%% file: Main.tex
\documentclass[11pt]{article}

\usepackage{jheppub} %

\usepackage[utf8]{inputenc}
\usepackage{amsmath,amsthm,amsfonts,amssymb,mathtools,bbm,mathdots,tensor,bigints}
\usepackage[sans]{dsfont}
\usepackage[mathscr]{eucal}
\usepackage[dvipsnames]{xcolor}
\usepackage[]{hyperref}
\hypersetup{colorlinks = true, urlcolor=MidnightBlue
, citecolor=red, linkcolor=PineGreen}%

\usepackage{tikz}
\usetikzlibrary{hobby}
\usetikzlibrary{decorations.markings,intersections,patterns,backgrounds}

\usepackage{epsfig,setspace}

\usepackage{tikz-cd}
\usepackage{relsize}
\usepackage{cancel}
\usepackage{caption}
\usepackage{subcaption}

\newtheorem*{thm}{Theorem}

\newcommand{\mscr}[1]{\mathscr{#1}}
\newcommand{\mcal}[1]{\mathcal{#1}}
\newcommand{\mfk}[1]{\mathfrak{#1}}
\newcommand{\mbb}[1]{\mathbb{#1}}
\newcommand{\mbf}[1]{\mathbf{#1}}
\newcommand{\mbs}[1]{\boldsymbol{#1}}
\newcommand{\msf}[1]{\mathsf{#1}}
\newcommand{\pcab}{\widehat{\mcal{P}}_{g,n}}
\newcommand{\newl}{\medskip\noindent}
\newcommand{\wh}[1]{\widehat{#1}}
\newcommand{\wt}[1]{\widetilde{#1}}

\newenvironment{eqaligned}[1][]
{\begin{equation}
    \begin{aligned}
    #1
} 
{
    \end{aligned}
\end{equation}
}

\newenvironment{eqgathered}[1][]
{\begin{equation}
    \begin{gathered}
    #1
} 
{
    \end{gathered}
\end{equation}
}

\title{\centering Hyperbolic Geometry of Superstring \\ Perturbation Theory}

\author{Seyed Faroogh Moosavian, Roji Pius}
\affiliation{Perimeter Institute for Theoretical Physics, Waterloo, ON N2L 2Y5, Canada}
\emailAdd{sfmoosavian@gmail.com}
\emailAdd{\quad rojipius@imsc.res.in}

\abstract{We explore the hyperbolic structure of the RNS formulation of perturbative  superstring theory. The aim is to provide a systematic method to explicitly compute on-shell and off-shell closed superstring amplitudes with an arbitrary number of external states and loops. Using hyperbolic geometry, we construct gluing-compatible off-shell string measures by giving a set of gluing-compatible local coordinates around external punctures and a gluing-compatible distribution of picture-changing operators. These amplitudes satisfy the required off-shell factorization property. This provides a formalism within which string-theory amplitudes can be computed explicitly once the corresponding string measures are expressed in terms of certain coordinates on Teichm\"uller space, the so-called Fenchel-Nielsen coordinates.}

\begin{document} 
\maketitle

\section{Introduction}
\label{sec:intro}
\input{Sections/Introduction.tex}

\section{Off-Shell Bosonic-String Amplitudes}
\label{Off-Shell Bosonic String Amplitudes}
\input{Sections/Off-Shell_Bosonic-String_Amplitudes.tex}

\section{Hyperbolic Geometry and Off-Shell Bosonic-String Amplitudes}
\label{H Off-Shell b Amplitudes}
\input{Sections/Hyperbolic_Geometry_and_Off-Shell_Bosonic-String_Amplitudes.tex}

\section{Off-Shell Superstring Amplitudes}
\label{Off-Shell Superstring Amplitudes}
\input{Sections/Off-Shell_Superstring_Amplitudes.tex}

\section{Hyperbolic Geometry and Off-Shell Superstring Amplitudes}
\label{HOff-Shell Superstring Amplitudes}

\input{Sections/Hyperbolic_Geometry_and_Off-Shell_Superstring_Ampliudes.tex}

\section{Discussion and Future Directions}
\label{future}
\input{Sections/Discussion_and_Future_Directions.tex}

\bigskip

{\bf Acknowledgement:} It is our pleasure to thank  Davide Gaiotto and Ashoke Sen  for detailed discussions and important comments on the draft. We are grateful to Sujay Ashok, Kevin Costello, Ted Erler, Thiago Fleury, Abhijit Gadde, Rajesh Gopakumar, Sitender Pratap Kashyap, Bernard Maskit, Shiraz Minwalla, Kasra Rafi, Kalyana Rama, Brad Safnuk, Mrityunjay Verma, Edward Witten, Scott Wolpert, and Barton Zwiebach for helpful discussions.  Research at Perimeter Institute is supported by the Government of Canada through Industry Canada and by the Province of Ontario through the Ministry of Research \& Innovation.

\appendix 

\section{Hyperbolic Surfaces and their Teichm\"uller Space}
\label{hyperbolic}
\input{Appendices/Hyperbolic_Surfaces_and_their_Teichmuller_Space.tex}

\section{Matrices for Fenchel-Nielsen Coordinates}
\label{FGFNcoordinates}
\input{Appendices/Matrices_for_Fenchel-Nielsen_Coordinates.tex}

\bibliography{References}
\bibliographystyle{JHEP}

\end{document}

%% file: Sections/Introduction.tex
One of the main powers of quantum field theory is that it comes with  a set of relatively simple rules for calculating measurable quantities.  For instance, it allows us to compute the quantities of interest, such as the scattering amplitudes, as a  perturbative expansion in the coupling parameters of the theory by evaluating the Feynman diagrams using  the Feynman rules. These rules are derived from the Lagrangian of theory and express amplitudes with arbitrary external states and loops as well-defined integrals. Therefore, we can consider this Feynman prescription as a calculable formulation of the perturbative quantum field theory.\par

\newl Similarly, in the conventional formulation of string theory,  scattering amplitudes are obtained  by evaluating the string diagrams which are the string theory analog of Feynman diagrams. Unlike in  quantum field theory, the rules for evaluating string diagrams are  not derived from a Lagrangian. Instead, the rules are given without referring to an underlying Lagrangian  \cite{Polyakov1981a, Polyakov1981b,FriedanMartinecShenker1986}.   By exploring the conformal structure of string theory, powerful covariant methods to compute  amplitudes in perturbative string theory were introduced in \cite{FriedanMartinecShenker1986}. Some aspects of the formalism that remained slightly opaque were elaborated recently \cite{Berera199301,Berera199406, Belopolsky1997a,Belopolsky1997b,Belopolsky1997c, Witten2012a,Witten2012b,Witten2012c,Witten201306,DonagiWitten201304,DonagiWitten201404,Sen201408,SenWitten201504,PiusRudraSen201410,Sen201411,Sen201501,Sen201508a,Sen201508b,PiusSen1604,Sen201610,Sen201606,Sen201607,Sen201609, deLacroixErbinKashyapSenVerma1703}. Therefore, we can safely say that  perturbative string theory provides a well-defined formal procedure for computing string amplitudes as an expansion in the string coupling.  However, we can consider this as a calculable formulation of the string theory only if it comes with a set of well-defined practical  rules for  evaluating  amplitudes.  Unfortunately, such a set of practical rules for computing amplitudes  is not readily available.
 
\newl  In this paper, we describe a practical procedure for performing computations in perturbative string theory by using hyperbolic geometry. Below, we spell out the main objectives of this paper followed by a brief summary of the results. 
 
 \subsection{Objectives}
 
 The path integral formulation of string theory requires summing over all possible worldsheets of the relativistic strings. The contribution of each worldsheet is weighted by a unimodular complex number whose phase is the area of the corresponding worldsheet. In his seminal papers \cite{Polyakov1981a, Polyakov1981b}, Polyakov showed that this summation over the two-dimensional surfaces reduces to two dimensional exactly solvable conformal field theory on Riemann surfaces. This connection was established by introducing a metric on each of the worldsheets. The net effect of this is to introduce an unphysical degree of freedom, the Liouville field, into the problem that disappears from the worldsheet action due to its Weyl invariance. Demanding the Weyl invariance at the quantum level makes sure that the Liouville field will not reappear in the action due to the quantum correction. Moreover, if we assume that the external states in the scattering process satisfy the classical on-shell condition, then the amplitudes are also independent of the specific choice of the metric on the worldsheet and only depend on its conformal class or equivalently its complex structure. Since the classical on-shell condition is assumed for the external states, we shall call such amplitudes as {\it the on-shell} string amplitudes.  
 
\paragraph{Bosonic String Amplitudes and Moduli Space of Riemann Surfaces:} As we have mentioned, a worldsheet admits complex structure (a holomorphic coordinate $z$) and hence is a Riemann surface with $n$ boundaries. The conventional formulation of the bosonic string theory  defines the $g$-loop contribution to a typical amplitude for the scattering of  $n$ asymptotic closed string states as a path integral over the embedding $X^{\mu}(z,\bar z)$ of the string worldsheet.  Using  the  state-operator correspondence, it is possible to replace the boundary states with  vertex operators in an appropriate conformal field theory on the genus $g$ Riemann surface. Let us denote the vertex operator representing a state carrying  quantum numbers $a_i$ and momentum $k^{\mu}_i$ by $V_{a_i,k_i}(z_i,\bar z_i)$. Then the momentum-space expression for a $g$-loop string amplitude for $n$ closed-string states having quantum numbers $a_i$ and momentum $k_i^{\mu}$ with $i$ running from $1$ to $n$ is given by,
\begin{equation}\label{nloop}
A^{(n)}_g(a_1,k_1;...;a_n,k_n)=\bigintsss dg_{ab}dX^\mu e^{-S_p[ g_{ab},X^{\mu}]}\prod_{i=1}^n\int d^2z_i\sqrt{-g(z_i)}~V_{a_i,k_i}(z_i),\nonumber\\
\end{equation}
where $dg_{ab}$ is the measure for integration over all (conformally-inequivalent) metrics, and $S_p$ is the Polyakov action for a free relativistic string propagating through space-time.  The path integral \eqref{nloop} can be made well-defined by carrying out the gauge fixing by introducing the Fadeev-Popov  ghosts. Redundant transformations that must be fixed  are  the Weyl transformations and the reparametrizations that keep the locations of punctures unaffected. This can be achieved by restricting the path integral over all possible metrics on the Riemann surface to a set of metrics that are not related to each other by the above-mentioned gauge transformations. Any equivalence class of metrics defined by the above transformations must contribute once to the path integral. Different conformal classes correspond to Riemann surfaces with different complex structures. As it is shown by Riemann himself and can be seen via the index theorem, the number of parameters specifying each class is finite. These parameters form a non-compact space called the moduli space of Riemann surfaces. Therefore, the gauge-fixed path integral over the metrics on the Riemann surface is reduced to integration over the moduli space of Riemann surfaces. For questions of IR behavior of amplitudes, one considers a natural compactification of this space \cite{DeligneMumford1969a}. The compactification can be obtained by adding the so-called Deligne-Mumford compactification locus, associated with the Riemann surfaces with nodes, to $\mathcal{M}_{g,n}$. Hence the $g$-loop on-shell amplitudes in bosonic string theory  with $n$ external closed-string states can be considered  as the integral of a suitable differential form of degree $6g-6+2n$ over the compactified  moduli space of genus-$g$ Riemann surfaces with $n$ punctures $\overline{\mathcal{M}}_{g,n}$ :
\begin{equation}
A^{(n)}_g(a_1,k_1;\ldots;a_n,k_n)=\bigintsss_{\overline{\mathcal{M}}_{g,n}}\Omega_{6g-g+n}. \label{amplitude as an integral of form over moduli space}
\end{equation}
Therefore, in order to obtain  a practical prescription for evaluating amplitudes in string theory, we need to understand the geometry of the moduli space of Riemann surfaces and  specify the rules for  performing integration over it.  Unfortunately, there is no known simple description of the moduli space. This is mainly due to the following fact. Characterizing the space of  Riemann surfaces requires varying the complex structure by infinitesimal deformation of the metric on the Riemann surface.  But the infinitesimal deformations only lead us to the Teichm\"uller space of  Riemann surfaces  instead of  the moduli space.  The Teichm\"uller space is the space of Riemann surfaces whose metrics are not related by  diffeomorphisms that  can be continuously connected to the identity. The moduli space of  genus-$g$ Riemann surfaces with $n$ distinguished punctures is  obtained by identifying the points in the corresponding  Teichm\"uller space $\mathcal{T}_{g,n}$ that are related by the action of the  mapping class group Mod$_{g,n}$:
	\begin{equation}
	\mathcal{M}_{g,n}\equiv\frac{\mathcal{T}_{g,n}}{\mathrm{Mod}_{g,n}}.
	\end{equation}
Mod$_{g,n}$ is  the group of diffeomorphisms on a genus-$g$ Riemann surface with $n$ distinguished punctures that can not be continuously connected to the identity transformation.  The action of mapping class group on Teichm\"uller space is very complicated and as a result,  the geometry and the topology of the moduli space are also very complicated. Therefore, finding an explicit fundamental domain of the mapping-class group in the Teichm\"uller space is a highly nontrivial task. This makes explicit integration of a function (or more precisely a form) over the moduli space  of an arbitrary Riemann surface a daunting task.

\newl Riemann surfaces are distinguished by their complex structure. On the other hand, The  Uniformization Theorem of Poincar\'e and Koebe \cite{Abikoff198110} asserts that every genus-$g$ Riemann surface $\mathcal{R}_{g,n}$ with $n$ distinguished punctures subject to the constraint $2g+n\ge 3$, can be obtained by the proper discontinuous action of a Fuchsian group $\Gamma$ on the Poincar\'e upper half plane $\mathbb{H}$: 
\begin{equation}
\mathcal{R}_{g,n}\simeq \frac{\mathbb{H}}{\Gamma} \label{Uniformization Theorem}.
\end{equation}
The Fuchsian group $\Gamma$ is a subgroup of the automorphic group of the Poincar\'e upper half plane. Riemann surfaces obtained this way are hyperbolic and have constant curvature $-1$ everywhere. This implies the following correspondence:
\begin{equation}
\text{Hyperbolic Structure} \qquad \Longleftrightarrow \qquad  \text{Complex Structure}.
\end{equation}
This means that for any surface with $2g+n\ge 3$, any conformal class of metrics contains a unique hyperbolic metric. Therefore, the moduli space of a Riemann surface which is essentially the classifying space for the conjugacy classes of the distinguished complex structures on it can naturally be identified  with the moduli space of hyperbolic surfaces. 

\newl A convenient parametrization of Teichm\"uller space is provided by the so-called Fenchel-Nielsen coordinates \cite{FenchelNielsen2002,ImayoshiTaniguchi1992,Hubbard2006}.  The basic idea behind the Fenchel-Nielsen  parametrization is that every compact hyperbolic surface of genus $g$  with  $n$ punctures  can be obtained by gluing $2g-2+n$ pairs of pants. By varying the parameters of this construction, every compact hyperbolic metric can be obtained. At each gluing site, there are two parameters: the length of the boundaries $\ell$ and the twist parameter $\tau_i$. Since there are $3g-3+n$ gluing sites for  a Riemann surface of genus $g$  with  $n$ punctures, the  Fenchel-Nielsen coordinates  of the corresponding Teichm\"uller space $\mathcal{T}_{g,n}$ is given by $(\tau_i,\ell_i),~i=1,\ldots,3g-3+n$. A specific value of the Fenchel-Nielsen coordinates corresponds to a particular Riemann surface $\mcal{R}$. As we mentioned above, it is also possible to construct $\mcal{R}$ by considering the quotient of a fundamental domain in the upper half-plane by the action of a Fuchsian group $\Gamma$. It turns out that the generators of $\Gamma$ can be expressed in terms of the associated Fenchel-Nielsen coordinates \cite{Maskit2001}. For more details, see Appendix \ref{FGFNcoordinates}. 

\newl  Other than having this very convenient parametrization, there is another advantage of using hyperbolic geometry. The only known explicit computation of an integral over the moduli space $\mathcal{M}_{g,n}$, for generic values of $g$ and $n$, is the computation of the Weil-Petersson (WP) volumes of the moduli space of the bordered  hyperbolic surfaces, due to Mirzakhani \cite{Mirzakhani200610}. A prominent role in these developments is played by Fenchel-Nielsen coordinates. 

\newl Due to this outstanding progress and the above-mentioned convenient properties, one may wonder whether one can use hyperbolic geometry to provide a framework for computing scattering amplitudes in string theory. This forms the main aim of this work.

\newl \noindent{\bf Objective 1}: {\it Expressing the bosonic-string amplitudes as an integral over an explicit region in Teichm\"uller space of hyperbolic surfaces.}

\paragraph{Superstring Amplitudes and Picture-Changing Operators:} Amplitudes in superstring theory  are the super analogs of the  amplitudes in the bosonic string theory. Instead of all possible Riemann surfaces with  punctures, we need to sum over all possible super-Riemann surfaces with Neveu-Schwarz (NS) and/or Ramond (R) punctures \cite{BaranovSchwarz198510,Friedan1986a,RoslySchwarzVoronov1988a}. More precisely,  $g$-loop closed superstring amplitudes of  $n_{\text{R}}$ asymptotic closed superstring  states from the R sector and $n_{\text{NS}}$ asymptotic closed superstring states from the NS sector is defined as the path integral over the embedding $X^{\mu}(z,\theta;\bar z,\bar\theta)$ of the superstring worldsheet. Such worldsheets are $\mscr{N}=1$ super-Riemann surface with $n_{\text{R}}$ boundaries having R boundary conditions and $n_{\text{NS}}$ boundaries having  NS boundary conditions. Here $(z,\theta)$ is the super-holomorphic coordinate on  the surface. Gauge redundancies of this path integral are the super-reparametrizations that do not move the punctures and the super-Weyl transformations. Like the bosonic-string theory, we must gauge fix the path integral to get rid of these gauge redundancies. The gauge-fixing  restricts the path integral to the space of inequivalent superconformal or super-complex structures. Again using the index theorem, it can be shown that the number of parameters that describes each class is finite. The parameters describing these classes involve both commuting and  anti-commuting variables.  Therefore, the gauge-fixed path integral reduces to integration over the super-moduli space of $\mscr{N}=1$ super-Riemann surfaces \cite{Witten2012a,Witten2012b,Witten2012c,Witten201306}.\par

\newl Another formulation of superstring amplitudes is the so-called picture-changing formalism in which superstring amplitudes (after summation over possible spin structures) are written as integrals over the moduli space of ordinary Riemann surfaces \cite{FriedanMartinecShenker1986}. One first integrates out odd super moduli at the cost of introducing the so-called picture-changing operators (PCO) on the worldsheet. The naive  picture-changing prescription assumes that their locations on the surface do not matter as long as we impose the following condition.  For a genus-$g$ Riemann surface with  $n_{\text{NS}}$ NS punctures and $2n_{\text{R}}$ R punctures, denoted by $\mathcal{R}_{g,n_{\text{NS}},2n_{\text{R}}}$, we must insert $2g-2+n_{\text{NS}}+n_{\text{R}}$ number of PCOs on arbitrary locations on the Riemann surface. Consider such a surface with one degenerating cycle. Assume that cutting along this cycle separates the surface into two  surfaces $\mathcal{R}_{g_1,n^1_{\text{NS}},2n^1_{R}}$ and $\mathcal{R}_{g_2,n^2_{\text{NS}},2n^2_{\text{R}}}$. Then  PCOs distribution on the degenerate surface must be such that each component surface has $2g_i-2+n^i_{\text{NS}}+n^i_{\text{R}},~i=1,2$ number of PCOs.  If there are more than one degenerating curve then a similar condition must be true with respect to all of the  subsurfaces. Then the $g$-loop superstring amplitude of $n_{\text{NS}}$ NS states and $2n_{\text{R}}$ R states can be defined as an integral over an appropriately-constructed form $\Omega_{6g-g+2n}$ on the moduli space of genus-$g$ Riemann surface with $n=n_{\text{NS}}+2n_{\text{R}}$ punctures:
\begin{equation*}
A^{(n_{\text{NS}},n_{\text{R}})}_g(a^{\text{NS}}_i,k^{\text{NS}}_i;a^{\text{R}}_j,k^{\text{R}}_j)=\bigintsss_{\overline{\mathcal{M}}_{g,n}}\Omega_{6g-g+2n} \left(z_1(m),\ldots,z_{2g-2+n_{\text{NS}}+n_{\text{R}}}(m)\right),
\end{equation*}
where $z_i(m)$ denotes the position of $i^{th}$ PCO on the Riemann surface corresponding to the point in the moduli space  represented by the coordinate $m$. The set $\left(z_1(m),\ldots,z_{2g-2+n_{\text{NS}}+n_{\text{R}}}(m)\right)$ provides a distribution of PCOs over the entire region of moduli space that satisfies the above-mentioned property. $a_i^{\text{NS}},k_i^{\text{NS}}$ denote the quantum numbers and momentum vector of the $i^{th}$ NS external state and  $a_j^{\text{R}},k_j^{\text{R}}$ denote the quantum numbers and momentum vector of the $j^{th}$ R external state.

\newl Superstring amplitudes defined using the naive picture-changing formalism are not well-defined due to the presence of  unphysical singularities. The origin of these singularities can be traced back to the fact that replacing the odd coordinates of the supermoduli space with PCOs is not possible in a smooth way over the entire supermoduli space. The choice of the locations of PCOs corresponds to a particular choice of a gauge for the worldsheet gravitino field \cite{VerlindeVerlinde198702}. It is known that choosing a globally-valid gauge for the gravitino field is impossible \cite{Witten2012a,Witten2012b,Witten2012c,Witten201306, DonagiWitten201304,DonagiWitten201404}. In the picture-changing formalism, this breakdown of the global choice of the gauge for the  gravitino field shows up as the spurious singularities of the integration measure of a typical superstring amplitude. The presence of spurious singularities means that the  computation of amplitudes in  superstring theory  using PCOs has to be based on piecing together local descriptions. The vertical integration prescription introduced in \cite{Sen201408} and systematically developed in \cite{SenWitten201504} provides an efficient method to piece together the local descriptions (see also \cite{ErlerKonopkaSachs201312}). However, in order to implement the vertical integration prescription, it is essential to know the explicit region over which a superstring measure is supposed to be integrated.

\newl {\bf Objective 2}:  {\it  Expressing the superstring amplitudes as an integration over an explicit region in Teichm\"uller space of hyperbolic surfaces in terms of Fenchel-Nielsen coordinates and describe a distribution of PCOs on the worldsheets that respects the above-mentioned factorization property.}

\newl Achieving this goal would make the implementation of vertical integration possible. The resulting superstring amplitudes are free from spurious poles.

\paragraph{Off-Shell Superstring Amplitudes and 1PI Approach:} The conventional formulation of string perturbation theory requires imposing the classical on-shell condition on the external states and  Weyl invariance.  Demanding the classical on-shell condition for external states and the Weyl invariance at the quantum level have significant consequences which we explain below.

\newl  {\small \it Mass Renormalization and Classical On-Shell Condition}: Consider a string theory amplitude corresponding to the scattering of $n$-external states representing particles carrying momenta $k_1, \ldots, k_n$ and other discrete quantum numbers $a_1, \ldots, a_n$ with tree level masses $m_{a_1} , \ldots, m_{a_n}$. The momenta $k_i$ are required to satisfy the tree-level on-shell condition $k^2_i = -m^2_{a_i}$. The conventional formulation of string theory yields the result for what in a quantum field theory can be called `truncated Green's function on classical mass shell'
\begin{equation}\label{ek4}
    \begin{aligned}
        R^{(n)}_{a_1\ldots a_n}(k_1,\ldots k_n) 
        &\equiv \lim_{k_i^2 \to -m_{a_i}^2} F^{(n)}_{a_1\ldots a_n} (k_1, \ldots k_n),
         \\
        F^{(n)}_{a_1\ldots a_n}  (k_1, \ldots k_n)&\equiv
        G^{(n)}_{a_1\ldots a_n} (k_1,\ldots k_n)\prod_{i=1}^n (k_i^2 + m_{a_i}^2),
    \end{aligned}
\end{equation}
where $G^{(n)}_{a_1\ldots a_n} (k_1,\ldots k_n)$ corresponds to the momentum-space Green's function in the quantum field theory.  This expression is similar to the expression for the S-matrix elements in a quantum field theory but it has significant differences.

\newl To define S-matrix elements in quantum field theory, we need to first consider the two-point function $G^{(2)}_{ab}(k, k')$  for all sets of fields whose tree-level masses are all equal to $m$ described by the matrix
\begin{equation} \label{ek6}
G^{(2)}_{ab}(k, k') = (2\pi)^{D+1} \delta^{(D+1)}(k+k') \, Z^{1/2}(k)_{ac} (k^2 + M_p^2)_{cd}^{-1} 
(Z^{1/2}(-k))_{db}^T,
\end{equation}
where $M_p^2$ is the mass$^2$ matrix and $Z^{1/2}(k)$ is the wave-function renormalization matrix, the latter being free from poles near $k^2+m^2\simeq 0$.  The sum over $c,d$ is restricted to states which have the same tree-level mass $m$ as the states labeled by the indices $a,b$. $D+1$ is the total number of non-compact space-time dimensions. We can diagonalize $M_p^2$ and absorb the diagonalizing  matrices into the wave-function renormalization factor $Z^{1/2}(k)$ to express  $M_p^2$ as a diagonal matrix.  These eigenvalues, which we shall denote by $m_{a,p}^2$, are the squares of the physical masses.  Then S-matrix elements are defined by
\begin{equation} \label{ek5}
S^{(n)}_{a_1\ldots a_n}(k_1,\ldots k_n) =
\lim_{k_i^2 \to -m_{a_i,p}^2} G^{(n)}_{b_1\ldots b_n}
(k_1,\ldots k_n)\prod_{i=1}^n \left\{Z_i^{-1/2}(k_i)_{a_i, b_i} (k_i^2 + m_{a_i,p}^2)\right\},
\end{equation}
where $m_{a_i,p}$ is the physical mass of the $i$-th particle, defined as the location of the pole in the two-point Green's function $G^{(2)}$ as a function of $-k^2$. $Z_i(k_i)_{ a_i,b_i}$ is the residue at this pole.

\newl At the tree level $Z=1$, $M_p^2=m^2 \, I$, where $I$ is the identity matrix, and hence $R^{(n)}$ and $S^{(n)}$  agree.  However, in general $R^{(n)}$ and $S^{(n)}$ are different. While $S^{(n)}$ defined in \eqref{ek5} is the physically-relevant quantity, the conventional formulation of string theory directly computes $R^{(n)}$ defined in \eqref{ek4}. This is serious trouble for states receiving mass renormalization. Assume that an external state with the quantum number $a_i$ and the tree-level mass $m_{a_i}$ receives mass renormalization. Then beyond one loop, the radiative corrections introduce a series of $\frac{1}{k_i^2+m_{a_i}^2}$. At the same time string theory requires $k_i^2+m_{a_i}^2=0$. This makes the on-shell amplitudes involving  the states receiving mass renormalization ill-defined beyond the tree level. 

\newl {\small \it Dynamical Shift of Vacuum and Conformal Invariance}: The background spacetimes that avoid the Weyl anomaly satisfy a set of classical equations. However, there are  situations in which  quantum corrections modify such backgrounds.  Consider an $\mathcal{N}=1$ supersymmetric compactification of string theory down to 3+1 dimensions, where we have $U(1)$ gauge fields with Fayet-Iliopoulos (FI) terms generated at one  loop \cite{DineSeibergWitten198701,AtickDixonSen198702,DineIchinoseSeiberg198711,GreenSeiberg198804}.  It is possible to ensure that only one gauge field has the FI term by choosing a suitable linear combination of these gauge fields. Typically, there are massless scalars $\phi_i$ charged under $U(1)$ gauge fields. The FI term generates a potential of the form
\begin{equation} \label{e1}
{1\over g_s^2}\, \left( \sum_i q_i \phi_i^* \phi_i - C\, g_s^2\right)^2,
\end{equation}
where $q_i$ is the charge carried by $\phi_i$, $C$ is a numerical constant that determines the coefficient of the FI term, and $g_s$ is the string coupling constant. $C$ could be positive or negative and $q_i$s for different fields could have different signs. If we expand the potential in powers of $\phi_i$ around the perturbative vacuum $\phi_i=0$, it is clear that some of these scalars can become tachyonic. The form of effective potential suggests that the correct procedure to compute physical quantities is to shift the corresponding fields so that we have a new vacuum where $\sum_i q_i \langle\phi_i^*\rangle \langle\phi_i\rangle = C\, g_s^2$, and quantize string theory around this new background. However since classically the $C\, g_s^2$ term is absent from this potential \eqref{e1}, this new vacuum is not a solution to the classical equations of motion. As a result, on-shell  methods \cite{Belopolsky1997a,Belopolsky1997b,Belopolsky1997c,D'HokerPhong2002a,D'HokerPhong2002b,D'HokerPhong2002c,D'HokerPhong2002d,D'HokerPhong2002e,D'HokerPhong2002f,D'HokerPhong2002g,Witten2012a,Witten2012b,Witten2012c,Witten201306}, which require that we begin with a conformally-invariant worldsheet theory, is not suitable for carrying out a systematic perturbation expansion around this new background.

\newl It is possible to cure the singularities associated with mass renormalization and dynamical modification of vacuum by defining appropriate off-shell string amplitudes \cite{PiusRudraSen201311,PiusRudraSen201401,PiusRudraSen201410,Sen201408,SenWitten201504,Sen201411,Sen201501,Sen201508a,Sen201508b,Sen201606}. The off-shell amplitudes are defined by relaxing the classical on-shell condition on external states. The classical on-shell condition, in the worldsheet conformal field theory language, is equivalent to choosing integrated vertex operators having conformal dimensions $(1,1)$. Therefore, in order to construct off-shell amplitudes we must use vertex operators with arbitrary conformal dimensions.
 
\newl  Consider a reference coordinate system $z$ on a Riemann surface. Let $z_i$ be the location of the $i$-th puncture in the $z$-coordinate system and $w_i$ be the local coordinate system around the $i$-th puncture, related to $z$ by some functional relation $z=f_i(w_i)$. It must be such that $w_i=0$ maps to $z=z_i$: $f_i(0)=z_i$. Then the contribution to the $n$-point off-shell amplitude from genus-$g$  Riemann surfaces can be expressed as
\begin{equation}
\bigintsss_{\overline{\mathcal{M}}_{g,n}}  \left\langle \prod_{i=1}^n f_i\circ V_i(0) \, \times \, 
\hbox{ghost insertions}\right\rangle\,,
\end{equation}
where $f\circ V(0)$ denotes the conformal transformation of the vertex operator $V$ by the function $f(w)$, the correlator $\langle\ldots\rangle$ is evaluated in the reference 
$z$-coordinate system and $\int_{\overline{\mathcal{M}}_{g,n}}$ denotes the integration over the compactified moduli space of Riemann surfaces of genus $g$ with $n$ punctures. A detailed description of how to construct the integration measure for a given choice of local coordinate system will be explained later. The off-shell amplitudes defined this way depend on the choice of the local coordinate system $w_i$ but  are independent of the choice of the reference coordinate system $z$.

\newl It is important to  make sure that physical quantities extracted from the off-shell amplitudes defined this way are independent of the choice of local coordinates. It is shown \cite{PiusRudraSen201311,PiusRudraSen201401,Sen201408,SenWitten201504} that the physical quantities like the renormalized masses of physical states and S-matrix elements are independent of the choice of local coordinates  if we work with a class of local coordinates satisfying the following  properties:

\begin{itemize}

\item 
The local coordinate system is  symmetric in all punctures, i.e. the function $f_i(w)$ should depend on $i$ only via the location $z_i$ of the puncture.

\item The choice of local coordinates is continuous over the entire region of moduli space. 

\item Riemann surfaces that can be obtained by gluing two or more Riemann surfaces are declared to  contribute to the 1PR amplitudes\footnote{we elaborate in the main body of the paper}. The gluing relation is
\begin{equation} \label{egluing}
w_1 w_2 = e^{-s+i\theta}, \, \quad 0\le \theta < 2\pi, \quad 0\le s<\infty\,,
\end{equation}
where  $w_1$ and $w_2$ are the local coordinates at the punctures used for gluing. We shall call the corresponding
Riemann surfaces {\it 1PR Riemann surfaces}. The rest of the Riemann surfaces are declared as {\it 1PI Riemann surfaces}. The choice of local coordinates on a 1PR surface must be the one induced from the local coordinates of the glued 1PI surfaces. However, the choice of local coordinates on the 1PI surfaces is arbitrary. 
\end{itemize}
We shall call local coordinates satisfying the above criteria  
{\it gluing-compatible local coordinates}. In the case of the superstring amplitudes, we also need to impose a similar restriction on the distribution of PCOs.  On 1PI Riemann surfaces, we are allowed to choose an arbitrary  distribution of PCOs. However, the distribution of PCOs on 1PR surfaces is the one induced from the surfaces that are being glued. 

\newl  Note that the off-shell string amplitudes $\Gamma_{a_1\ldots a_n}^{(n)}(k_1,\ldots,k_n)$ for $n$ external states with quantum numbers $a_i$, momenta  $k_i$, and tree level masses $m_i$, do not compute the off-shell Green's function  $G_{a_1\ldots a_n}^{(n)}(k_1,\ldots,k_n)$ of the string theory. They are however related as 
  \begin{align}\label{offgreen}
  \Gamma_{a_1\ldots a_n}^{(n)}(k_1,\ldots,k_n)=G_{a_1\ldots a_n}^{(n)}(k_1,\ldots,k_n)\prod_{i=1}^n(k_i^2+m_i^2).
  \end{align}
Scattering amplitudes defined following LSZ-prescription using  off-shell amplitudes defined this way  are unitary and satisfy cutting rules \cite{PiusSen1604,Sen201610,Sen201606,Sen201607}.

\newl {\bf Objective 3}:  {\it Describing an explicit construction of the 1PI region inside the moduli space of  Riemann surfaces in terms of Fenchel-Nielsen coordinates and also describe a gluing compatible choice of local coordinates around the punctures. Also, provide a systematic method for distributing PCOs over the entire region of moduli space in a gluing-compatible fashion.  Then construct the superstring measure using the Beltrami differentials associated with the Fenchel-Nielsen vector fields.}

 \subsection{Summary of the Results}
  
Let us now summarize the main results of this work. 
 
\subsubsection*{Bosonic-String Measure and Fenchel-Nielsen Coordinates} 
Assume that the Fenchel-Nielsen  coordinates $(\tau_i,\ell_i),~i=1,\ldots,3g-3+n$ are defined with respect to the pair of pants decomposition $\{C_i\},~i=1,\ldots,3g-3$, where $C_i$ denote a simple closed geodesic\footnote{A simple geodesic is a geodesic that does not intersect itself.} on the hyperbolic surface. The tangent space at a point in the Teichm\"uller space is spanned by the Fenchel-Nielsen  vector fields
 \begin{equation*}
 \left\{\frac{\partial}{\partial \tau_i},\frac{\partial }{\partial \ell_i}\right\},\qquad i=1,\ldots,3g-3+n.
 \end{equation*} 
These vector field $\frac{\partial}{\partial \tau_i}$ is  the  twist vector field $t_{C_i}$ associated with the curve $C_i$.  The twist field $t_{\alpha}$, where $\alpha$ is a simple closed geodesic,  generates a flow in $\mathcal{T}_{g,n}$. This flow can be understood as the operation of  cutting the hyperbolic surface along $\alpha$ and attaching the boundaries after rotating one boundary relative to the other by some amount $\delta$. The magnitude $\delta$ parametrizes the flow in  $\mathcal{T}_{g,n}$. The vector field $\frac{\partial}{\partial \ell_i}$ is the dual to $\frac{\partial}{\partial \tau_i}$ with respect to the Weil-Petersson symplectic form on the Teichm\"uller space \cite{Wolpert198203,Wolpert198302,Wolpert198508}.

\newl Using the Beltrami differentials that represent the vector fields $\{\frac{\partial}{\partial \tau_i},\frac{\partial }{\partial \ell_i}\},~i=1,\ldots,3g-3+n$ we can write the off-shell bosonic-string measure as 
\begin{equation}\label{pformpgnbos1}
\Omega_{6g-6+2n}^{(g,n)}(|\Phi\rangle)=(2\pi \mathrm{i})^{-(3g-3+n)}\langle\mathcal{R}|B_{6g-6+2n}|\Phi\rangle,
\end{equation}
where
\begin{equation*}\label{opevormbos1}
B_{6g-6+2n}\left[\frac{\partial}{\partial \ell_1},\frac{\partial}{\partial \tau_1},\ldots,\frac{\partial}{\partial \ell_{3g-3+n}},\frac{\partial}{\partial \tau_{3g-3+n}}\right]\equiv\prod\limits_{a=1}^{3g-3+n}b(t_{\tau_a})\delta\tau_a\,b(t_{\ell_a})\delta\ell_a,
\end{equation*}
and 
\begin{equation}\label{bv1}
\begin{aligned}
    b(t_{\tau_i})&=\bigintsss_{\mathcal{F}} d^2z\left(b_{zz}\mathbf{t}_{\tau_i}+b_{\bar z\bar z}\bar{\mathbf{t}}_{\tau_i}\right),
    \\
    b(t_{\ell_i})&=\bigintsss_{\mathcal{F}} d^2z\left(b_{zz}\mathbf{t}_{\ell_i}+b_{\bar z\bar z}\bar{\mathbf{t}}_{\ell_i}\right).
\end{aligned}
\end{equation}
Here $\mathcal{F}$ denotes the fundamental region in the upper half-plane representing the surface $\mathcal{R}$. $\mathbf{t}_{\tau_a}$ is the representative of the vector field $\frac{\partial}{\partial \tau_a}$ and $\mathbf{t}_{\ell_a}$ is the representative of the vector field $\frac{\partial}{\partial \ell_a}$. $\mathbf{t}_{\tau_a}$ and $\mathbf{t}_{\ell_a}$ deform the hyperbolic structure of the surface by infinitesimal amount $\delta\tau_a$ and $\delta\ell_a$, respectively, and $\langle \mathcal{R}|$ is the surface state associated with the surface $\mathcal{R}$. This state represents the state that is created on the boundaries of the   disc $D_i, ~i=1,\ldots,n$ around the  punctures having unit radius measured in the local coordinates around the punctures  by performing the functional integral over the fields of the conformal field theory on $\mathcal{R}-\sum_iD_i$.  The state
\begin{equation*}
    |\Phi\rangle=|\Psi_1\rangle\otimes\ldots\otimes |\Psi_n\rangle\in\mathcal{H}^{\otimes n},
\end{equation*}
where $\mathcal{H}$ is the Hilbert space of the worldsheet conformal field theory.  Then $\langle\mathcal{R}|B_{6g-6+2n}|\Phi\rangle$ describes the $n$-point correlation function on $\mathcal{R}$ with the vertex operator for $|\Psi_i\rangle$ inserted at the $i^{th}$ puncture using the gluing compatible-local coordinate system  around that puncture.

\subsubsection*{Gluing-Compatible Local Coordinates and Hyperbolic Metric} 

The requirement of gluing-compatibility demands that the choice of local coordinates on a 1PR Riemann surface must be the same as the local coordinates induced on it from the component surfaces. At first sight, the local coordinates around the punctures that are induced from the hyperbolic metric on the surface seem to be a gluing-compatible choice. This is true  for surfaces with nodes, constructed by the gluing of hyperbolic surfaces. However, it turns out that  when we construct the plumbing family of hyperbolic surfaces without nodes, the resulting surface does not have constant curvature $-1$ everywhere. This implies that the induced local coordinates on the glued surface are not the local coordinates defined using the hyperbolic metric on it. Therefore, the induced local coordinate system from the hyperbolic metric is not gluing-compatible.

\newl In the simplest form of degeneration, two parts of a degenerate Riemann surface are connected by a single long cylinder. Assume that we are gluing surfaces with hyperbolic metrics on them. In this case, the cylinder connecting the component surfaces has a curve on it where the curvature is accumulated. This implies that the glued surface is not hyperbolic. This is the consequence of the following facts: 1) in the plumbing fixture construction, one glues two surfaces by identifying curves around the punctures, 2) the glued surface is hyperbolic only if the component surfaces are glued through a geodesic, 3) however there are no geodesic curves in the neighborhood of punctures. According to the Uniformization Theorem, any surface subject to the condition $2g+n\ge 3$ can be made hyperbolic. In fact, we can make it hyperbolic by performing a Weyl transformation. The correct Weyl factor can be obtained by solving the so-called {\it curvature-correction equation} \cite{Wolpert199002,ObitsuWolpert200704,MelroseZhu201606}.  To describe this equation, consider a compact Riemann surface having  metric $ds^2$ with the Gauss curvature $\mathbf{C}$. The conformally-equivalent  metric $e^{2f}ds^2$ has constant curvature $-1$, if $f$ satisfy the curvature-correction equation
  \begin{equation}\label{constantcurvatureq1}
  Df-e^{2f}=\mathbf{C},
  \end{equation}
where $D$ denotes the Laplace-Beltrami operator on the surface.

\newl The analysis can be generalized to a surface with an arbitrary number of nodes. The result provides the relation between induced local coordinates on the glued surface and the local coordinates defined using the hyperbolic metric. We can then find a gluing-compatible choice of local coordinates. Before specifying this choice, it is important to describe the 1PI and 1PR regions of the moduli space. The 1PI region inside the moduli space consists of surfaces with no closed geodesics on them with lengths less than or equal to $c_*$, the collar constant \cite{Randol197910,Wolpert1987a}. The 1PR region inside the moduli space consists of surfaces with at least one closed geodesic on them with a length less than $c_*$. The length of the collar constant $c_*$ is assumed to be infinitesimal compared to any geodesic on the surface having a finite length.

\newl We can now specify a gluing-compatible choice of local coordinates found by solving the curvature-correction equation to second-order in $c_*$. For this, let us introduce another infinitesimal parameter $\epsilon$. In order to define local coordinates in the 1PI region, we divide it into subregions. Let us denote the subregion in the 1PI region consists of surfaces with $m$ simple closed geodesics of length between $c_*$ and $(1+\epsilon)c_*$ by $\mathbf{R}_m$. For surfaces belong to the subregion $\mathbf{R}_0$, we choose the local coordinate around the $i^{th}$ puncture to be $e^{\frac{\pi^2}{c_*}}w_i$. In terms of $w_i$, the hyperbolic metric in the neighborhood of the puncture takes the following form
\begin{equation}
 d s^2_{hyp}=\left(\frac{|dw_i|}{|w_i|\ln|w_i|}\right)^2, \qquad  i=1,\ldots,n. 
\end{equation}
For surfaces belong to the region $\mathbf{R}_{m}$ with $m\ne 0$, we choose the local coordinates around the $i$\textsuperscript{th} puncture to be  $e^{\frac{\pi^2}{c_*}}\widetilde w_{i,m}$, where $\widetilde w_{i,m}$ is obtained by solving the following equation
\begin{equation}
\left(\frac{|d\widetilde{w}_{i,m}|}{|\widetilde{w}_{i,m}|\mathrm{ln}|\widetilde{w}_{i,m}|}\right)^2=\left(\frac{|dw_i|}{|w_i|\mathrm{ln}|w_i|}\right)^2\left(1-\sum_{j=1}^mf(l_j)E^0_j(w_i)\right).
\end{equation}
$f(x)$ is a smooth function of $x$ such that $f(c_*)=\frac{c_*^2}{3}$ and $f((1+\epsilon)c_*)=0$, and $E^0_{j}(w_i)$ is the leading term in the Eisenstein series defined with respect to the $j$\textsuperscript{th} node on the surface.

\newl The 1PR region consists of surfaces with  $m$ simple closed geodesics of length  $0\le l_j<c_* ,~j=1,\ldots,m$. We choose the local  coordinates around the $i$\textsuperscript{th} puncture to be $e^{\frac{\pi^2}{c_*}}\widehat{w}_{i,m}$ where $\widehat{w}_{i,m}$ is obtained by solving the following equation
\begin{equation}
\left(\frac{|d\widehat{w}_{i,m}|}{|\widehat{w}_{i,m}|\mathrm{ln}|\widehat{w}_{i,m}|}\right)^2=\left(\frac{|dw_i|}{|w_i|\mathrm{ln}|w_i|}\right)^2\left(1-\sum_{j=1}^m\frac{l_j^2}{3}E^0_j(w_i)\right).
\end{equation} 
Note that different choices of the function $f$ and different values for the parameters $c_*$ and  $\epsilon$ give different choices of gluing-compatible local coordinates. It is shown in \cite{PiusRudraSen201311,PiusRudraSen201401,Sen201408,SenWitten201504}  that all such choices of local coordinates give the same value for the physical quantities. Therefore, the function $f$ and the arbitrary parameters $c_*$ and $\epsilon$ will not appear in the measurable quantities as long as  we restrict our computations to the second order in $c_*$. 

\newl In principle, it is possible to solve the curvature-correction equation to arbitrary orders of $c_*$. One can then define the corresponding coordinates by a similar procedure.

\subsubsection*{Integration of Bosonic-String Measures}

The off-shell bosonic-string measure that we constructed using Fenchel-Nielsen coordinates of Teichm\"uller space is an object which lives in the moduli space of hyperbolic surfaces because it is invariant under mapping class group (MCG) by construction. The action of MCG on the Teichm\"uller space is highly nontrivial and, as a result, the geometry of moduli space is considerably more difficult than the corresponding Teichm\"uller space. This means that in order to evaluate string amplitudes we need a special method for integrating an MCG-invariant function over the moduli space. Below, we describe such a method.

\newl Consider a hyperbolic surface $\mcal{R}$ with $n$ borders having hyperbolic lengths $L_i,~i=1,\ldots,n$. Assume that we have the following identity,
  \begin{equation}\label{eq:an identity on the hyperbolic surface R}
  \sum_{\alpha\in\overline{\text{Mod}}\cdot \gamma} f_i(l_{\alpha})=L_i,
  \end{equation}
  where $f_i$ are real functions of the hyperbolic length $l_{\alpha}$ of the curve $\alpha$ on the hyperbolic surface, which is an MCG image of the curve $\gamma=\sum_{i=1}^k\gamma_i$. Here, $\gamma_i$s are simple closed curves on the $\mathcal{R}$ with hyperbolic length $l_{\gamma_i}$. $\overline{\text{Mod}}\cdot \gamma$ denotes the set of inequivalent MCG images of the curve $\gamma$. We also assume that cutting along curves $\gamma$ produces $s$ disconnected bordered surfaces $\mathcal{R}^j(\gamma),~j=1,\ldots,s$. The Mirzakhani-McShane identity for bordered hyperbolic surfaces \cite{McShane199105,McShane199805,Mirzakhani200610} is an example of identities of the form  \eqref{eq:an identity on the hyperbolic surface R}.
  
\newl Consider an MCG-invariant function $H(\mbs{\ell},\mbs{\tau})$ defined on the bordered surface $\mathcal{R}$ with borders having lengths $\mathbf{L}=(L_1,\ldots,L_n)$. $(\mbs{\ell},\mbs{\tau})$ denotes the Fenchel-Nielsen coordinates collectively, defined with respect to a pair of pants decomposition of $\mathcal{R}$. Since $H(\mbs{\ell},\mbs{\tau})$ is invariant under MCG transformations, we are free to choose any pair of pants decomposition of the surface in order to define Fenchel-Nielsen coordinates. Interestingly, the geometry of the moduli space and the structure of the identity \eqref{eq:an identity on the hyperbolic surface R} allow us to reduce the integration of $H(\mbs{\ell},\mbs{\tau})$ over the moduli space to the integration over the lower-dimensional moduli spaces as follows:
\begin{equation}
\hspace*{-.2cm}
    \begin{aligned}
	\bigintssss_{\mathcal{M}_{g,n}(\mathbf{L})}dV~H(\mbs{\ell},\mbs{\tau})&=\frac{1}{L_i}\sum_{\alpha\in \mathrm{Mod}.\gamma}\int_{\mathcal{M}_{g,n}(\mathbf{L})}dV~f_i({\ell}_{\gamma})H(\mbs{\ell},\mbs{\tau})
	\\
	&=\frac{1}{L_i|\mathrm{Sym}(\gamma)|}\int_{\mathcal{M}_{g,n}(\mathbf{L})^{\gamma}}dV~f_i({\ell}_{\gamma})H(\mbs{\ell},\mbs{\tau})
	\\
	&=\frac{1}{L_i|\mathrm{Sym}(\gamma)|}\int_{\mathbb{R}_+^k}d{\ell}_{\gamma_1}\ldots d{\ell}_{\gamma_k} \int_{0}^{\wh{\ell}_{\gamma_1}}\ldots  \int_{0}^{\wh{\ell}_{\gamma_k}} d\tau_{\gamma_1}\ldots  d\tau_{\gamma_k}
	\\ &\times\int_{\mathcal{M(R}^1(\gamma))}dV(\mathcal{R}^1(\gamma))\ldots\int_{\mathcal{M(R}^s(\gamma))}dV(\mathcal{R}^s(\gamma))~f_i(\ell_{\gamma})H(\mbs{\ell},\mbs{\tau}).
    \end{aligned}
\end{equation}
$\tau_{\gamma_i}$ is the twist along $\gamma_i$, and $\wh{\ell}_i\equiv 2^{-M_{\gamma_i}}\ell_{\gamma_i}$ with $M_{\gamma_i}=1$ if $\gamma_i$ bounds a torus with one boundary component, and $M_{\gamma_i}=0$ otherwise. In the second step, we absorbed the MCG images of the curve $\gamma$ to the integration by enlarging the region of integration from the moduli space $\mathcal{M}(\mathcal{R})$ to a certain covering space $\mathcal{M}(\mathcal{R})^{\gamma}$. The covering space $\mathcal{M}(\mathcal{R})^{\gamma}$ can be obtained from the Teichm\"uller space by taking the quotient with the subgroup of the full mapping class group that does not act on the curve $\gamma$. In the third step, we used the fact that the space $\mathcal{M}(\mathcal{R})^{\gamma}$ can be obtained from the moduli spaces of component surfaces, obtained by cutting the surface $\mathcal{R}$ along the curves $\gamma_i, ~i=1,\ldots,k$, and $k$ number of infinite cones formed by $(\ell_i,\tau_i)$ for $i=1,\ldots,k$. Now, we can repeat the same procedure for integration over $\mathcal{M(R}^j(\gamma))$. In this way, the integral can be written over an explicit domain, as we will explain in Section \ref{integration over ms}.
  
\subsubsection*{Superstring Amplitudes and the PCOs}
    
Off-shell superstring amplitudes can be obtained by constructing the off-shell superstring measure  and then integrating it over the moduli space of hyperbolic surfaces.\footnote{Note that off-shell string measures are sections of a certain bundle over the moduli space. These measures are thus dependent on the coordinates of the corresponding moduli space.} Like in the bosonic-string theory, using the Beltrami differentials associated with Fenchel-Nielsen vector fields in the tangent space of Teichm\"uller space and a choice of gluing compatible local coordinates around the punctures, we construct the off-shell superstring measure. However, the construction of the off-shell superstring measure is more involved due to the presence of PCOs. For obtaining consistent off-shell superstring amplitudes, we need to impose the gluing-compatibility requirement for the  PCOs distribution as well as for the local coordinates. A gluing-compatible choice of PCOs distribution can be found using the Mirzakhani-McShane identity as described in \cite{Pius201808}.
       
\newl Although the gluing-compatible choice of PCOs distribution provides off-shell superstring amplitudes that is consistent with off-shell factorization properties still it might be ill-defined due to the presence of unphysical singularities. These unwanted singularities appear when the PCOs distribution satisfies certain global conditions on the moduli space. It is not clear that the proposed distribution of PCOs avoids these conditions. In the case that it does, one can follow the vertical integration prescription, proposed in \cite{Sen201408} and later elaborated in \cite{SenWitten201504}, to avoid such singularities.\footnote{There is a generalization of the vertical integration procedure by Erler and Konopka developed in \cite{ErlerKonopka201710}.}

\subsection{Organization of the Paper}

This paper is organized as follows. In Section \ref{Off-Shell Bosonic String Amplitudes}, we discuss the formal construction of off-shell amplitudes in bosonic-string theory. In Section \ref{H Off-Shell b Amplitudes}, we describe the explicit construction of off-shell amplitudes in bosonic-string theory using hyperbolic surfaces. In Section \ref{Off-Shell Superstring Amplitudes}, we explain the general construction of off-shell superstring amplitudes that are free from infrared and spurious divergences.  In Section \ref{HOff-Shell Superstring Amplitudes}, we describe an explicit construction of superstring amplitudes with off-shell external states in terms of hyperbolic surfaces. We end the main part of the paper in Section \ref{future} by mentioning some interesting directions that require further investigation. In Appendix \ref{hyperbolic}, we discuss basic facts about hyperbolic surfaces and their Teichm\"uller space. The Appendix \ref{FGFNcoordinates} is devoted to the description of a systematic algorithm for expressing the generators of a Fuchsian group in terms of the Fenchel-Nielsen coordinates of the associated surface.

\newl {\bf Note added:} After the appearance of this and the subsequent works \cite{MoosavianPius201706,MoosavianPius201708}, the applications of hyperbolic geometry in string theory have been explored in several works including \cite{CostelloZwiebach201909,Cho201912,Firat202102}.

%% file: Sections/Off-Shell_Bosonic-String_Amplitudes.tex
In this section, we shall  review the general construction of off-shell amplitudes in bosonic-string theory. 
  
\subsection{The Worldsheet Theory}

The bosonic string theory is formulated in terms of a conformal field theory (CFT) defined on a Riemann surface. This CFT consists of two sectors, the matter and the ghost sector. The total CFT has a central charge $(0,0)$. The matter CFT has  a central charge $(26,26)$ and the central charge of the reparametrization ghosts CFT is $(-26,-26)$. The ghost system is composed of the anti-commuting fields $b,c,\bar b,\bar c$.   The  Hilbert space of CFT is denoted by $\mathcal{H}$. We denote by $\mathcal{H}_0$ a  subspace of $\mathcal{H}$ defined as
  \begin{equation}\label{h0bosonic}
  |\Psi\rangle\in \mathcal{H}_0,\qquad\mathrm{if}\qquad(b_0-\bar b_0)|\Psi\rangle=0,\quad(L_0-\bar L_0)|\Psi\rangle =0,
  \end{equation}
   where $\bar L_n$ and $L_n$ denote the total Virasoro generators in the left and the right-moving sectors of the worldsheet theory.  Physical states that appear as external states in the S-matrix elements computation belong to the subspace  $\mathcal{H}_1$  of $\mathcal{H}_0$, consisting of states with the ghost-number $2$ and satisfy the extra condition
    \begin{equation}\label{h1bosonic}
  |\Psi\rangle\in \mathcal{H}_1,\qquad\mathrm{if}\qquad |\Psi\rangle\in \mathcal{H}_0,\quad (b_0+\bar b_0)|\Psi\rangle=0.
  \end{equation}
We shall denote the BPZ-conjugate of the state $|\phi_i\rangle\in \mathcal{H}_0$  by $\langle \phi_i^c|$ satisfying
\begin{equation}\label{BPZconjugateb}
\begin{gathered}
    \begin{aligned}
    \langle\phi_i^c|\phi_j\rangle&=\delta_{ij},
    \\
    \langle\phi_j|\phi_i^c\rangle&=(-1)^{n_{\phi_i}}\delta_{ij},
    \end{aligned}
    \\
    \sum_i|\phi_i\rangle\langle \phi^c_i|=(-1)^{n_{\phi_i}}\sum_i|\phi_j^c\rangle\langle\phi_j|=1,
\end{gathered}
\end{equation}
where $n_{\phi_i}$ is the ghost number of the state $|\phi_i\rangle$. 

\subsection{Off-Shell Bosonic-String Amplitudes}\label{sec:offshell bosonic-string amplitudes}

The $g$-loop on-shell amplitudes with $n$ external states in bosonic-string theory are obtained by  integrating an appropriate $6g-6+2n$ real dimensional differential form over  $\overline{\mathcal{M}}_{g,n}$, the compactified moduli space of Riemann surfaces. This differential form is constructed by computing the CFT correlator of unintegrated vertex operators with conformal dimension $(1,1)$ corresponding to the external states satisfying the tree-level on-shell condition.  However, generic states of string theory undergo mass  renormalization. States having masses different from the tree-level masses are mapped to vertex operators having conformal dimensions different from $(1,1)$. Therefore,  for generic states in string theory, on-shell amplitudes defined using  vertex operators with conformal dimension $(1,1)$ do not compute S-matrix elements beyond tree level. This forces us to consider the off-shell amplitudes constructed using vertex operators of arbitrary conformal dimensions. 

\newl Since  on-shell amplitudes are defined using vertex operators of conformal dimension $(1,1)$, they do not depend on spurious data like the choice of local coordinates around the punctures at which vertex operators are inserted. This means that the integration measure of an on-shell amplitude is a genuine differential form on $\mathcal{M}_{g,n}$. But off-shell amplitudes defined using vertex operators with arbitrary conformal dimensions do depend on the choice of local coordinates around the punctures. Therefore, we can not consider the integration measure of an off-shell amplitude as a genuine differential form on $\overline{\mathcal{M}}_{g,n}$. Instead, we need to  think of it as a differential form defined on a section of a larger space $\mathcal{P}_{g,n}$. This space is  defined as a fiber bundle over $\overline{\mathcal{M}}_{g,n}$. The fiber direction of  $\mathcal{P}_{g,n}$ corresponds to possible choices  of the local coordinates around the $n$ punctures of a genus-$g$ Riemann surface. If we restrict ourselves to the states that belong to the Hilbert space $\mathcal{H}_0$, then we can consider the differential form of our interest as defined on a section of space $\widehat{\mathcal{P}}_{g,n}$. This space is smaller compared to $\mathcal{P}_{g,n}$.\footnote{We mode out the phases of local coordinates. Thus we essentially consider the quotient of $\mathcal{P}_{g,n}$ by the group generated by rotations of local coordinates. Therefore, the resulting space, $\widehat{\mathcal{P}}_{g,n}$, is smaller than $\mathcal{P}_{g,n}$.} We can understand  $\widehat{\mathcal{P}}_{g,n}$ as a base space of the fiber bundle $\mathcal{P}_{g,n}$ with the fiber direction corresponds to the phases of local coordinates.

\subsubsection*{Tangent Vectors of $\mathcal{P}_{g,n}$}

In order to construct a differential form on a space, we need  to first study its tangent space at a generic point. Since we are interested in constructing a differential form on a section of  $\mathcal{P}_{g,n}$, we need to study the  tangent space of $\mathcal{P}_{g,n}$ associated with deformations of the punctured Riemann surface and/or the choice of local coordinates around the punctures. We shall do this using the idea of Schiffer variation \cite{SchifferSpencer,Gardiner197501,Zwiebach199206,Sen201408}.
   \begin{figure}
\begin{center}
\usetikzlibrary{backgrounds}
\begin{tikzpicture}[scale=.45]

\draw[black,   thick] (0,2) to[curve through={(1,2.2)..(3,3)..(8,0) .. (3,-3) ..(0,-2)  .. (-3,-3)..(-8,0)..(-3,3)..(-1,2.2)}] (0,2);
\draw[black,  thick] (2.5,0) to[curve through={(3.5,1)..(5,1)}] (6,0);
\draw[black,  thick] (2.3,.2) to[curve through={(3.5,-.75)..(5,-.75)}] (6.2,.2);
\draw[black,  thick] (-2.5,0) to[curve through={(-3.5,1)..(-5,1)}] (-6,0);
\draw[black,  thick] (-2.3,.2) to[curve through={(-3.5,-.75)..(-5,-.75)}] (-6.2,.2);
\draw[black,  very thick] (-5,2) ellipse (.06 and .06);
\draw[black,  very thick] (-5,-2) ellipse (.06 and .06);
\draw[black,  very thick] (7,0) ellipse (.06 and .06);
\draw[black,  very thick] (5,2.1) ellipse (.5 and 1);
\draw[black,  very thick] (5,-1.95) ellipse (.5 and 1.1);
\draw[black,  very thick] (-3.2,1.95) ellipse (.5 and 1);
\draw[black,  very thick] (-3.2,-1.85) ellipse (.5 and 1.1);
\draw[black,  very thick] (0,0) ellipse (.5 and 2);
\draw[black,  very thick] (-7,0) ellipse (1 and .75);
\draw[gray,  very thick] (-5,2.05) ellipse (.7 and .7);
\draw[gray,  very thick] (-5,-1.95) ellipse (.7 and .7);
\draw[gray,  very thick] (7,0) ellipse (.7 and .7);
\draw  node[above] at (-5,3) {$D_1$};
\draw  node[below] at (-5,-3) {$D_2$};
\draw  node[right] at (8,0) {$D_3$};
\draw  node[below right] at (-7.5,2.3) {$P_1$};
\draw  node[above right] at (-7.5,-2.3) {$P_2$};
\draw  node[above ] at (1.5,-.7) {$P_4$};
\draw  node[above ] at (-1.5,-.7) {$P_3$};
\draw  node[below ] at (6.5,-.7) {$P_5$};
\end{tikzpicture}
\end{center}

\caption{A pair of pants decomposition of a genus-2 surface with 3 punctures. $P_i$ denotes a pair of pants with local coordinates $z_i$ inside $P_i$ for $i=1,\ldots,5$. $D_i$ denotes the disc around the $i$\textsuperscript{th} puncture with unit radius with respect to the local coordinate  $w_i$ defined around the $i$\textsuperscript{th} puncture for $i=1,2,3$.}
\label{Riemann surface coordinates}
\end{figure}
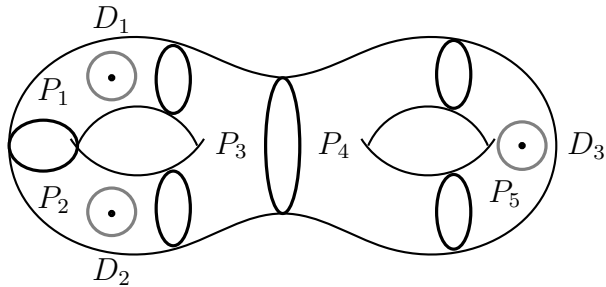
To elucidate the idea of Schiffer variation, consider a Riemann surface $\mathcal{R}\in\mathcal{P}_{g,n}$. This means that $\mathcal{R}$ is a genus-$g$ Riemann surface with $n$ punctures and a specific choice of local coordinates around the punctures. We shall denote the local coordinate around the $i$\textsuperscript{th} puncture by $w_i$  and the disc  around the $i$\textsuperscript{th} puncture by $D_i$ for $i=1,\ldots,n$. $D_i$ is defined by all $w_i$s that satisfy $|w_i|\le 1$. Now, consider a pair of pants decomposition of $\mathcal{R}-\sum_i D_i$  by choosing $3g-3+n$ homotopically non-trivial disjoint curves on it. This gives $2g-2+n$ number of pairs of pants denoted by $P_i,~i=1,\ldots,2g-2+n$, see Figure \ref{Riemann surface coordinates}. Denote the coordinate inside $P_i$ by $z_i$. Assume that the $i^{th}$ disc $D_i$ shares its boundary $|w_i|=1$ with the $j$\textsuperscript{th} pair of pants $P_j$. Also, assume that the $k$\textsuperscript{th} pair of pants $P_k$ shares a boundary with the $m$\textsuperscript{th} pair of pants $P_m$. Then we have the following transition functions 
\begin{equation}\label{transitionf1}
\begin{aligned}
z_j&=f_{i}(w_i),\quad&&\text{analytic for}&&\quad P_j\cap{D}_i,\quad\text{can have  singularities elsewhere},
\\
z_k&=f_{km}(z_m),\quad&&\text{analytic for}&&\quad P_k\cap{P}_m,\quad\text{can have singularities elsewhere}.
\end{aligned}
\end{equation}
The Schiffer variation generates all deformations of $\mathcal{P}_{g,n}$ by varying the transition functions associated with the discs around the punctures $f_i(w_i),~i=1,\ldots,n$ by keeping all other transition functions $f_{km}(z_m)$ fixed. We can generate such variations by keeping the coordinates $z_k$ inside the pair of pants $P_k,~k=1,\ldots,2g-2+n$ fixed and changing the coordinates inside the disc $D_i$ from $w_i\to w^{\epsilon}_i$ for $i=1,\ldots,n$. This change of coordinates deforms the transition function associated with the disc $D_i$ around the $i$\textsuperscript{th} puncture as follows
\begin{equation}\label{deformfi}
f_i^{\epsilon}(w_i)=f_i(w_i)-\epsilon v^{(i)}(z_j),~v^{(i)}(z_j)=f'_i(w_i)v^{(i)}(w_i).
\end{equation}
Here, we assumed that the boundary of $D_i$ is shared with the pair of pants $P_j$. The form of $v^{(i)}(w_i)$ can be obtained from the fact that $f^{\epsilon}_i(w_i^{\epsilon})=z_k=f_i(w_i)$. Then, the tangent vector of $\mathcal{P}_{g,n}$ is given by
 \begin{equation}\label{tangentvpgn}
\vec{v}(z)=(v^{(1)}(z),\ldots,v^{(n)}(z)).
 \end{equation}
$z$ denotes the coordinate on $\mathcal{R}-\sum_i D_i$. The behavior of $v^{(i)}(z)$ on $\mathcal{R}$ determines the kind of deformations it induces on $\mathcal{P}_{g,n}$

\begin{itemize}
\item {\it $\vec{v}(z)$ is a null vector}: if it is holomorphic everywhere except possibly at the punctures.

\item {\it $\vec{v}(z)$ deforms the local coordinates around punctures}: if it is holomorphic inside $D_i,~i=1,\ldots,n$ which vanishes at the punctures and it does not holomorphically extend into $\mathcal{R}-\sum_i D_i$.

\item {\it $\vec{v}(z)$ moves the punctures}: if it is holomorphic inside $D_i,~i=1,\ldots,n$ and is nonvanishing at the punctures and also it does not holomorphically extend into $\mathcal{R}-\sum_i D_i$.

\item {\it $\vec{v}(z)$ generates deformations in the intrinsic moduli of $\mathcal{R}$ which is not associated with punctures}: if it has poles at one or more punctures and also it does not holomorphically extend into $\mathcal{R}-\sum_i D_i$. $3g-3$ of such vector fields with poles of order $q=1,\ldots,3g-3$ at any of the punctures generate the complete set of moduli deformations of $\mathcal{R}$.
\end{itemize}

\subsubsection*{Differential Forms on $\widehat{\mathcal{P}}_{g,n}$}

Consider $p$ tangent vectors $V_1,\ldots,V_p$ of $\mathcal{P}_{g,n}$ and let $\vec{v}_1,\ldots,\vec{v}_p$ be the corresponding $n$-tuple vector fields. We can construct an operator-valued $p$-form $B_p$, whose contraction with the tangent vectors $V_1,\ldots,V_p$ is given by 
\begin{equation}\label{opevormbos}
B_p[V_1,\ldots,V_p]=b(\vec{v}_1)\ldots b(\vec{v}_p),
\end{equation}
where $b(\vec{v})$ is defined as 
\begin{equation}\label{bv}
b(\vec{v})=\sum_i\oint dw_iv^{(i)}(w_i)b^{(i)}(w_i)+\sum_i\oint d\bar w_i\bar v^{(i)}(\bar w_i)\bar b^{(i)}(\bar w_i).
\end{equation}
$b,\bar b$ denote the anti-ghost fields. We define the $p$-form on $\mathcal{P}_{g,n}$ as
\begin{equation}\label{pformpgnbos}
\Omega_p^{(g,n)}(|\Phi\rangle)=(2\pi \mathrm{i})^{-(3g-3+n)}\langle\mathcal{R}|B_p|\Phi\rangle.
\end{equation}
Here, $|\Phi\rangle$ is some element of $\mathcal{H}^{\otimes n}$ with ghost number 
$$n_{\Phi}=p+6-6g.$$
$\langle \mathcal{R}|$ is the surface-state associated with the surface $\mathcal{R}$, which describes the state that is created on the boundaries of $D_i$ by performing a functional integral over the fields of the CFT on $\mathcal{R}-\sum_iD_i$.  Inner product between $\langle\mathcal{R}|$ and a state $|\Psi_1\rangle\otimes\ldots\otimes |\Psi_n\rangle\in\mathcal{H}^{\otimes n}$ 
\begin{equation}\label{innerprcft}
\langle\mcal{R}|(|\Psi_1\rangle\otimes\ldots\otimes |\Psi_n\rangle),
\end{equation} 
describes the $n$-point correlation function on $\mathcal{R}$ with the vertex operator for $|\Psi_i\rangle$ inserted at the $i^{th}$ puncture using the local coordinate system $w_i$ around that puncture. It is clear that $\Omega_p^{(g,n)}(|\Phi\rangle)$ is a $p$-form on $\mathcal{P}_{g,n}$. Remember that a $p$-form on a space generates a number when contracted with $p$ tangent vectors of this space and this number is antisymmetric under the exchange of any pair of tangent vectors. Since anti-ghost fields $b,\bar b$ are anti-commuting, $\Omega_p^{(g,n)}(|\Phi\rangle)$ also has this property and, hence, $\Omega^{(g,n)}_p$ is a $p$-form on $\mathcal{P}_{g,n}$.

\newl It is argued in \cite{Nelson198902} that constructing consistent off-shell string measures requires us to restrict $|\Phi\rangle$ to be an element of $\mathcal{H}_0^{\otimes n}$. In other words, $|\Phi\rangle$ has to satisfy the following conditions
\begin{equation}\label{eq:conditions on off-shell states b_0 L_0}
(b_0-\bar b_0)|\Phi\rangle=(L_0-\bar L_0)|\Phi\rangle=0.
\end{equation}
These conditions mean that the state $|\Phi\rangle$ is insensitive to the phase of the chosen local coordinates. As a result, the $p$-form $\Omega^{(g,n)}_p$ is independent of the phase of local coordinates around the punctures. In order to check this claim, note that a tangent vector that generates a phase rotation for the local coordinates has the following nonvanishing components, $v^{(i)}(w_i)=w_i$ and $\bar v^{(i)}(\bar w_i)=-\bar w_i$. It is clear that such vectors do not change $\Omega_p^{(g,n)}(|\Phi\rangle)$ if $|\Phi\rangle\in \mathcal{H}_0$. This is because for  $\vec{v}=(0,\ldots,w_i,\ldots,0)$, $b(\vec{v})=b^{(i)}_0-\bar b^{(i)}_0$ and $T(\vec{v})=L^{(i)}_0-\bar L^{(i)}_0$, where the superscripts denote that the modes  $b_0,L_0$ and their complex conjugates are defined with respect to the local coordinates inside the $i$\textsuperscript{th} puncture. Here 
\begin{equation}\label{tv}
T(\vec{v})=\sum_i\oint dw_iv^{(i)}(w_i)T^{(i)}(w_i)+\sum_i\oint d\bar w_i\bar v^{(i)}(\bar w_i)\bar T^{(i)}(\bar w_i).
\end{equation}
Since $(b_0-\bar b_0)|\Phi\rangle=(L_0-\bar L_0)|\Phi\rangle=0$, the  change in  $\Omega_p^{(g,n)}(|\Phi\rangle)$ due to the phase rotation of local coordinates vanishes. Therefore, consistent off-shell string measures are $p$-forms on $\widehat{\mathcal{P}}_{g,n}$, where $\widehat{\mathcal{P}}_{g,n}$ is the space obtained by identifying points in ${\mathcal{P}}_{g,n}$ parametrizing Riemann surfaces whose chosen local coordinates are related by phase rotations.

\subsubsection*{Differential Form on Sections of $\widehat{\mathcal{P}}_{g,n}$
}
Let us discuss the construction of tangent vectors on a section of $\widehat{\mathcal{P}}_{g,n}$ where choosing a section corresponds to a specific choice of local coordinates. Changing along a section corresponds to changing the moduli parameters while keeping the local coordinates around the punctures fixed. Therefore, in order to construct a $p$-form on a section of $\widehat{\mathcal{P}}_{g,n}$, we need to consider only those tangent vectors that give rise to variations of moduli parameters.

\newl Consider a genus-$g$ Riemann surface $\mcal{R}$ with $n$ punctures. Let us divide $\mcal{R}$ into $2g-2+2n$ patches. $n$ patches are obtained by considering small disks $D_i$s around the $n$ punctures. The remaining $2g-2+n$ patches are pairs of pants obtained by pair-of-pants decomposition of $\mcal{R}-\sum_iD_i$. We denote the local coordinate on the $m$\textsuperscript{th} patch by $z_m$. Since there are $3g-3+2n$ interfaces between these patches ($n$ interfaces between the disks and the remaining pairs of pants, and $3g-3+n$ interfaces between the pairs of pants). We thus have $3g-3+2n$ number of transition functions. As a result, tangent vectors of $\pcab$ constructed using the idea of Schiffer variation will have $3g-3+2n$ number of components.

\newl Denote the real coordinates of the moduli space of genus-$g$ Riemann surface with $n$ punctures by $(\mfk{t}^{(1)},\ldots,\mfk{t}^{(6g-6+2n)})$. Assume that the $m$\textsuperscript{th} and $n$\textsuperscript{th} coordinate patches have a non-empty intersection along the contour $C_{mn}$ runs between them. For the $(3g-3+2n)$-tuple vector field  $\vec{v}_{k}=(0,\ldots,0,\frac{\partial z_m}{\partial \mfk{t}^{(k)}}\big|_{z_n},0,\ldots,0)$ corresponding to the variation in the moduli $\mfk{t}^{(k)}$, $b(\vec{v}_k)$ defined in \eqref{bv} is given by
\begin{equation}\label{bvgen}
b(\vec{v}_{k})=\sum_{(mn)}\bigointssss_{C_{mn}}\left(dz_m\left.\frac{\partial z_m}{\partial \mfk{t}^{(k)}}\right|_{z_n}b_{z_mz_m}-d\bar z_{m}\left.\frac{\partial \bar z_m}{\mfk{t}^{(k)}}\right|_{z_n}b_{\bar z_m\bar z_m}\right),
\end{equation} 
where the sum over $(mn)$ denotes the summation over the number of overlaps between coordinate patches. Then, the change in the local coordinates $z_m$ and $z_n$ under a change in the moduli $\mfk{t}^{(k)}$ are related as follows \cite{Polchinskivol011998}
\begin{equation}\label{modtvar}
\left.\frac{\partial z_m}{\partial \mfk{t}^{(k)}} \right|_{z_n}=v_{km}^{z_m}-\left.\frac{\partial z_m}{\partial z_n}\right|_{\mfk{t}}v_{kn}^{z_n}=v_{km}^{z_m}-v_{kn}^{z_m},\qquad v_{km}^{z_m}\equiv \frac{d z_m}{d \mfk{t}^{(k)}}.
\end{equation}
From \eqref{modtvar}, we see that 
 \begin{equation}\label{bvgen1}
b(\vec{v}_{k})=\sum_{m=1}^{2g-2+2n}\bigointssss_{C_{m}}\left(dz_mv^{z_m}_{km}b_{z_mz_m}-d\bar z_{m}v^{\bar z_m}_{km}b_{\bar z_m\bar z_m}\right).
\end{equation} 
Using Stokes Theorem, we get
 \begin{equation}\label{bvgen2}
b(\vec{v}_{k})=\bigintssss d^2z\left(b_{zz}\mu_{k\bar z}^z+b_{\bar z\bar z}\mu_{kz}^{\bar z}\right).
\end{equation} 
 Here $\mu_k$ denotes the Beltrami differential associated with the moduli $\mfk{t}^{(k)}$ which is related to the coordinate changes via the infinitesimal version of the Beltrami equation
\begin{equation}\label{beltrami}
    \begin{aligned}
 \mu_{kz_m}^{\bar z_m}&=\partial_{z_m}v_{km}^{\bar z_m},
 \\
 \mu_{k\bar z_m}^{z_m}&=\partial_{\bar z_m}v_{km}^{ z_m}.
    \end{aligned}
\end{equation}
 
\subsubsection*{Gluing-Compatible Integration Cycle}

As we have explained above, off-shell bosonic-string amplitudes are given by integrating top-degree  differential forms on a section of $\widehat{\mathcal{P}}_{g,n}$. These forms are built using states $|\Phi\rangle\in\mathcal{H}_1^{\otimes n}$, defined in \eqref{h1bosonic}, representing the off-shell external states. In particular, the ghost number of $|\Phi\rangle$ is $2n$. Therefore, the degree of relevant differential form is $p=6g-6+2n$, which matches with the dimension of $\mathcal{M}_{g,n}$. However, since it depends on the choice of local coordinates around the punctures, it is not a genuine top-form on $\mathcal{M}_{g,n}$. Contracting $\Omega_{6g-6+2n}^{(g,n)}(|\Phi\rangle)$ with a tangent vector $\vec{v}$, generating deformations of local coordinate without varying the surface $\mathcal{R}$, gives a non-zero number. Therefore, $\Omega_{6g-6+2n}^{(g,n)}(|\Phi\rangle)$ is sensitive to the choice of section of $\pcab$. 

\newl On the other hand, physical quantities that can be extracted from off-shell amplitudes like the renormalized masses and the S-matrix elements must be independent of the choice of a section of $\pcab$ \cite{Sen201408}. This is true only if we  impose a condition on the choice, known as the {\it gluing-compatibility} \cite{PiusRudraSen201311,PiusRudraSen201401}. To describe this condition, consider two Riemann surfaces $\mathcal{R}_1$ and $\mathcal{R}_2$. $\mathcal{R}_1$ is a genus-$g_1$ surface with $n_1$ punctures and $\mathcal{R}_2$ is a genus $g_2$ surface with $n_2$ punctures. Denote the collection of pairs of pants in a pair-of-pants decomposition of $\mathcal{R}_1$ by $\{P^{(1)}_k\}$ and the discs around punctures by $D_1^{(1)},\ldots,D_{n_1}^{(1)}$. Similarly denote the collection of pairs of pants in a pairs-of-pants decomposition of $\mathcal{R}_2$ by $\{P^{(2)}_k\}$ and the discs around punctures by $D_1^{(2)},\ldots,D_{n_2}^{(2)}$. We can glue discs $D_i^{(1)}$ and  $D_j^{(2)}$ using the gluing relation:
\begin{equation}
w_i^{(1)}w_j^{(2)}=e^{-s+\mathrm{i}\theta},\qquad 0\leq s<\infty,\quad 0\leq\theta<2\pi. \label{plumb}
\end{equation}
This will produce another surface $\mathcal{R}$ with genus $g=g_1+g_2$ and $n=n_1+n_2-2$ punctures. Surfaces that can be constructed this way belong to the region near the boundary of $\mathcal{M}_{g,n}$. This part of $\mathcal{M}_{g,n}$ can be parametrized by the coordinates of $\mathcal{M}_{g_1,n_1}$, $\mathcal{M}_{g_2,n_2}$ and $(s,\theta)$. The gluing-compatibility condition requires that the section of the fiber bundle $\widehat {\mathcal{P}}_{g,n}$ over this region of moduli space be chosen such that the relationship between coordinates of $\{P_k^{(1)}\}$ and $D_1^{(1)},\ldots,\cancel{D_i^{(1)}},\ldots, D^{(1)}_{n_1}$ depends only on the moduli of $\mathcal{M}_{g_1,n_1}$ and not on the moduli of $\mathcal{M}_{g_2,n_2}$ and $(s,\theta)$. Similarly  the relation between coordinates of $\{P_k^{(2)}\}$ and $D_1^{(2)},\ldots,\cancel{D_j^{(2)}},\ldots, D^{(2)}_{n_2}$ depends only on the moduli of $\mathcal{M}_{g_2,n_2}$ and not on the moduli of $\mathcal{M}_{g_1,n_1}$ and $(s,\theta)$. Also, the dependence of these relations on the moduli of $\mathcal{M}_{g_l,n_l}$ must be the one induced from the choice of section $\widehat{\mathcal{P}}_{g_l,n_l}$ for $l=1,2$. 

\newl A choice of a section of $\wh{\mcal{P}}_{g,n}$ satisfying the gluing-compatible requirement explained above provides a well-defined off-shell bosonic-string measure associated with $g$-loop contribution to a scattering process involving  $n$ arbitrary external states. It is well-defined in the sense that the physical results do not depend on the choice of local coordinates around the punctures. This concludes our brief recap of the construction of off-shell bosonic-string measures.

%% file: Sections/Hyperbolic_Geometry_and_Off-Shell_Bosonic-String_Amplitudes.tex
In this section, we shall explicitly construct off-shell  amplitudes in bosonic-string theory using hyperbolic geometry following the general construction described in section \ref{Off-Shell Bosonic String Amplitudes}. For a quick introduction to the theory of hyperbolic surfaces see Appendix \ref{hyperbolic}. Standard references on the subject are \cite{ImayoshiTaniguchi1992,Hubbard2006,KeenLakic2007}.

\subsection{Degenerating Family of Hyperbolic Surfaces}
In this section, we explain how to construct degenerating families of hyperbolic surfaces. Before explaining the details, let us set the stage first.

\newl A Riemann surface is called hyperbolic if it is equipped with a hyperbolic metric, i.e. a metric with constant curvature $-1$ everywhere on the surface. 
One important advantage of using a hyperbolic metric is that Riemann surfaces with nodes obtained by the gluing of hyperbolic surfaces, as we explain in detail below, are again hyperbolic.  This suggests that if we choose local coordinates around the punctures as the one induced from the hyperbolic metric on the surface, at least on the  complete degeneration limit, where the gluing parameter vanishes and a node is developed, this choice will match with local coordinates induced from the component surfaces. This is an essential constraint satisfied by a gluing-compatible section of $\widehat{\mathcal{P}}_{g,n}$.  We shall argue in the rest of this section that  away from the degeneration locus also we can  define  a consistent choice of local coordinate that leads to a gluing-compatible section of $\widehat{\mathcal{P}}_{g,n}$.

\newl A hyperbolic surface can be represented  as a quotient of the upper half-plane $\mathbb{H} $ by a Fuchsian group.  A puncture on a hyperbolic surface corresponds to the fixed point of a parabolic element of the Fuchsian group acting on the upper half-plane $\mathbb{H}$. For a puncture $p$, there is a natural local coordinate $w$ with $w(p) = 0$, and the hyperbolic metric around the puncture is locally given by \cite{Wolpert201001}
 \begin{equation}
 ds^2=\left(\frac{|dw|}{|w|\mathrm{ln}|w|}\right)^2.
 \end{equation}
Let $z$ be the coordinate on the upper half-plane. Then, the distinguished local coordinate is
		\begin{equation}\label{local coordinate for the cusp at infinity}
		w = e^{2\pi\mathrm{i}z}, 
		\end{equation}
	for the puncture corresponding to the parabolic element whose fixed point is at infinity $z_{\infty}=i\infty$ on the upper half-plane $\mathbb{H}$. As required this choice of local coordinate  is invariant under the translation, $z \to z + 1$, which represents  the action of the generator of the corresponding parabolic element.  In terms of the coordinate $z$, the metric around the puncture takes the form 
		\begin{equation}
		ds^2=\frac{dzd\bar{z}}{(\mathrm{Im}z)^2},
		\end{equation}
		which is the  hyperbolic metric for the Poincar\'e upper half-plane $\mathbb{H}$, as it should be.  The local canonical coordinate around the puncture $w$ given by \eqref{local coordinate for the cusp at infinity} is unique modulo a phase factor.  If the fixed point of the parabolic element corresponding to the cusp is at a finite point $x$  on the real axis of the upper half-plane, then the local coordinate is given by 
		\begin{equation}
		w = e^{-\frac{2\pi\mathrm{i}}{z-x}}. \label{local coordinate for cusp at finite real value $x$}
		\end{equation}
Before using the proposed choice of local coordinates for constructing off-shell amplitudes, it is important to ensure that it satisfies the gluing-compatibility  requirement.  For this, we should first answer the following question: {\it What are the local coordinates induced around the punctures on the surface obtained via gluing of hyperbolic surfaces?} To proceed, let us analyze the metric on hyperbolic surfaces obtained via the gluing of hyperbolic surfaces. 

\subsubsection*{Gluing Hyperbolic Surfaces}
 
The degenerating families of Riemann surfaces are readily given by cut-and-paste constructions in hyperbolic geometry, following Fenchel and Nielsen \cite{FenchelNielsen2002,ImayoshiTaniguchi1992,Hubbard2006}. In the following, we shall discuss the relationship between the cut-and-paste construction and the gluing construction of degenerating families of hyperbolic surfaces. For more details on the cut-and-paste construction and Fenchel-Nielsen coordinates, see Appendix \ref{hyperbolic}.
  
\begin{figure}\centering
\begin{tikzpicture}[scale=.65]
\draw[line width=1pt] (-3,1) .. controls (-2.25,1) and (-1.75,.75)  ..(-1,0);
\draw[line width=1pt] (-3,-2) .. controls(-2.25,-2) and (-1.75,-1.75)  ..(-1,-1);
\draw[line width=1pt] (-3,.4) .. controls(-1.75,.5)  and (-1.25,-1.2) ..(-3,-1.4);
\draw[line width=1pt,color=gray] (-3,.7) ellipse (.115 and .315);
\draw[line width=1pt,color=gray] (-3,-1.7) ellipse (.115 and .315);
\draw[line width=.75pt,color=gray] (0,-0.5) ellipse (.115 and .155);
\draw[line width=1pt,color=black] (0.35,-0.5) ellipse (.085 and .115)node  [below ] {$\gamma_2$};
\draw[line width=1pt] (-1,0) .. controls(-.7,-.25) ..(1,-0.5)node[above] {$p$};
\draw[line width=1pt] (-1,-1) .. controls(-.75,-.75)  ..(1,-0.5);
\draw[line width=1pt] (4,0) .. controls(5,1) and (6,1) ..(7,0);
\draw[line width=1pt] (4,-1) .. controls(5,-2) and (6,-2) ..(7,-1);
\draw[line width=1pt] (7,0) .. controls(7.2,-.15)  ..(8,-.15);
\draw[line width=1pt] (7,-1) .. controls(7.2,-.85)  ..(8,-.85);
\draw[line width=1pt,color=gray] (8,-.5) ellipse (.15 and .35);
\draw[line width=1pt] (4.75,-.75) .. controls(5.25,-1.15) and (5.75,-1.15) ..(6.25,-.75);
\draw[line width=1pt] (4.75,-.35) .. controls(5.25,0.1) and (5.75,0.1) ..(6.25,-.35);
\draw[line width=1pt] (4.75,-.35) .. controls(4.65,-.475) and (4.65,-.625) ..(4.75,-.75);
\draw[line width=1pt] (6.25,-.35) .. controls(6.35,-.475) and (6.35,-.625) ..(6.25,-.75);
\draw[line width=1pt] (4,0) .. controls(3.5,-.4) ..(2,-0.5)node[above] {$q$};
\draw[line width=1pt] (4,-1) .. controls(3.5,-.65)  ..(2,-0.5);
\draw[line width=.75pt,color=gray] (3.35,-0.525) ellipse (.1 and .145);
\draw[line width=1pt,color=black] (3,-0.525) ellipse (.075 and .115)node  [below  ] {$\gamma_1$};
\draw (-3.75,.5) node[above ] {$\beta_2$}  (-3.75,-2) node[above] {$\beta_1$} (8.2,-.2)node [below right ] {$\beta_3$};
\draw[->,line width =1pt] (9.5,-.5)--(10.5,-.5);
\draw[line width=1pt] (12.65,1) .. controls (14.15,.8) and (13.85,.35)  ..(14.65,0);
\draw[line width=1pt] (12.65,-2) .. controls(14.5,-1.8) and (13.85,-1.35)  ..(14.65,-1);
\draw[line width=1pt] (12.65,.4) .. controls(13.85,.5)  and (14.40,-1.2) ..(12.65,-1.4);
\draw[line width=1pt,color=gray] (12.65,.7) ellipse (.115 and .315);
\draw[line width=1pt,color=gray] (12.65,-1.7) ellipse (.115 and .315);
\draw[line width=.75pt,color=gray] (15.65,-0.52) ellipse (.115 and .2);
\draw[line width=1pt,color=black] (16,-0.525) ellipse (.085 and .115)node  [below ] {$\gamma$};
\draw[line width=1pt] (14.65,0) .. controls(14.9,-.15) ..(16.05,-0.4);
\draw[line width=1pt] (14.65,-1) .. controls(14.9,-.85)  ..(16.05,-0.65);
\draw[line width=1pt] (17,0) .. controls(18,1) and (19,1) ..(20,0);
\draw[line width=1pt] (17,-1) .. controls(18,-2) and (19,-2) ..(20,-1);
\draw[line width=1pt] (20,0) .. controls(20.2,-.15)  ..(21,-.15);
\draw[line width=1pt] (20,-1) .. controls(20.2,-.85)  ..(21,-.85);
\draw[line width=1pt,color=gray] (21,-.5) ellipse (.15 and .35);
\draw[line width=1pt] (17.75,-.75) .. controls(18.25,-1.15) and (18.75,-1.15) ..(19.25,-.75);
\draw[line width=1pt] (17.75,-.35) .. controls(18.25,0.1) and (18.75,0.1) ..(19.25,-.35);
\draw[line width=1pt] (17.75,-.35) .. controls(17.65,-.475) and (17.65,-.625) ..(17.75,-.75);
\draw[line width=1pt] (19.25,-.35) .. controls(19.35,-.475) and (19.35,-.625) ..(19.25,-.75);
\draw[line width=1pt] (17,0) .. controls(16.5,-.4) ..(16.05,-0.4);
\draw[line width=1pt] (17,-1) .. controls(16.5,-.65)  ..(16.05,-0.65);
\draw[line width=.75pt,color=gray] (16.35,-0.54) ellipse (.1 and .15);
\draw[line width=1pt,color=black] (16,-0.525) ellipse (.075 and .115);
\draw (12.15,.5) node[above ] {$\beta_2$}  (12.15,-2) node[above ] {$\beta_1$} (21.2,-.2)node [below right ] {$\beta_3$};
\end{tikzpicture}

\caption{The gluing of two surfaces at punctures $p$ and $q$.}
\label{plumbing1}
\end{figure}
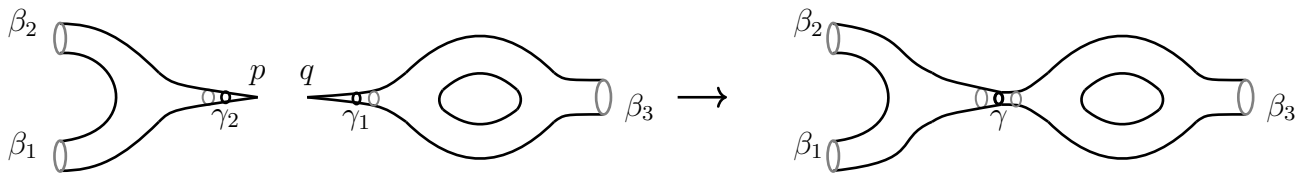

\paragraph{The Gluing Construction:} Consider $\overline{\mathcal{M}}_g$, the Deligne-Mumford stable curve compactification of the  moduli space of genus-$g$ Riemann surfaces \cite{DeligneMumford1969a}. Let us denote the compactification divisor of $\overline{\mathcal{M}}_g$ by $\mathcal{D}$. A point of $\mathcal{D}$ represents a Riemann surface $\mathcal{R}$ with nodes. By definition, a neighborhood of a node $\mfk{n}$ of $\mathcal{R}$ is complex-analytically isomorphic to either $\{|w^{(1)}|<\epsilon\}$ or
   \begin{equation}
 U\equiv \{w^{(1)}w^{(2)}=0|~|w^{(1)}|,|w^{(2)}|<\epsilon\},
 \end{equation}
where $w^{(1)}$ and $w^{(2)}$ are the local coordinates around the  two sides of $\mfk{n}$. To move away from the compactification divisor, let us consider the following family of surfaces fibered over a disk with complex coordinate $t$ 
   \begin{equation}
   \left\{w^{(1)}w^{(2)}=t|~|w^{(1)}|,|w^{(2)}|<\epsilon,|t|<\epsilon\right\}.
   \end{equation}
   We can identify $U$ with the fiber at $t=0$. A deformation of $\mathcal{R}\in \overline{\mathcal{M}}_g$ which opens the node is given by varying the parameter $t$. 
   
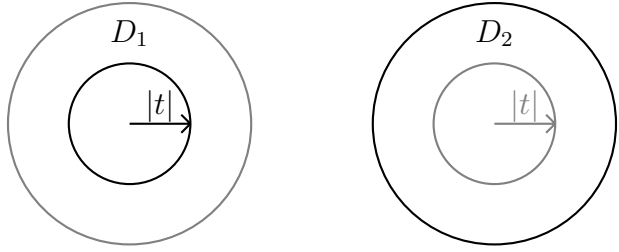
\begin{figure}
\begin{center}
\begin{tikzpicture}[scale=.8]
	\draw[thick,color=black] (3,-2.5) ellipse (2 and 2)(3,-1)node{$D_2$};%
	\draw[thick,color=gray] (3,-2.5) ellipse (1 and 1); %
	\draw[thick,color=gray](3,-2.5)--(4,-2.5)(3.85,-2.35)--(4,-2.5)(3.85,-2.65)--(4,-2.5)(3.5,-2.2)node{$|t|$};
	
	\draw[thick,color=gray] (-3,-2.5) ellipse (2 and 2);%
	\draw[thick,color=black] (-3,-2.5) ellipse (1 and 1); %
	\draw[thick](-3,-2.5)--(-2,-2.5)(-2.15,-2.35)--(-2,-2.5)(-2.15,-2.65)--(-2,-2.5)(-2.5,-2.2)node{$|t|$}(-3,-1)node{$D_1$};	
\end{tikzpicture}
\end{center}
\caption{Two annuli with inner radius $|t|$ and outer radius 1 obtained by removing a disc of radius $|t|$ from $D_1$ and $D_2$ where $t$ is a complex parameter.} \label{Fig.5}
\end{figure}

\newl Consider a Riemann surface $\mathcal{R}_0\in \mathcal{D}\subset \overline{\mathcal{M}}_g$ with $m$ nodes denoted by $\mfk{n}_1,\ldots,\mfk{n}_m$. For the node $\mfk{n}_i$, the punctures $p_i$ and $q_i$ of $\mathcal{R}_0-\{\mfk{n}_1,\ldots,\mfk{n}_m\}$ are paired. Let 
   \begin{equation}
   \begin{aligned}
       U_i^1&\equiv\left\{|w_i^1|<1\right\}, &\qquad i&=1,\ldots, m,
       \\
       \quad U_i^2&\equiv\left\{|w_i^2|<1\right\}, &\qquad i&=1,\ldots, m,
   \end{aligned}
   \end{equation}
   be the disjoint neighborhoods of the punctures $p_i$ and $q_i$, respectively. Here, $w^{(1)}_i$ and $w^{(2)}_i$ with $w_i^{(1)}(p_i)=0$ and $w_i^{(2)}(q_i)=0$ are the local coordinates around the two sides of the node $\mfk{n}_i$. Consider an open set $\mathcal{V}\subset \mathcal{R}_0$ disjoint from the set $U_i^1,U_i^2$ which support the Beltrami differentials $\{\mu_a\}$. Beltrami differentials span the tangent space of the Teichm\"uller space of $\mathcal{R}_0-\{\mfk{n}_1,\ldots,\mfk{n}_m\}$. The dimension of this space is $3g-3-m$. Given 
   \begin{equation}
       s\equiv (s_1,\ldots,s_{3g-3-m})\in \mathbb{C}^{3g-3-m},
   \end{equation}
   for a neighborhood of the origin, the sum  $\mu(s)=\sum_js_j\mu_j$  is a solution $\omega^{\mu(s)}$ of the Beltrami equation. Assume that the surface $\omega^{\mu(s)}(\mathcal{R}_0)=\mathcal{R}_s$ is a quasiconformal deformation of $\mathcal{R}_0$.  Then, we shall parametrize the opening of nodes as follows. The map $\omega^{\mu(s)}$ is conformal on $U_i^1$ and $U_i^2$ and therefore $w_i^{(1)}$ and $w_i^{(2)}$ serve as local coordinates for $\omega^{\mu(s)}(U_i^1),~\omega^{\mu(s)}(U_i^2)\subset \omega^{\mu(s)}(\mathcal{R}_0)$. Given 
   \begin{equation}
       t=(t_1,\ldots,t_m)\in \mathbb{C}^m,\qquad |t_i|<1,
   \end{equation}
   we construct the family of surface $\mathcal{R}_{t,s}$, parametrized by $s$ and $t$ as follows. We first remove the discs $\{0<|w_i^{(1)}|\leq |t_i|\}$ and $\{0<|w^{(2)}_i|\leq |t_i|\}$ from $\mathcal{R}_s$ (see Figure \ref{Fig.5}), and then attach $\{|t_i|<|w^{(1)}_i|< 1\}$ to $\{|t_i|<|w^{(2)}_i|< 1\}$ by identifying $w^{(1)}_i$ and $\frac{t_i}{w^{(2)}_i}$. The construction is complex, and the tuple $(t,s)$ parametrizing $\mathcal{R}_{t,s}$ provides a local complex coordinate chart. 

   \newl The above construction can be easily generalized to surfaces with punctures.   
     
\begin{figure}
\begin{center}
\usetikzlibrary{backgrounds}
\begin{tikzpicture}[scale=.6]

\draw[<->,line width=.25pt](-3,0)--(11,0);
\draw[->,line width=.25pt](4,0)--(4,5);

\draw[line width=1pt] (1,0) .. controls (2,3) and (6,3)  ..(7,0);
\draw[line width=1pt] (-1,0) .. controls (.5,5) and (7.5,5)  ..(9,0);

\draw[line width=1pt, color=gray] (1.25,3)node[above left]{$\pi-l$}--(4,0) -- (6.75,3)node[above right]{$l$};
\end{tikzpicture}
\end{center}

\caption{The fundamental domain of a collar is the region between the annulus and the wedge.}
\label{collar1}
\end{figure}
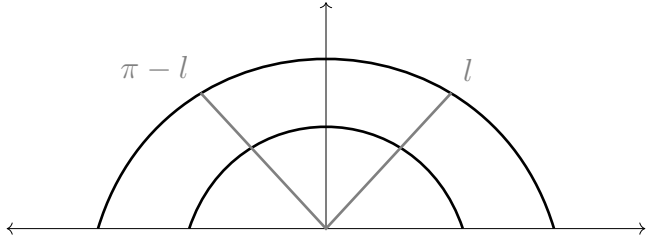

\paragraph{Gluing of Hyperbolic Surfaces:} We now discuss the gluing of Riemann surfaces equipped with a hyperbolic metric. Consider a hyperbolic surface $\mathcal{R}$.  For a geodesic $\alpha$ on $\mathcal{R}$ of length $l_{\alpha}$, a neighborhood with area $2l_{\alpha}\cot\frac{l_{\alpha}}{2}$ is called the collar around $\alpha$ \cite{Keen1974,Keen197410,Matelski197610,ChavelFeldman197812,Randol197910,Wolpert1987a}.  The standard collar around the geodesic $\alpha$  is the  collection of points whose hyperbolic distance from $\alpha$ is less than  $w(\alpha)$, which is defined through
	\begin{equation}
	\sinh w(\alpha)\sinh\frac{l_{\alpha}}{2}=1.
	\end{equation}
The standard collar can be described as a quotient of the upper half-plane $\mathbb{H}$. To describe this quotient space, consider the transformation $z\to e^{l_{\alpha}}z$ which is represented by the following matrix in PSL$(2,\mathbb{R})$
\begin{equation}
		M=\left(\begin{array}{cc}e^{\frac{l_{\alpha}}{2}} & 0 \\0 & e^{-\frac{l_{\alpha}}{2}}\end{array}\right).
\end{equation}
It is clear from this form that it generates a cyclic subgroup of PSL$(2,\mathbb{R})$. We shall denote this subgroup by $\Gamma_{\alpha}$. The fundamental domain is given by a strip in $\mathbb{H}$ (see Figure \ref{collar1}). If we quotient $\mathbb{H}$ with $z\to e^{l_{\alpha}}z$ relation, we identify the two sides of the strip. This gives a cylinder that is topologically  an annulus which has an induced hyperbolic structure from $\mathbb{H}$. The simple closed geodesic of this hyperbolic annulus has hyperbolic length $l_{\alpha}$. The standard collar is the  quotient of the wedge $\left\{\frac{l_{\alpha}}{2}<\mathrm{arg} z<\pi-\frac{l_{\alpha}}{2}\right\}$ by $\Gamma_{\alpha}$ (see Figure \ref{collar1}).

\newl The standard collar or  hyperbolic annulus can be constructed via the gluing of two discs $D_1=\{|y_1|<1\}$ and $D_2=\{|y_2|<1\}$ and by endowing a hyperbolic metric on it.   The  gluing locus 
 \begin{equation}\label{plumbingfiber}
 \mathcal{F}=\left\{y_1y_2=t\,\Big|~|y_1|,|y_2|,|t|<1\right\},
 \end{equation}
 is a complex manifold fibered  over the disk $D=\{|t|<1\}$. The $t=0$ fibre is singular. It is the union of the discs $D_1$ and $D_2$ joined at the origin. To obtain a hyperbolic metric, we need to remove the origin from $D_1$ and $D_2$. Each of the punctured disks has a complete hyperbolic metric given by
 \begin{equation}\label{hypemetricd}   
 ds^2_0=\left(\frac{|dy_1|}{|y_1|\ln|y_1|}\right)^2, \qquad \{0<|y_1|<1\}\cup \{0<|y_2|<1\}.
 \end{equation} 
The  $t\neq 0$ fibre is the annuli $\{|t|<|y_1|<1\}$ with complete hyperbolic metric
\begin{equation}\label{hype}
  ds_t^2=\text{sinc}^{-2}\left(\frac{\pi\ln|y_1|}{\ln|t_j|} \right)ds_0^2,
\end{equation}
 in the $j$\textsuperscript{th} collar of $\mathcal{R}_t$, where sinc is the normalized sinc function given by $\text{sinc}(x)\equiv\frac{\text{sin}(\pi x)}{\pi x}$. For small $|t|$, we have the following expansion of the hyperbolic metric on the punctured disc \cite{Wolpert199002}
 \begin{equation}
ds^2_t= \left(1+\frac{1}{3}\Theta^2+\frac{1}{15}\Theta^4+\ldots\right)ds^2_0, \qquad \{|t|<|y_1|<1\},
 \end{equation} 
  where $\Theta\equiv\pi\frac{\ln |y_1|}{\ln|t|}$. The limit $t\longrightarrow 0$ of $ds_t^2$ is real analytic in $(\ln\frac{1}{|t|})^{-1}$ and furthermore it is uniform for $|y_1|,|y_2|>\epsilon$ \cite{Wolpert199002}. The closed geodesic in the $t$-fiber is $|y_1|=|t|^{\frac{1}{2}}$ and has the length
  \begin{equation}\label{eq:length of the core geodesics}
  l=-\frac{2\pi^2}{\mathrm{ln}|t|}.
  \end{equation}
It is known that  there exists a positive constant $c_*$, known as the {\it collar constant}, such that if the length $l$ of a geodesic $\gamma$ on $\mathcal{R}$ is less than or equal to $c_*$, then the standard collar embeds isometrically about $\gamma$ \cite{Keen1974,Keen197410,Matelski197610,ChavelFeldman197812,Randol197910,Wolpert1987a}. We shall call a geodesic whose length is at most $c_*$ a {\it short geodesics}. Therefore, whenever the length of a simple geodesic along which the cut-and-paste construction can be done becomes less than the collar constant, we can replace the collar around this short geodesic with the hyperbolic annulus. This allows us to find the needed bridge between the gluing and the cut-and-paste construction.

\newl The disc $\mathscr{D}=\{|t|<1\}$ can be thought of as the moduli space for the hyperbolic annulus. The Riemannian Weil-Petersson (WP) metric on $\mathscr{D}$ is given by \cite{Wolpert1987a}
  \begin{equation} \label{WP metric on the moduli space of the Riemann surfaces}
  ds^2_{\text{WP}}=-\frac{2\pi^3|dt|^2}{|t|^2(\mathrm{ln}|t|)^3}.
  \end{equation}
The Fenchel-Nielsen coordinates $(\ell,\tau)$ for the moduli space of hyperbolic annulus as \cite{Wolpert1987a}
\begin{equation} \label{ltrelation}
    \begin{aligned}
    \ell&=-\frac{2\pi^2}{\mathrm{ln}|t|},
    \\
    \frac{2\pi\tau}{\ell}&=\mathrm{arg}~t=\mathrm{Im(ln}~t).
    \end{aligned}
\end{equation}
If we use these relations, then (\ref{WP metric on the moduli space of the Riemann surfaces}) can be written as $d\ell\wedge d\tau$. This perfectly agrees with Wolpert's formula for the WP volume form on the moduli space of hyperbolic surfaces obtained via the WP symplectic form \cite{Wolpert198508}.
 
\newl This identification can be generalized to the case of a hyperbolic surface with $m$ disjoint short geodesics. The collar neighborhood  of each of the short geodesics can be interpreted as a hyperbolic annulus.  Given $\epsilon>0$, we have the following estimate for the length $l_i$ of the simple closed geodesic of the $t_i$\textsuperscript{th} annulus \cite{Wolpert1987a}
   \begin{equation}
   \left|\frac{2\pi^2}{l_i}-\ln\frac{1}{|t_i|}\right|<\epsilon,\qquad i=1,\ldots,m.
   \end{equation}
   Therefore,  we see that the degeneration of a hyperbolic metric is associated to the formation of  wide collars about short geodesics. \par

\newl From the above discussion, it is tempting to claim that a hyperbolic surface near the boundary of the moduli space can be understood as a surface obtained by the gluing of Riemann surfaces with a hyperbolic metric. Putting differently, one would claim that a metric on a surface obtained by gluing two hyperbolic surfaces is again hyperbolic. In the following, we see that this is not quite the case and there is an accumulation of curvature on certain regions of the resulting surface. We then show how this issue can be resolved following \cite{Wolpert199002,ObitsuWolpert200704}.

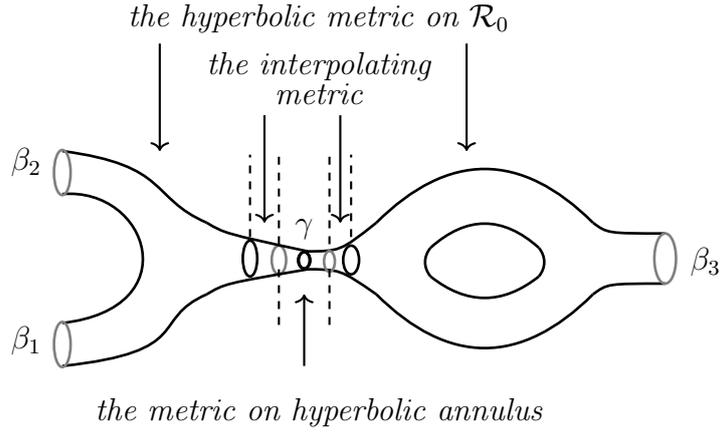
\begin{figure}
\begin{center}
\usetikzlibrary{backgrounds}
\begin{tikzpicture}[scale=.95]
\draw[line width=1pt] (12.65,1) .. controls (14.15,.8) and (13.85,.35)  ..(14.65,0);
\draw[line width=1pt] (12.65,-2) .. controls(14.5,-1.8) and (13.85,-1.35)  ..(14.65,-1);
\draw[line width=1pt] (12.65,.4) .. controls(13.85,.5)  and (14.40,-1.2) ..(12.65,-1.4);
\draw[line width=1pt,color=gray] (12.65,.7) ellipse (.115 and .315);
\draw[line width=1pt,color=gray] (12.65,-1.7) ellipse (.115 and .315);
\draw[line width=1pt,color=gray] (15.65,-0.52) ellipse (.1 and .2);
\draw[line width=1pt] (16.65,-0.52) ellipse (.11 and .2);
\draw[line width=1pt,color=black] (16,-0.525) ellipse (.085 and .115);
\draw[line width=1pt] (14.65,0) .. controls(14.9,-.15) ..(16.05,-0.4);
\draw[line width=1pt] (14.65,-1) .. controls(14.9,-.85)  ..(16.05,-0.65);
\draw[line width=1pt] (17,0) .. controls(18,1) and (19,1) ..(20,0);
\draw[line width=1pt] (17,-1) .. controls(18,-2) and (19,-2) ..(20,-1);
\draw[line width=1pt] (20,0) .. controls(20.2,-.15)  ..(21,-.15);
\draw[line width=1pt] (20,-1) .. controls(20.2,-.85)  ..(21,-.85);
\draw[line width=1pt,color=gray] (21,-.5) ellipse (.15 and .35);
\draw[line width=1pt] (17.75,-.75) .. controls(18.25,-1.15) and (18.75,-1.15) ..(19.25,-.75);
\draw[line width=1pt] (17.75,-.35) .. controls(18.25,0.1) and (18.75,0.1) ..(19.25,-.35);
\draw[line width=1pt] (17.75,-.35) .. controls(17.65,-.475) and (17.65,-.625) ..(17.75,-.75);
\draw[line width=1pt] (19.25,-.35) .. controls(19.35,-.475) and (19.35,-.625) ..(19.25,-.75);
\draw[line width=1pt] (17,0) .. controls(16.5,-.4) ..(16,-0.4);
\draw[line width=1pt] (17,-1) .. controls(16.5,-.65)  ..(16,-0.65);
\draw[line width=1pt,color=gray] (16.35,-0.54) ellipse (.075 and .145);
\draw[line width=1pt] (15.25,-0.5) ellipse (.1 and .25);
\draw[line width=1pt,color=black] (16,-0.525) ellipse (.075 and .115);
\draw[thick,style=dashed](16.35,-0.3) --(16.35,1);
\draw[thick,style=dashed](16.65,-0.2)--(16.65,.95);
\draw[thick,style=dashed](15.65,-0.3) --(15.65,1);
\draw[thick,style=dashed](15.25,-0.2)--(15.25,.95);
\draw[thick,style=dashed](16.35,-0.65) --(16.35,-1.5);
\draw[thick, style=dashed](15.65,-0.65)--(15.65,.-1.5);
\draw (16.2,2.5) node[above]{\it the hyperbolic metric on $\mathcal{R}_0$};
\draw[thick,->](18.25,2.5)--(18.25,1);
\draw[thick,->](14,2.5)--(14,1);
\draw (16.2,1.5) node[above]{\it $\substack{\text{\fontsize{12}{12}\selectfont the interpolating} \\ \text{\fontsize{12}{12}\selectfont metric}}$};
\draw[thick,->](16.5,1.5)--(16.5,0);
\draw[thick,->](15.45,1.5)--(15.45,0);
\draw (16.2,-3) node[above]{\it the metric on hyperbolic annulus};
\draw[thick,->](16,-2)--(16,-1);
\draw (12.15,.5) node[above ] {$\beta_2$}  (12.15,-2) node[above ] {$\beta_1$} (21.2,-.2)node [below right ] {$\beta_3$}  (16,0.2) node  [below ] {$\gamma$};
\end{tikzpicture}
\end{center}

\caption{The grafted metric on the surface obtained via the gluing.}
\label{plumbing2}
\end{figure}

\paragraph{Hyperbolic Metric on the Glued Family:} Consider a hyperbolic surface $\mcal{R}_0$ with $m$ nodes. Using gluing construction, we can construct a family of non-degenerate Riemann surfaces $\mcal{R}_t$ parametrized by $m$-tuple $t\equiv (t_1,\ldots,t_m)$ from the degenerate hyperbolic surface $\mcal{R}_0$. By removing the $m$ nodes from $\mcal{R}_0$, we get a hyperbolic surface $\mcal{\widehat{R}}_0$. On $\widehat{\mcal{R}}_0$, we have a pair of punctures $p_i,q_i$  instead of the node $\mfk{n}_i$ for $i=1,\ldots,m$. The surface $\widehat{\mathcal{R}}_0$ is a finite union of Riemann surfaces with punctures. We denote the local coordinates around the punctures $p_i$ and $q_i$ induced from the hyperbolic metric by $w^{(1)}_i$ and $w^{(2)}_i$ with the property that $w^{(1)}_i(p_i)=0$ and $w^{(2)}_i(q_i)=0$. Let us analyze the metric on $\mcal{R}_t$. In terms of the local coordinates $w^{(1)}_i$ and $w^{(2)}_i$, the hyperbolic metric on $\widehat{\mathcal{R}}_0$ has the following local expression 
 \begin{equation}\label{eq:hyperbolic metric around the punctures}
ds^2=\left(\frac{|d\zeta|}{|\zeta|\ln|\zeta|}\right)^2,\qquad \zeta=w^{(1)}_i,w^{(2)}_i.
 \end{equation}
Using \eqref{eq:hyperbolic metric around the punctures}, it is straightforward to check that there is no geodesic around the punctures. Applying the gluing construction to these punctures identifies the curves $w_i^{(1)}=\sqrt{t}$ and $w_i^{(2)}=\sqrt{t}$ in the neighborhood of the punctures. We can obtain a hyperbolic surface by gluing two hyperbolic surfaces only if the component surfaces glued along a geodesic. However, the curves identified by the gluing construction are not geodesics. Therefore, the resulting surface $\mcal{R}_t$ cannot be a hyperbolic surface. In fact, one can check that the metric on $\mcal{R}_t$ has constant curvature $-1$ everywhere except near the $w_i^{(1)}=w_i^{(2)}=\sqrt{t}$. Therefore, the collar does not have a hyperbolic metric on it. However, we can make the metric on the collar in $\mcal{R}_t$ hyperbolic by performing a conformal transformation. In terms of local coordinates $w^{(1)}_i$ and $w^{(2)}_i$, the resulting metric on the $i$\textsuperscript{th} collar takes the following form 
  \begin{equation}\label{hype1}
 ds_t^2=\text{sinc}^{-2}\left(\frac{\pi\ln|\zeta|}{\ln|t_i|} \right)ds_0^2
		\qquad~ \zeta=w^{(1)}_i,w^{(2)}_i. 
  \end{equation}
As a result of this conformal transformation, the metric on the collars of $\mcal{R}_t$ becomes hyperbolic. The metric away from the collars remains hyperbolic as before the transformation. However, these two metrics do not match with each other in the tails of collars. Therefore, we interpolate between the two choices at the two ends of the collars.  The result is a smooth grafted metric $ds_{\textbf{gf}}^2$ for $\mathcal{R}_t$, see Figure \ref{plumbing2}. The grafted metric has curvature $-1$ everywhere except at the tails of collars on $\mathcal{R}_t$.

\newl Since we have found a smooth metric on the glued surface, we can make it hyperbolic by performing an appropriate conformal transformation. The proper conformal transformation that does this job can be found by solving the so-called {\it curvature-correction equation} \cite{Wolpert199002,ObitsuWolpert200704,MelroseZhu201606}. To describe this equation, consider a compact Riemann surface with the metric $ds^2$ and the Gauss curvature\footnote{In two dimensions, the Gaussian curvature is half of the Ricci curvature of the surface.} $\mathbf{ C}$.  Then, the  metric $e^{2f}ds^2$ on this surface has constant curvature $-1$ if
  \begin{equation}\label{constantcurvatureq}
  Df-e^{2f}=\mathbf{ C},
  \end{equation}
  where $D$ is the Laplace-Beltrami operator on the surface. In order to find the  hyperbolic metric on $\mathcal{R}_t$, we need to specify the grafted metric properly. For this, let us introduce a positive constant $b_*$ and a negative constant $a_0$. The grafted metric $ds_{\textbf{gf}}^2$ on $\mathcal{R}_t$ is defined as follows:
\begin{itemize}
 \item  In the region complement to the region in $\mathcal{R}_t$ described by
    \begin{equation}
   \mathcal{F}_{b_*}=\left\{\left(w_i^{(1)},w_i^{(2)},t_i\right)\left|~w_i^{(1)}w_i^{(2)}=t_i,|w_i^{(1)}|,|w_i^{(2)}|<b_*;~i=1,\ldots,m\right\}\right.,
   \end{equation}
   we use the hyperbolic metric $ds^2$ on $\widehat{\mathcal{R}}_0$ restricted to $\mathcal{R}_{b_*}$. The surface $\mathcal{R}_{b_*}$ is obtained from $\widehat{\mathcal{R}}_0$ by removing the punctured discs $\{0<|w_i^{(1)}|\leq b_*\}$ about $p_i$ and $\{0<|w_i^{(2)}|\leq b_*\}$ about $q_i$ for $i=1,\ldots,m$.

 \item In the region in  $ \mathcal{F}_{b_*}$ that is complement to the  collar bands described by 
    \begin{equation}
 e^{a_0}b_*\leq |w_i^{(1)}|\leq b_*,\quad e^{a_0}b_*\leq |w_i^{(2)}|\leq b_*;~i=1,\ldots,m,
   \end{equation}
   we use the metric $ds_t^2$  on the hyperbolic annulus given in \eqref{hype1}.

 \item In the collar bands $\{e^{a_0}b_*\leq |w_i^{(j)}|\leq b_*\}$ for $j=1,2$, we use the geometric interpolation of the two metrics $ds^2$ and $ds_t^2$ given by 
    \begin{equation}\label{interpmetric}
   ds_{\textbf{gf}}^2=(ds^2)^{1-\eta}(ds_t^2)^{\eta} \quad \mathrm{with} \quad \eta=\eta\left(\ln\left(\frac{|w^{(j)}_i|}{b_*}\right)\right). 
   \end{equation}
   Here, $\eta(a)$ is a smooth function which is one for $a\leq a_0<0$ and zero for $a\geq 0$.
  \end{itemize} 
Now we shall discuss the leading correction to the grafted metric needed for  making it a hyperbolic metric on the glued family \cite{ObitsuWolpert200704}. Let us denote the Fuchsian group associated with the Riemann surface $\mcal{R}$ with $\Gamma$. Assume that  $\Gamma$ has the stabilizer of infinity  $\Gamma_{\infty}$ (i.e. the  group generated by the transformation $z\to z+1$). We can then consider the relative Poincar\'e series 
 \begin{equation}\label{eisenstein}
 E(z;2)=\sum_{A\in \Gamma/\Gamma_{\infty}}(\mathrm{Im}  A (z))^2,
 \end{equation}
 where $z$ is the coordinate on $\mathbb{H}$. The function $(\mathrm{Im} z)^2$ on $\mathbb{H}$ is an eigenfunction of the hyperbolic Laplacian with eigenvalue 2. This seres, known as the {\it Eisenstein series},  converges locally-uniformly on $\mathbb{H}$.  It has the expansion
 \begin{equation}\label{eisentseincusp}
 E(z;2)=(\mathrm{Im} z)^2+\wh{e}(z),
 \end{equation}
 with  $\wh{e}(z)$ bounded as $\mcal{O}((\mathrm{Im} z)^{-1})$ for large values of $\mathrm{Im} z$. The quotient space $\{\mathrm{Im} (z)>1\}/\Gamma_{\infty}$ embeds in $\mathbb{H}/\Gamma$. This region in the upper half-plane that corresponds to a neighborhood of the puncture with hyperbolic area $1$ associated with $\Gamma_{\infty}$ on the hyperbolic surface $\mathcal{R}=\mathbb{H}/\Gamma$  is known as the {\it cusp region} for this puncture in $\mathbb{H}$. Cusp regions for distinct punctures are disjoint.
 
\newl Remember that $\widehat{\mcal{R}}_0$ is a finite union of Riemann surfaces with punctures, and each component has an associated Fuchsian group. Therefore, we can define an Eisenstein series with respect to a puncture. This Eisenstein series lives on the component surface that contains the puncture with respect to which we define the Eisenstein series. It is useful to consider a special truncation of these Eisenstein series for a given choice of $\eta$ and parameters $b_*,a_0$ and $t$.

\newl {\small\bf Truncated Eisenstein Series:} {\it The special truncation  $E^{\#}$ of the Eisenstein series is given by the following modification in the cusp regions
 \begin{itemize}
 \item In the cusp region associated with the puncture that is used for the gluing that produces $\mcal{R}_t$ from $\widehat{\mcal{R}}_0$, and for $\mathrm{Im} z>1$, we define
 \begin{equation}\label{trnceisenstein1}
 \begin{aligned}
     E^{\#}(z;2)&\equiv[1-\eta(-2\pi\mathrm{Im}\,z-\mathrm{ln}~b_*)](\mathrm{Im}\,z)^2
     \\
     &+\left[1-\eta\left(-2\pi\mathrm{Im}\,z+\mathrm{ln}\left(\frac{b_*}{|t|}\right)+a_0\right)\right]\wh{e}(z).
 \end{aligned}
 \end{equation}
\item In the cusp region associated with other punctures, and for $\mathrm{Im} z>1$, we define
 \begin{equation}\label{trnceisenstein2}
 E^{\#}(z;2)\equiv \left[1-\eta(-2\pi\ln\,z+\ln\left(\frac{b_*}{|t|}\right)+a_0)\right]E(z;2).
 \end{equation}
 \end{itemize}}
 
Let us extend these truncated Eisenstein series to define $E^{\dagger}$, the {\it melding of the Eisenstein series}, on $\mcal{R}_t$. For this, we first extend the definition of $E^{\#}$ by zero on the components of $\widehat{\mathcal{R}}_0$ that do not contain the puncture used to define $E$. Then, we define $E^{\dagger}$ on the glued surfaces $\mcal{R}_t$ as follows. Away from the collars in $\mcal{R}_t$, $E^{\dagger}$ is the same as the non-zero $E^{\#}$ in that region. On the $i$\textsuperscript{th} collar of $\mcal{R}_t$ on the overlap $\{|t|/b_*<|w^{(1)}|<b_*\}\cap\{|t|/b_*<|w^{(2)}|<b_*\}$, $E^{\dagger}$ is defined as the sum of $E^{\#}$ at $w^{(1)}_i$ and $E^{\#}$ at $w^{(2)}_i=w^{(1)}_i/t$.

\newl As we mentioned earlier, $(\mathrm{Im}(z))^2$ is the eigenfunction of the hyperbolic Laplacian with eigenvalue $2$. The contribution of the truncation of the Eisenstein series to the hyperbolic metric can be determined by analyzing the quantity $(D_{\bf{gf}}-2)E^{\dagger}$ on the collar in which $D_{\bf{gf}}$ is the Laplacian in the grafted metric. In the complement of the collar, the grafted metric is the hyperbolic metric and $E^\dagger=E$ and $(D_{\bf{gf}}-2)E^{\dagger}$ becomes $(D_{\bf hp}-2)E$ for $D_{\bf hp}\equiv y^2\left(\frac{\partial^2}{\partial x^2}+\frac{\partial^2}{\partial y^2}\right)$ where $z\equiv x+\mathrm{i} y$. By the definition of $E$ given in (\ref{eisenstein}), this quantity is zero. However, this quantity is non-zero on the collar and can be used to determine the contribution to the hyperbolic metric on the collar from the grafted metric defined on the collar band. It can be shown that \cite{ObitsuWolpert200704}:
		\begin{equation}\label{grafted Laplacian acts on the melting}
		(D_{\bf{gf}}-2)E^{\dagger}(z)=-\frac{1}{4\pi}\Lambda +\mathcal{O}\left(\left(-\ln|t|\right)^{-1}\right),
\end{equation}
where
\begin{equation}
    \Lambda\equiv\frac{\partial}{\partial a}\left(a^4\frac{\partial}{\partial a}\eta(a)\right), \qquad a\equiv\ln|w^{(1)}|,\ln|w^{(2)}|.
\end{equation}
Finally, we can use this result to find the relation between the hyperbolic metric on the glued family and the grafted metric. Assume that $t=(t_1,\ldots,t_m)\in\mathbb{C}^{m}$ and $|t_i|<b_*^4,~i=1,\ldots,m$ are the gluing parameters for a surface $\mathcal{R}_t$. To find the degenerate expansion of the hyperbolic metric on $\mathcal{R}_t$, one should use the curvature-correction equation (\ref{constantcurvatureq}). Using this approach, the hyperbolic metric on the surface $\mathcal{R}_t$ has the following expansion \cite{Wolpert199002,ObitsuWolpert200704}:
		\begin{equation}
		ds_{\bf hp}^2=ds_{\textbf{gf}}^2\left(1+2(D_{\bf{gf}}-2)^{-1}(1+{\bf C}_{\mathrm{graft}})+\mathcal{O}\left(\parallel 1+{\bf C}_{\mathrm{graft}}\parallel\right)^2\right) \label{hyperbolicintermsgaussiancurvature},
		\end{equation}
in which $\parallel\cdot\parallel$ is an appropriate norm and ${\bf C}_{\mathrm{graft}}$ is the Gaussian curvature of the grafted metric given by \cite{Wolpert199002}:
		\begin{equation}
		{\bf C}_{\text{graft}}=-1-\frac{\epsilon^2}{6}\Lambda+\mathcal{O}(\epsilon^4),\qquad \epsilon\equiv\frac{\pi}{\ln|t|}.
		\end{equation}
		$\Lambda$ is given in (\ref{grafted Laplacian acts on the melting}). Using this relation, the hyperbolic metric on $\mathcal{R}_t$ can be expanded in terms of the grafted metric \cite[Theorem $4$]{ObitsuWolpert200704} as follows

\newl {\small\bf The Expansion of Hyperbolic Metric on $\mathcal{R}_{t}$}: {\it Given a choice of $b_*<1$ and a cut-off function $\eta$, for all small $t$, the hyperbolic metric $ds^2_{\bf hp}$ of the  Riemann surface $\mathcal{R}_{t}$, obtained by the gluing of the $m$ pairs of annuli, has the following expansion}
		\begin{equation}\label{graftedhyp}
		ds_{\bf hp}^2=ds_{\bf gf}^2\left\{1+\frac{4\pi^4}{3}\sum_{i=1}^m(\ln|t_i|)^{-2}\left(E^{\dagger}_{i,1}+E^{\dagger}_{i,2}\right)+\sum_{i=1}^m\mathcal{O}\left((\ln|t_i|)^{-3}\right)\right\}.
		\end{equation}
		{\it The functions $E^{\dagger}_{i,1}$ and $E^{\dagger}_{i,2}$ are the melding of the Eisenstein series $E(\cdot;2)$ associated to the pair of cusps plumbed to form the $i$\textsuperscript{th} collar.}
		
\newl This expansion for the hyperbolic metric on $\mathcal{R}_{t}$ can be written in terms of the $i$\textsuperscript{th} collar geodesic  $l_{i}=-\frac{2\pi^2}{\ln|t_i|}$ computed in $ds_t^2$ metric:
		\begin{equation}\label{graftedhyp2}
		ds_{\bf hp}^2=ds_{\bf gf}^2\left(1+\sum_{i=1}^m\frac{l_i^{2}}{3}\left(E^{\dagger}_{i,1}+E^{\dagger}_{i,2}\right)+\sum_{i=1}^m\mathcal{O}\left(l_i^3\right)\right). 
		\end{equation}
		We can use this expansion to compute the lengths of an arbitrary simple closed geodesics on $\mcal{R}_t$ as follows
		\begin{itemize}
			\item
			The hyperbolic length of the geodesic in the $i$\textsuperscript{th} collar  is given by
			\begin{equation}\label{collargeodesic}
			l_i^{(\bf{hp})}=l_i+\mathcal{O}\left(l_i^3\right). 
			\end{equation}
			Note that there is no correction of order $l_i^2$ to the length of the $i${\textsuperscript{th}} collar geodesic.
			
			\item The  length of a simple closed geodesic $\alpha$,  disjoint from the collars, is given by
			\begin{equation}
			\widetilde{l}_{\alpha}=l_{\alpha}+\sum_{i=1}^m\frac{l_i^2}{6}\bigintssss_{\alpha}ds\,\left(E^\dagger_{i,1}+E^\dagger_{i,2}\right)+\sum_{i=1}^m\mathcal{O}\left(l_i^3\right).
			\end{equation} 
			In this formula, $\widetilde{l}_{\alpha}$ and $l_\alpha$ denote the hyperbolic lengths of $\alpha$ on $\mcal{R}_t$ and $\mcal{R}_0$, respectively. Away from the collars, $E_{i,1}^\dagger(z,2)=E_{i,1}(z,2)$, so we can write:
			\begin{equation}
			\widetilde{l}_{\alpha}=l_{\alpha}+\sum_{i=1}^m\frac{l_i^2}{6}\bigintssss_{\alpha}ds\,\left(E_{i,1}+E_{i,2}\right)+\sum_{i=1}^m\mathcal{O}\left(l_i^3\right).  \label{lengthawayc}
			\end{equation}
	\end{itemize}

\subsection{Gluing-Compatibility and Hyperbolic Geometry}\label{sec:gluing-compatibility and hyperbolic geometry}
Before specifying a gluing-compatible section, let us first define the 1PI and 1PR regions of the moduli space $\mathcal{M}_{g,n}$:

\newl{\small\bf 1PI and 1PR Regions of the Moduli Space:} {\it the region inside the moduli space consisting of hyperbolic surfaces having no simple closed geodesics with length less than or equal to the collar constant $c_*$ is declared as the 1PI region of the moduli space. The region complement to the 1PI region is declared as the 1PR region.}

\newl Here, we assume that $c_*$ is a small constant that is much less than the hyperbolic length of any simple closed geodesic on the surface  that is  homotopically-inequivalent to the simple closed geodesic on the collars. The rationale behind this assumption will be explained below. With these definitions of 1PI and 1PR regions of the moduli space, {\it we choose the local coordinates as the ones that are induced from the hyperbolic metric on the surface with the scaling factor of $e^{\frac{\pi^2}{c_*}}$}. Let us explain the need for the scaling factor. The 1PR region of the moduli space should be constructed using the gluing in which the gluing parameter runs between $0$ and $1$. On the other hand, it is clear from  \eqref{eq:length of the core geodesics} that our definition of the 1PR region of the moduli space consists of all Riemann surfaces satisfying
\begin{equation}
\left\{w_i^{(1)}w_i^{(2)}=t_i~|~|t_i|\leq |w_i^{(1)}|,|w_i^{(2)}|<1,~|t_i|<e^{-\frac{2\pi^2}{c_*}},~i=1,\ldots,m\right\}.
\end{equation}
However we can make the gluing parameter to run from $0$ to $1$ by rescaling the local coordinates $w_i^{(1)}$ and $w_i^{(2)}$ as follows
\begin{equation}\label{rescaled coordinate}
\widetilde{w}_i^{(1)}\equiv e^{\frac{\pi^2}{c_*}}w_i^{(1)}\qquad \widetilde{ w}_i^{(2)}\equiv e^{\frac{\pi^2}{c_*}}w_i^{(2)}, \qquad i=1,\ldots,m.
\end{equation}
In terms of these rescaled coordinates, the family of surfaces in the 1PR region is constructed by the following gluing relation
\begin{equation}
\left\{\widetilde{w}_i^{(1)}\widetilde{w}_i^{(2)}=\widetilde{t}_i~|~|\widetilde {t}_i|\leq|\widetilde{w}^{(1)}_i|,|\widetilde{ w}^{(2)}_i|<e^{\frac{\pi^2}{c_*}},~|\widetilde{t}_i|<1,~i=1,\ldots,m\right\}.
\end{equation}

\newl Note that this definition does not fix the phases of local coordinates. Such a phase ambiguity will not affect the definition of off-shell amplitudes due to the condition \eqref{eq:conditions on off-shell states b_0 L_0} satisfied by the off-shell states. This choice provides a gluing-compatible section of $\pcab$ only if the gluing construction applied to one or more hyperbolic surfaces gives a hyperbolic metric on the resulting surface. As we have explained above, this can be done by performing an appropriate conformal transformation. 

\newl We are in a position to check the validity of the proposed choice of local coordinates. We proposed that the local coordinate around a puncture to be $\widetilde{w}=\exp{\left(\frac{\pi^2}{c_*}\right)}w$,  where $w$ is induced from the hyperbolic metric on the surface. In terms of the local coordinate $w$, the hyperbolic metric around the puncture takes the form $\left(\frac{|dw|}{|w|\ln|w|}\right)^2$. One of the main features of a gluing-compatible choice of local coordinates is that each component of the surface in the 1PR region is independent of the gluing parameters and moduli parameters of other components. Equation \eqref{graftedhyp2} shows that the proposed choice of local coordinates does not satisfy this criterion and as such does not provide a gluing-compatible section of $\pcab$. Let us elaborate on this point. By construction, the local coordinates induced from the grafted metric $ds_{\bf gf}^2$ depend only on the moduli parameters associated with the component surfaces to which the punctures belong. To obtain a hyperbolic metric $ds_{\bf hp}^2$ from the grafted metric, we should multiply with a Weyl factor that depends on the gluing parameters. The net effect is that the local coordinates induced from the hyperbolic metric will depend on the gluing parameters, and hence they will not be a gluing-compatible choice of local coordinates. It is thus clear that we can obtain a choice of local coordinates that do not depend on the gluing parameters by multiplying the hyperbolic metric with the inverse of the Weyl factor obtained by solving the curvature-correction equation \eqref{constantcurvatureq}, the factor that is multiplied with $ds_{\bf gf}^2$ in \eqref{graftedhyp2}. 

\newl Based on this observation, we can modify the proposed choice of local coordinates and make it gluing-compatible as long as we keep the parameter $c_*$ very small, which can be explained as follows. Exactly at the degeneration locus, the local coordinates on the glued surface match with local coordinates induced from the hyperbolic metric. Otherwise, the difference between them is of order $l_i^2$, where $l_i$ is the length of the simple closed geodesic on the $i$\textsuperscript{th} collar. When the length $l_i$ is very small compared to the lengths of other simple closed geodesics on the surface, we can neglect the higher-order corrections. For a gluing-compatible section of $\pcab$, it is important that the choice of local coordinates in the 1PR region matches the local coordinates induced from the component surfaces. Remember that we defined the 1PR region by considering hyperbolic surfaces that have simple closed geodesics less than $c_*$. By demanding $c_*$ to be very small, we are making sure that the difference between the coordinates induced from the hyperbolic metric and the ones induced from the metrics on component surfaces through gluing is very small. In other words, we can use \eqref{graftedhyp2} and ignore all higher-order corrections beyond the order $l_i^2$. Moreover, it is clear from \eqref{collargeodesic} that for small $c_*$, the definition of 1PI and 1PR regions given in section \ref{sec:gluing-compatibility and hyperbolic geometry} remains the same.

\newl Putting together everything we have explained so far, we shall now describe a choice of local coordinates that is gluing-compatible. As in the case of naive local coordinates induced from the hyperbolic metric, this choice of local coordinates is unique up to a phase ambiguity. To specify the choice of local coordinates, let us introduce another infinitesimal parameter $\epsilon$. In order to define local coordinates in the 1PI region, we divide it into subregions. Let us denote the subregion in the 1PI region consists of surfaces with $m$ simple closed geodesics of length between $c_*$ and $(1+\epsilon)c_*$ by $\mathbf{R}_m$. For surfaces belong to the subregion $\mathbf{R}_0$, we choose the local coordinate around the $i$\textsuperscript{th} puncture to be $\exp{\left(\frac{\pi^2}{c_*}\right)}w_i$. In terms of $w_i$, the hyperbolic metric in the neighborhood of the puncture takes the following form
\begin{equation}
ds^2_{\bf hp}=\left(\frac{|dw_i|}{|w_i|\ln|w_i|}\right)^2, \qquad  i=1,\ldots,n. 
\end{equation}
For surfaces belong to the region $\mathbf{R}_{m}$ with $m\ne 0$, we choose the local coordinates around the $i$\textsuperscript{th} puncture to be  $\exp{\left(\frac{\pi^2}{c_*}\right)}\widetilde w_{i,m}$, where $\widetilde w_{i,m}$ is obtained by solving the following equation
\begin{equation}\label{eq:metric on the 1PI and near its boundary}
\left(\frac{|d\widetilde{w}_{i,m}|}{|\widetilde{w}_{i,m}|\mathrm{ln}|\widetilde{w}_{i,m}|}\right)^2=\left(\frac{|dw_i|}{|w_i|\mathrm{ln}|w_i|}\right)^2\left(1-\sum_{j=1}^mf(l_j)E^0_j(w_i)\right).
\end{equation}
$f(x)$ is a smooth function of $x$ such that $f(c_*)=\frac{c_*^2}{3}$ and $f((1+\epsilon)c_*)=0$, $E^0_{j}(w_i)$ is the leading term in the Eisenstein series defined with respect to the $j$\textsuperscript{th} node on the surface. For more explanation about the expansion of the Eisenstein series around a puncture, see appendix of \cite{Pius201808}. 

\newl Finally, the 1PR region consists of surfaces with $m$ simple closed geodesics of length  $0\le l_j<c_* ,~j=1,\ldots,m$. We choose the local  coordinates around the $i$\textsuperscript{th} puncture to be $\exp^{\left(\frac{\pi^2}{c_*}\right)}\widehat{w}_{i,m}$ where $\widehat{w}_{i,m}$ is obtained by solving the following equation
\begin{equation}\label{eq:metric in the 1PR region}
\left(\frac{|d\widehat{w}_{i,m}|}{|\widehat{w}_{i,m}|\mathrm{ln}|\widehat{w}_{i,m}|}\right)^2=\left(\frac{|dw_i|}{|w_i|\mathrm{ln}|w_i|}\right)^2\left(1-\sum_{j=1}^m\frac{l_j^2}{3}E^0_j(w_i)\right).
\end{equation} 
Note that different choices of $f$ and different values for the parameters $c_*$ and  $\epsilon$ give different choices of gluing-compatible local coordinates. On the other hand, the condition for the physically-relevant quantities to be independent of these choices is the condition of gluing-compatibility \cite{PiusRudraSen201311,PiusRudraSen201401,Sen201408,SenWitten201504}. Therefore, the function $f$ and the arbitrary parameters $c_*$ and $\epsilon$ will not appear in the measurable quantities as long as  we restrict our computations to the second order in $c_*$. 

\newl We can  generalize the procedure to higher order in $c_*$, say to the $n$\textsuperscript{th} order in $c_*$.  {\it The idea behind choosing the  local coordinates around the punctures is that it should be the same as the coordinate induced from the grafting metric inside the 1PR region of the moduli space. }  Therefore, we need to  find the explicit solution for the curvature-correction equation \eqref{graftedhyp2} to $n$\textsuperscript{th} order in $c_*$. This will relate the hyperbolic metric with the grafting metric that is valid up to $n$\textsuperscript{th} order in $c_*$. The length of the core geodesic can also receive correction beyond the second order. Therefore, beyond the second order, we need to correct both the definition of the 1PR region and the choice of local coordinates to make it gluing compatible. The algorithm for finding a gluing compatible choice of local coordinate around the punctures up to $c_*^n,~n\ge 3$ is as follows:
  
		\begin{enumerate}
			\item[(1)] We first solve the curvature-corrections equation (\ref{constantcurvatureq}) to $n$\textsuperscript{th} order in $c_*$. Using this solution, we can find the correction to the hyperbolic metric $ds_{\bf hp}^2$ up to the order $n$ from (\ref{hyperbolicintermsgaussiancurvature}).
			
			\item[(2)]  We then compute the correction to the lengths of the core geodesics in collars using the hyperbolic metric $ds_{\bf hp}^2$  expressed in terms of the grafted metric.
			
			\item[(3)] We next define the 1PI region inside the moduli space as the region consists of hyperbolic surfaces with no simple geodesics of length less than $C_*$. $C_*$ is the length of the geodesic on the collars computed using $ds_{\bf hp}^2$ on the family of hyperbolic surfaces obtained with gluing parameters $|t_i|=\exp{\left(-\frac{2\pi^2}{c_*}\right)}$. The 1PR region is the complement of the 1PI region. 
			
			\item[(4)] The next step is to define the local coordinates on surfaces inside the 1PR regions. We choose it to be $\exp{\left(\frac{\pi^2}{c_*}\right)}\widehat{w}_{i,m}^{(n)},~i=1,\ldots,n$, where $\wh{w}_{i,m}^{(n)}$ is the natural local coordinate around the punctures induced from the grafted metric, represented in terms of the hyperbolic metric.
			
			\item[(5)] In order to define the local coordinates around the punctures on surfaces inside the 1PI region,  we divide the 1PI region into subregions. Let us denote the subregion in the 1PI region consists of surfaces with $m$ simple closed geodesics of length between $C_*$ and $(1+\epsilon)C_*$ by $\mathbf{R}^{(n)}_m$. For the surfaces belong to the subregion $\mathbf{R}^{(n)}_0$, we choose the local coordinate around the $i$\textsuperscript{th} puncture to be $\exp{\left(\frac{\pi^2}{c_{*}}\right)}w_i$. For the surfaces belong to the subregion $\mathbf{R}^{(n)}_m$ for $m\ne 0$, we choose the local coordinate around the $i$\textsuperscript{th} puncture to be $\exp{\left(\frac{\pi^2}{c_{*}}\right)}\widetilde{w}_{i,m}^{(n)}$. $\widetilde{w}_{i,m}^{(n)}$ is induced from the metric interpolating between the grafted and hyperbolic metrics.

		\end{enumerate}
		
		   Note again that choosing different interpolating functions $f$ and various values for the parameter $c_*$ and  $\epsilon$ give different choices of gluing compatible local coordinates. As such, they will not appear in the measurable quantities as long as we restrict them to the same order in $c_*$ up to which we have solved the curvature-correction equation.

\subsection{Integrating a Bosonic-String Measure over the Moduli Space}

To determine an off-shell bosonic-string amplitude, we need to construct a gluing-compatible section of $\pcab$. This section then needs to be integrated over the relevant moduli space to determine the corresponding off-shell bosonic-string amplitude. In the previous section, we determined a gluing-compatible choice of local coordinates using hyperbolic geometry. In this section, we explain how one can integrate this gluing-compatible section of $\pcab$ over $\overline{\mathcal{M}}_{g,n}$, considered as the moduli space of hyperbolic surfaces, using (an extension of) the method developed by Mirzakhani \cite{Mirzakhani200603}.

\subsubsection{Integration Measure on a Section of $\widehat{\mathcal{P}}_{g,n}$} 

The construction of string measure on a gluing compatible section of $\widehat{\mathcal{P}}_{g,n}$ requires choosing a parametrization of the moduli space of hyperbolic surfaces and computing the Beltrami differentials associated with them, as explained in Section \ref{sec:offshell bosonic-string amplitudes}. The Fenchel-Nielsen parametrization provides a convenient choice of global coordinates for the Teichm\"uller space of hyperbolic surfaces \cite{FenchelNielsen2002,ImayoshiTaniguchi1992,Hubbard2006}. Although this parameterization  is not well suited for explicitly describing the moduli space, as explained in  Section \ref{integration over ms}, the integration over the moduli space can be efficiently performed using these coordinates. Therefore, we choose to work with them. Below, we shall discuss the construction of Beltrami differentials associated with the Fenchel-Nielsen coordinates and describe the explicit procedure for writing the integration measure for the off-shell bosonic-string theory.

\newl The basic idea behind the Fenchel-Nielsen parametrization is that every hyperbolic metric on a Riemann surface can be obtained by piecing together the metric from simple subdomains. The basic building block of a compact hyperbolic surface is  a  pair of pants. This construction is based on the following observations. 
\begin{itemize}
	
	\item[1)]  The lengths $\lambda_1,\lambda_2$ and $\lambda_3$ of alternating sides of a right hexagon in the hyperbolic plane have any value between $(0,\infty)$.  If the hexagon is doubled across the unlabeled sides, then we obtain a pair of pants $P$ with geodesic boundary lengths $l_j=2\lambda_j,~j=1,2,3$. 
	
	\item[2)]  Given a pair of pants $P$ with boundaries $\alpha_j,~j=1,2,3$ and a pair of pants $Q$ with boundaries $\beta_j,~j=1,2,3$, $P$ can be glued to $Q$ along the boundaries $\alpha_1$ and $\beta_1$ if both of them have the same hyperbolic length. The resulting surface is a geometric sum, that has a hyperbolic metric whose restriction to each component pair of pants is the original hyperbolic metric of that component. The proof of this claim is based on the observation that the geometry in a tubular neighborhood of a simple closed geodesic of length $l$ is completely determined by $l$ \cite{Wolpert1987a}.
\end{itemize}

By gluing $2g-2+n$ pairs of pants, a compact surface of genus $g$  and $n$ boundaries can be constructed. Assume that the boundary components of the surface are curves with lengths $L_i,~i=1,\ldots,n$. When all $L_i=0,~i=1,\ldots,n$, we obtain a Riemann surface with $n$ punctures. By varying the parameters of this construction, every compact hyperbolic metric can be obtained. At each gluing site, there are two parameters. At the gluing site where we glue the boundary $\alpha_1$ of the pair of pants $P$ with boundary $\beta_1$ of the pair of pants $Q$, these parameters are the  length $l(\alpha_1)=l(\beta_1)=\ell$ of the boundaries $\alpha_1$ and $\beta_1$ and the twist parameter $\tau$.  The twist parameter is defined as follows. Let $p_1$ be a point on boundary $\alpha_1$ and point $q_1$ be a point on boundary $\beta_1$. Compared to other points on $\alpha_1$, the point $p_1$ is assumed to have the minimum hyperbolic distance from the boundary $\alpha_2$. Similarly,  point $q_1$ has the minimum hyperbolic distance from the boundary $\beta_2$. The twist parameter $\tau$ is the distance between $p_1$ and $q_1$ along $\alpha_1\sim\beta_1$.  Then the parameters $(\tau_j,\ell_j),~1\leq j\leq 3g-3+n,~\tau_j\in \mathbb{R},~l_j\in \mathbb{R}^+$ for a fixed pair of pants decomposition $\mathcal{P}$ endows the Teichm\"uller space $\mathcal{T}_{g,n}$ of the genus-$g$ Riemann surfaces with $n$ boundary components with a global real-analytic coordinates. There is a natural symplectic form known as the Weil-Petersson (WP) symplectic form whose explicit form in terms of the Fenchel-Nielsen coordinates is \cite{Wolpert198508}:
\begin{equation}\label{WolpertMagic}
	\omega_{\text{WP}}=\sum_{i=1}^{3g-3+n}d\ell_j\wedge d\tau_j. 
\end{equation}
Consider the Fenchel-Nielsen  coordinates $(\tau_i,\ell_i),~i=1,\ldots,3g-3+n$  defined with respect to a pants decomposition of a hyperbolic surface  $\mathcal{R}$  using $3g-3+n$  curves $\{C_i\},~i=1,\ldots,3g-3+n$, where $C_i$ denotes a simple closed geodesic on the  surface. For $i\neq j$, the curves $C_i$ and $C_j$ are disjoint and non-homotopic.  The tangent space at a point in  the Teichm\"uller space is spanned by the Fenchel-Nielsen vector fields $\left\{\frac{\partial}{\partial \tau_i},\frac{\partial }{\partial \ell_i}\right\},~i=1,\ldots,3g-3+n$. $\frac{\partial}{\partial \tau_i}$ is  the  twist vector field $t_{C_i}$ associated with the curve $C_i$.  The twist field $t_{\alpha}$, for a simple closed geodesic $\alpha$,  generates a flow on  $\mathcal{T}_{g,n}$. This flow can be understood as the operation of  cutting the hyperbolic surface along $\alpha$ and attaching the boundaries after rotating one boundary relative to the other by some amount $\delta$. The magnitude $\delta$ parametrizes the flow on $\mathcal{T}_{g,n}$. 

\newl The universal cover of a hyperbolic surface $\mathcal{R}$ is the upper half-plane $\mathbb{H}$ endowed with the hyperbolic metric. Assume that the uniformizing group is a finitely-generated Fuchsian group $\Gamma$ and suppose that the simple closed geodesic $\alpha$  corresponds to the element 
$$A=\left(\begin{array}{cc}a & b \\c & d\end{array}\right),$$  in $ \Gamma$. Let $\langle A\rangle$ denote the infinite cyclic group generated by $A$. Then, the  Beltrami differential corresponding to the twist vector field $t_{\alpha}$ is given by \cite{Wolpert198203}  
\begin{equation}\label{beltramitwist}
	\mathbf{t }_{\alpha}=\frac{\mathrm{i}}{\pi}(\mathrm{Im}z)^2\overline{ \Theta}_{\alpha},
\end{equation}
where $\Theta_{\alpha}$ is the relative Poincar\'e series 
\begin{equation}\label{poicares}
	\begin{gathered}
	    \Theta_{\alpha}\equiv\sum_{B\in \langle A\rangle \backslash  \Gamma}\msf{W}_{B^{-1}AB},
        \\
        \msf{W}_A\equiv\frac{(a+d)^2-4}{\Big(cz^2+(d-a)z-b\Big)^2}.
	\end{gathered}
\end{equation}
$z$ is the coordinate on the upper half-plane. Therefore, we conclude that {\it the Beltrami differential for the Fenchel-Nielsen  vector field $\frac{\partial}{\partial \tau_i}$ is given by $\mathbf{t}_{C_i}$.} The Beltrami differential for the Fenchel-Nielsen  vector field $\frac{\partial}{\partial l_i}$ can also be constructed by noting the that $\frac{\partial}{\partial l_i}$ is dual to the twist vector field $\frac{\partial}{\partial \tau_i}$  with respect to the WP symplectic form \cite{Wolpert198302}. We denote the  Beltrami differential for the Fenchel-Nielsen  coordinate vector field $\frac{\partial}{\partial l_i}$ as $\mathbf{ l}_{C_i}$.

\newl Putting these facts together, the off-shell bosonic-string measure is given by
\begin{equation}\label{eq:the bosonic-string measure}
	\Omega_{6g-6+2n}^{(g,n)}(|\Phi\rangle)=(2\pi \mathrm{i})^{-(3g-3+n)}\langle\mathcal{R}|B_{6g-6+2n}|\Phi\rangle,
\end{equation}
where
\begin{equation}\label{eq:the expression for B}
	B_{6g-6+2n}\left[\frac{\partial}{\partial \ell_1},\frac{\partial}{\partial \tau_1},\ldots,\frac{\partial}{\partial \ell_{3g-3+n}},\frac{\partial}{\partial \tau_{3g-3+n}}\right]=\prod_{i=1}^{3g-3+n}b(\mathbf{t}_{C_i})b(\mathbf{l}_{ C_i}),
\end{equation}
and 
\begin{alignat}{1}\label{eq:the expressions for b(v)}
	b(\mathbf{t}_{C_i})&=\int_{\mathcal{F}} d^2z\left(b_{zz}\mathbf{t}_{C_i}+b_{\bar z\bar z}\overline{\mathbf{t}}_{C_i}\right),\nonumber
	\\
	b(\mathbf{l}_{C_i})&=\int_{\mathcal{F}} d^2z\left(b_{zz}\mathbf{l}_{C_i}+b_{\bar z\bar z}\overline{\mathbf{l}}_{C_i}\right).
\end{alignat}
Here $\mathcal{F}$ denotes the fundamental domain in the upper half-plane for the action of the Fuchsian group $\Gamma$ associated with the Riemann surface $\mathcal{R}$.  The state $\langle\mathcal{R}|$ is the surface state associated with $\mathcal{R}$ and the state  $|\Phi\rangle=|\Psi_1\rangle\otimes\ldots\otimes |\Psi_n\rangle\in\mathcal{H}^{\otimes n}$ represents the tensor product of the off-shell states inserted at the punctures. The quantity  $\langle\mathcal{R}|B_{6g-6+2n}|\Phi\rangle$ describe the $n$-point correlation function on $\mathcal{R}$ with the off-shell vertex operators for the states $|\Psi_i\rangle, i=1,\ldots,n$, inserted at the punctures using the gluing compatible choice of local coordinates  around the punctures described in Section \ref{sec:gluing-compatibility and hyperbolic geometry}.

\subsubsection{Integration over the Moduli Space}\label{integration over ms}
In order to obtain off-shell bosonic-string amplitudes, we need to integrate the off-shell bosonic-string measure over $\overline{\mathcal{M}}_{g,n}$, the compactified moduli space  of genus $g$  Riemann surface with $n$ punctures. $\mathcal{M}_{g,n}$  is the quotient of the Teichm\"uller space  $\mathcal{T}_{g,n}$ with respect to the action of the mapping class group (MCG) $\text{Mod}_{g,n}$. Although the Fenchel-Nielsen parametrization provides a simple description of the Teichm\"uller space of a hyperbolic surface, they fail to provide a convenient description of the corresponding moduli space. The reason is that the form of the action of  the mapping class group on  Fenchel-Nielsen coordinates and the explicit description of moduli space are not known  explicitly in general  \cite{HatcherThurston198003,Hatcher199906}.   As a result, integrating a mapping class group invariant function (or more precisely a mapping-class-group-invariant differential form)\footnote{Note that by construction, well-defined objects on the moduli space are those on the corresponding Teichm\"uller space that are invariant under the action of the mapping-class group.} over the moduli space of hyperbolic surfaces is not a  straightforward task. In this section, we shall discuss a way to bypass this difficulty using an extension of the method introduced by Mirzakhani \cite{Mirzakhani200610}.

\newl The basic idea of \cite{Mirzakhani200610} is as follows.  The integration over the moduli space can be lifted to an appropriate covering space, and the pull-back of the WP volume form defines a volume form in the covering space. The most important feature of  this covering space is that it can be decomposed  into a product of lower-dimensional moduli spaces  and a set of two-dimensional  infinite cones. These infinite cones are parametrized by some pairs $(l_i,\tau_i)$ of Fenchel-Nielsen coordinates. We can again lift the integration over the lower dimensional moduli spaces to obtain another set of lower dimensional moduli spaces and infinite cones. In the end, we will end up with integration over a set of infinite cones with an explicit domain of integration. 

\newl Let us explain how one can unwind an integral and lift it to an appropriate covering space. Consider the space $M$ and a covering space $N$ thereof given by $\pi: N\longrightarrow M$. If $dv_{M}$ is a volume form on $M$, then $dv_N\equiv\pi^{-1}*(dv_M)$ defines a volume form on $N$.  Suppose that $f$ is a smooth function on $N$, then the push-forward $\pi_*f$ of $f$ to $M$ using the covering map $\pi$ is
\begin{equation}\label{coveri}
	(\pi_*f)(x)\equiv\sum_{y\in \pi^{-1}\{x\}}f(y).
\end{equation}
Here, we used the fact that the value of the push-forward of a function $f$ at a point $x$ of $M$ can be obtained by the summation over the values of the function $f$ at all points in the fiber of the point $x$ in $N$. This relation defines a smooth function on $M$. Then, the integral of this push-forward function on $M$ can be lifted to $N$ using
\begin{equation}\label{covint}
	\bigintssss_{M}dv_M~(\pi_*f)~=\bigintssss_{N}dv_N~f.
\end{equation}
Therefore, if we have a covering of the moduli space that has a well-defined region in the Teichm\"uller space, then  the integration of a function on the moduli space can be performed by expressing this function as a push-forward of a function in the covering space. In the remaining part of this section, we shall explain that it is indeed possible to find such a covering of the moduli space and express the MCG-invariant functions as a push-forward of a function defined in the covering space.

\paragraph{Relevant Covering Spaces of the Moduli Space:} To keep the discussion general, we consider bordered hyperbolic surfaces, which have boundaries of non-zero lengths. We denote a genus-$g$ hyperbolic surface with $n$ boundaries satisfying $2g+n\ge 3$ by $\mathcal{R}_{g,n}(\mbf{L})$ where $\mathbf{L}\equiv\{L_1,\ldots,L_n\}$ denotes the collection of the lengths of boundaries. The corresponding punctured surface is obtained by sending the lengths of all boundaries to zero. 

\newl The moduli space of $\mathcal{R}_{g,n}(\mbf{L})$, which we denote by $\mathcal{M}_{g,n}(\mathbf{L})$, can be obtained from its Teichm\"uller space, denoted as $\mathcal{T}_{g,n}(\mathbf{L})$, by identifying different points which are related by the action of all mapping class group elements, which is, in general, a complicated action. However, the following covering spaces of the moduli space have a comparatively nicer structure. Assume that $\gamma$ is a simple closed curve on $\mathcal{R}_{g,n}(\mathbf{L})$. Consider a space  $\mathcal{M}_{g,n}(\mathbf{L})^{\gamma}$ which is obtained from $\mathcal{T}_{g,n}(\mathbf{L})$ by identifying different points that are related by the subgroup of the mapping class group that keeps $\gamma$ fixed. The space $\mathcal{M}_{g,n}(\mathbf{L})^{\gamma}$ is larger than $\mathcal{M}_{g,n}(\mathbf{L})$ because it is obtained by taking  a quotient  of $\mathcal{T}_{g,n}(\mathbf{L})$ by a smaller group. The most important feature of this space is that it can be thought of as a product of lower dimensional moduli spaces and infinite cones \cite{Mirzakhani200610} where the infinite cones correspond to  twists about the curve $\gamma$ with values between $0$ and the length $\ell_{\gamma}$  of the curve $\gamma$.\footnote{As we explain below, there is a situation when the range of $\tau$ is between $0$ and $\frac{\ell_{\gamma}}{2}$.} The lower dimensional moduli spaces are the moduli spaces of surfaces obtained by cutting $\mathcal{R}_{g,n}(\mbf{L})$ along $\gamma$. 

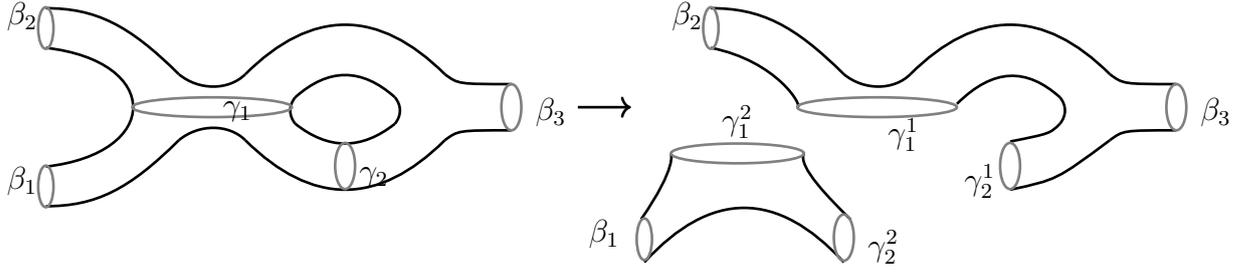
\begin{figure}
	\begin{center}
		\usetikzlibrary{backgrounds}
		\begin{tikzpicture}[scale=.875]
		\draw[line width=1pt] (1,1) .. controls (1.75,1) and (2.25,.75)  ..(3,0);
		\draw[line width=1pt] (1,-2) .. controls(1.75,-2) and (2.25,-1.75)  ..(3,-1);
		\draw[line width=1pt] (1,.4) .. controls(2.75,.2)  and (2.75,-1.2) ..(1,-1.4);
		\draw[gray, line width=1pt] (1,.7) ellipse (.115 and .315);
		\draw[gray, line width=1pt] (1,-1.7) ellipse (.115 and .315);
		\draw[line width=1pt] (3,0) .. controls(3.3,-.25) and (3.75,-.25) ..(4,0);
		\draw[line width=1pt] (3,-1) .. controls(3.3,-.75) and (3.75,-.75) ..(4,-1);
		\draw[line width=1pt] (4,0) .. controls(5,1) and (6,1) ..(7,0);
		\draw[line width=1pt] (4,-1) .. controls(5,-2) and (6,-2) ..(7,-1);
		\draw[line width=1pt] (7,0) .. controls(7.2,-.15)  ..(8,-.15);
		\draw[line width=1pt] (7,-1) .. controls(7.2,-.85)  ..(8,-.85);
		\draw[gray, line width=1pt] (8,-.5) ellipse (.15 and .35);
		\draw[line width=1pt] (4.75,-.75) .. controls(5.25,-1.15) and (5.75,-1.15) ..(6.25,-.75);
		\draw[line width=1pt] (4.75,-.35) .. controls(5.25,0.1) and (5.75,0.1) ..(6.25,-.35);
		\draw[line width=1pt] (4.75,-.35) .. controls(4.65,-.475) and (4.65,-.625) ..(4.75,-.75);
		\draw[line width=1pt] (6.25,-.35) .. controls(6.35,-.475) and (6.35,-.625) ..(6.25,-.75);
		\draw[line width =1pt, color=gray] (3.5,-.5) ellipse (1.2 and .15);
		\draw[line width =1pt, color=gray] (5.5,-1.39) ellipse (.15 and .35);
		\draw (0.25,.5) node[above right] {$\beta_2$}  (0.25,-2) node[above right] {$\beta_1$} (8.2,-.2)node [below right ] {$\beta_3$}  (3.5,-.25) node  [below right ] {$\gamma_1$} (5.55,-1.25) node  [below right ] {$\gamma_2$};
		\draw[->,line width =1pt] (9,-.5)--(9.8,-.5);
		\draw[line width=1pt] (11,1) .. controls (11.75,1) and (12.25,.75)  ..(13,0);
		\draw[line width=1pt] (11,.4) .. controls(12,.2)  and (12.4,-.5) ..(12.3,-.5);
		\draw[line width=1pt] (9.95,-2.2) .. controls(10.5,-1.4)  and (10.4,-1.2) ..(10.4,-1.2);
		\draw[line width=1pt] (12.4,-1.2) .. controls(12.3,-1.4)  and (12.8,-1.9) ..(13.1,-2.2);
		\draw[line width=1pt] (10,-2.85) .. controls(11,-1.75) and (12,-1.75)  ..(13,-2.85);
		\draw[gray, line width=1pt] (11,.7) ellipse (.115 and .315);
		\draw[gray, line width=1pt] (10,-2.5) ellipse (.115 and .315);
		\draw[line width=1pt] (13,0) .. controls(13.3,-.25) and (13.75,-.25) ..(14,0);
		\draw[line width=1pt] (14,0) .. controls(15,1) and (16,1) ..(17,0);
		\draw[line width=1pt] (15.5,-1.75) .. controls(16.2,-1.6)  ..(17,-1);
		\draw[line width=1pt] (17,0) .. controls(17.2,-.15)  ..(18,-.15);
		\draw[line width=1pt] (17,-1) .. controls(17.2,-.85)  ..(18,-.85);
		\draw[gray, line width=1pt] (18,-.5) ellipse (.15 and .35);
		\draw[line width=1pt] (15.5,-1.02) .. controls(16.1,-.9) ..(16.25,-.75);
		\draw[line width=1pt] (14.7,-.45) .. controls(15.25,0.1) and (15.75,0.1) ..(16.25,-.35);
		\draw[line width=1pt] (16.25,-.35) .. controls(16.35,-.475) and (16.35,-.625) ..(16.25,-.75);
		\draw[line width =1pt, color=gray] (13.5,-.5) ellipse (1.2 and .15);
		\draw[line width =1pt, color=gray] (11.4,-1.2) ellipse (1 and .15);
		\draw[line width =1pt, color=gray] (15.5,-1.39) ellipse (.15 and .35);
		\draw[line width =1pt, color=gray] (13,-2.475) ellipse (.15 and .35);
		\draw (10.25,.5) node[above right] {$\beta_2$}  (9,-2.8) node[above right] {$\beta_1$} (18.2,-.2)node [below right ] {$\beta_3$}  (13.5,-.5) node  [below right ] {$\gamma^1_1$} (14.65,-1.2) node  [below right ] {$\gamma^1_2$} (11,-.3) node  [below right ] {$\gamma^2_1$} (13.2,-2.2) node  [below right ] {$\gamma^2_2$};
		\end{tikzpicture}
	\end{center}
	
	\caption{Cutting the surface along the curve $\gamma=\gamma_1+\gamma_2$.}
	\label{cutting1}
\end{figure}

\newl Let us elaborate on the above discussion. Assume that $\gamma$ is a multi-curve on $\mathcal{R}_{g,n}(\mbf{L})$
\begin{equation}\label{curvg}
	\gamma=\sum_{i=1}^k \gamma_i.
\end{equation}
$\{\gamma_1,\ldots,\gamma_k\}$ are disjoint non-homotopic  simple closed curves on $\mathcal{R}_{g,n}(\mbf{L})$. Denote the mapping class group of $\mathcal{R}_{g,n}(\mbf{L})$ by $\text{Mod}_{g,n}(\mbf{L})$. For $h\in \text{Mod}_{g,n}(\mbf{L})$, let $h\cdot\gamma=\sum_{i=1}^k h\cdot\gamma_i$. Assume that $\mathcal{O}_{\gamma}$ be the set of homotopy classes of elements of the set $\text{Mod}_{g,n}(\mbf{L})\cdot\gamma$
\begin{equation}
	\mathcal{O}_{\gamma}\equiv\left\{\left[\alpha\right]\left|~\alpha=h\cdot\gamma,~~[\alpha]\sim[\beta] \iff \alpha\mathrm{ ~and}~ \beta ~\mathrm{are ~homotopic}\right.\right\}.
\end{equation}
The covering space $\mathcal{M}_{g,n}(\mathbf{L})^{\gamma}$ is defined as follows:
\begin{equation}
	\mathcal{M}_{g,n}(\mathbf{L})^{\gamma}\equiv\left\{(\mathcal{R}_{g,n}(\mbf{L}),\eta)| X\in \mathcal{M}_{g,n}(\mathbf{L}),~\eta=\sum_{i=1}^k\eta_i\in \mathcal{O}_{\gamma}, \eta_i\mathrm{ ~ is~a ~closed ~geodesics~on~}\mathcal{R}_{g,n}(\mbf{L})\right\}.\nonumber \label{mgng}
\end{equation}
For any set $A$ of homotopy classes of simple closed curves on $\mcal{R}_{g,n}(\mbf{L})$, define the stabilizer of $A$, Stab($A$), by
\begin{equation}\label{stabg}
	\mathrm{Stab}(A)\equiv\{h\in \text{Mod}_{g,n}(\mbf{L})|~h\cdot A=A\}\subset \text{Mod}_{g,n}(\mbf{L}).
\end{equation}
The symmetry group of $\gamma$, Sym($\gamma$), is defined by
\begin{equation}
	\mathrm{Sym}(\gamma)\equiv\mathrm{Stab}(\gamma)/\cap_{i=1}^k\mathrm{Stab}(\gamma_i)\label{symg}.
\end{equation}
Denoting the Dehn twist along $\gamma$ by $\phi_{\gamma}$, one can then define the following subgroup of $\text{Mod}_{g,n}(\mbf{L})$
\begin{equation}\label{ggamma}
	G_{\gamma}\equiv\bigcap_{i=1}^k~ \mathrm{Stab}(\gamma_i)\subset \text{Mod}_{g,n}(\mbf{L}).
\end{equation}
$G_{\gamma}$ is generated by the set of Dehn twists $\phi_{\gamma_i},~i=1,\ldots,k$, and  the elements of the mapping class group of $\mcal{R}_{g,n}(\gamma)$. The surface $\mcal{R}_{g,n}(\gamma)$ denote the surface $\mcal{R}_{g,n}(\mbf{L})-U_{\gamma}$, where $U_{\gamma}$ is an open set homeomorphic to $\bigcup_{i=1}^k(0,1)\times \gamma_i$ around $\gamma$. Cutting the surface $\mcal{R}_{g,n}(\mbf{L})$ along each of the curves $\gamma_i$ produces two additional boundary components. Therefore, the surface $\mcal{R}_{g,n}(\gamma)$, which is obtained by cutting $\mcal{R}_{g,n}(\mbf{L})$ along all the curves in $\gamma$, is a possibly disconnected surface with $n+2k$ boundary components and $s$ connected components. Let us denote the connected component of $\mcal{R}_{g,n}({\gamma})$ by $\mathcal{R}^i(\gamma),~i=1,\ldots,s$, see figures \ref{cutting1} and \ref{cutting3}. The space $\mathcal{M}_{g,n}(\mathbf{L})^{\gamma}$ is then identified with \cite{Mirzakhani200603}
\begin{equation}\label{mgnlg} 
	\mathcal{M}_{g,n}(\mathbf{L})^{\gamma}\equiv\mathcal{T}_{g,n}(\mathbf{L})/G_{\gamma}. 
\end{equation}

\newl On the other hand, $\mathcal{T}_{g,n}(\mbf{L})$ admits the following factorization \cite{Mirzakhani200610,Wolpert201001}:
\begin{equation}
	\mathcal{T}_{g,n}(\mbf{L})=\prod_{j=1}^s\mathcal{T(R}^j(\gamma))\times \prod_{\gamma_i}\mathbb{R}_{>0}\times \mathbb{R}.  \label{teichmullerfact}
\end{equation}
$\mcal{T}(\mcal{R}^j(\gamma))$ is the Teichm\"uller space of $\mcal{R}^j(\gamma)$. Denoting the mapping class group of $\mcal{R}^j(\gamma)$ by $\text{Mod}(\mcal{R}^j(\gamma))$, we then have \cite{Mirzakhani200610,Wolpert201001}:
\begin{equation}\label{mgnlg1}
	\mathcal{M}_{g,n}(\mathbf{L})^{\gamma}=\prod_{j=1}^s\mathcal{T(R}^j(\gamma))/\text{Mod}(\mcal{R}^j(\gamma))\prod_{\gamma_i}(\mathbb{R}_{>0}\times \mathbb{R})/\mathrm{Dehn}_*(\gamma_i).
\end{equation}
Here $\mathrm{Dehn}_*(\gamma_i)$ is generated by a half-twist if the curve bounds a torus with a single boundary.\footnote{The reason for the half-twist in the case of a torus with a boundary component is that this surface has a non-trivial automorphism called hyperelliptic involution.} Otherwise, it is generated by a simple twist. It acts only on the variables $\tau_j$ with the range $0\le\tau_{\gamma_j}\le\ell_{\gamma_j}/2$ if $\gamma_j$ bounds a torus with a single boundary or otherwise with the range $0\le\tau_{\gamma_j}\le\ell_{\gamma_j}$. Using the factorization \eqref{mgnlg1} of $\mathcal{M}_{g,n}(\mathbf{L})^{\gamma}$,  the volume form of $\mathcal{M}_{g,n}(\mbf{L})$ can be decomposed as:
\begin{equation}
	dV=\prod_{i=1}^sdV(\mathcal{R}^i(\gamma))\times \prod_{\gamma_j}d\ell_{\gamma_j}\wedge d\tau_{\gamma_j}, \label{mggammavoel}
\end{equation}
where $dV(\mathcal{R}^j(\gamma))$ denotes the volume form of $\mathcal{M}(\mcal{R}^j(\gamma))$, the moduli space of surfaces $\mathcal{R}^j(\gamma)$. We can perform the integration over $\mathcal{M}_{g,n}(\mathbf{L})^{\gamma}$ by first performing the integration over $\prod_{j=1}^s\mathcal{M(R}^j(\gamma))$ for fixed values of the lengths of curves $\gamma_j$. We then integrate over $d\tau_{\gamma_j}$, on appropriate ranges as described above, followed by integration over $d\ell_{\gamma_j}$ for $j=1,\ldots s$ from $0$ to $\infty$. 

\begin{figure}
	\begin{center}
		\usetikzlibrary{backgrounds}
		\begin{tikzpicture}[scale=.875]
		\draw[line width=1pt,->] (.5,.6)--(5,3);
		\draw[line width=1pt,->] (.5,.6) --(5,-1.4);
		\draw[gray, line width=1.5pt, style=dashed] (3.65,.75) ellipse (.25 and 1.5);
		\draw[ line width=1pt] (4.75,.8) ellipse (.25 and 2.07);
		\draw[->](3.65,.75)--(3.65,2.25);
		\draw (3.75,1) node[above right] {$\frac{\ell_{\gamma}}{2\pi}$} ;
		\draw[->](.5,.75)--(3.65,2.4);
		\draw (2,2) node[above right] {$\ell_{\gamma}$} ;
		\end{tikzpicture}
	\end{center}
	
	\caption{A cone with infinite length representing the covering space  $\mathcal{M}_{0,4}(\mbf{L})^\gamma$. The thick gray circle with radius $\frac{\ell_{\gamma}}{2\pi}$ corresponds to the range of the twist parameter associated with the curve $\gamma$ with length $\ell_{\gamma}$.}
	\label{cone}
\end{figure}
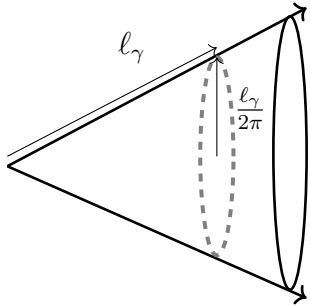

\hypertarget{IntegrationOfMCGInvariantFunctionOverModuliSpace}{\paragraph{Integration of MCG-Invariant Functions over the Moduli Space:}} An  off-shell bosonic-string measure for a typical scattering process is an object living in the moduli space. It is invariant under the mapping class group transformations. In this subsection, we shall describe a method for performing the integration of an MCG-invariant function over the moduli space.

\newl To be able to lift the integration over the moduli space to a covering space, we need a covering map. For this, assume that we have the following identity:
\begin{equation}\label{mcshane}
	\sum_{\alpha\in \overline{\text{Mod}}{\cdot}\gamma}f_i({\ell}_{\alpha}(\mathcal{R}_{g,n}(\mbf{L})))=L_i.
\end{equation}
$f_i$ are real functions of the hyperbolic length ${\ell}_{\alpha}(\mathcal{R}_{g,n}(\mbf{L}))$ of the curve $\alpha$. The curve $\alpha$ is an image of the curve $\gamma$ under the mapping class group. Also, $\overline{\text{Mod}}\cdot \gamma$ denotes the set of inequivalent MCG images of the curve $\gamma$. The Mirzakhani-McShane identity, which we shall explain below, is an example of these types of identities known generically as identities on hyperbolic surfaces. We also assume that cutting along curves $\gamma_i,~i=1,\ldots,k$ provides $s$ disconnected bordered hyperbolic surfaces. Let then $H(\mbs{\ell},\mbs{\tau})$ denote an MCG-invariant function defined on the bordered surface $\mathcal{R}_{g,n}(\mbf{L})$, where $(\mbs{\ell},\mbs{\tau})$ denotes the collection of Fenchel-Nielsen coordinates defined with respect to a pants decomposition of $\mathcal{R}_{g,n}(\mbf{L})$. By MCG-invariance, we mean
	\begin{equation}
		\forall \mfk{m}\in\text{Mod}_{g,n}(\mbf{L})\quad\Longrightarrow\quad H(\mfk
  {m}\cdot\mbs{\ell},\mfk{m}\cdot\mbs{\tau})=H(\mbs{\ell},\mbs{\tau}). \label{MCGInvariance}
\end{equation}
Since $H(\mbs{\ell},\mbs{\tau})$ is invariant under all MCG transformations{,} we are free to choose any pair of pants decomposition of the surface to define the Fenchel-Nielsen coordinates. We can then integrate $H(\mbs{\ell},\mbs{\tau})$ over $\mathcal{M}_{g,n}(\mathbf{L})$ as follows
\begin{equation}\label{funinte}
\hspace*{-.2cm}
    \begin{aligned}
	\bigintssss_{\mathcal{M}_{g,n}(\mathbf{L})}dV~H(\mbs{\ell},\mbs{\tau})&=\frac{1}{L_i}\sum_{\alpha\in \mathrm{Mod}.\gamma}\int_{\mathcal{M}_{g,n}(\mathbf{L})}dV~f_i({\ell}_{\gamma})H(\mbs{\ell},\mbs{\tau})
	\\
	&=\frac{1}{L_i|\mathrm{Sym}(\gamma)|}\int_{\mathcal{M}_{g,n}(\mathbf{L})^{\gamma}}dV~f_i({\ell}_{\gamma})H(\mbs{\ell},\mbs{\tau})
	\\
	&=\frac{1}{L_i|\mathrm{Sym}(\gamma)|}\int_{\mathbb{R}_+^k}d{\ell}_{\gamma_1}\ldots d{\ell}_{\gamma_k} \int_{0}^{\wh{\ell}_{\gamma_1}}\ldots  \int_{0}^{\wh{\ell}_{\gamma_k}} d\tau_{\gamma_1}\ldots  d\tau_{\gamma_k}
	\\ &\times\int_{\mathcal{M(R}^1(\gamma))}dV(\mathcal{R}^1(\gamma))\ldots\int_{\mathcal{M(R}^s(\gamma))}dV(\mathcal{R}^s(\gamma))~f_i(\ell_{\gamma})H(\mbs{\ell},\mbs{\tau}).
    \end{aligned}
\end{equation}
$\tau_{\gamma_i}$ is the twist along $\gamma_i$, and $\wh{\ell}_i\equiv 2^{-M_{\gamma_i}}\ell_{\gamma_i}$ with $M_{\gamma_i}=1$ if $\gamma_i$ bounds a torus with one boundary component, and $M_{\gamma_i}=0$ otherwise. In general, the twist parameter along $\gamma_i$ can be between 0 and $\ell_{\gamma_i}$. In the case of a simple geodesic $\gamma_i$ separating off a torus with one boundary, Stab($\gamma_i$) contains a half twist and so $\tau_{\gamma_i}$ varies over the range $\{0\leq \tau_{\gamma_i}\leq \frac{\ell_{\gamma_i}}{2}\}$. The reason is that every Riemann surface $\mathcal{R}\in \mathcal{M}_{1,1}(\mathbf{L})$ comes with an elliptic involution, but when $(g,n)\neq (1,1)$, a generic point in $\mathcal{M}_{g,n}(\mathbf{ L})$ does not have any non-trivial automorphism fixing the boundary components set-wise. We can then repeat the same procedure for the integration over $\mathcal{M(R}^j(\gamma))$.  We shall continue this procedure until we are left with  integrations over the infinite cones parametrized by $(\ell_i,\tau_i)$ for $i=1,\ldots,3g-3+n$. The final integrand would be a product of some number of the function $f_i$, appearing in \eqref{mcshane}, with the function $H(\mbs{\ell},\mbs{\tau})$ integrated over the following domain:
\begin{alignat}{4}
	0&\le\tau_i\le \wh{\ell}_i,
	\qquad &&i=1,\ldots,3g-3+n,
	\nonumber \\
	0&\le\ell_i< \infty,\qquad &&i=1,\ldots,3g-3+n.
\end{alignat}
To be able to do explicit integration, we have to specify the form of identities in \eqref{mcshane}.

\hypertarget{MirzakhaniMcShaneIdentity}{\paragraph{The Mirzakhani-McShane Identity:}}
  
 The Mirzakhani-McShane identity provides a function defined on the Teichm\"uller space such that the sum of its values over the elements of each orbit of $\text{Mod}_{g,n}(\mbf{L})$ is a constant independent of the orbit \cite{McShane199105,McShane199805,Mirzakhani200610}. 
 Before stating the identity, let us discuss some aspects of the infinite simple geodesic rays on a hyperbolic pair of pants.

\begin{figure}
\begin{center}
\usetikzlibrary{backgrounds}
\begin{tikzpicture}[scale=1.85]
\draw[line width=1pt] (1,1) .. controls (1.75,1) and (2.25,.25)  ..(3.5,.4);
\draw[line width=1pt] (1,-2) .. controls(1.75,-2) and (2.25,-1.25)  ..(3.5,-1.2);
\draw[line width=1.5pt,style=dashed] (3.7,-.2) .. controls (2.5,-.15) and (1.75,-.2)  ..(1.7,-.35);
\draw[line width=1.5pt] (3.3,-.4) .. controls (2.5,-.45) and (1.75,-.35)  ..(1.7,-.35);
\draw[line width=1.5pt] (3.35,0) .. controls (2.5,0) and (1.9,.75)  ..(1.7,.8);
\draw[line width=1.5pt,style=dashed] (1.7,.8) .. controls (1.5,.7) and (1.3,0.1)  ..(1.5,0.05);
\draw[line width=1.5pt] (1.4,.92) .. controls (1.65,.85) and (1.7,0.1)  ..(1.5,0.05);
\draw[line width=1.5pt,style=dashed] (1.4,.92) .. controls (1.2,.7) and (1.15,0.3)  ..(1.3,0.2);
\draw[line width=1.5pt] (1.25,.95) .. controls (1.5,.9) and (1.5,0.1)  ..(1.3,0.2);
\draw[line width=1.5pt,style=dashed] (1.25,.95) .. controls (1.1,.7) and (1.11,0.5)  ..(1.15,0.3);
\draw[line width=1.5pt] (1.05,1) .. controls (1.25,.9) and (1.3,0.5)  ..(1.15,0.3);
\draw[line width=1pt] (1,.4) .. controls(2,-.2)  and (2,-.8) ..(1,-1.4);
\draw[line width=1.5pt] (3.6,.1)--(3.7,.1);
\draw[line width=1.5pt] (3.65,-.65)--(3.75,-.65);
\draw[line width=1.5pt] (3.3,-.9)--(3.4,-.9);
\draw[gray, line width=1.5pt] (1,.7) ellipse (.115 and .315);
\draw[gray,line width=1.5pt] (1,-1.7) ellipse (.115 and .315);
\draw[gray, line width=1.5pt] (3.5,-.4) ellipse (.2 and .8);
\draw (0.25,.5) node[above right] {$\beta_2$}  (0.25,-2) node[above right] {$\beta_3$}  (3.6,.5)node [above right ] {$\beta_1$}  (3.35,-.15)  node[above right] {$z_1$}(3.7,0)  node[above right] {$z_2$} (3.7,-.8)  node[above right] {$y_2$} (3.35,-1.1)  node[above right] {$y_1$} (3.25,-.6)  node[above right] {$w_1$}  (3.7,-.4)node [above right ] {$w_2$} ;
\end{tikzpicture}
\end{center}
\caption{The complete geodesics that are disjoint from $\beta_2,\beta_3$ and are orthogonal to $\beta_1$.}
\label{complete geodesics}
\end{figure}
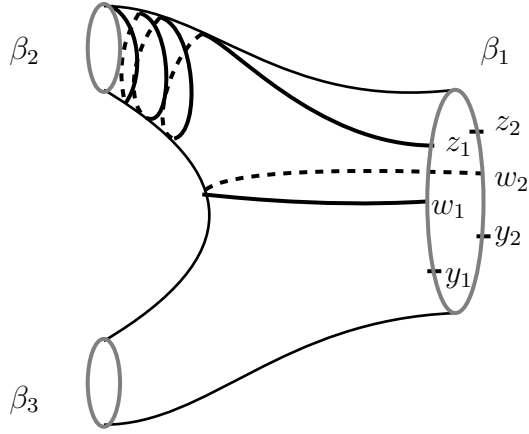

\newl Consider the unique hyperbolic pair of pants $P(x_1,x_2,x_3)$ with geodesic boundary curves $\{\beta_1,\beta_2,\beta_3\}$ such that $l_{\beta_i}(P)=x_i,~i=1,2,3$. The boundary curves are allowed to have vanishing lengths. $P(x_1,x_2,x_3)$ is obtained by pasting two copies  of the (unique) hyperbolic hexagons,  whose edges are  geodesics  that meet perpendicularly with non-adjacent sides of length $\frac{x_1}{2},\frac{x_2}{2}$ and $\frac{x_3}{2}$, along the three remaining sides. Thus  $P(x_1,x_2,x_3)$ admits a reflection  involution symmetry $\mathcal{J}$ which interchanges the two hexagons. On such a hyperbolic pair of pants, there are $5$ complete geodesics disjoint from $\beta_1,\beta_2$, and $\beta_3$. Two of these geodesics meet $\beta_1$, say at $z_1$ and $z_2$, and spiral around $\beta_2$, as in Figure \ref{complete geodesics}.  Similarly, the other two geodesics meet $\beta_1$, at say $y_1$ and $y_2$, and spiral around $\beta_3$. The fifth geodesic is the common simple geodesic perpendicular to $\beta_1$ meeting $\beta_1$ at two points, say $w_1$ and $w_2$. We have
\begin{equation}
    \begin{aligned}
        \mathcal{J}(y_1)=y_2, \qquad \mathcal{J}(z_1)=z_2, \qquad \mathcal{J}(w_1)=w_2.
    \end{aligned}
\end{equation}
The geodesic length  of the interval between $z_1$ and $z_2$ along the boundary  $\beta_1$ containing $w_1$ and $w_2$ is given by \cite{Mirzakhani200610}
    \begin{equation}\label{rinterval}
    \mathcal{D}(x_1,x_2,x_3)=x_1-\mathrm{ln}\left(\frac{\mathrm{cosh}(\frac{x_2}{2})+\mathrm{cosh}(\frac{x_1+x_3}{2})}{\mathrm{cosh}(\frac{x_2}{2})+\mathrm{cosh}(\frac{x_1-x_3}{2})}\right).
    \end{equation}
Twice times the length of the geodesic between the two geodesics perpendicular to $\beta_1$ and spiraling around $\beta_2$ and $\beta_3$ is given by \cite{Mirzakhani200610}
     \begin{equation}\label{einterval}
    \mathcal{E}(x_1,x_2,x_3)=2\mathrm{ln}\left(\frac{e^{\frac{x_1}{2}}+e^{\frac{x_2+x_3}{2}}}{e^{-\frac{x_1}{2}}+e^{\frac{x_2+x_3}{2}}}\right).
    \end{equation}
We say three isotopy classes of connected simple closed curves $(\alpha_1,\alpha_2,\alpha_3)$ on $\mcal{R}_{g,n}(\mbf{L})$ bound a pair of pants if there exists an embedded pair of pants such that its boundaries are $\alpha_1,\, \alpha_2$, and $\alpha_3$. Here $\alpha_i$ can have length zero. To state the Mirzakhani-McShane identity, the following definitions are useful
  
\begin{itemize}
  
  \item For $1\leq i\leq n$, let $\mathcal{F}_i$ be the set of unordered pairs of isotopy classes of simple closed curves $\{\alpha_1,\alpha_2\}$ bounding a pair of pants with $\beta_i$ such that $\alpha_1,\alpha_2\notin \partial(\mcal{R}_{g,n}(\mbf{L}))$, as in Figures \ref{cutting3} and \ref{cutting2}. 
  
  \item For $1\leq i\neq j \leq n$, let $\mathcal{F}_{i,j}$ be the set of isotopy classes of simple closed curves $\gamma$ that together with $\beta_i$ and $\beta_j$ bounding a pair of pants, as in Figure \ref{cutting4}. 
  \end{itemize} 
  
\begin{figure}
\begin{center}
\usetikzlibrary{backgrounds}
\begin{tikzpicture}[scale=.9]
\draw[line width=1pt] (1,1) .. controls (1.75,1) and (2.25,.75)  ..(2.75,.2);
\draw[line width=1pt] (1,-2) .. controls(1.75,-2) and (2.25,-1.75)  ..(3,-1);
\draw[line width=1pt] (1,.4) .. controls(1.7,0)  and (1.7,-1) ..(1,-1.4);
\draw[gray, line width=1pt] (1,.7) ellipse (.115 and .315);
\draw[gray,line width=1pt] (1,-1.7) ellipse (.115 and .315);
\draw[gray, line width=1pt] (3,-1) .. controls(3.3,-.75) and (3.75,-.75) ..(4,-1);
\draw[line width=1pt] (4,0.2) .. controls(5,1) and (6,1) ..(7,0);
\draw[line width=1pt] (2.75,0.2) .. controls(3,0.05) and (3.1,0.2) ..(3,1.5);
\draw[line width=1pt] (4,0.2) .. controls(3.8,0.05) and (3.6,0.2) ..(3.7,1.5);
\draw[line width=1pt] (4,-1) .. controls(5,-2) and (6,-2) ..(7,-1);
\draw[line width=1pt] (7,0) .. controls(7.2,-.15)  ..(8,-.15);
\draw[line width=1pt] (7,-1) .. controls(7.2,-.85)  ..(8,-.85);
\draw[gray, line width=1pt] (8,-.5) ellipse (.15 and .35);
\draw[gray, line width=1pt] (3.35,1.5) ellipse (.35 and .15);
\draw[line width=1pt] (4.75,-.75) .. controls(5.25,-1.1) and (5.75,-1.1) ..(6.25,-.75);
\draw[line width=1pt] (4.75,-.35) .. controls(5.25,0.1) and (5.75,0.1) ..(6.25,-.35);
\draw[line width=1pt] (4.75,-.35) .. controls(4.65,-.475) and (4.65,-.625) ..(4.75,-.75);
\draw[line width=1pt] (6.25,-.35) .. controls(6.35,-.475) and (6.35,-.625) ..(6.25,-.75);
\draw[line width =1.2pt, color=gray]  (2.8,-.48) ellipse (.15 and .65);
\draw[line width =1.2pt, color=gray] (4.15,.-.46) ellipse (.15 and .68);
\draw[line width=1pt] (2.2,-.5) ellipse (.25 and .4);

\draw (0.25,.5) node[above right] {$\beta_2$}  (0.25,-2) node[above right] {$\beta_3$} (8.2,-.2)node [below right ] {$\beta_4$}  (3.2,1.5)node [above right ] {$\beta_1$} (2,-1.5) node  [above right ] {$\alpha_1$} (4.3,-1.5) node  [above right ] {$ \alpha_2$};
\end{tikzpicture}
\end{center}

\caption{Cutting the surface along $\alpha_1+\alpha_2$ produces a pair of pants, a genus-1 surface with 3 borders and a genus-1 surface with 2 borders.}
\label{cutting3}
\end{figure}
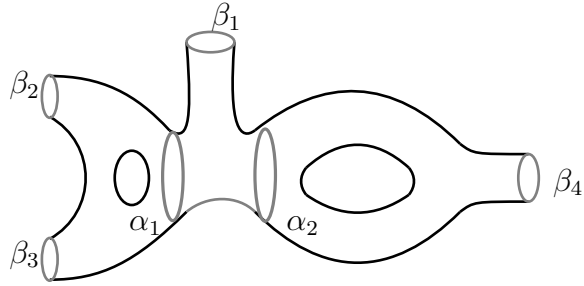
  
  \begin{figure}
\begin{center}
\usetikzlibrary{backgrounds}
\begin{tikzpicture}[scale=.9]
\draw[line width=1pt] (1,1) .. controls (1.75,1) and (2.25,.75)  ..(2.75,.2);
\draw[line width=1pt] (1,-2) .. controls(1.75,-2) and (2.25,-1.75)  ..(3,-1);
\draw[line width=1pt] (1,.4) .. controls(1.7,0)  and (1.7,-1) ..(1,-1.4);
\draw[gray,line width=1pt] (1,.7) ellipse (.115 and .315);
\draw[gray,line width=1pt] (1,-1.7) ellipse (.115 and .315);
\draw[line width=1pt] (3,-1) .. controls(3.3,-.75) and (3.75,-.75) ..(4,-1);
\draw[line width=1pt] (4,0.2) .. controls(5,1) and (6,1) ..(7,0);
\draw[line width=1pt] (2.75,0.2) .. controls(3,0.05) and (3.1,0.2) ..(3,1.5);
\draw[line width=1pt] (4,0.2) .. controls(3.8,0.05) and (3.6,0.2) ..(3.7,1.5);
\draw[line width=1pt] (4,-1) .. controls(5,-2) and (6,-2) ..(7,-1);
\draw[line width=1pt] (7,0) .. controls(7.2,-.15)  ..(8,-.15);
\draw[line width=1pt] (7,-1) .. controls(7.2,-.85)  ..(8,-.85);
\draw[gray, line width=1pt] (8,-.5) ellipse (.15 and .35);
\draw[gray, line width=1pt] (3.35,1.5) ellipse (.35 and .15);
\draw[line width=1pt] (4.75,-.75) .. controls(5.25,-1.1) and (5.75,-1.1) ..(6.25,-.75);
\draw[line width=1pt] (4.75,-.35) .. controls(5.25,0.1) and (5.75,0.1) ..(6.25,-.35);
\draw[line width=1pt] (4.75,-.35) .. controls(4.65,-.475) and (4.65,-.625) ..(4.75,-.75);
\draw[line width=1pt] (6.25,-.35) .. controls(6.35,-.475) and (6.35,-.625) ..(6.25,-.75);
\draw[line width =1.2pt, color=gray] [rotate around={160:(3.7,-.2)}]  (3.7,-.2) ellipse (1 and .15);
\draw[line width =1.2pt, color=gray] (5.5,.35) ellipse (.15 and .4);
\draw[line width=1pt] (2.2,-.5) ellipse (.25 and .4);

\draw (0.25,.5) node[above right] {$\beta_2$}  (0.25,-2) node[above right] {$\beta_3$} (8.2,-.2)node [below right ] {$\beta_4$}  (3.2,1.5)node [above right ] {$\beta_1$} (3.5,-.3) node  [below right ] {$\alpha_1$} (5.55,.5) node  [below right ] {$\alpha_2$};
\end{tikzpicture}
\end{center}

\caption{Cutting the surface along $\alpha_1+\alpha_2$ produces a pair of pants and genus-1 surface with 5 borders.}
\label{cutting2}
\end{figure}
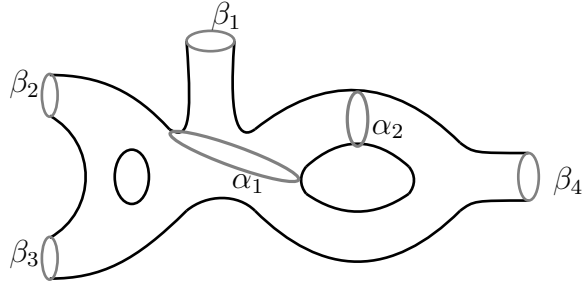

  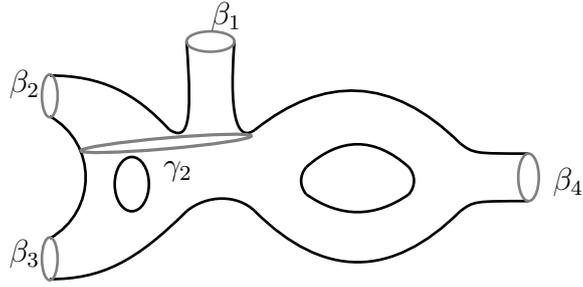
\begin{figure}
\begin{center}
\usetikzlibrary{backgrounds}
\begin{tikzpicture}[scale=.9]
\draw[line width=1pt] (1,1) .. controls (1.75,1) and (2.25,.75)  ..(2.75,.2);
\draw[line width=1pt] (1,-2) .. controls(1.75,-2) and (2.25,-1.75)  ..(3,-1);
\draw[line width=1pt] (1,.4) .. controls(1.7,0)  and (1.7,-1) ..(1,-1.4);
\draw[gray, line width=1pt] (1,.7) ellipse (.115 and .315);
\draw[gray, line width=1pt] (1,-1.7) ellipse (.115 and .315);
\draw[line width=1pt] (3,-1) .. controls(3.3,-.75) and (3.75,-.75) ..(4,-1);
\draw[line width=1pt] (4,0.2) .. controls(5,1) and (6,1) ..(7,0);
\draw[line width=1pt] (2.75,0.2) .. controls(3,0.05) and (3.1,0.2) ..(3,1.5);
\draw[line width=1pt] (4,0.2) .. controls(3.8,0.05) and (3.6,0.2) ..(3.7,1.5);
\draw[line width=1pt] (4,-1) .. controls(5,-2) and (6,-2) ..(7,-1);
\draw[line width=1pt] (7,0) .. controls(7.2,-.15)  ..(8,-.15);
\draw[line width=1pt] (7,-1) .. controls(7.2,-.85)  ..(8,-.85);
\draw[gray, line width=1pt] (8,-.5) ellipse (.15 and .35);
\draw[gray, line width=1pt] (3.35,1.5) ellipse (.35 and .15);
\draw[line width=1pt] (4.75,-.75) .. controls(5.25,-1.1) and (5.75,-1.1) ..(6.25,-.75);
\draw[line width=1pt] (4.75,-.35) .. controls(5.25,0.1) and (5.75,0.1) ..(6.25,-.35);
\draw[line width=1pt] (4.75,-.35) .. controls(4.65,-.475) and (4.65,-.625) ..(4.75,-.75);
\draw[line width=1pt] (6.25,-.35) .. controls(6.35,-.475) and (6.35,-.625) ..(6.25,-.75);
\draw[line width =1.2pt, color=gray] [rotate around={5:(2.7,.0)}]  (2.7,.0) ellipse (1.25 and .08);
\draw[line width=1pt] (2.2,-.6) ellipse (.25 and .4);

\draw (0.25,.5) node[above right] {$\beta_2$}  (0.25,-2) node[above right] {$\beta_3$} (8.2,-.2)node [below right ] {$\beta_4$}  (3.2,1.5)node [above right ] {$\beta_1$} (2.5,-.1) node  [below right ] {$\gamma_2$};
\end{tikzpicture}
\end{center}

\caption{Cutting the surface along $\gamma_2$ produces a pair of pants and a genus-2 surface with 3 borders.}
\label{cutting4}
\end{figure}
We can now state the Mirzakhani-McShane identity for $\mathcal{R}_{g,n}(\mathbf{L})$ as follows \cite{Mirzakhani200610} 
  \begin{equation}\label{gmidentity}
  \sum_{\{\alpha_1,\alpha_2\}\in \mathcal{F}_1}\mathcal{D}(L_1,l_{\alpha_1(\mathcal{R})},l_{\alpha_2(\mathcal{R})})+\sum_{i=2}^n\sum_{\gamma\in \mathcal{F}_{1,i}}\mathcal{E}(L_1,L_i,l_{\gamma}(\mathcal{R}))=L_1.
  \end{equation}
A useful special case is where the boundary $\beta_1$ tends to a puncture $p_1$ 
    \begin{equation}\label{gmidentityp}
  \sum_{\{\alpha_1,\alpha_2\}\in \mathcal{F}_1}\frac{1}{1+e^{\frac{l_{\alpha_1}(\mathcal{R})+l_{\alpha_2}(\mathcal{R})}{2}}}+\sum_{i=2}^n\sum_{\gamma\in \mathcal{F}_{1,i}}\frac{1}{2}\left(\frac{1}{1+e^{\frac{l_{\gamma}(\mathcal{R})+L_i}{2}}}+\frac{1}{1+e^{\frac{l_{\gamma}(\mathcal{R})-L_i}{2}}}\right)=\frac{1}{2}.
  \end{equation}

\paragraph{Computing the Volume of Moduli Spaces:} Let us now explain the simplest possible application of these results, which is the computation of the WP volumes of moduli spaces. We explain the illustrative examples of the WP volumes of $\mathcal{M}_{1,1}(\mathcal{L}=0)$ and $\mathcal{M}_{0,4}(\mathbf{L})$.

   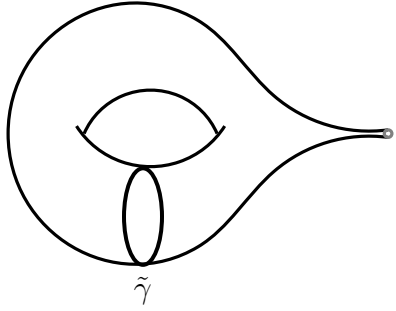
\begin{figure}
\begin{center}
\usetikzlibrary{backgrounds}
\begin{tikzpicture}[scale=.5]
\draw[black, very thick] (14,-.1) to[curve through={(11,-1)..(9,-3)..(4,0)..(9,3)..(11,1)}] (14,.1);
\draw[black, very thick] (9.5,0) to[curve through={(8.5,1)..(7,1)}] (6,0);
\draw[black, very thick] (9.7,.2) to[curve through={(8.5,-.75)..(7,-.75)}] (5.8,.2);
\draw[gray, very thick](14,0) ellipse (.1 and .1);
\draw[ line width=1.5pt](7.55,-2.2) ellipse (.5 and 1.28);
\draw (7.55,-3.5) node[below] {$\tilde\gamma$} ;
\end{tikzpicture}
\end{center}

\caption{The curve $\widetilde\gamma$ on the one-punctured torus.}
\label{Torus1p}
\end{figure}

\begin{figure}
\begin{center}
\usetikzlibrary{backgrounds}
\begin{tikzpicture}[scale=.875]
\draw[line width=1pt,->] (.5,.6)--(5,3);
\draw[line width=1pt,->] (.5,.6) --(5,-1.4);
\draw[gray, line width=1.5pt, style=dashed] (3.65,.75) ellipse (.25 and 1.5);
\draw[ line width=1pt] (4.75,.8) ellipse (.25 and 2.07);
\draw[->](3.65,.75)--(3.65,2.25);
\draw (3.75,1) node[above right] {$\frac{\ell_{\tilde\gamma}}{2\pi}$} ;
\draw[->](.5,.75)--(3.65,2.4);
\draw (2,2) node[above right] {$\ell_{\tilde\gamma}$} ;
\end{tikzpicture}
\end{center}

\caption{The cone with the infinite length representing the covering space  $\mathcal{M}_{1,1}^{\widetilde{\gamma}}$. The thick gray circle with radius $\frac{\ell_{\widetilde\gamma}}{2\pi}$ corresponds to the range of the twist parameter associated with $\widetilde\gamma$ having length $\ell_{\widetilde\gamma}$}.
\label{conet}
\end{figure}
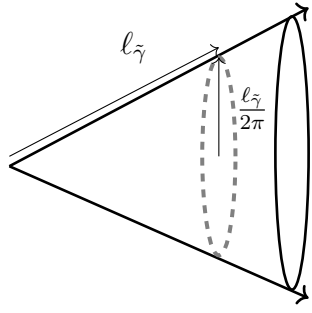

\newl {\small \bf The WP Volume of $\mathcal{M}_{1,1}$:} Let us compute the volume of $\mcal{M}_{1,1}(L=0)$, the moduli space of one-punctured torus $\mcal{R}_{1,1}$. The WP symplectic form is given by $d\ell\wedge d\tau$, where $\ell$ and $\tau$ are the Fenchel-Nielsen coordinates. We then have
\begin{equation}\label{volm11}
\mathrm{Vol}(\mathcal{M}_{1,1})=\bigintssss_{\mathcal{M}_{1,1}}d\ell\wedge d\tau.
\end{equation}
The McShane identity for $\mcal{R}_{1,1}$ is given by \cite{McShane199105,McShane199805}
\begin{equation}\label{mcshaneoneptorus}
\sum_{\gamma\in \mathcal{F}_{1}}\frac{1}{1+e^{\ell_{\gamma}}}=\frac{1}{2}.
\end{equation}
Any curve in $\mathcal{F}_1$ can be obtained from a single representative geodesic say $\widetilde\gamma$ by an action of the mapping class group, see Figure \ref{Torus1p}. We can then compare the McShane identity for the one-punctured hyperbolic torus with \eqref{mcshane} and make the following identification
\begin{equation}
\left.\frac{\partial }{\partial L_{1}}f(\ell_{\widetilde \gamma})\right|_{L_1=0}=\frac{2}{1+e^{\ell_{\widetilde \gamma}}}.
\end{equation}
Following the integration prescription discussed above, let us insert the McShane identity inside the volume integration \eqref{volm11}
\begin{equation}\label{1volm11}
    \begin{aligned}
\mathrm{Vol}(\mathcal{M}_{1,1})&=\sum_{\gamma\in \mathrm{Mod}\cdot\widetilde{\gamma}}\bigintssss_{\mathcal{M}_{1,1}}\frac{2}{1+e^{\ell_{ \gamma}}}d\ell_{\gamma}\wedge d\tau_{\gamma}
\\
&=\bigintssss_{\mathcal{M}^{\widetilde{\gamma}}_{1,1}}\frac{2}{1+e^{\ell_{ \tilde\gamma}}}d\ell_{\widetilde{\gamma}}\wedge d\tau_{\widetilde{\gamma}}.
    \end{aligned}
\end{equation}
The covering space $\mathcal{M}^{\widetilde \gamma}_{1,1}$ of $\mathcal{M}_{1,1}$ is given by 
\begin{equation}\label{m11cover}
\mathcal{M}^{\widetilde \gamma}_{1,1}=\frac{\mathcal{T}_{1,1}}{\mathrm{Stab}(\widetilde\gamma)}=\left\{(\ell_{\widetilde\gamma},\tau_{\widetilde\gamma})\,\,|\,\,0\leq \tau\leq \ell_{\widetilde\gamma} \right\}.
\end{equation}
This identification is needed because the full twist around the curve $\widetilde\gamma$ is a Dehn twist around it. Therefore, the space $\mathcal{M}^{\widetilde \gamma}_{1,1}$ can be identified with an infinite cone, as illustrated in Figure \ref{conet}. Finally, we get
\begin{equation}\label{2volm11}
\mathrm{Vol}(\mathcal{M}_{1,1})=\bigintssss_{0}^{\infty}d\ell_{\widetilde{\gamma}}\int_0^{\ell_{\widetilde{\gamma}}}d\tau_{\widetilde{\gamma}}\frac{2}{1+e^{\ell_{\widetilde{\gamma}}}}=\frac{\pi^2}{6}.
\end{equation}

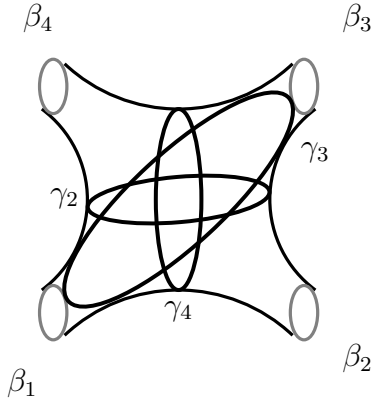
\begin{figure}
	\begin{center}
		\usetikzlibrary{backgrounds}
		\begin{tikzpicture}[scale=.6]
		
		\draw[black, very thick] (20,-2) to[curve through={(21,0)}] (20,2);
		\draw[black, very thick] (26,-2) to[curve through={(25,0)}] (26,2);
		\draw[black,very  thick] (20.5,-3) to[curve through={(23,-2)}] (25.5,-3);
		\draw[black,very  thick] (20.5,3) to[curve through={(23,2)}] (25.5,3);
		\draw[gray,very  thick](20.25,-2.5) ellipse (.3 and .6);
		\draw[gray,very  thick](20.25,2.5) ellipse (.3 and .6);
		\draw[gray,very  thick](25.75,-2.5) ellipse (.3 and .6);
		\draw[gray,very  thick](25.75,2.5) ellipse (.3 and .6);
		
		\draw  node[below] at (23,-2) {$\gamma_4$};
		\draw  node[below] at (20.5,.5) {$\gamma_2$};
		\draw  node[below] at (26,1.5) {$\gamma_3$};
		\draw  node[below right] at (19,-3.5) {$\beta_1$};
		\draw  node[left] at (27.5,-3.5) {$\beta_2$};
		\draw  node[left] at (27.5,4) {$\beta_3$};
		\draw  node[left] at (20.5,4) {$\beta_4$};
		\draw[line width=1.5pt](23,0) ellipse (.5 and 2);
		\draw[ line width=1.5pt, rotate around={95:(23,0)}](23,0) ellipse (.5 and 2);
		\draw[ line width=1.5pt, rotate around={133:(23,0)}](23,0) ellipse (1 and 3.3);
		\end{tikzpicture}
	\end{center}
	
	\caption{The curves $\gamma_2,\gamma_3$ and $\gamma_4$ on the sphere with four borders.}
	\label{sphere4borders}
\end{figure}

\paragraph{\small\bf The WP Volume of $\mathcal{M}_{0,4}(\mathbf{L})$:} Let us next compute the WP volume of $\mcal{M}_{0,4}(\mbf{L})$, the moduli space of the hyperbolic surface $\mcal{R}_{0,4}(\mathbf{L})$. The borders $\{\beta_1,\beta_2,\beta_3,\beta_4\}$ have lengths $\mathbf{ L}=(L_1,L_2,L_3,L_4)$. The Mirzakhani-McShane identity for $\mcal{R}_{0,4}(\mathbf{L})$ is given by 
  \begin{equation}\label{gmidentityo4}
\sum_{i=2}^4\sum_{\gamma\in \mathcal{F}_{1,i}}\mathcal{E}(L_1,L_i,\ell_{\gamma})=L_1.
  \end{equation}
Proceeding as the previous example, the WP volume of $\mathcal{M}_{0,4}(\mathbf{L})$ is
\begin{equation} \label{vol04}
    \begin{aligned}
\mathrm{Vol}(\mathcal{M}_{0,4}(\mathbf{ L}))&=\frac{1}{L_1}\sum_{i=2}^4\sum_{\gamma\in \mathcal{F}_{1,i}}\int_{\mathcal{M}_{0,4}(\mathbf{ L})}\mathcal{E}(L_1,L_i,\ell_{\gamma})~d\ell_{\gamma}\wedge d\tau_{\gamma}\nonumber\\
&=\frac{1}{L_1}\sum_{i=2}^4\int_{\mathcal{M}_{0,4}^{\gamma_i}(\mathbf{ L})}\mathcal{E}(L_1,L_i,\ell_{\gamma_i})~d\ell_{\gamma_i}\wedge d\tau_{\gamma_i}\nonumber\\
&=\frac{1}{L_1}\sum_{i=2}^4\int_{0}^{\infty}d\ell_{\gamma_i} \int_0^{\ell_{\gamma_i}}d\tau_{\gamma_i}~\mathcal{E}(L_1,L_i,\ell_{\gamma_i})\nonumber\\
&=2\pi^2+\frac{1}{2}(L_1^2+L_2^2+L_3^2+L_4^2).    
    \end{aligned}
\end{equation}
The curves $\gamma_2,\gamma_3$ and $\gamma_4$ are shown in Figure \ref{sphere4borders}.

%% file: Sections/Off-Shell_Superstring_Amplitudes.tex
We now consider off-shell amplitudes in supersymmetric string theories.  To avoid cluttering, we shall discuss heterotic string theory whose holomorphic part is the same as the type-II superstring theory and its antiholomorphic part is similar to the bosonic string theory compactified on 16-dimensional integer, even, self-dual lattice \cite{GrossHarveyMartinecRohm1985a,GrossHarveyMartinecRohm1985b}. In this section, we shall  review the general construction of the off-shell amplitudes in the superstring theory.   
  
\subsection{The Worldsheet Theory}\label{sworlsheet}
  
The world-sheet theory of heterotic string theory at the tree level contains the matter field theory with central charge $(26,15)$, and the ghost system of total central charge $(-26,-15)$ containing the anti-commuting fields $b,c,\bar b,\bar c$ and the commuting $\beta,\gamma$ ghosts. The $(\beta,\gamma)$  system can be bosonized  \cite{FriedanMartinecShenker1986}, by replacing it with a system involving a scalar field $\phi$ with background charge $Q=-2$ and two conjugate anti-commuting fields $\xi$ and $\eta$ of conformal dimensions $0$ and $1$, respectively. The bosonization prescription reads
\begin{equation}\label{bgpxe}
\beta(z)=\partial_z \xi(z)e^{-\phi(z)}, \qquad\gamma(z)=\eta(z)e^{\phi(z)}.
\end{equation}
Reversing this prescription we get the following identifications:
 \begin{equation}
 \begin{aligned}
     \xi(z)&=\mathrm{ H}(\beta(z)),&\qquad \eta(z)&=\partial_z\gamma(z)\delta(\gamma(z)), 
     \\
     e^{\phi(z)}&=\delta(\beta(z)),&\qquad e^{-\phi(z)}&=\delta(\gamma(z)).
 \end{aligned}
 \end{equation}
where $\mathrm{H}$ denotes the Heaviside step function. Therefore, the complete set of bosonization relations are
\begin{equation}
\begin{aligned}
    \gamma(z)&=\eta e^{\phi(z)},&\qquad \delta(\gamma(z))&=e^{-\phi(z)},
    \\
    \beta(z)&=\partial_z \xi(z) e^{-\phi(z)},&\qquad  \delta(\beta(z))&=e^{\phi(z)}.
\end{aligned}
\end{equation}
When the bosonic superconformal ghost system, i.e. $(\beta,\gamma)$ system, is bosonized, we are left with a choice among different Bose sea levels or a picture number \cite{FriedanMartinecShenker1986}. Each choice of a picture number leads to a different inequivalent representation of the ghost vacuum. However, each choice leads to a different description of the same physical states \cite{FriedanMartinecShenker1986}. To go from one description to the other, we need an operation, which is called the picture-changing operation. As is clear from the name, it changes the picture number of the bosonic superconformal ghost vacuum and also the primaries, i.e. the vertex operators, of the superconformal field theory build upon that vacuum. The picture-changing operation is done using the picture-changing operator (PCO) $\chi(z)$ which is given by:
\begin{equation}\label{hpart}
\begin{aligned}
    \chi(z)&=\{Q_B,\xi(z)\}=\oint dw~ j_{B}(w)\xi(z),
    \\
j_{B}(z)&=c(z)\Big(T_{m}(z)+T_{\beta,\gamma}(z)\Big)+\gamma(z)T_F(z)+b(z)c(z)\partial c(z)-\frac{1}{4}\gamma(z)^2b(z).
\end{aligned}
\end{equation}
Here $Q_B$ denotes the world-sheet BRST charge, $j_B$ denotes the BRST current and  $T_F(z)$ denote the superpartner of the matter stress tensor $T_m(z)$. The picture-changing operator is a BRST-invariant dimension-zero primary operator having picture number one. Being BRST-exact, it might naively seem that the picture-changing operator is trivial once acted on physical states. The origin of its non-trivial action comes from the fact that $\xi(z)$ does not commute with $\eta_0$, which enters the definition of states appearing in the S-matrix computation, as we explain below. 

\newl Let us denote the total Hilbert space of the world-sheet theory  by $\mathcal{H}=\mathcal{H}^{\text{NS}}\otimes\mathcal{H}^{\text{R}}$, where $\mathcal{H}^{\text{NS}}$ denote the Nevue-Schwarz (NS) sector and $\mathcal{H}^{\text{R}}$ denote the Ramond (R) sector. We  denote by $\mathcal{H}_0=\mathcal{H}_0^{\text{NS}}\otimes\mathcal{H}_0^{\text{R}}$, a  subspace of $\mathcal{H}$ defined as
\begin{equation}\label{h0super}
 \begin{gathered}
     |\Psi\rangle\in \mathcal{H}_0, \quad \mathrm{if} \qquad(b_0-\bar b_0)|\Psi\rangle=0,\quad(L_0-\bar L_0)|\Psi\rangle =0,\quad\eta_0|\Psi\rangle=0,
     \\
    \text{picture number $(|\Psi\rangle)$}=
      \begin{cases}
        -1, & \quad  |\Psi\rangle\in \mathcal{H}_0^{\text{NS}},
        \\
        -\frac{1}{2}, &\quad  |\Psi\rangle\in \mathcal{H}_0^{\text{R}}.
      \end{cases}
 \end{gathered}
\end{equation}
 Physical states that will appear as the external states in the S-matrix computation belong to the  subspace  $\mathcal{H}_1$  of $\mathcal{H}_0$ containing states with ghost number $2$ satisfying the additional condition
    \begin{equation}\label{h1super}
  |\Psi\rangle\in \mathcal{H}_1,\qquad\text{if}\qquad |\Psi\rangle\in \mathcal{H}_0,\qquad (b_0+\bar b_0)|\Psi\rangle=0.
  \end{equation}
This concludes the discussion of the world-sheet theory and its Hilbert space. 
 
 \subsection{Off-Shell Superstring Amplitudes}
  
  The construction of the off-shell  amplitudes in superstring theory is  similar to the construction of the off-shell amplitudes in the bosonic string theory. For instance, in the superstring theory also we need to choose gluing-compatible local coordinates around the punctures at which the vertex operators are inserted.  However, there are additional complications that we need to address for off-shell superstring amplitudes. These complications are arising from the following fact. In order to construct the genus $g$ contribution to off-shell amplitudes in superstring theory with $n_{\text{NS}}$ number of NS external states having the picture number $-1$  and $2n_{\text{R}}$ number of R external states having the picture number $-\frac{1}{2}$,  we need to insert $2g-2+n_{\text{NS}}+n_{\text{R}}$ PCOs on the genus $g$ Riemann surface with $n=n_{\text{NS}}+2n_{\text{R}}$ punctures. Inserting these operators on the worldsheet introduces the following additional issues compared to the case of bosonic string theory
  \begin{itemize}
  \item[(1)] For consistency, on the 1PR region of the moduli space, we need to distribute PCOs on the surface in a way that is consistent with the PCO distribution which is induced from gluing by the relation \eqref{plumb}.
  
  \item[(2)] PCOs may collide with each other or with vertex operators. These collisions  introduce singularities into the corresponding amplitude which have no physical interpretation. Moreover, even if there are no such collisions, the presence of PCOs can introduce spurious singularities in the amplitude when the locations of PCOs on the surface satisfy certain global conditions.
  \end{itemize}
  
  As a result, in order to define off-shell superstring amplitudes, we need to define an integration cycle judiciously such that it avoids the occurrence of spurious singularities. 
  
  \newl Therefore,  like in  bosonic string theory, the integration measure of the off-shell amplitudes in the superstring theory is not a genuine differential form on the moduli space $\mathcal{M}_{g,n}$. Instead, we need to think of the integration measure of an off-shell superstring amplitude involving $n_{\text{NS}}$ number of NS states and $2n_{\text{R}}$ number of R states as a differential form defined on  a section of a larger space $\mathcal{P}_{g,n_{\text{NS}},2n_{\text{R}}}$. This space is a fiber bundle over the moduli space  $\mathcal{M}_{g,n}$ whose fiber directions correspond to different choices of the local coordinates around the punctures where the vertex operators are inserted and the locations of $2g-2+n_{\text{NS}}+2n_{\text{R}}$ number of PCOs. If we restrict ourselves to  the states that belong to the Hilbert space $\mathcal{H}_0$, then we can consider the differential form of our interest as a section of a space $\wh{\mathcal{P}}_{g,n_{\text{NS}},2n_{\text{R}}}$ that is smaller compared to $\mathcal{P}_{g,n_{\text{NS}},2n_{\text{R}}}$. We can understand  $\wh{\mathcal{P}}_{g,n_{\text{NS}},2n_{\text{R}}}$ as the base space of the fiber bundle $\mathcal{P}_{g,n_{\text{NS}},2n_{\text{R}}}$ with the fiber direction corresponds to the phases of local coordinates. 
  
  \subsubsection*{A Gluing-Compatible Integration Cycle}
  
  In superstring theory, the gluing-compatibility of the integration cycle refers to  choosing the local coordinates and the PCO distribution that respect gluing by the relation \eqref{plumb}. Inside the 1PI region of moduli space, we are free to choose  local coordinates and the PCOs distribution arbitrarily. The definition of a gluing-compatible choice of local coordinates is the same as described for the bosonic-string amplitudes in Section \ref{sec:offshell bosonic-string amplitudes}. Therefore, we shall discuss only the meaning of the gluing-compatibility requirement on the distribution of PCOs on the worldsheet.
  
  \newl Consider the situation where the genus $g$ Riemann surface with $n_{\text{NS}}$ punctures having NS-sector states and $2n_{\text{R}}$ punctures having R-sector states degenerate into a genus $g_1$ Riemann surface with $n^1_{\text{NS}}$ punctures having NS-sector states and $2n^1_{\text{R}}$ punctures having R-sector states and a genus $g_2$ Riemann surface with $n^2_{\text{NS}}$ punctures with NS-sector states and $2n^2_{\text{R}}$ punctures having R-sector states. There can be two possible types of degenerations: 
  \begin{itemize}
\item  The degeneration where an NS-sector state propagates along the tube connecting the component Riemann surfaces: In this case, the genera of the degenerate surfaces are related by the constraint $g=g_1+g_2$. Furthermore, 
$n_{\text{NS}}=n_{\text{NS}}^1+n_{\text{NS}}^2-2$ and $n_{\text{R}}=n^1_{\text{R}}+n^2_{\text{R}}$. We need to distribute   $2g-2+n_{\text{NS}}+n_{\text{R}}$ number of PCOs on the surfaces belong to the inside of the 1PR region of the moduli space such that $2g_1-2+n^1_{\text{NS}}+n^1_{\text{R}}$ number of  PCOs are on the genus $g_1$ component surface and $2g_2-2+n^2_{\text{NS}}+n^2_{\text{R}}$ number of PCOs are on the genus $g_2$ component surface.

\item The degeneration where an R-sector state propagates along the tube connecting the  component Riemann surfaces: In this case, the genera of the degenerate surfaces are related by the constraint $g=g_1+g_2$. Furthermore, $n_{\text{NS}}=n_{\text{NS}}^1+n_{\text{NS}}^2$ and $n_{\text{R}}=n^1_{\text{R}}+n^2_{\text{R}}-1$. We need to distribute $2g-2+n_{\text{NS}}+n_{\text{R}}$ number of PCOs on surfaces belong to the inside of the 1PR region of the moduli space such that $2g_1-2+n^1_{\text{NS}}+n^1_{\text{R}}$ number of  PCOs are on the genus $g_1$ component surface and $2g_2-2+n^2_{\text{NS}}+n^2_{\text{R}}$ number of PCOs are on the genus $g_2$ component surface. The remaining  PCO has to be distributed on a homotopically non-trivial cycle on the tube connecting the two components as follows
\begin{equation}\label{pcocycle}
\chi_0=\bigointssss\frac{dz}{z}\chi(z).
\end{equation}
\end{itemize}

There is a reason for distributing the extra PCO on a homotopically non-trivial cycle on the tube connecting the two components, instead of placing it at a point on the tube. Inserting the extra PCO at a point on the tube will not allow us to interpret  the integration over the moduli of the tube with an R-sector state propagating along it  as a propagator. This is due to the fact that the $b_0$ mode of the reparametrization ghost field does not commute with the picture-changing operator. On the other hand, if we work within the Hilbert space  $\mathcal{H}_0$, then the PCO distributed over a cycle commute with $b_0, L_0$, and $\bar b_0, \bar L_0$. This means that  smearing the PCO on a cycle on the tube allows us to interpret the integration over the moduli of the tube with the R-sector state propagating along it as a propagator \cite{Sen201501}.

\subsubsection*{Singularities Associated with PCOs}
  
 As we discussed in Section \ref{sworlsheet}, the super-reparametrization ghost system $(\beta,\gamma)$ can be replaced  with $(\phi,\xi,\eta)$ system. Let us analyze the following  correlation function involving $\phi,\xi,$ and $\eta$ fields on a genus $g$ Riemann surface \cite{VerlindeVerlinde198702}
 \begin{align}\label{phietxi}
& \left\langle \prod_{i=1}^{n+1}\xi(x_i)\prod_{j=1}^n\eta(y_i)\prod_{k=1}^me^{q_k}\phi(z_k)\right\rangle_{\alpha,\beta}\nonumber\\
=&\frac{\prod_{j=1}^n\Theta[^{\alpha}_{\beta}](-\vec y_j+\sum\vec x-\sum \vec y+\sum q \vec z-2\vec \Delta)}{\prod_{j=1}^{n+1}\Theta[^{\alpha}_{\beta}](-\vec x_j+\sum\vec x-\sum \vec y+\sum q \vec z-2\vec \Delta)}\frac{\prod_{i_1<i_2}E(x_{i_1},x_{i_2})\prod_{j_1<j_2}E(y_{j_1},y_{j_2})}{\prod_{i<j}E(x_{i},y_{j})\prod_{k<l}E(z_k,z_l)^{q_kq_l}\prod_k\sigma(z_k)^{2q_k}},\nonumber\\
&\sum_{k=1}^mq_k=2g-2.
 \end{align}
Here $\Theta[^{\alpha}_{\beta}]$ is the genus $g$ theta function having characteristic $(\alpha,\beta)$ corresponding  to a specific  choice of spin structure, $E(x,y)$ is  the prime form and $\sigma(z)$ is a $\frac{g}{2}$ differential representing the conformal anomaly of the ghost system. $\sum \vec x, \sum \vec y$ and $\sum q \vec z$ denote $\sum_{i=1}^{n+1}\vec x_i,\sum_{j=1}^{n}\vec y_j$ and $\sum_{k=1}^mq_k\vec z_k$, respectively, with 
  \begin{equation}
  \vec x=\bigintssss_{P_0}^x\vec \omega,
  \end{equation}
  where $\omega_i$ stands for  the Abelian differentials on the surface and $P_0$ is an arbitrary point of the surface.   $\vec \Delta$ is the Riemann class vector characterizing the divisor of zeroes of the theta function. Components of the  Riemann class vector  are given by 
  \begin{equation}\label{RiemannClass}
  \Delta_k=\pi\mathrm{i}+\frac{\tau_{kk}}{2}-\frac{1}{2\pi\mathrm{i}}\sum_{j\neq k}\int_{A_j}\omega_j(P)\int_{P_0}^P\omega_k,
  \end{equation}
  where $\tau$ denotes the period matrix and $A_j$ denote the $j^{th}$ A-cycle on the Riemann surface.  The dependence of the Riemann class on $P_0$ will cancel the dependence of $\vec x$  on $P_0$ to make the theta function independent of $P_0$.

\newl There are two kinds of singularities that appear in the  correlation function \eqref{phietxi} :
\begin{itemize}
\item  The prime form $E(x,y)$ has a simple zero at $x=y$. Therefore, the prime forms appearing in the denominator introduce  poles in the correlation function. These poles  correspond to the collision of operators. 

\item  It is known that the Theta function on a genus $g$ Riemann surface has zeroes. Therefore,  the factor $\prod_{j=1}^{n+1}\Theta[^{\alpha}_{\beta}](-\vec x_j+\sum\vec x-\sum \vec y+\sum q \vec z-2\vec \Delta)$ in the denominator also introduces poles in the correlation function. 
 \end{itemize}

Using the identifications \eqref{bgpxe} and operator product expansions of $\eta$ and $\xi$ fields, it is possible to see that this factor in the denominator becomes independent of the locations of the $\beta$'s and $\gamma$'s \cite{Lechtenfeld198911,Morozov199001}. As a result,  the  poles associated with Theta functions do not depend on the locations of insertion of $\beta$ and $\gamma$ fields. Note that the vertex operators and the BRST charge are constructed using the $\beta\gamma$-system, but the picture changing operator contains $\partial\xi,\eta$ and $e^{q\phi}$ factors that can not be expressed as polynomials of $\beta$ and $\gamma$.  This suggests that these poles originate from picture-changing operators in the correlation function. Also, the locations of these poles are functions of the locations of PCOs in the correlation function.

\subsubsection*{A Prescription for Avoiding Spurious Poles}
  
The picture-changing formulation of superstring theory requires inserting  a certain number of PCOs on the worldsheet to define amplitudes. Since PCOs do not correspond to any physical excitations, the singularities introduced by them in the  measure of superstring amplitudes have no physical meaning, and hence are called {\it spurious singularities}. Amplitudes must be free from spurious singularities and, therefore, we need to construct integration cycles that avoid the occurrence of these singularities. Such integration cycles can be constructed systematically by using the {\it vertical integration} prescription  introduced in \cite{Sen201408} and elaborated in \cite{SenWitten201504} (and generalized in \cite{ErlerKonopka201710}). This prescription is based on the following observations

\begin{itemize}
	\item The locations of spurious poles are functions of the positions of PCOs.
 
	\item Changing the position of one PCO at a time depends only on the initial and final locations of that PCO and not on the path. Also this path  is allowed to pass through spurious poles. Such a path will not introduce any singularity. 
\end{itemize} 

We shall now briefly review this prescription  \cite{Sen201408,SenWitten201504}.

The spurious singularity specifies a real codimension-two locus on the worldsheet. The Riemann surface has real dimension two. Therefore, the locations on the Riemann surface which introduce spurious singularity into the integrand form a finite set of points on the surface. We thus need to choose only those sections in ${\wh{\mathcal{P}}_{g,n_{\text{NS}},2n_{\text{R}}}}$ as integration cycle for superstring amplitudes  that do not contain locations of PCOs leading to spurious singularities. However, in general, it is not possible to find such a continuous section over the whole moduli space. Therefore, we need to split the whole moduli space into  different regions and choose sections of ${\wh{\mathcal{P}}_{g,n_{\text{NS}},2n_{\text{R}}}}$ over each region. Each choice must avoid the occurrence of spurious poles. But in a generic situation, these sections will not smoothly join together along the boundaries of different regions. In the vertical integration prescription, a systematic procedure is proposed for the gluing of these sections along their boundaries such that spurious singularities are avoided over each region. This will determine a formal spurious-pole-free integration cycle in $\wh{\mathcal{P}}_{g,n_{\text{NS}},n_{\text{R}}}$. 
 
\paragraph{A Warm-Up Example:} In order to elucidate  the general construction of a spurious-pole-free integration cycle, let us consider the case  with  only one PCO. For simplicity, assume that $\dim_{\mathbb{R}}\left(\mathcal{M}_{g,n}\right)=q=2$, where $n=n_{\text{NS}}+2n_{\text{R}}$. Let us denote by $\wh{\mathcal{P}}^{d}_{g,n_{\text{NS}},2n_{\text{R}}}$ the subspace of $\wh{\mathcal{P}}_{g,n_{\text{NS}},2n_{\text{R}}}$ where only positions of PCOs that avoid spurious singularities are allowed. We shall denote the position of the PCO by $a$ and the point in the moduli space by $m$. Hence, the corresponding off-shell superstring measure takes the following form:
\begin{equation}
\omega_{q}(m,a)=\left\langle\left(\mathcal{X}(a)-\partial_{a}\xi(a)da\right)\wedge\mathcal{O}\right\rangle_q. \label{ExaOme}
\end{equation}
This form is defined on a section of $\wh{\mathcal{P}}^{d}_{g,n_{\text{NS}},2n_{\text{R}}}$. If this section is defined by the function 
\begin{equation}
s:\mathcal{M}_{g,n}\longrightarrow\wh{\mathcal{P}}^{d}_{g,n_{\text{NS}},2n_{\text{R}}},
\end{equation}
 the scattering amplitude $\mathcal{A}_{g,n_{\text{NS}},2n_{\text{R}}}$ is defined by integrating the pull-back of this form to $\mathcal{M}_{g,n}$ under $s$ over  $\mathcal{M}_{g,n}$: 
\begin{equation}
\mathcal{A}_{g,n_{\text{NS}},2n_{\text{R}}}\equiv\bigintssss_{\mathcal{M}_{g,n}}s^{*}(\omega_n)=\bigintssss_{\mathcal{M}_{g,n}} \omega_{n}(m,s(m)).
\end{equation}
\begin{figure}
	\begin{center}
		\begin{tikzpicture}
		\node[right=3pt] (0,0) {$p_{ijk}$};
		\node at (-2,-.2) {$T_i$};
		\node at (0,2) {$T_j$};
		\node at (2,-.2) {$T_k$};
		\draw[line width=2pt] (0,0) -- node[align=center,below,xshift=-.1cm]{$B_{ij}$} (-2,2); 
		\draw[line width=2pt] (0,0) -- node[align=center,below,xshift=.2cm]{$B_{jk}$} (2,2);
		\draw[line width=2pt] (0,0) -- node[align=center,left]{$B_{ki}$} (0,-2);
		\end{tikzpicture}
	\end{center}
	\caption{Triangulation of the two-dimensional moduli space} \label{Tri2Dim}
\end{figure}
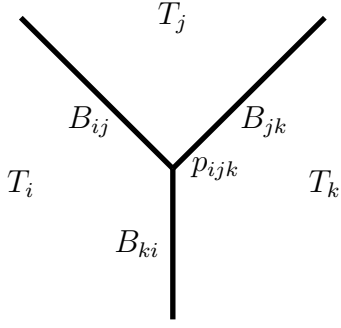
\noindent In general, there exist no global section $s:\mathcal{M}_{g,n}\longrightarrow\wh{\mathcal{P}}^{d}_{g,n_{\text{NS}},2n_{\text{R}}}$. The general idea for circumventing this difficulty is to find a fine tiling of $\mathcal{M}_{g,n}$ and define a local section for each of the tiles in the tiling. Such fine tiling can be given by triangulation of the moduli space.  We denote a triangulation by $T\equiv\bigcup_{i=1}^{\mathrm{\# ~of ~triangles}}T_i$, in which $T_i$'s are triangles  of the triangulation (see Figure \ref{Tri2Dim}).  The scattering amplitude can receive the following three kinds of contributions:
\begin{enumerate}
	\item[(1)]  A set of local sections of $\wh{\mathcal{P}}^{d}_{g,n_{\text{NS}},2n_{\text{R}}}$ are defined on the triangles, by construction. The local section of $\wh{\mathcal{P}}^{d}_{g,n_{\text{NS}},2n_{\text{R}}}$ defined on $T_i$ is given by the following map
	\begin{equation}
	s_i : T_i\longrightarrow \wh{\mathcal{P}}^{d}_{g,n_{\text{NS}},2n_{\text{R}}},\qquad i=1,\ldots,\mathrm{\# ~of ~triangles}.
	\end{equation}
	These sections avoid spurious singularities. Then, the contribution from $T_i$ to the scattering amplitudes is given by:
	\begin{equation}
	\mathcal{A}^{(i)}_{g,n_{\text{NS}},2n_{\text{R}}}\equiv\int_{T_i}s_i^{*}(\omega_{q})=\int_{T_i} \omega_{q}(m,s_i(m)),\quad i=1,\ldots,\mathrm{\# ~of ~triangles}. \label{ConTri}
	\end{equation}
 
	\item[(2)]  In general, the local sections $s_i$ and $s_j$ do not agree on $B_{ij}$. Therefore, we need to consider the appropriate correction factor from $B_{ij}$ to the scattering amplitude, which corresponds to making the sections $s_i$ and $s_j$  agree on $B_{ij}$. We denote the correction factors from various common boundaries of the triangles by $\mathcal{A}^{(ij)}_{g,n_{\text{NS}},2n_{\text{R}}},~~i,j=1,\ldots,\mathrm{\# ~of ~triangles}$.
	\item{(3)} In general, even after making the sections coincide on the boundaries  $B_{ij}$, $B_{jk}$ and $B_{ki}$,   that meet at the vertex of a triangle, the corresponding sections may not agree on the intersection of these boundaries denoted by the points $p_{ijk}$. Therefore, we need to consider the appropriate correction factor from $p_{ijk}$. We denote the correction factors from various common points of the various boundaries of the triangles by $\mathcal{A}^{(ijk)}_{g,n_{\text{NS}},2n_{\text{R}}},~~i,j,k=1,\ldots,\mathrm{\# ~of ~triangles}$. In general, there can be more than three triangles that share the same vertex. We denote the correction factors from the common vertex of $n$ triangles by $\mathcal{A}^{(i_1,\ldots,i_n)}_{g,n_{\text{NS}},2n_{\text{R}}},~~i_1,\ldots,i_n=1,\ldots,\mathrm{\# ~of ~triangles}$. 
	
\end{enumerate}

\begin{figure}
	\begin{center}
		\begin{tikzpicture} 
		
		\coordinate (1) at (-2,0);
		\coordinate (2) at (2,0);
		\coordinate (3) at (0,-3.9);
		
		\node[below] at (0,-3.9) {$m^*\in B_{ij}$};
		\node[above,yshift=.1cm] at (-2,0) {$s_i(m^*)$};
		\node[above,yshift=.1cm] at (2,0) {$s_j(m^*)$};
		\filldraw (0,-3.9) circle (1.5pt);
		\filldraw (-2,0) circle (1.5pt);
		\filldraw (2,0) circle (1.5pt);
		
		\draw[line width=1pt] (-3,0) ellipse  (60pt and 39pt) node[left]{$s_i(m)$};
		\draw[line width=1pt] (3,0) ellipse  (60pt and 39pt) node[right]{$s_j(m)$};
		\draw [line width=1pt,style=dashed] (-2,0) to [out=20,in=160] (0,0) node[above]{$P_{ij}(m,v)$} to [out=-20,in=-160] (2,0);
		
		\draw[line width=2pt] (-6,-5) -- (-4,-3) -- (5,-3) -- (3,-5) -- node[align=center,below]{$\mathcal{M}_{g,n}$} cycle;
		\draw[line width=1pt] (-3,-4.5) -- (2,-3.5);
		
		\draw[line width=1pt] (3) -- (1);
		\draw[line width=1pt] (3) -- (2);
		
		\end{tikzpicture}
	\end{center}
	\caption{Vertical segment for a two-dimensional moduli space. $s_i(m)$ and $s_j(m)$ are sections over $T_i$ and $T_j$. The definition of a vertical segment involves the choice of a curve $P_{ij}(m,v)$ in $\mathcal{R}(m^*)$ that connects sections $s_i(m^*)$ and $s_j(m^*)$ over $m^*\in B_{ij}$. For a fixed $m^*\in B_{ij}$, as the parameter $v$ changes over the interval $\left[0,1\right]$, the curve connects the two sections $s_i(m^*)$ and $s_j(m^*)$ over $m^*$.} \label{VerSegExa}
\end{figure}
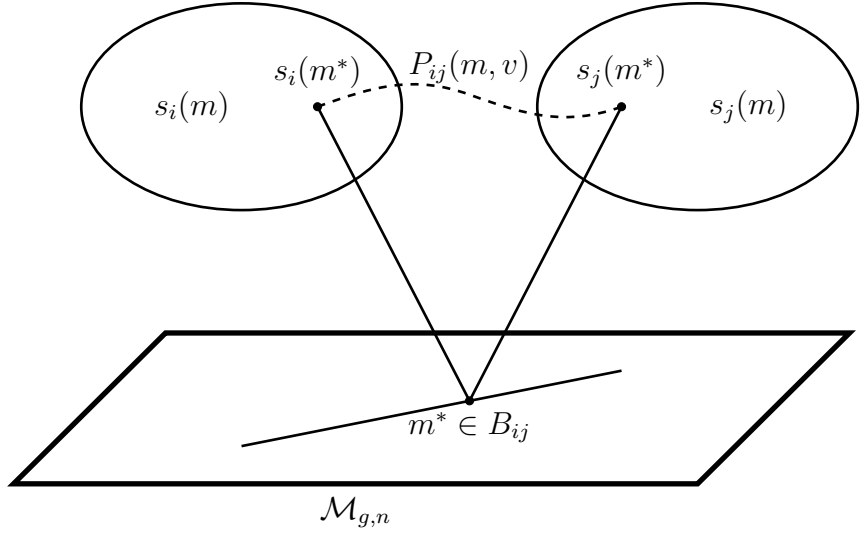

To find the full amplitude $\mathcal{A}_{g,n_{\text{NS}},2n_{\text{R}}}$, we need to find the  expressions for $\mathcal{A}^{(ij)}_{g,n_{\text{NS}},2n_{\text{R}}}$ and $\mathcal{A}^{(i_1,\ldots,i_n)}_{g,n_{\text{NS}}2,n_{\text{R}}}$'s. The determination of these expressions is done by the method of vertical integration. We explore each piece separately:

\begin{itemize}
	\item {\small\it Determining $\mathcal{A}^{(ij)}_{g,n_{\text{NS}},2n_{\text{R}}}$}: The expression for $\mathcal{A}^{(ij)}_{g,n_{\text{NS}},2n_{\text{R}}}$ can be obtained by choosing a vertical segment $\mathcal{V}_{ij}$ over $B_{ij}$. To construct this segment, we choose a point $m\in B_{ij}$ and a curve $P_{ij}(m,v)$ that connects $s_i(m)$ and $s_j(m)$ in $\mathcal{R}(m)\in\wh{\mathcal{P}}^{d}_{g,n_{\text{NS}},2n_{\text{R}}}$. The parameter $v\in\left[0,1\right]$ labels a position along the curve:
	\begin{equation}
	P_{ij}(m,v) : [0,1]\longrightarrow \wh{\mathcal{P}}^{d}_{g,n_{\text{NS}},2n_{\text{R}}}.
	\end{equation}
	The vertical segment can be parametrized as follows
	\begin{equation}
	\mathcal{V}_{ij}\equiv\left\{(m,v)\left|~m\in B_{ij},~v\in\left[0,1\right]\right.\right\}.
	\end{equation}
	To get the correction factor from the boundary $B_{ij}$, $\omega_{q}$ should be integrated over the path $P_{ij}(m,v)$ that connects the sections $s_i(m)$ and  $s_j(m)$ over the $B_{ij}$ instead of the sections themselves (see Figure \ref{VerSegExa}). A point on $P_{ij}(m,v)$  is given by a value of the parameter $v$. Therefore, the form depends on $(m,v)$. From \eqref{ExaOme}, we can first integrate over $v$:

    \begin{eqaligned}\label{DoubleIntersection}
    \mathcal{A}^{(ij)}_{g,n_{\text{NS}},2n_{\text{R}}}&=\int_{\mathcal{V}_{ij}}\omega_{q}(m,v)=\int_{m\in B_{ij}}\int_{v\in\left[0,1\right]}\left\langle\left(\mathcal{X}(m,v)-\partial_{v}\xi(m,v)dv\right)\wedge\mathcal{O}\right\rangle_q
    \\
    &=\int_{m\in B_{ij}}\left\langle\left(\xi(s_i(m))-\xi(s_j(m))\wedge \mathcal{O}\right)\right\rangle_{q-1}.
    \end{eqaligned}
    
    \item {\small\it Determining $\mathcal{A}^{(i_1,\ldots,i_n)}_{g,n_{\text{NS}},n_{\text{R}}}$ for $n\ge 3$}: Let us consider the simplest case, namely $n=3$. The contribution $\mathcal{A}^{(i,j,k)}_{g,n_{\text{NS}},n_{\text{R}}}$ is present only if the vertical segments $\mathcal{V}_{ij}$ over $B_{ij}$, $\mathcal{V}_{jk}$ over $B_{jk}$ and $\mathcal{V}_{ki}$ over $B_{ki}$ do not match over $p_{ijk}$. However, \eqref{DoubleIntersection} is independent of the choice of the path $P_{ij}(m,v)$. Hence, it is independent of the choice of the vertical segment. Therefore, we can always choose the curves $P_{ij}(m,v)$, $P_{jk}(m,v)$ and $P_{ki}(m,v)$ such that vertical segments $\mathcal{V}_{ij}$ over $B_{ij}$, $\mathcal{V}_{jk}$ over $B_{jk}$ and $\mathcal{V}_{ki}$ over $B_{ki}$ match over the triple intersection point $p_{ijk}$. Therefore, there is no contribution from $p_{ijk}$. Similarly, all $\mathcal{A}^{(i_1,\ldots,i_n)}_{g,n_{\text{NS}},n_{\text{R}}}$ can make to vanish for $n\ge 4$.
\end{itemize}

Adding  all the contributions together with appropriate signs gives the final expression for a scattering process involving $n_{\text{NS}}$ external NS states and $2n_{\text{R}}$ R external states whose associate moduli space has (real) two-dimensional moduli space and needs only one PCO insertion. The sign factor can be fixed by specifying the orientations of $B_{ij}$ \cite{SenWitten201504}.

\begin{figure}
	\begin{center}
		\begin{tikzpicture}
		\draw[line width=2pt] (1.5,-1) -- (2,1); 
		\draw[line width=2pt] (1.5,-1) -- (2,-1.5); 
		\draw[line width=2pt] (2,1) -- (-1,2);
		\draw[line width=2pt] (-1,2) -- (-1,-1);
		\draw[line width=2pt] (-1,-1) -- (-1.5,-1.5);
		\draw[line width=2pt] (-1,-1) -- (1.5,-1);
		\draw[line width=2pt] (2,1) -- (3,3);
		\draw[line width=2pt] (1.75,0) -- (5,1);
		\draw[line width=2pt] (-1,2) -- (-3,4);
		\draw[line width=2pt] (-3,4) -- (-4,-1);
		\draw[line width=2pt] (-4,-1) -- (-1,0);
		\draw[line width=2pt] (-3.8,0) -- (-5,3);
		\draw[line width=2pt] (-3,4) -- (3,3);
		\draw[line width=2pt] (-4,-1) -- (-4.5,-1.5);
		\end{tikzpicture}
	\end{center}
	\caption{The dual triangulation of the two-dimensional moduli space. Each codimension-one face (i.e. an edge) is shared by two polyhedra and each codimension-two face (i.e. a vertex) is shared by three polyhedra.} \label{DTin2D}
\end{figure}
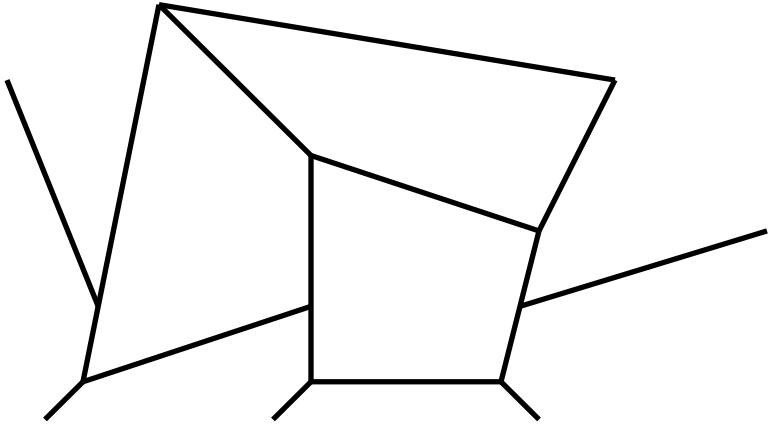

\paragraph{The General Vertical Integration Prescription:} We shall now explain a systematic procedure for finding the continuous integration cycles that avoid spurious poles, for the case of higher dimensional moduli spaces with more than one PCO.  An appropriate tiling of the moduli space can be obtained by considering its dual triangulation. By definition, a {\it dual triangulation $\Upsilon$ of an $n$-dimensional manifold} is given by gluing together the $n$-dimensional polyhedra along their boundary faces. The faces of an $n$-dimensional polyhedron have codimensions $1\le k\le n$. The gluing should be in such a way that every codimension-$k$ face of a polyhedron in $\Upsilon$ is contained in exactly $k+1$ polyhedra in $\Upsilon$ (see Figure \ref{DTin2D}). The latter property of a dual triangulation gives better control over the number of polyhedra that have a common codimension-$k$ face. Therefore, it would be easier to find the correction factors for these faces. Hence, we assume that we have a reasonably-fine dual triangulation $\Upsilon$ of the moduli space.

\newl Denote the codimension-$k$ face that is shared by the codimension zero faces $\mathcal{M}_0^{\alpha_0},\ldots,\mathcal{M}_0^{\alpha_k}$ by $\mathcal{M}_k^{\alpha_0\ldots \alpha_k}$. We can pick an orientation of the $\mathcal{M}_k^{\alpha_0\ldots \alpha_k}$ as follows:
\begin{equation}
\partial\mathcal{ M}_k^{\alpha_0\ldots \alpha_k}=-\sum_{\beta}\mathcal{M}_{k+1}^{\alpha_0\ldots \alpha_k\beta},    \label{orientm}
\end{equation}
where the sum runs over all codimension 0 faces $\mathcal{M}_0^{\beta}$, distinct from $\mathcal{M}_0^{\alpha_0},\ldots,\mathcal{M}_0^{\alpha_k}$, that have nonempty intersection with $\mathcal{M}_k^{\alpha_0\ldots \alpha_k}$.  From \eqref{orientm}, it is clear that the orientation of $\mathcal{M}_k^{\alpha_0\ldots \alpha_k}$  changes sign under $\alpha_i \leftrightarrow \alpha_j$ for any pair $(i,j)$.

\newl To define a well-defined integration cycle, we must remove the locus of spurious singularities. As we saw, these loci correspond to the bad locations of PCOs on the worldsheet. Therefore, we must remove these loci. The information about the locations of  PCOs is encoded in $\wh{\mathcal{P}}_{g,n_{\text{NS}},2n_{\text{R}}}$. If we choose a particular  Riemann surface, i.e. a point in $\mathcal{M}_{g,n},~n=n_{\text{NS}}+2n_{\text{R}}$, the fiber direction of $\wh{\mathcal{P}}_{g,n_{\text{NS}},2n_{\text{R}}}$, in addition to the information about the choice of local coordinates around the punctures up to phases, includes the information about the locations of PCOs. Therefore, we need to remove those points of $\wh{\mathcal{P}}_{g,n_{\text{NS}},2n_{\text{R}}}$ corresponding to  bad locations of  PCOs on the worldsheet which lead to spurious singularities. For this reason, denote by $\wh{\mathcal{P}}^d_{g,n_{\text{NS}},2n_{\text{R}}}$ the subspace of $\wh{\mathcal{P}}_{g,n_{\text{NS}},2n_{\text{R}}}$ in which, in each fiber, one deletes the bad points (points which causes  spurious singularities) at which the PCO should not be inserted. Make $\Upsilon$ to be fine enough so that the projection map $\phi:\wh{\mathcal{P}}^d_{g,n_{\text{NS}},2n_{\text{R}}}\to \mathcal{M}_{g,n}$ has a gluing-compatible section $s^{\alpha}$ over each of the polyhedra $\mathcal{M}_0^{\alpha}$. This means that along the section $s^{\alpha}$, we will not encounter any spurious singularity.

\newl Further conditions have to be imposed on $s^{\alpha}$. Since $\Upsilon$ is a dual triangulation, a codimension-$k$ face of the polyhedron $\mathcal{M}_0^{\alpha}$ is shared with  $k+1$ polyhedra. The $k+1$ sections over these $k+1$ polyhedra come with $k+1$ allowed PCO distributions, denoted by $a^0,\ldots,a^k$, that are free from spurious poles. Each $a^{\alpha}$ stands for a set of $K=2g-2+n_{\text{NS}}+n_{\text{R}}$ points $(z_1^{\alpha},\ldots, z_K^{\alpha})$ where $z_i^{\alpha}(m)\in \mathcal{R}(m)$. Here $\mathcal{R}(m)$ denotes the Riemann surface that corresponds to the point $m$ in  $\mathcal{M}_{g,n}$.  Then, at each point that belongs to the codimension-$k$ face, we can have $(k+1)^K$ possible arrangement of PCOs $(z_1,\ldots,z_K)$ where each $z_i$ can take values $z_i^{\alpha_0},\ldots,z_i^{\alpha_k}$. The additional condition on the section $s^{\alpha}$ is that {\it the PCO distribution associated with each $s^{\alpha}$ should be such that at each point belonging to the codimension-$k$ face, all $(k+1)^K$ possible arrangement of PCOs $(z_1,\ldots,z_K)$ should be free from spurious singularities.} We shall call this condition on the sections $s^{\alpha}$ as {\it spurious-pole-=free condition}. This also means that the distribution of PCOs over the Riemann surface may not go over from one polyhedron $\mathcal{M}_0^{\alpha}$ to $\mathcal{M}_0^{\beta}$ smoothly, there is a discontinuity in the distribution.

\newl The gap in the integration cycle caused by this discontinuity in the PCO distribution can be filled by providing compensating integration cycles.  Let $\mathcal{M}_1^{\alpha\beta}$ be the codimension-$1$ face shared by the codimension-$0$ faces $\mathcal{M}_0^{\alpha}$ and $\mathcal{M}_0^{\beta}$. Then on $\mathcal{M}_1^{\alpha\beta}$, we need to choose a path $P_{\alpha\beta}$ from the PCO locations $(z_1^{\alpha},\ldots,z_K^{\alpha})$ to $(z_1^{(\beta)},\ldots,z_K^{(\beta)})$. If we denote by $\widetilde{\Xi}(m)$ the product $\widetilde{\mathcal{R}}(m)\times\ldots\times \widetilde{\mathcal{R}}(m)$ of $K$ copies of $\widetilde{\mathcal{R}}(m)$, where $\widetilde{\mathcal{R}}(m)$ is the universal cover of $\mathcal{R}(m)$, then $P_{\alpha\beta}$ can be regarded as a path in  $\widetilde{\Xi}$ from the PCO locations on $\mathcal{M}^{\alpha}_0$ to the PCO locations on $\mathcal{M}_0^{\beta}$. Once a path $P_{\alpha\beta}$ has been chosen in this way, we will choose $P_{\beta\alpha}$ to be  $-P_{\alpha\beta}$, i.e. the same path traversed in the opposite direction.  The paths will be constructed by moving PCOs one at a time from an initial location $z_j^{(\alpha)}$ (for some $j$) to a final location $z_j^{(\beta)}$.

\newl For a codimension-$2$ face $\mathcal{M}_2^{\alpha\beta\gamma}$, there are three sets of PCO data: $(z_1^{\alpha},\ldots,z_K^{\alpha}), (z_1^{\beta},\ldots,z_K^{\beta})$, and $(z_1^{\gamma},\ldots,z_K^{\gamma})$. We now consider $3^K$ PCO configurations $(z_1,\ldots,z_K)$ with $z_i$ taking values $z^{(\alpha)}_i, z^{(\beta)}_i$ or $z^{(\gamma)}_i$ for each $i$. Then, we have the  path $P_{\alpha\beta}$ from the origin $(z_1^{\alpha},\ldots,z_K^{\alpha})$ to the point $(z_1^{\beta},\ldots,z_K^{\beta})$, the path $P_{\beta\gamma}$ is a path from $(z_1^{\beta},\ldots,z_K^{\beta})$ to $(z_1^{\gamma},\ldots,z_K^{\gamma})$ and $P_{\gamma\alpha}$ is represented by a path  from $(z_1^{\gamma},\ldots,z_K^{\gamma})$ to $(z_1^{\alpha},\ldots,z_K^{\alpha})$. Together they form a closed path in $\widetilde{\Xi}(m)$. Next, we need to choose a subspace $P_{\alpha\beta\gamma}$ of $\widetilde{\Xi}(m)$, satisfying the following properties:
\begin{itemize}
	\item The boundary of $P_{\alpha\beta\gamma}$ is given by 
	\begin{equation}
	\partial P_{\alpha\beta\gamma}\simeq-P_{\alpha\beta}-P_{\beta\gamma}-P_{\gamma\alpha},   \label{pkalpha1}
	\end{equation}
	where $\simeq$ means that the boundary of $P_{\alpha\beta\gamma}$ can be regarded as a collection of $2$-dimensional subspaces of $\Xi$, which is the product $\mathcal{R}(m)\times\ldots\times \mathcal{R}(m)$ of $K$ copies of $\mathcal{R}(m)$, whose corner points agree with those of the right-hand side of \ref{pkalpha1}. However, the hypercubes themselves may not be identical since we might have used different choices for constructing the faces of various dimensions from the given corner points and might even have used different representatives for some of the PCO locations on $\widetilde{\mathcal{R}}(m)$.
	
    \item $P_{\alpha\beta\gamma}$ is made of a collection of rectangles whose vertices are  points of $\widetilde{\Xi}$ with coordinates $z_i^{\alpha},z_i^{\beta}$ or $z_i^{\gamma}$ and along each rectangle only two coordinates of $\widetilde{\Xi}$ vary. 
	
    \item $P_{\alpha\beta\gamma}$  is chosen to be antisymmetric under  the exchange of any two pairs of its subscripts. 
	
\end{itemize}

We can continue this procedure for higher codimensions. Given a codimension-$k$ face $\mathcal{M}_k^{\alpha_0\ldots \alpha_k}$ shared by $k+1$ codimension zero faces $\mathcal{M}_0^{\alpha_0},\ldots,\mathcal{M}_0^{\alpha_k}$, we can represent the PCO locations determined by the sections $s^{\alpha_0},\ldots,s^{\alpha_k}$ as  points in $\widetilde{\Xi}$, with the  $i$-th PCO location as $z_i^{\alpha_s}$.  The analysis at the previous step would have determined the $(k-1)$-dimensional subspaces $P_{\alpha_0\ldots \alpha_{k-1}}, P_{\alpha_0\ldots \alpha_{k-2}\alpha_k}$, etc., each of which can be represented as $(k-1)$-dimensional subspace of  $\widetilde{\Xi}$ composed of a union of hypercuboids with vertices given by a point in $X^{(k+1)}$ in which 
\begin{equation}
    \begin{gathered}
        X^{(k+1)}\equiv\{(m;a^{0},\ldots,a^{k})|~m\in\mathcal{M}_{g,n},
        \\ 
	\text{and} ~\left\{a^{\alpha}\right\}_{\alpha=1}^k~ \mathrm{are}~ k+1~\text{spurious-pole-free PCO distribution}\},
    \end{gathered}
\end{equation}
and in each hypercuboids only $k-1$ of the coordinates of $\widetilde{\Xi}$ vary. We now have to choose a $k$-dimensional subspace $P_{\alpha_0\ldots\alpha_k}$ of $\widetilde{\Xi}$ satisfying the condition
\begin{equation}
\partial P_{\alpha_0\ldots\alpha_k}\simeq-\sum_{i=0}^k(-1)^{k-i}P_{\alpha_0\ldots \alpha_{i-1}\alpha_{i+1}\ldots \alpha_k}. \label{pkalpha}
\end{equation}
Here, we have the same interpretation for $\simeq$ as above. We don't have to continue this forever. Because it is clear that we must have $k\leq n$ since $\mathcal{M}_k$ has codimension-$k$. Also we must have $k\leq K$ since $P_{\alpha_0\ldots\alpha_k}$ has dimension $k$. Typically, we always have $K\leq n$ and hence $k\leq K$ is the bound.

\newl Once we have chosen all the $P_{\alpha_0\ldots\alpha_k}$ via this procedure, we can formally construct a continuous integration cycle in $\wh{\mathcal{P}}^d_{g,n_{\text{NS}},2n_{\text{R}}}$ as follows. First, for each codimension-zero face $\mathcal{M}_0^{\alpha}$, the section $s^{\alpha}$ gives a section of $\wh{\mathcal{P}}^d_{g,n_{\text{NS}},2n_{\text{R}}}$ over $\mathcal{M}_0^{\alpha}$. Let us call this $\Sigma_{\alpha}$. In a generic situation, $s^{\alpha}$ and $s^{\beta}$ do not match at the boundary $\mathcal{M}_1^{\alpha\beta}$ separating $\mathcal{M}_0^{\alpha}$ and $\mathcal{M}_0^{\beta}$, leaving a gap in the integration cycle between $\Sigma_{\alpha}$ and $\Sigma_{\beta}$.  We fill these gap by including, for each $\mathcal{M}_1^{\alpha\beta}$, a cycle $\Sigma_{\alpha\beta}$ obtained by fibering $P_{\alpha\beta}$ on $\mathcal{M}_1^{\alpha\beta}$ with gluing-compatible coordinates around the punctures.  However since on the codimension-2 face $\mathcal{M}_2^{\alpha\beta\gamma}$,  $P_{\alpha\beta}, P_{\beta\gamma}$ and $P_{\gamma\alpha}$ enclose a non-zero subspace of $\Xi$, the subspaces $\Sigma_{\alpha\beta},\Sigma_{\beta\gamma}$ and $\Sigma_{\gamma\alpha}$ will not meet. This gap will have to be filled by the space $\Sigma_{\alpha\beta\gamma}$ obtained by fibering $P_{\alpha\beta\gamma}$ over $\mathcal{M}_2^{\alpha\beta\gamma}$ with a set of gluing-compatible local coordinates around the punctures.  Proceeding this way, we include all subspaces $\Sigma_{\alpha_0\ldots\alpha_k}$ obtained by fibering $P_{\alpha_0\ldots\alpha_k}$ over $\mathcal{M}_k^{\alpha_0\ldots\alpha_k}$ with a gluing-compatible local coordinates choice. This formally produces a continuous integration cycle in $\wh{\mathcal{P}}^d_{g,n_{\text{NS}},2n_{\text{R}}}$.  We shall call the segments $\Sigma_{\alpha_0\ldots\alpha_k}$ for $k\geq1$ the vertical segments. Note that the vertical segments are allowed to pass through the spurious poles.

\paragraph{The Integration Measure:} The construction of superstring measures is  similar to  those of bosonic-string measures, except for the additional contributions coming from the PCOs. A typical superstring measure is given by
\begin{equation}
\Omega_{p}^{g,n_{\text{NS}},2n_{\text{R}}}=(2\pi \mathrm{i})^{-(3g-3+n)}\langle\mcal{R}|\mathcal{B}_p|\Phi\rangle,~~~~~\mathcal{B}_p=\mathop{\sum_{r=0}^p}_{r\leq J}K^{(r)}\wedge B_{p-r}, \label{supstringmeasure}
\end{equation}
where $J=2g-2+n_{\text{NS}}+n_{\text{R}}$, $B_p$ is the same as \eqref{opevormbos}, $|\Phi\rangle$ is some element of the Hilbert space $\mathcal{H}^{\otimes n}$ of the world-sheet superconformal field theory with the ghost number $n_{\Phi}=p+6-6g$. $\langle \mcal{R}|$ is the surface state and  the inner product between $\langle\mcal{R}|$ and a state $|\Psi_1\rangle\otimes\ldots\otimes |\Psi_n\rangle\in\mathcal{H}^{\otimes n}$  describes the $n$-point correlation function on $\mathcal{R}$ with the vertex operators for $|\Psi_i\rangle$ are inserted at punctures using the local coordinate system $w_i$ around that puncture. $K^{(r)}$ is an operator-valued $r$-forms  along the fiber direction of $\wh{\mathcal{P}}^d_{g,n_{\text{NS}},n_{\text{R}}}$ that corresponds to positions of the PCOs. It is defined as the $r$-form component of 
\begin{equation} 
\mathbf{ K}=\left(\chi(z_1)-\partial \xi(z_1)dz_1\right)\wedge\ldots\wedge\left(\chi(z_{J})-\partial \xi(z_{J})dz_{J}\right).
\end{equation}
The contributions from the codimension-zero faces $\mathcal{M}_0^{\alpha}$ are straightforward to describe; we simply pull back the form $\Omega_{d}^{g,n_{\text{NS}},2n_{\text{R}}},~ d=6g-6+2n_{\text{NS}}+4n_{\text{R}}$  to $\mathcal{M}_0^{\alpha}$ using the section $s^{\alpha}$ and integrate it over $\mathcal{M}^{\alpha}$. We write $\mu_p^{\alpha}(m)=(s^{\alpha})^*(\Omega_{p}^{g,n_{\text{NS}},2n_{\text{R}}})$, so the contribution of $\mathcal{M}^{\alpha}_0$ to the scattering amplitude is given by $\int_{\mathcal{M}^{\alpha}_0}\mu_d^{\alpha}$, which is free  from spurious singularities.

\newl $\Omega_{p}^{g,n_{\text{NS}},2n_{\text{R}}}$ given by \eqref{supstringmeasure} has an important property. Let us consider  some $k$-dimensional region $P$ of $\Xi(m)$, constructed using the procedure described above. Suppose that along $P$, the PCO locations $z_{i_1},\ldots,z_{i_k}$ vary, keeping the other PCO locations fixed. Suppose further that along the edges of $P$ on which $z_i$ varies, its limits are $u_i$ and $v_i$. Then we have 
\begin{equation}
\begin{gathered}
    \int_P \Omega_{d}^{g,n_{\text{NS}},2n_{\text{R}}}=(2\pi \mathrm{i})^{-(3g-3+n)}\langle\Sigma|\widetilde{\mathcal{B}}_{d-k}|\Phi\rangle,
    \\
    \widetilde{\mathcal{B}}_{d-k}=\mathop{\sum_{r=0}}_{r\leq J-k}\widetilde{K}^{(r)}\wedge B_{d-r},   \label{omn1}
\end{gathered}
\end{equation}
where the overall sign has to be fixed from the orientation of the subspace $P$ and  $\widetilde{K}^{(r)}$ is the $r$-form component of 
\begin{equation}
\widetilde{\mathbf{K}}= \prod_{s=1}^k\left(\xi(u_{i_s})-\xi(v_{i_s})\right)\prod_{j=1;j\neq i_1,\ldots,i_k}^J(\chi(z_i)-\partial \xi(z_i)dz^i).
\end{equation}
Since the set $(u_i,v_i,z_i)$ takes values from the set $(z_i^{\alpha_0},\ldots,z_i^{\alpha_J})$, the result is free from spurious singularities as long as the sections over each polyhedron satisfy the spurious pole free condition even if the subspace $P$ contains spurious poles. The full amplitude may then be expressed as 
\begin{equation}
\sum_{k=0}^J(-1)^{k(k+1)/2}\sum_{\{\alpha_0,\ldots,\alpha_k\}}\int_{\mathcal{M}_k^{\alpha_0\ldots\alpha_k}}\int_{P_{\alpha_0\ldots\alpha_k}}\Omega_{d}^{g,n_{\text{NS}},2n_{\text{R}}}.   \label{fullampc}
\end{equation}
We can thus summarize the procedure to construct the off-shell superstring amplitudes as follows:
	\begin{itemize}
		\item Consider a fixed codimension-$k$ face $\mathcal{M}_k^{\alpha_0\dots\alpha_k}$ in the dual triangulation of the moduli space and the $k$-dimensional fiber $P_{\alpha_0\dots\alpha_k}$ over it;
  
		\item Find the contribution to the amplitude from this fiber using \eqref{omn1};
  
		\item Sum over all codimension-$k$ faces $\mathcal{M}_k^{\alpha_0\dots\alpha_k}$ in the dual triangulation of the moduli space. The summation takes into account the orientations of the codimension-$k$ faces represented by the $(-1)^{k(k+1)/2}$ factor in \eqref{fullampc}.
\end{itemize}

\hypertarget{RegulatingIRDivergences}{\subsubsection*{Regulating the Infrared Divergences}}
  
The integration cycle for an off-shell amplitude has to be treated carefully near the boundary of the moduli space where the surface  degenerates. This is because  the off-shell amplitudes  can have infrared divergences from these regions. These regions  correspond to $s\to \infty$ limit of \eqref{plumb}. There are two kinds of infrared divergences that can appear in off-shell amplitudes

 \begin{itemize}
 \item {\it Generic degeneration}: referred to a degeneration in which the state that propagates  through a degenerating cycle carries a generic off-shell momentum.
 
 \item {\it Special degeneration}:  referred to a degeneration in which the state that propagates  through a degenerating cycle is forced to have zero or on-shell momentum.
 \end{itemize}
  
  We can regulate the infrared divergences caused by the generic degenerations by making the analytic continuation  $s\to \mathrm{i}s$ and including a damping factor $e^{-\epsilon s}$ in the integral as $s\to\infty$ \cite{Berera199301,Berera199406}. The infrared divergences caused by the special  degenerations are regulated by restricting the integral over $s$ integral by some upper cut-off \cite{Witten2012a,Witten2012b,Witten2012c,Witten201306}.

%% file: Sections/Hyperbolic_Geometry_and_Off-Shell_Superstring_Ampliudes.tex
In this section, we shall describe an explicit construction of off-shell superstring amplitudes  using hyperbolic surfaces. This construction is similar to the construction of  off-shell bosonic-string amplitudes explained in Section \ref{H Off-Shell b Amplitudes}. Hence, we shall restrict our discussion  to additional details.

\subsection{Decomposition of Off-Shell Superstring Amplitudes}
In this section, we provide a method to decompose an off-shell superstring measure using identities on hyperbolic surfaces. 

\paragraph{Gluing-Compatibility and  the Mirzakhani-McShane Decomposition:} The gluing-compatibility requirement in the case of off-shell superstring amplitudes refers to the gluing-compatible choice of  local coordinates around the punctures where the vertex operators are inserted and the gluing-compatible distribution of PCOs on the surface. We define the 1PR and 1PI regions of the moduli space in the same way as we define in the section \ref{sec:gluing-compatibility and hyperbolic geometry}.  Below, we shall describe a systematic method for making the distribution of PCOs and the choice of local coordinates around the punctures on the Riemann surface gluing-compatible. To this purpose,  let us discuss an important property of the Mirzakhani-McShane identity.

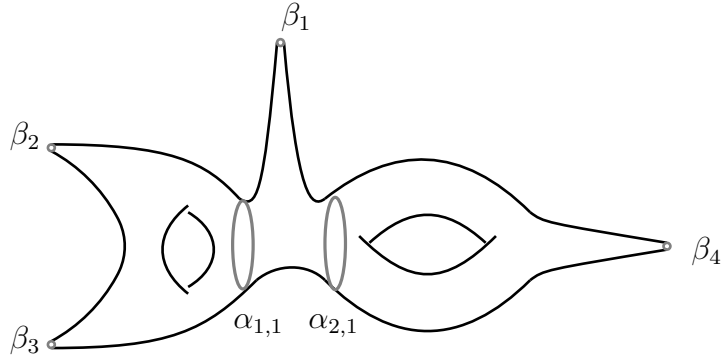
\begin{figure}
\begin{center}
\usetikzlibrary{backgrounds}
\begin{tikzpicture}[scale=.9]
\draw[line width=1pt] (0,1) .. controls (1.75,1) and (2.25,.75)  ..(2.75,.2);
\draw[line width=1pt] (0,-2) .. controls(1.75,-2) and (2.25,-1.75)  ..(3,-1);
\draw[line width=1pt] (0,.9) to[curve through={(.95,-0.05)..(1,-.15)..(1,-.85)..(.95,-.95)}] (0,-1.9);
\draw[gray, line width=1pt] (0,.95) ellipse (.05 and .05);
\draw[gray,line width=1pt] (0,-1.95) ellipse (.05 and .05);
\draw[line width=1pt] (3,-1) .. controls(3.3,-.75) and (3.75,-.75) ..(4,-1);
\draw[line width=1pt] (4,0.2) .. controls(5,1) and (6,1) ..(7,0);
\draw[line width=1pt] (2.75,0.2) .. controls(3,0.05) and (3.1,0.2) ..(3.3,2.5);
\draw[line width=1pt] (4,0.2) .. controls(3.8,0.05) and (3.6,0.2) ..(3.4,2.5);
\draw[line width=1pt] (4,-1) .. controls(5,-2) and (6,-2) ..(7,-1);
\draw[line width=1pt] (7,0) .. controls(7.2,-.15)  ..(9,-.45);
\draw[line width=1pt] (7,-1) .. controls(7.2,-.85)  ..(9,-.55);
\draw[gray, line width=1pt] (9,-.5) ellipse (.05 and .05);
\draw[gray, line width=1pt] (3.35,2.5) ellipse (.05 and .05);
\draw[line width=1pt] (4.5,-.35) .. controls(5.25,-1.1) and (5.75,-1.1) ..(6.5,-.35);
\draw[line width=1pt] (4.65,-.45) .. controls(5.25,0.1) and (5.75,0.1) ..(6.35,-.45);
\draw[line width=1pt] (2,0) .. controls(2.5,-.3) and (2.5,-.8) ..(2,-1.1);
\draw[line width=1pt] (2,0.1) .. controls(1.5,-.3) and (1.5,-.8) ..(2,-1.2);
\draw (0,.75) node[above left] {$\beta_2$}  (0,-2.25) node[above left] {$\beta_3$} (9.2,-.2)node [below right ] {$\beta_4$}  (3.2,2.5)node [above right ] {$\beta_1$} (2.5,-2) node  [above right ] {$\alpha_{1,1}$} (3.6,-2) node  [above right ] {$ \alpha_{2,1}$};
\draw[line width =1.2pt, color=gray]  (2.8,-.48) ellipse (.15 and .65);
\draw[line width =1.2pt, color=gray] (4.15,.-.46) ellipse (.15 and .68);
\end{tikzpicture}
\end{center}

\caption{Cutting the surface along the curve $\alpha_{1,1}+\alpha_{2,1}$ produces a pair of pants with one puncture and 2 borders, a genus 1 surface with 2 punctures and one border, and a genus 1 surface with one puncture and one border.}
\label{cutting31}
\end{figure}

\begin{figure}
\begin{center}
\usetikzlibrary{backgrounds}
\begin{tikzpicture}[scale=.9]
\draw[line width=1pt] (0,1) .. controls (1.75,1) and (2.25,.75)  ..(2.75,.2);
\draw[line width=1pt] (0,-2) .. controls(1.75,-2) and (2.25,-1.75)  ..(3,-1);
\draw[line width=1pt] (0,.9) to[curve through={(.95,-0.05)..(1,-.15)..(1,-.85)..(.95,-.95)}] (0,-1.9);
\draw[gray, line width=1pt] (0,.95) ellipse (.05 and .05);
\draw[gray,line width=1pt] (0,-1.95) ellipse (.05 and .05);
\draw[line width=1pt] (3,-1) .. controls(3.3,-.75) and (3.75,-.75) ..(4,-1);
\draw[line width=1pt] (4,0.2) .. controls(5,1) and (6,1) ..(7,0);
\draw[line width=1pt] (2.75,0.2) .. controls(3,0.05) and (3.1,0.2) ..(3.3,2.5);
\draw[line width=1pt] (4,0.2) .. controls(3.8,0.05) and (3.6,0.2) ..(3.4,2.5);
\draw[line width=1pt] (4,-1) .. controls(5,-2) and (6,-2) ..(7,-1);
\draw[line width=1pt] (7,0) .. controls(7.2,-.15)  ..(9,-.45);
\draw[line width=1pt] (7,-1) .. controls(7.2,-.85)  ..(9,-.55);
\draw[gray, line width=1pt] (9,-.5) ellipse (.05 and .05);
\draw[gray, line width=1pt] (3.35,2.5) ellipse (.05 and .05);
\draw[line width=1pt] (4.5,-.35) .. controls(5.25,-1.1) and (5.75,-1.1) ..(6.5,-.35);
\draw[line width=1pt] (4.65,-.45) .. controls(5.25,0.1) and (5.75,0.1) ..(6.35,-.45);
\draw[line width=1pt] (2,0) .. controls(2.5,-.3) and (2.5,-.8) ..(2,-1.1);
\draw[line width=1pt] (2,0.1) .. controls(1.5,-.3) and (1.5,-.8) ..(2,-1.2);
\draw (0,.75) node[above left] {$\beta_2$}  (0,-2.25) node[above left] {$\beta_3$} (9.2,-.2)node [below right ] {$\beta_4$}  (3.2,2.5)node [above right ] {$\beta_1$} (3.5,-.4) node  [below right ] {$\alpha_{1,2}$} (5.55,.5) node  [below right ] {$ \alpha_{2,2}$};
\draw[line width =1.2pt, color=gray] [rotate around={160:(3.7,-.2)}]  (3.7,-.2) ellipse (1 and .15);
\draw[line width =1.2pt, color=gray] (5.5,.35) ellipse (.15 and .4);
\end{tikzpicture}
\end{center}

\caption{Cutting the surface along the curve $\alpha_{1,2}+\alpha_{2,2}$ produces a pair of pants with one puncture and 2 borders and a genus 1 surface with 3 punctures and 2 borders.}
\label{cutting21}
\end{figure}
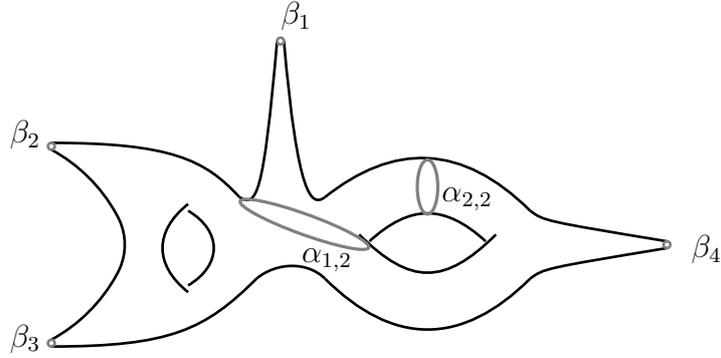

\begin{figure}
\begin{center}
\usetikzlibrary{backgrounds}
\begin{tikzpicture}[scale=.9]
\draw[line width=1pt] (0,1) .. controls (1.75,1) and (2.25,.75)  ..(2.75,.5);
\draw[line width=1pt] (0,-2) .. controls(1.75,-2) and (2.25,-1.75)  ..(3,-1);
\draw[line width=1pt] (0,.9) to[curve through={(.95,-0.05)..(1,-.15)..(1,-.85)..(.95,-.95)}] (0,-1.9);
\draw[gray, line width=1pt] (0,.95) ellipse (.05 and .05);
\draw[gray,line width=1pt] (0,-1.95) ellipse (.05 and .05);
\draw[line width=1pt] (3,-1) .. controls(3.3,-.75) and (3.75,-.75) ..(4,-1);
\draw[line width=1pt] (4,0.5) .. controls(5,1) and (6,1) ..(7,0);
\draw[line width=1pt] (2.75,0.5) .. controls(3,0.35) and (3.1,1) ..(3.3,2.5);
\draw[line width=1pt] (4,0.5) .. controls(3.8,0.35) and (3.6,0.2) ..(3.4,2.5);
\draw[line width=1pt] (4,-1) .. controls(5,-2) and (6,-2) ..(7,-1);
\draw[line width=1pt] (7,0) .. controls(7.2,-.15)  ..(9,-.45);
\draw[line width=1pt] (7,-1) .. controls(7.2,-.85)  ..(9,-.55);
\draw[gray, line width=1pt] (9,-.5) ellipse (.05 and .05);
\draw[gray, line width=1pt] (3.35,2.5) ellipse (.05 and .05);
\draw[line width=1pt] (4.5,-.35) .. controls(5.25,-1.1) and (5.75,-1.1) ..(6.5,-.35);
\draw[line width=1pt] (4.65,-.45) .. controls(5.25,0.1) and (5.75,0.1) ..(6.35,-.45);
\draw[line width=1pt] (2,0) .. controls(2.5,-.3) and (2.5,-.8) ..(2,-1.1);
\draw[line width=1pt] (2,0.1) .. controls(1.5,-.3) and (1.5,-.8) ..(2,-1.2);
\draw (0,.75) node[above left] {$\beta_2$}  (0,-2.25) node[above left] {$\beta_3$} (9.2,-.2)node [below right ] {$\beta_4$}  (3.2,2.5)node [above right ] {$\beta_1$} (0.2,.3) node  [below right ] {$\gamma_2$};
\draw[line width =1.2pt, color=gray] [rotate around={8:(2.4,.2)}]  (2.4,.2) ellipse (1.5 and .09);
\end{tikzpicture}
\end{center}

\caption{Cutting the surface along $\gamma_2$ produces a pair of pants with 2 punctures and one border and a genus-2 surface with 2 punctures and one border.}
\label{cutting41}
\end{figure}
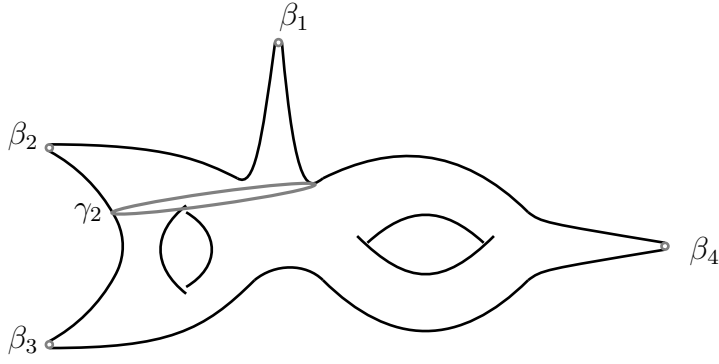

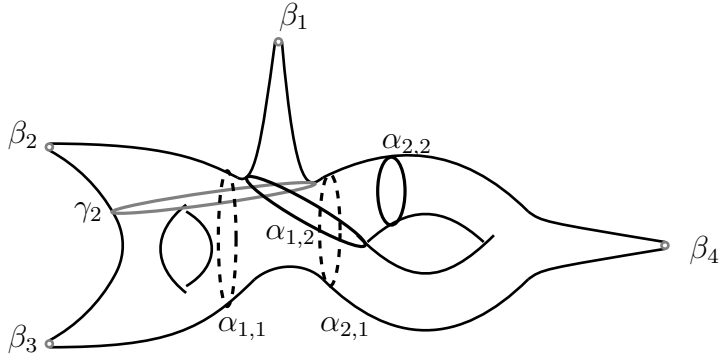
\begin{figure}
\begin{center}
\usetikzlibrary{backgrounds}
\begin{tikzpicture}[scale=.9]
\draw[line width=1pt] (0,1) .. controls (1.75,1) and (2.25,.75)  ..(2.75,.5);
\draw[line width=1pt] (0,-2) .. controls(1.75,-2) and (2.25,-1.75)  ..(3,-1);
\draw[line width=1pt] (0,.9) to[curve through={(.95,-0.05)..(1,-.15)..(1,-.85)..(.95,-.95)}] (0,-1.9);
\draw[gray, line width=1pt] (0,.95) ellipse (.05 and .05);
\draw[gray,line width=1pt] (0,-1.95) ellipse (.05 and .05);
\draw[line width=1pt] (3,-1) .. controls(3.3,-.75) and (3.75,-.75) ..(4,-1);
\draw[line width=1pt] (4,0.5) .. controls(5,1) and (6,1) ..(7,0);
\draw[line width=1pt] (2.75,0.5) .. controls(3,0.35) and (3.1,1) ..(3.3,2.5);
\draw[line width=1pt] (4,0.5) .. controls(3.8,0.35) and (3.6,0.2) ..(3.4,2.5);
\draw[line width=1pt] (4,-1) .. controls(5,-2) and (6,-2) ..(7,-1);
\draw[line width=1pt] (7,0) .. controls(7.2,-.15)  ..(9,-.45);
\draw[line width=1pt] (7,-1) .. controls(7.2,-.85)  ..(9,-.55);
\draw[gray, line width=1pt] (9,-.5) ellipse (.05 and .05);
\draw[gray, line width=1pt] (3.35,2.5) ellipse (.05 and .05);
\draw[line width=1pt] (4.5,-.35) .. controls(5.25,-1.1) and (5.75,-1.1) ..(6.5,-.35);
\draw[line width=1pt] (4.65,-.45) .. controls(5.25,0.1) and (5.75,0.1) ..(6.35,-.45);
\draw[line width=1pt] (2,0) .. controls(2.5,-.3) and (2.5,-.8) ..(2,-1.1);
\draw[line width=1pt] (2,0.1) .. controls(1.5,-.3) and (1.5,-.8) ..(2,-1.2);
\draw (0,.75) node[above left] {$\beta_2$}  (0,-2.25) node[above left] {$\beta_3$} (9.2,-.2)node [below right ] {$\beta_4$}  (3.2,2.5)node [above right ] {$\beta_1$} (0.2,.3) node  [below right ] {$\gamma_2$}(2.3,-1.4) node  [below right ] {$\alpha_{1,1}$}(3.8,-1.4) node  [below right ] {$\alpha_{2,1}$}(3,-.05) node  [below right ] {$\alpha_{1,2}$}(4.7,1.3) node  [below right ] {$\alpha_{2,2}$};
\draw[line width =1.2pt, color=gray] [rotate around={8:(2.4,.2)}]  (2.4,.2) ellipse (1.5 and .09);
\draw[line width =1.2pt, style=dashed] (4.1,-.275) ellipse (.15 and .825);
\draw[line width =1.2pt, style=dashed] (2.6,-.4) ellipse (.13 and 1);
\draw[line width =1.2pt, color=black] (5,.3) ellipse (.2 and .5);
\draw[line width =1.2pt, color=black] [rotate around={60:(3.75,0)}] (3.75,0) ellipse (.15 and 1);
\end{tikzpicture}
\end{center}
\caption{All of the curves $\gamma_2, \alpha_{1,1}+\alpha_{2,1}, \alpha_{1,2}+\alpha_{2,2}$ intersect each other.}
\label{cutting42}
\end{figure}

\paragraph{A Property of the Mirzakhani-McShane Decomposition:} The Mirzakhani-McShane identity for the bordered hyperbolic surface  $\mathcal{R}$ is given by 
  \begin{equation}\label{gmidentity1}
  \sum_{\{\alpha_1,\alpha_2\}\in \mathcal{F}_1}\mathcal{D}(L_1,l_{\alpha_1(\mathcal{R})},l_{\alpha_2(\mathcal{R})})+\sum_{i=2}^n\sum_{\gamma\in \mathcal{F}_{1,i}}\mathcal{E}(L_1,L_i,l_{\gamma}(\mathcal{R}))=L_1.
  \end{equation}
For details about the summation and different quantities appearing in this identity see \hyperlink{MirzakhaniMcShaneIdentity}{The Mirzakhani-McShane Identity}.  Note that the function $\mathcal{D}(x,y,z)$ given by 
\begin{equation}\label{DR1}
   \mathcal{D}(x,y,z)=2~\mathrm{ln}\left( \frac{e^{\frac{x}{2}}+e^{\frac{y+z}{2}}}{e^{\frac{-x}{2}}+e^{\frac{y+z}{2}}}\right),
      \end{equation}
vanishes in the limits $y\to \infty$ keeping $x,z$ fixed and $z\to \infty$ keeping $x,y$ fixed. Similarly, the function $\mathcal{E}(x,y,z)$ given by 
\begin{equation}\label{DR2}
   \mathcal{E}(x,y,z)=x-\mathrm{ln}\left( \frac{\mathrm{cosh}(\frac{y}{2})+\mathrm{cosh}(\frac{x+z}{2})}{\mathrm{cosh}(\frac{y}{2})+\mathrm{cosh}(\frac{x-z}{2})}\right),
   \end{equation}
vanishes in the limit $z\to \infty$ keeping $x,y$ fixed. This limiting behavior has an interesting consequence. Consider a $g$-loop superstring amplitude  with $n_{\text{NS}}$ external NS states and $2n_{\text{R}}$ external R states. The superstring measure corresponding to this process is given by the differential form $\Omega_d^{g,n_{\text{NS}},2n_{\text{R}}}(z_1,\ldots,z_K)$, where $d=6g-6+2(n_{\text{NS}}+2n_{\text{R}})$  is the degree of the form. $\{z_1,\ldots,z_K\}$ denote to the locations of  $K=2g-2+n_{\text{NS}}+n_{\text{R}}$ number of  the PCOs on the surface.  If we forget the subtle issues coming from the spurious singularities, the amplitude $A_{g,n_{\text{NS}},2n_{\text{R}}}$ of such a process can be computed by integrating $\Omega_d^{g,n_{\text{NS}},2n_{\text{R}}}(z_1,\ldots,z_K)$ over $\mathcal{M}_{g,n}$, the moduli space of genus-$g$ surfaces with $n=n_{\text{NS}}+2n_{\text{R}}$ punctures
 \begin{equation}\label{samp}
A_{g,n_{\text{NS}},2n_{\text{R}}}=\int_{\mathcal{M}_{g,n}}\Omega_d^{g,n_{\text{NS}},2n_{\text{R}}}(z_1,\ldots,z_K).
 \end{equation}
Let us insert the Mirzkhani-McShane identity for punctured surfaces
    \begin{equation}\label{gmidentityp1}
  \sum_{\{\alpha_1,\alpha_2\}\in \mathcal{F}_1}\frac{2}{1+e^{\frac{l_{\alpha_1}+l_{\alpha_2}}{2}}}+\sum_{i=2}^n\sum_{\gamma\in \mathcal{F}_{1,i}}\frac{2}{1+e^{\frac{l_{\gamma}}{2}}}=1,
  \end{equation}
inside the integration measure over the moduli space. Following the discussion in \hyperlink{IntegrationOfMCGInvariantFunctionOverModuliSpace}{Integration of MCG-Invariant Functions over the Moduli Space}, we obtain
\begin{eqaligned}\label{sampM}
A_{g,n_{\text{NS}},2n_{\text{R}}}&=\sum_j\int_{\mathcal{M}^{\alpha_{1,j}+\alpha_{2,j}}_{g,n}}dV\left[\frac{2}{1+e^{\frac{l_{\alpha_{1,j}}+l_{\alpha_{2,j}}}{2}}}\right]\Omega_d^{g,n_{\text{NS}},2n_{\text{R}}}(z_1,\ldots,z_K)
\\
&+\sum_{i=2}^n\int_{\mathcal{M}^{\gamma_i}_{g,n}}dV\left[\frac{2}{1+e^{\frac{l_{\gamma_i}}{2}}}\right]\Omega_d^{g,n_{\text{NS}},2n_{\text{R}}}(z_1,\ldots,z_K),
\end{eqaligned}
where the sum over $j$ denotes the summation over distinct pair of curves $\{\alpha_{1,j},\alpha_{2,j}\}$ in $\mathcal{F}_1$ that can not be related via mapping class group action and $\gamma_i$ is a representative curve from the set $\mathcal{F}_{1,i}$. $\mathcal{M}^{\alpha_{1,j}+\alpha_{2,j}}_{g,n}$ denotes the covering space of $\mathcal{M}_{g,n}$ associated with the curve $\alpha_{1,j}+\alpha_{2,j}$. $\mathcal{M}^{\gamma_i}_{g,n}$ denotes the covering space of $\mathcal{M}_{g,n}$ associated with the curve $\gamma_i$. These spaces have been defined in \eqref{mgnlg1}. Note that each term in the decomposition \eqref{sampM} can be associated with a particular boundary of the moduli space. Assume that the curve $\alpha_{1,j}+\alpha_{2,j}$ degenerates. In that situation,  $\{\alpha_{1,j},\alpha_{2,j}\}$ degenerate, the hyperbolic length of at least one of the curves in the pair $\{\alpha_{1,l},\alpha_{2,l}\}$ with $l\neq j$ and all curves $\gamma_i$ in the second sum  becomes infinite. Similarly, if  any of the curves $\gamma_i$ in the second summation degenerates then the hyperbolic length of another curve $\gamma_j$ in the sum with $i\neq j$ and at least one of the curves in all of the pairs $\{\alpha_{1,i},\alpha_{2,i}\}$ in the first sum becomes infinite. This means that when the surface degenerates with  both of the curves in the pair of curves $\{\alpha_{1,j},\alpha_{2,j}\}$   having vanishingly-small hyperbolic lengths, the  finite contribution comes only from the term corresponds to this pair of curves in the right-hand side of equation \ref{sampM}.  Other terms provide only exponentially suppressed contributions (see Figures \ref{cutting31}, \ref{cutting21}, \ref{cutting41}, and \ref{cutting42}).

\paragraph{Maximal Decomposition of Off-Shell Superstring Amplitudes:} Let us decompose the superstring amplitude $A_{g,n_{\text{NS}},2n_{\text{R}}}$ by inserting the Mirzakhani-McShane identity inside the integration measure
\begin{eqaligned}\label{sampM1}
A_{g,n_{\text{NS}},2n_{\text{R}}}&=\sum_j\int_{\mathcal{M}^{\alpha_{1,j}+\alpha_{2,j}}_{g,n}}\left[\frac{2}{1+e^{\frac{l_{\alpha_{1,j}}+l_{\alpha_{2,j}}}{2}}}\right]\Omega_{d}^{g,n_{\text{NS}},2n_{\text{R}}}(z_1,\ldots,z_K)
\\
&+\sum_{i=2}^n\int_{\mathcal{M}^{\gamma_i}_{g,n}}\left[\frac{2}{1+e^{\frac{l_{\gamma_i}}{2}}}\right]\Omega_d^{g,n_{\text{NS}},2n_{\text{R}}}(z_1,\ldots,z_K),
\end{eqaligned}
where the pair of curves  $\{\alpha_{1,j},\alpha_{2,j}\}$ together with a puncture form a pair of pants. The curve $\gamma_i$ together with two of the punctures form a pair of pants. Typical examples of such curves are shown in Figures \ref{cutting31} and \ref{cutting21}. A typical example of the curve $\gamma_i$ is shown in Figure \ref{cutting41}. The covering spaces $\mathcal{M}^{\alpha_{j,1}+\alpha_{j,2}}_{g,n}$ and $\mathcal{M}^{\gamma_i}_{g,n}$ are products of  lower dimensional moduli spaces and cones. There is one cone for cutting along $\gamma_i$ and two cones for cutting along $\alpha_{1,j}$ and $\alpha_{2,j}$. The lower dimensional moduli spaces are the moduli space of the surfaces obtained by cutting the original surface along $\alpha_{1,j}$ and $\alpha_{2,j}$ for the first term and along $\gamma_i$ for the second term.
   
\newl Consider the following term in the above decomposition of the amplitude
\begin{equation}\label{sampM1t}
 \int_{\mathcal{M}^{\alpha_{1,j}+\alpha_{2,j}}_{g,n}}\left[\frac{2}{1+e^{\frac{l_{\alpha_{1,j}}(\mathcal{R})+l_{\alpha_{2,j}}(\mathcal{R})}{2}}}\right]\Omega_d^{g,n_{\text{NS}},2n_{\text{R}}}(z_1,\ldots,z_K).
 \end{equation}
Cutting the surface along $\alpha_{1,j}$ and  $\alpha_{2,j}$ produces the following three components: 1) a pair of pants whose boundary elements are a puncture $\beta_1$ and two borders $\alpha_{1,j}$ and $\alpha_{2,j}$, 2) a genus-$g_j$ surface $\mathcal{R}_{\alpha_{1,j}}$ with $n_j$ punctures and one border $\alpha_{1,j}$, and 3) a genus-$\widehat g_j=g-g_j$ surface $\mathcal{R}_{\alpha_{2,j}}$ with $\widehat n_j=n-n_j$ punctures and one border $\alpha_{2,j}$. For an illustration see Figure \ref{cutting31}. Following the discussion in \hyperlink{IntegrationOfMCGInvariantFunctionOverModuliSpace}{Integration of MCG-Invariant Functions over the Moduli Space}, we can write this term as 

\begin{eqaligned}\label{sampM2t}
& \int_{\mathbb{R}_+^2}dl_{\alpha_{1,j}} dl_{\alpha_{2,j}}\int_0^{ 2^{-M_{\alpha_{1,j}}}l_{\alpha_{1,j}}} \int_0^{2^{-M_{\alpha_{2,j}}}l_{\alpha_{2,j}}} d\tau_{\alpha_{1,j}}d\tau_{\alpha_{2,j}}
\\
&\int_{\mathcal{M}(\mathcal{R}_{\alpha_{1,j}})}dV(\mathcal{R}_{\alpha_{1,j}})\int_{\mathcal{M}(\mathcal{R}_{\alpha_{1,j}})}d V(\mathcal{R}_{\alpha_{2,j}})~\left[\frac{2 \Omega_d^{g,n_{\text{NS}},2n_{\text{R}}}(z_1,\ldots,z_K)}{1+e^{\frac{l_{\alpha_{1,j}}(\mathcal{R})+l_{\alpha_{2,j}}(\mathcal{R})}{2}}}\right].
\end{eqaligned}

We can now insert the Mirzakhani-McShane identity for the surfaces $\mathcal{R}_{\alpha_{1,j}}$  and $\mathcal{R}_{\alpha_{2,j}}$ inside the integration over $\mathcal{M}(\mathcal{R}_{\alpha_{1,j}})$ and $\mathcal{M}(\mathcal{R}_{\alpha_{2,j}})$. This provides a further decomposition of the amplitude. We can continue this procedure for every term in the decomposition \eqref{sampM1}. Since there is a finite number of steps, we eventually end up with only a set of pairs of pants as component surfaces. We shall call the resulting decomposition as {\it the maximal decomposition of the amplitude}. \par 

\newl {\it The most striking feature of the maximal decomposition is that each term in the decomposition corresponds to a specific pants decomposition of the surface. Due to the limiting behavior of the functions appearing in the Mirzakhani-McShane identity, it is clear that when the length of any curve $\gamma$ tends to zero, only those terms in the maximal decomposition whose associated pants decomposition do not contain any curve intersecting with $\gamma$ will be finite.} This property of the maximal decomposition can be used to get an effective description of a gluing-compatible choice of local coordinates and PCOs distribution. Consider the boundary of the 1PI region corresponding to $m$ simple closed geodesics $\gamma_1,\ldots, \gamma_m$ with length $c_*$. Near this boundary, the finite contribution to the amplitude comes from those terms in the maximal decomposition associated with the pants decomposition that contains all of the curves $\gamma_1,\ldots, \gamma_m$. On the other hand, a consistent off-shell superstring amplitude must be gluing-compatible. Combining these two observations, we conclude that the gluing-compatibility must only be applied only for those terms that remain finite after the degenerations of the curves $\gamma_1,\ldots,\gamma_m$. This means that the choice of local coordinates around the punctures and the distribution of PCOs on the surface must be gluing-compatible only for these terms.

\paragraph{Infrared Regulators  and Maximal Decompositions:} Off-shell superstring amplitudes can have infrared divergences coming from the boundary region of the moduli space where the surface degenerates. We shall regulate them by properly choosing the integration cycle in that region. Consider  the boundary region of moduli space corresponding to the  degeneration of a specific curve $\gamma$. Based on the observations that we made above, it is not difficult to see that we need to modify the integration cycle for only those terms in  the maximal decomposition that are associated with the degeneration of curve $\gamma$. For these terms, the integration cycle contains an integration over the hyperbolic length $l_{\gamma}$ of the curve $\gamma$. 

\newl Consider the two types of degenerations defined in \hyperlink{RegulatingIRDivergences}{Regulating the Infrared Divergences}. For generic degeneration near the boundary, we shall deform the integration cycle by changing $l_{\gamma} \to -\mathrm{i}l_{\gamma}$ for $l_{\gamma}\leq c_*$. We also need to multiply the measure with a factor of $\exp{\left(-\frac{2\pi^2}{l_\gamma}\varepsilon\right)}$, where $\varepsilon$ is a small constant. This can be seen from \eqref{ltrelation}, which specifies the relation between the hyperbolic length of the simple closed geodesic on the degenerating collar and the gluing parameter. For the special degeneration, we shall place a lower cut $\frac{2\pi^2}{\Delta}<c_*$, where $\Delta$ is an upper cut-off for the integral over $s$, for the integration over $l_{\gamma}$.

\subsection{Gluing-Compatiblity and Hyperbolic Geometry}
We now describe a gluing-compatible choice of local coordinates around punctures and a gluing-compatible PCO distribution. Together, they provide a well-defined off-shell superstring measure, which might still contain spurious poles. These singularities could then be avoided using the vertical integration prescription. 

\subsubsection*{A Gluing-Compatible Choice of Local Coordinates}
Consider the modified $k$\textsuperscript{th} term in the maximal decomposition of the amplitude. For this term, we use the measure defined using the following coordinates around the punctures 
  \begin{itemize}
 \item  For $ l_{\mathcal{C}^j_k}\geq (1+\epsilon)c_*$ for $j=1,\ldots,3g-3+n$, we choose $e^{\frac{\pi^2}{c_*}}w_{i,m}$ as the local coordinate around the $i$\textsuperscript{th} puncture, where $w_i$ is induced from the hyperbolic metric. 
 
\item Near the  boundary of the 1PI region, where the curves, $\{\widetilde{\mathcal{C}}^k_j,~j=1,\ldots,m\}$, from pair of pants decomposition $\mathcal{P}_k$ degenerate, we choose an annular region $c_*\leq l_{\widetilde{\mathcal{C}}^k_j}<(1+\epsilon)c_*,\quad j=1,\ldots,m $. Inside  this annular regions  use the local  coordinates around the punctures to be $\exp^{\left(\frac{\pi^2}{c_*}\right)}\widetilde{w}_{i,m}$ where $\widetilde{w}_i$ is the local coordinates induced from the  metric \eqref{eq:metric on the 1PI and near its boundary}.

\item Inside the 1PR region defined by   $ l_{\widetilde{\mathcal{C}}^k_j},<c_* ,~j=1,\ldots,m$, we choose the local  coordinates around the punctures to be$\exp^{\left(\frac{\pi^2}{c_*}\right)}\widehat{w}_{i,m}$ where $\widehat{w}_i$ is the local coordinates induced from the  metric \eqref{eq:metric in the 1PR region}.
  \end{itemize}

This concludes our discussion of a choice of gluing-compatible local coordinates around the punctures for defining off-shell superstring measures. We now turn to the distribution of PCOs. 

\subsubsection*{A Gluing-Compatible Distribution of PCOs} 

We shall utilize the above-mentioned property of the maximal decomposition of the superstring amplitude to distribute the PCOs on the worldsheet in a gluing-compatible fashion. This has been constructed in \cite{Pius201808}. We shall discuss the net effect of such a gluing-compatible choice of PCOs distribution.

\newl Consider the $k$\textsuperscript{th} term in the maximal decomposition which is associated with the pants decomposition $\mathcal{P}_k=\left\{\mathcal{C}_k^j,~j=1,\ldots,3g-3+n\right\}$. $\mathcal{C}_k^j$ are non-homotopic simple closed geodesics on the surface with lengths $l_{\mcal{C}_k^j}$. The PCOs distribution described in \cite{Pius201808} has two main features: 1) we need to extend the region of integration to the whole Teichm\"uller space, and 2) the integration measure is multiplied with the following factor
\begin{equation}
\prod\limits_{j=1}^{3g-3+n}\text{sinc}^2\left(\tau_{\mcal{C}_{k}^j}\right),
\end{equation}
where $\tau_{\mcal{C}_{k}^j}$ is the twist parameter along $\mcal{C}_{k}^j$. Within this modified term, we distribute the PCOs as follows. For simplicity, let us assume that all of the off-shell external states are from the NS sector with picture number $-1$. We, therefore, need to distribute $2g-2+n_{\text{NS}}$ PCOs on the surface, where $n_{\text{NS}}$ denotes the number of external states. For this term, let us use the following choice of PCOs distribution. 
\begin{itemize}
 \item  For $ l_{\mathcal{C}_k^j}\geq (1+\epsilon)c_*$ for $j=1,\ldots,3g-3+n_{\text{NS}}$, we distribute one PCO on each of the pairs of pants in the decomposition $\mcal{P}_k$. For each pair of pants, we consider its decomposition into two hexagons. We place one PCO in the center of each hexagon. The PCOs distribution on the corresponding pair of pants is given by the average of PCOs on the two hexagons. 
 
\item  In the complement region, we need to modify the PCOs distribution in order to compensate for the violation of gluing-compatibility due to the difference between the hyperbolic metric and the metric induced from the gluing by the relation \eqref{plumb}. 
\end{itemize}

The off-shell superstring amplitudes defined this  way may contain spurious poles. We thus need to construct an integration cycle that avoids these unwanted divergences. We should again work with each term in the maximal decomposition. Within each term, we can perform vertical integration to make this term free from spurious divergences. As a result, the full superstring amplitude is free from spurious divergences.

%% file: Sections/Discussion_and_Future_Directions.tex
In this paper, we described an explicit construction of off-shell amplitudes in bosonic-string and superstring theories by exploring the hyperbolic geometry of surfaces and moduli spaces thereof. Following the basic ideas in this paper, a calculable formulation of closed bosonic-string and superstring field theories can be constructed \cite{MoosavianPius201706,MoosavianPius201708,Pius201808}. We conclude by collecting some interesting future directions for further investigations

\newl It is known that amplitudes in two-dimensional topological gravity and topological string theory obey certain recursion relations \cite{Witten199001, Kontsevich199206,Mirzakhani200603,DijkgraafVerlindeVerlinde199011}.  It is interesting to check whether some  limits of bosonic-string or superstring amplitudes have a recursive structure. It would be exciting if the field theory limit of superstring amplitudes has such a property. In the case of topological gravity, it is known that the underlying reason for such a recursive structure is the Mirzakhani-McShane identity and the symplectic reduction of the Wolpert form on the moduli space \cite{Wolpert198508,Mirzakhani200603}. This suggests that  by studying the implication of symplectic reduction for the Beltrami differentials associated with the Fenchel-Nielsen vector fields that we used to define string measures, it might be possible to check whether there exists a limit of amplitudes with a recursive structure. 

\newl In order to perform explicit computations of scattering amplitudes in superstring theory following the ideas developed in this paper, we need to express the correlation functions of relevant conformal field theories on a Riemann surface in terms of Fenchel-Nielsen coordinates. It is possible to express the generators of a Fuchsian group in terms of Fenchel-Nielsen coordinates \cite{Maskit199905,Maskit2001}. On the other hand, the correlation function of relevant CFTs (at least in the flat background spacetime) on a surface $\mathcal{R}$ can be written in terms of well-defined quantities like Selberg's zeta functions, which can be expressed through generators of the Fuchsian group corresponding to $\mathcal{R}$ \cite{Kierlanczyk198609,DHokerPhong198612,Sarnak198703}. This might provide a way to express correlation functions in terms of FN coordinates. What we have explored in this paper, together with the progress in this direction would make explicit computations in superstring theory possible.

%% file: Appendices/Hyperbolic_Surfaces_and_their_Teichmuller_Space.tex
In this appendix, we briefly review the theory of hyperbolic surfaces and their Teichm\"uller space. More rigorous discussions on this can be found in  \cite{ImayoshiTaniguchi1992,Hubbard2006,KeenLakic2007}. 

\subsection{Basics of Hyperbolic Geometry}
\newl A hyperbolic surface is a Riemann surface with  a metric whose curvature is $-1$. According to  the Uniformization Theorem \cite{Abikoff198110}, any Riemann surface with genus $g\geq 2$ can be made hyperbolic. For a genus one surface  at least one puncture and for a genus-zero surface at least three punctures are required to make them hyperbolic. The hyperbolic surface $\mathcal{R}$ can be uniformized as  $\mathbb{H}/\Gamma$ for some group $\Gamma$ of isometries of the hyperbolic plane $\mathbb{H}$, defined as 
\begin{equation}
\mathbb{H}=\{z:~\mathrm{Im}~z>0\}.
\end{equation}
The group $\Gamma$ is  the Fuchsian group associated with $\mathcal{R}$.  Each hyperbolic surface $\mathcal{R}$ inherits, by projection from $\mathbb{H}$, its own hyperbolic geometry. 

\newl An open set $\mathcal{F}$ of the upper half-plane $\mathbb{H}$ is a fundamental domain for $\Gamma$ if $\mathcal{F}$ satisfies the following three conditions:

\begin{enumerate}
\item $\msf{g}(\mathcal{F})\cap \mathcal{F}=\emptyset$ for every $\msf{g}\in \Gamma$ with $\msf{g}\not=id$;

\item If $\bar{\mathcal{F}}$ is the closure of $\mathcal{F}$ in $\mathbb{H}$, then $\mathbb{H}=\bigcup_{\msf{g}\in \Gamma}\msf{g}(\bar{\mathcal{F}})$;

\item The relative boundary $\partial\mathcal{ F}$ of $\mathcal{F}$ in $\mathbb{H}$ has measure zero with respect to the two-dimensional Lebesgue measure.
\end{enumerate}

\paragraph{\bf The Poincar\'e Upper Half-Plane:} The hyperbolic metric on  the upper half-plane is given by 
\begin{equation}\label{mDH}
 ds^2_{\bf P}=\frac{dzd\bar z}{(\mathrm{Im}z)^2}.
\end{equation}
The Gaussian curvature $\mbf{C}(h)$ of the Riemannian metric $k(z)^2|dz|^2,~(k(z)>0)$ is given by 
\begin{equation}\label{Gcurvatureh}
\mbf{C}(k)=-\frac{4}{h^2}\frac{\partial ^2\ln k}{\partial z\partial \bar z}.
\end{equation}
Therefore, for the hyperbolic metric, for which $k(z)=\frac{1}{\mathrm{Im}z}$, the Gaussian curvature $\mbf{C}(k)$ is a constant  with the value $-1$.  The isometries of the upper half-plane $\mathbb{H}$ with the hyperbolic metric  are of the form,
\begin{equation}\label{MH}
z\to \frac{az+b}{cz+d}, \qquad a,b,c,d\in \mathbb{R},\qquad ad-bc=1.
\end{equation}
Given the hyperbolic metric $ds^2_{\bf P}$, we define the hyperbolic length $l(C)$ of a curve $C$ on the upper half-plane $\mathbb{H}$ to be 
\begin{equation}\label{curvel}
l(\gamma)=\int_{C} ds_{\bf P}.
\end{equation}
The hyperbolic distance, $\rho(z,w)$, between two points $z$ and $w$ is defined to be the infimum of $l(C)$ taken over all  the curves $C$ in $\mathbb{H}$ that join $z$ to $w$. Such a curve is the geodesic connecting the  points $z$ and $w$ in the hyperbolic plane which is an arc of the circle or the line segment that passes through $z$ and $w$ and is orthogonal to the boundary of the hyperbolic plane.  This distance is known explicitly and is given by 
\begin{equation}\label{hdis}
\rho(z,w)=\mathrm{ln}\left(\frac{1+\tau(z,w)}{1-\tau(z,w)}\right)=2~\mathrm{tanh}^{-1}(\tau(z,w)),
\end{equation}
where $\tau(z,w)=\Big| \frac{z-w}{\bar z- w}\Big| $ in $\mathbb{H}$. \\

   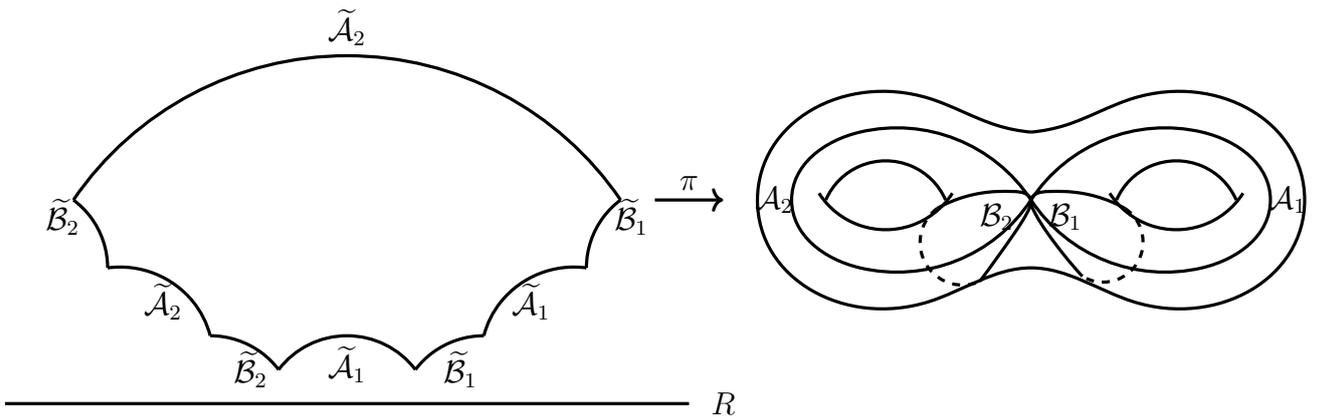
\begin{figure}\centering
   \hspace*{-.9cm}
\usetikzlibrary{backgrounds}
\begin{tikzpicture}[scale=.45]
\draw[black,very  thick] (0,2) to[curve through={(1,2.2)..(3,3)..(8,0) .. (3,-3) ..(0,-2)  .. (-3,-3)..(-8,0)..(-3,3)..(-1,2.2)}] (0,2);
\draw[black, very thick] (2.5,0) to[curve through={(3.5,1)..(5,1)}] (6,0);
\draw[black, very thick] (2.3,.2) to[curve through={(3.5,-.75)..(5,-.75)}] (6.2,.2);
\draw[black,very  thick] (-2.5,0) to[curve through={(-3.5,1)..(-5,1)}] (-6,0);
\draw[black,very  thick] (-2.3,.2) to[curve through={(-3.5,-.75)..(-5,-.75)}] (-6.2,.2);
\draw[very thick] (1.5,-2.25) to[curve through={(.5,-1)..(.05,-.25)..(0,0)..(.5,.25)..(1,.25)..}] (2.75,-.25);
\draw[very thick, style=dashed] (2.75,-.25) to[curve through={(3,-2)}] (1.5,-2.25);
\draw[very thick] (-1.5,-2.4) to[curve through={(-.5,-1)..(-.05,-.25)..(0,0)..(-.5,.25)..(-1,.25)..}] (-2.75,-.25);
\draw[very thick, style=dashed] (-2.75,-.25) to[curve through={(-3,-2)}] (-1.5,-2.4);
\draw[very thick] (0,0) to[curve through={(5,2)..(7,0)..(5,-2)}] (0,0) ;
\draw[very thick] (0,0) to[curve through={(-5,2)..(-7,0)..(-5,-2)}] (0,0);
\draw node at (7.5,0) {$\mathcal{A}_1$};
\draw node at (1,-.5) {$\mathcal{B}_1$};
\draw node at (-7.5,0) {$\mathcal{A}_2$};
\draw node at (-1,-.5) {$\mathcal{B}_2$};
\draw[very thick,->] (-11,0)--(-9,0);
\draw node at (-10,.5) {$\pi$};
\draw[very thick] (-30,-6)--(-10,-6);
\draw node at (-9,-6) {$R$};
\draw[black, very thick] (-22,-5) to[curve through={(-20,-4)}] (-18,-5);
\draw[black, very thick] (-24,-4) to[curve through={(-22.5,-4.5)}] (-22,-5);
\draw[black, very thick] (-27,-2) to[curve through={(-24.5,-3)}] (-24,-4);
\draw[black, very thick] (-28,0)to[curve through={(-27.5,-.5)}] (-27,-2) ;
\draw[black, very thick] (-28,0)to[curve through={(-20,4.25)}] (-12,0) ;
\draw[black, very thick] (-13,-2)to[curve through={(-12.5,-.5)}] (-12,0) ;
\draw[black, very thick] (-13,-2)to[curve through={(-15.5,-3)}] (-16,-4) ;
\draw[black,very  thick] (-16,-4) to[curve through={(-17.5,-4.5)}] (-18,-5);
\draw node [above] at (-20,4.25) {$\widetilde{\mathcal{A}}_2$};
\draw node [below,right] at (-15.5,-3) {$\widetilde{\mathcal{A}}_1$};
\draw node [below,left] at (-24.5,-3) {$\widetilde{\mathcal{A}}_2$};
\draw node [below,right ] at (-12.5,-.5) {$\widetilde{\mathcal{B}}_1$};
\draw node  [below] at(-20,-4) {$\widetilde{\mathcal{A}}_1$};
\draw node [below,left ] at (-27.5,-.5) {$\widetilde{\mathcal{B}}_2$};
\draw node [below,right] at (-17.5,-5) {$\widetilde{\mathcal{B}}_1$};
\draw node [below,left ] at (-22,-5) {$\widetilde{\mathcal{B}}_2$};
\end{tikzpicture}

\caption{The fundamental domain corresponding to a genus 2 surface. }
\label{Fuchsian Fundamental domain}
\end{figure}

\hypertarget{FuchsianUniformization}{\paragraph{The Fuchsian Uniformization:}} The Riemann surface $\mathcal{R}=\mathbb{H}/\Gamma$ can be considered as $\bar{\mathcal{F}}$ with points $\partial\mathcal{F}$ identified under the covering group $\Gamma$. Let $\pi:\mathbb{H}\to \mathcal{R}$ be the projection of $\mathbb{H}$ onto $\mathcal{R}=\mathbb{H}/\Gamma$. Since the Poincar\'e metric $ds^2_{\mbf P}$ is invariant under the action of $\Gamma$, we obtain a Riemannian metric $ds^2_{\mathcal{R}}$ on $\mathcal{R}$ which satisfies $\pi^*(ds^2_{\mathcal{R}})=ds^2_{\mbf P}$. We call  $ds^2_{\mathcal{R}}$ the hyperbolic metric on $\mathcal{R}$. Every $\msf{g}\in \Gamma$ corresponds to an element $[C_{\msf{g}}]$ of the fundamental group $\pi_1(\mathcal{R})$ of $\mathcal{R}$. In particular, $\msf{g}$ determines the free homotopy class of $C_{\msf{g}}$, where $C_{\msf{g}}$ is a representative of the class $[C_{\msf{g}}]$. We say that $\msf{g}$
covers the closed curve $C_{\msf{g}}$. 

\newl The axis $A_{\msf{g}}$ of a hyperbolic  element $\msf{g}$ is the geodesic on $\mathbb{H}$ that connects the fixed points of $\msf{g}$-action on $\mbb{H}$ given by 
\begin{equation}\label{gammaele}
\msf{g}(z)=\frac{az+b}{cz+d},\qquad a,b,c,d\in \mathbb{R},\qquad ad-bc=1.
\end{equation}
The closed curve $L_{\msf{g}}=A_{\msf{g}}/\langle \msf{g} \rangle$, the image on $\mathcal{R}$ of the axis $A_{\msf{g}}$ by the projection $\pi:\mathbb{H}\to \mathcal{R}$, is the unique geodesic with respect to the hyperbolic metric on $\mathcal{R}$ belonging to the free homotopy class of $C_{\msf{g}}$.  We call $L_{\msf{g}}$ the closed geodesic corresponding to $\msf{g}$, or to $C_{\msf{g}}$. Let $\msf{g}$ be a hyperbolic element of $\Gamma$ and $L_{\msf{g}}$ be the closed geodesic corresponding to $\msf{g}$. Then the hyperbolic length $l(L_{\msf{g}})$ of $L_{\msf{g}}$  satisfies,
\begin{equation}\label{lLgamma}
\mathrm{tr}^2(\msf{g}(z))=(a+d)^2=4\mathrm{cosh}^2\left(\frac{l(L_{\msf{g}})}{2}\right).
\end{equation}

\subsection{Teichm\"uller Space of Hyperbolic Surfaces}

In this section, we briefly explain the notion of Teichm\"uller space of a hyperbolic surface. The main references are \cite{ImayoshiTaniguchi1992,Hubbard2006,FarbMargalit2017}.

\newl Let $(g,n)$ denote the signature of a genus-$g$ surface with $n$ boundary components, which uniquely determines the topological classification, i.e. two surfaces with $(g_1,n_1)\ne(g_2,n_2)$ are topologically distinct. If one introduces additional structures such as a complex structure, then, within each topological class determined by $(g,n)$, there may be different surfaces. By definition, a Riemann surface is a one-dimensional complex manifold such that the transition functions of the charts are bi-holomorphic (i.e. the map and its inverse are holomorphic). The charts together with the transition functions between them define a complex structure on the surface. Two such complex structures are equivalent if there is a conformal (i.e complex analytic) map between them. The set of all such equivalent complex structures defines a {\it conformal class} in the set of all complex structures on the surface. However, all conformal classes are not equivalent i.e. the space of all complex structures on a Riemann surface of a given topological type is partitioned by the set of conformal classes. It turns out that there exists a continuum of conformal classes parameterized by a finite number of parameters. Due to the Uniformization Theorem, each conformal class of a Riemann surface contains a unique hyperbolic metric. Therefore, the classification space of all complex structures into conformal classes is the same as the classification space of all hyperbolic structures. Hence, we need a more precise definition of the classification space of hyperbolic structures. Intuitively, this space should classify all hyperbolic structures by some notion of equivalence relation. Therefore, we need to know the answers to two questions: {\it what is the definition of the hyperbolic structure?} and {\it what is the notion of equivalence of two hyperbolic structures?}. The answers are the followings \cite{FarbMargalit2017}
\begin{itemize}
\item {\small\it Hyperbolic structure:} A hyperbolic structure on a  surface $\mathcal{R}$ is a diffeomorphism $\phi:\mathcal{R}\longrightarrow S$, where $S$ is a surface with a finite-area hyperbolic metric and geodesic boundary components. The hyperbolic structure is denoted by the pair $(S,\phi)$.

\item {\small\it Equivalent hyperbolic structures:} Two hyperbolic structures on a  surface $\mathcal{R}$, given by  $\phi_1:\mathcal{R}\longrightarrow S_1$ and $\phi_2:\mathcal{R}\longrightarrow S_2$, are equivalent if there is an isometry $I:S_1\longrightarrow S_2$ such that the maps $I\circ \phi_1 : \mcal{R} \longrightarrow S_1$ and $\phi_2 : \mcal{R}\longrightarrow S_2$ are homotopic i.e. the map $I\circ S_1$ can be continuously deformed into $\phi_2$ by a homotopy map. The homotopies can move points in the boundary of $S_2$. In other words, the following diagram commutes up to homotopy:
\begin{equation}
\begin{tikzcd}
& \mcal{R} \arrow[swap]{dl}{\phi_1} \arrow{dr}{\phi_2}  \\ S_1  \arrow{rr}{I} && S_2
\end{tikzcd}
\end{equation}
Therefore, the relevant equivalence class is the homotopy-equivalence of the hyperbolic structures, i.e. equivalent hyperbolic structures can be deformed into each other by a homotopy map.
\end{itemize} 

In the same way that we can define the classification space of the conformal classes on a Riemann surface, we can define the classification space of the homotopy classes of the hyperbolic structures on a Riemann surface of a given topological type i.e. the space of the hyperbolic structures up to homotopies. This space is called {\it the Teichm\"uller space of the hyperbolic surfaces $\mcal{R}$ of a given topological type $(g,n)$} and it is denoted by $\mathcal{T}(\mcal{R})$.

\subsubsection*{Hyperbolic Pair of Pants}

Consider cutting the Riemann surface $\mathcal{R}$ which admits a hyperbolic metric by a family of mutually-disjoint simple closed geodesics on $\mathcal{R}$. Let $P$ be a compact connected component of the resulting union of subsurfaces. If $P$ contains no simple closed geodesic of $\mathcal{R}$, then $P$ should be homeomorphic to a planar region, say 
\begin{equation}\label{pairofpantsregion}
P_0=\left\{|z|<2\right\}-\left(\left\{|z+1|\leq \frac{1}{2}\right\}\cup\left\{|z-1|\leq \frac{1}{2}\right\} \right),
\end{equation}
which is simply a disc with two holes in it. Such a subsurface $P$, which we call a pair of pants,  can be considered as the smallest pieces that can be used for rebuilding $\mathcal{R}$.  Every boundary of $P$ is a simple closed geodesic on $\mathcal{R}$.

\newl Let us fix a pair of pants $P$ of $\mathcal{R}$ arbitrarily. Assume that the group $\Gamma$ is the Fuchsian model of the surface $\mathcal{R}$ acting on $\mathbb{H}$, and $\pi:\mathbb{H}\to \mathcal{R}=\mathbb{H}/\Gamma$ is the projection. Let $\wt{P}$ be a connected component of $\pi^{-1}(P)$. Denote by $\Gamma_{\wt{P}}$ the subgroup of $\Gamma$ consisting of all elements $\msf{g}$ of $\Gamma$ such that $\msf{g}(\wt{P})=\wt{P}$. Then $\Gamma_{\wt{P}}$ is a free group generated by two hyperbolic transformations, and $P=\wt{P}/\Gamma_{\wt{P}}$. Set $\wh{P}=\mathbb{H}/\Gamma_{\wt{P}}$, i.e. $\Gamma_{\wt{P}}$ is a Fuchsian model of $\wh{P}$. Then $\wh{P}$ is a surface obtained from $P$ by attaching a suitable doubly-connected region along each boundary component. Hence, $\wh{P}$ is again triply connected. Clearly, $P$ is considered as a subsurface of $\wh{P}$, which is a unique pair of pants of $\wh{P}$. In other words, $P$ is uniquely determined by $\Gamma_{\wt{P}}$. $P$ is called the Nielsen kernel of $\wh{P}$, and $\wh{P}$ is called the Nielsen extension of $P$. 

\paragraph{Complex Structure:} Let $L_1,L_2$ and $L_3$ be the boundary components of the pair of pants $P$ and let $\Gamma_0$ be a Fuchsian model of a domain $\mcal{F}$ acting on $\mathbb{H}$. Then $\Gamma_0$ is a free group generated by two hyperbolic transformations $\msf{g}_1$ and $\msf{g}_2$. We may assume that $\msf{g}_1$ and $\msf{g}_2$ cover $L_1$ and $L_2$, respectively (see \hyperlink{FuchsianUniformization}{The Fuchsian Uniformization} for the meaning of covering of a closed curve). It is possible to show that for an arbitrarily triple $(a_1,a_2,a_3)$ of positive numbers, there exists a triply connected planar Riemann surface $\mcal{F}$ such that $l(L_j)=a_j,~j=1,2,3.$ This is due to the fact that a hyperbolic hexagon with edges as geodesic arcs in the upper half-plane meeting perpendicularly at the corners is uniquely determined if we specify the lengths of three edges that do not meet at any corners of the hexagon. Let $D$ be the closed hyperbolic hexagon bounded by $\{C_j,L_j'\}_{j=1}^3$ corresponding to the triple $(a_1,a_2,a_3)$. Let $\eta_j, j=1,2,3,$ be the reflection with respect to $C_j$, i.e., anti-holomorphic automorphism of $\wh{\mathbb{C}}$  preserving $C_j$ point-wise.  Set 
\begin{equation}
    \msf{g}_1\equiv\eta_1\circ\eta_2,\qquad \msf{g}_2\equiv\eta_3\circ\eta_1.
\end{equation}
Then, $\msf{g}_1$ and $\msf{g}_2$ are hyperbolic elements of Aut$(\mathbb{H})$. Let $\Gamma_0$ be the group generated by $\msf{g}_1$ and $\msf{g}_2$. It is clear that $\mcal{F}=\mathbb{H}/\Gamma_0$ is triply connected, and that the unique pair of pants $P$ of $\mcal{F}$ is the interior of the set obtained by identifying the boundary of $D\cup\eta_1(D)$ under the action by $\Gamma_0$.

\newl The complex structure of a pair of pants $P$ is uniquely determined by the hyperbolic lengths of its ordered boundary components. In other words, the action of generators of $\Gamma_0=\langle\msf{g}_1,\msf{g}_2\rangle$\footnote{Here, the notation $\langle\ldots\rangle$ means the group generated by $\ldots$.} are uniquely determined by the values of $a_1,a_2$ and $a_3$. Let us assume that the action of $\msf{g}_1$ and $\msf{g}_2$ on $\mathbb{H}$ is given by 
\begin{align}\label{g1g2action}
\begin{gathered}
    \msf{g}_1(z)=\lambda^2z,~0<\lambda<1,
    \\ 
\msf{g}_2(z)=\frac{az+b}{cz+d},\qquad ad-bc=1,c>0,
\end{gathered}
\end{align}
 and, furthermore, assume that $1$ is the attractive fixed point of $\msf{g}_2$, or equivalently, $a+b=c+d,~0<-\frac{b}{c}<1$. Also, assume that 
 \begin{equation}
 (\msf{g}_3)^{-1}(z)=(\msf{g}_2\circ\msf{g}_1)(z)
 =\frac{\wt az+\wt b}{\wt cz+\wt d},\qquad \wt a\wt d-\wt b\wt c=1.
 \end{equation}
 Then one can show that 
 \begin{eqaligned}
    \left(\lambda+\frac{1}{\lambda}\right)^2&=4\mathrm{cosh}^2\left(\frac{a_1}{2}\right),
 \\
 (a+d)^2&=4\mathrm{cosh}^2\left(\frac{a_2}{2}\right),
 \\
 (\wt a+\wt d)^2&=4\mathrm{cosh}^2\left(\frac{a_3}{2}\right).
 \end{eqaligned}

\paragraph{Teichm\"uller Space:}	To define the Teichm\"uller space of a pair of pants $P$, we need the notion of a marked hyperbolic hexagon. A marked hyperbolic hexagon is a hexagon in the upper half-plane $\mathbb{H}$ with a distinguished vertex and ordered labeling $s_1,\ldots,s_6$ of its sides. The labeling must be in such a way that if we start from the distinguished vertex and move counterclockwise, the order of the sides is from $s_1$ to $s_6$. We denote the lengths of the sides by $l_i=\ell(s_i)$ in which $\ell(s_i)$ is the hyperbolic length of the side computed using the Poincar\'e metric $ds^2_{\bf P}$ on $\mathbb{H}$. We then have

\begin{thm}
    Let $a_1,a_2$ and $a_3$ denote the three positive real numbers. Then we have:
    
	\begin{enumerate}
		\item[(1)] There exists a marked hyperbolic hexagon in $\mathbb{H}$ such that $l_i=a_i$;
  
		\item[(2)] Any two such hyperbolic hexagons can be mapped into each other by an isometry in PSL$(2,\mathbb{R})$ such that the distinguished vertices are mapped into each other;
	\end{enumerate}
\end{thm}

This result basically says that {\it any positive triple $(a_1,a_2,a_3)$ specifies a unique hyperbolic hexagon up to} PSL$(2,\mathbb{R})$ action. If we consider two marked hyperbolic hexagons whose three non-adjacent sides have lengths $\frac{l_i}{2},~i=1,2,3$ and glue them along the other three sides, we get a pair of pants $P$ whose three geodesic boundaries, which we denote by $\gamma_i,~i=1,2,3$, have hyperbolic lengths $\ell_{\gamma_i}=l_i,~i=1,2,3$. Then we have the following result:

\begin{thm}
    The map $\psi : \mathcal{T}_{0,3}(P)\longrightarrow \mathbb{R}_+^{3}$ defined by $\psi(S)=(\ell_{\gamma_1}(S),\ell_{\gamma_2}(S),\ell_{\gamma_3}(S))$, is a homeomorphism.
\end{thm}

This basically means that the Teichm\"uller space of a pair of pants is determined by the lengths of its geodesic boundary components. Therefore, $\ell_{\gamma_i},~i=1,2,3$ can be thought of as coordinates on the Teichm\"uller space. We will explain below how the two facts that (1) any hyperbolic surface admits a pair of pants decomposition and (2) the coordinates on the Teichm\"uller space of a pair of pants is given by the hyperbolic lengths of its geodesic boundaries naturally, leads to a convenient global parametrization of Teichm\"uller space of a hyperbolic surface, the so-called Fenchel-Nielsen coordinates.

    \begin{figure}
\begin{center}
\usetikzlibrary{backgrounds}
\begin{tikzpicture}[scale=1]
\draw[line width=1pt] (1,1) .. controls (1.75,1) and (2.25,.75)  ..(2.75,.2);
\draw[line width=1pt] (1,-2) .. controls(1.75,-2) and (2.25,-1.75)  ..(3,-1);
\draw[line width=1pt] (1,.4) .. controls(1.7,0)  and (1.7,-1) ..(1,-1.4);
\draw[gray,line width=1pt] (1,.7) ellipse (.115 and .315);
\draw[gray,line width=1pt] (1,-1.7) ellipse (.115 and .315);
\draw[line width=1pt] (3,-1) .. controls(3.3,-.75) and (3.75,-.75) ..(4,-1);
\draw[line width=1pt] (4,0.2) .. controls(5,1) and (6,1) ..(7,0);
\draw[line width=1pt] (2.75,0.2) .. controls(3,0.05) and (3.1,0.2) ..(3,1.5);
\draw[line width=1pt] (4,0.2) .. controls(3.8,0.05) and (3.6,0.2) ..(3.7,1.5);
\draw[line width=1pt] (4,-1) .. controls(5,-2) and (6,-2) ..(7,-1);
\draw[line width=1pt] (7,0) .. controls(7.2,-.15)  ..(8,-.15);
\draw[line width=1pt] (7,-1) .. controls(7.2,-.85)  ..(8,-.85);
\draw[gray,line width=1pt] (8,-.5) ellipse (.15 and .35);
\draw[gray,line width=1pt] (3.35,1.5) ellipse (.35 and .15);
\draw[line width=1pt] (4.75,-.75) .. controls(5.25,-1.1) and (5.75,-1.1) ..(6.25,-.75);
\draw[line width=1pt] (4.75,-.35) .. controls(5.25,0.1) and (5.75,0.1) ..(6.25,-.35);
\draw[line width=1pt] (4.75,-.35) .. controls(4.65,-.475) and (4.65,-.625) ..(4.75,-.75);
\draw[line width=1pt] (6.25,-.35) .. controls(6.35,-.475) and (6.35,-.625) ..(6.25,-.75);
\draw[line width =1.4pt, ]  (2.85,-.48) ellipse (.15 and .65);
\draw[line width =1.4pt, ] (4.05,-.42) ellipse (.15 and .62);
\draw[line width =1.4pt] (2.25,0.25) ellipse (.1 and .35);
\draw[line width =1.4pt] (2.25,-1.25) ellipse (.1 and .35);
\draw[line width =1.4pt] (1.75,-.55) ellipse (.225 and .1);
\draw[line width =1.4pt] (5.5,0.375) ellipse (.1 and .365);
\draw[line width =1.4pt] (5.5,-1.365) ellipse (.1 and .35);
\draw[line width=1pt] (2.2,-.5) ellipse (.25 and .4);

\draw (0.25,.5) node[above right] {$b_2$}  (0.25,-2) node[above right] {$b_3$} (8.2,-.2)node [below right ] {$b_4$}  (3.2,1.7)node [above right ] {$b_1$} ;
\end{tikzpicture}
\end{center}

\caption{A pairs of pants decomposition of a genus-2 surface with four boundaries.}
\label{pair of pants decomposition}
\end{figure}
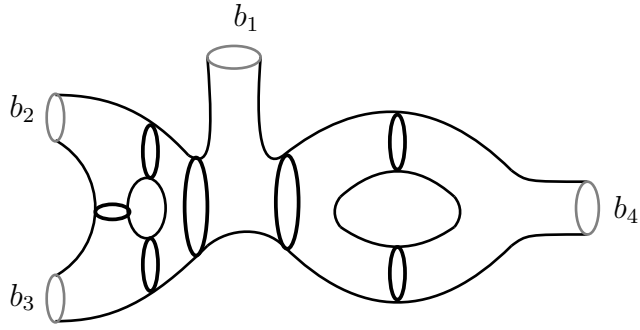

\paragraph{Fenchel-Nielsen Coordinates:}

Let $\mathcal{R}$ be a closed Riemann surface of topological type $(g,n)$. $\mathcal{R}$ can be cut along mutually disjoint simple closed geodesics with respect to the hyperbolic metric $ds_{\mathcal{R}}^2$ on $\mathcal{R}$. When there are no more closed geodesics of $\mathcal{R}$ contained in the remaining open set, then every piece would be a pair of pants $P$. The complex structure of each pair of pants of $\mathcal{R}$ is uniquely determined by a triple of the hyperbolic lengths of boundary geodesics of it, as we explain below. Then $\mathcal{R}$ can be reconstructed by gluing all resulting pairs of pants along their boundaries. Hence, we can consider, as  a system of coordinates for the Teichm\"uller space $\mathcal{T}(\mathcal{R})$, the pair of the set of lengths of all geodesics used in the above decomposition into pants and the set of the so-called twisting parameters used to glue the pieces. Such a system of coordinates is called Fenchel-Nielsen coordinates of $\mathcal{T}(\mcal{R})$. 

\newl Let us explain the precise definition of these coordinates. Consider the pants decomposition $\mathcal{P}=\{P_1,\ldots,P_{2g-2+n}\}$ of $\mathcal{R}$ consisting of all connected components of $\mathcal{R}-\cup_{j=1}^{3g-3+n}L_j$, corresponding to a system of decomposing curves $\mathcal{L}=\{L_1,\ldots,L_{3g-3+n}\}$. This is a set of mutually-disjoint simple closed geodesics on $\mathcal{R}$. Assume that $t$ is a coordinate for the Teilchm\"uller space $\mathcal{F}$ of the corresponding Fuchsian group. We denote the hyperbolic length $l(L_j(t))$ of $L_j(t)$ simply by $l_j(t)$ which we call the geodesic length function of $L_j$. It is known that every geodesic length function $l_j(t)$ is real-analytic on $\mathcal{F}$. For every $j$, let $P_{j,1}$ and $P_{j,2}$ be the two pairs of pants $\mathcal{P}$ having $L_j$ as a boundary component. Here we allow the case where $P_{j,1}=P_{j,2}$. Every pair of pants $P$ has an anti-holomorphic automorphism $J_P$ of order two. Moreover, the set $F_{J_P}=\{z\in P|J_P(z)=z\}$ of all fixed points of $J_P$ consists of three geodesics $\{D_j\}_{j=1}^3$ in $P$ satisfying the condition: For every $j~(j=1,2,3), D_j$ has the endpoints on, and is orthogonal to, both $L_j$ and $L_{j+1}$, where $L_4=L_1$. Take a fixed point of $J_k$ on $L_j$ for each $P_{j,k}~(k=1,2)$, and denote it by $c_{j,k}$.  Let $T_{j}(t)$ be the oriented arc on $L_j(t)$, which lies on $\mathcal{R}_t$ corresponds to the Fuchsian group $\Gamma_t$, from $c_{j,1}(t)$ to $c_{j,2}(t)$. Since $L_j(t)$ has the natural orientation determined from that of $L_j$, we can define the signed hyperbolic length $\tau(j)(t)$ of $T_j(t)$. Then we call $\theta_j(t)=2\pi\frac{\tau_j(t)}{l_j(t)}$, which is well defined modulo $2\pi$,  the twisting parameter with respect to $L_j$. Then the Fenchel-Nielsen coordinates of $\mathcal{T}(\mathcal{R})$ associated with the pants decomposition $\mathcal{P}$ is defined as $(l_1(t),\ldots,l_{3g-3+n}(t),\theta_1(t),\ldots,\theta_{3g-3+n}(t))$.

\subsection{Relation between Different Pants Decompositions}

   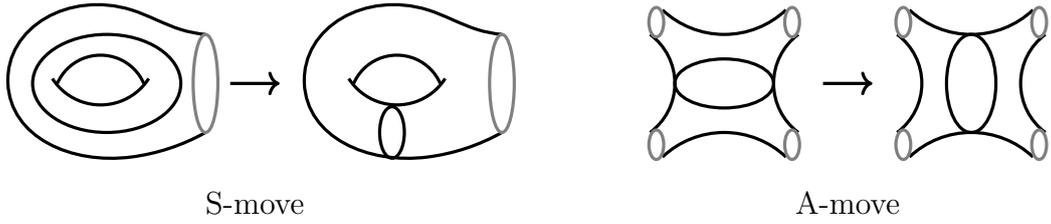
\begin{figure}
\begin{center}
\usetikzlibrary{backgrounds}
\begin{tikzpicture}[scale=.325]
\draw[black, very thick] (0,-2) to[curve through={(-3,-3)..(-8,0)..(-3,3)..(-1,2.2)}] (0,2);
\draw[black, very thick] (-2.5,0) to[curve through={(-3.5,1)..(-5,1)}] (-6,0);
\draw[black, very thick] (-2.3,.2) to[curve through={(-3.5,-.75)..(-5,-.75)}] (-6.2,.2);
\draw[gray, very thick](0,0) ellipse (.5 and 2);
\draw  node[below] at (2,-4) {S-move};
\draw[ very thick](-4,0) ellipse (3 and 2);
\draw[very thick,->] (1,0)--(3,0);
\draw[black, very thick] (12,-2) to[curve through={(9,-3)..(4,0)..(9,3)..(11,2.2)}] (12,2);
\draw[black, very thick] (9.5,0) to[curve through={(8.5,1)..(7,1)}] (6,0);
\draw[black, very thick] (9.7,.2) to[curve through={(8.5,-.75)..(7,-.75)}] (5.8,.2);
\draw[gray, very thick](12,0) ellipse (.5 and 2);
\draw[ very thick](7.55,-2) ellipse (.5 and 1.05);

\draw[black,very  thick] (18,-2) to[curve through={(19,0)}] (18,2);
\draw[black,very  thick] (24,-2) to[curve through={(23,0)}] (24,2);
\draw[black,very  thick] (18.5,-3) to[curve through={(21,-2)}] (23.5,-3);
\draw[black, very thick] (18.5,3) to[curve through={(21,2)}] (23.5,3);
\draw  node[below] at (26,-4) {A-move};
\draw[ gray,very thick](18.25,-2.5) ellipse (.3 and .6);
\draw[gray, very thick](18.25,2.5) ellipse (.3 and .6);
\draw[ gray,very thick](23.75,-2.5) ellipse (.3 and .6);
\draw[gray, very thick](23.75,2.5) ellipse (.3 and .6);
\draw[ very thick](21,0) ellipse (2 and 1);
\draw[very thick,->] (25,0)--(27,0);

\draw[black,very  thick] (28,-2) to[curve through={(29,0)}] (28,2);
\draw[black, very thick] (34,-2) to[curve through={(33,0)}] (34,2);
\draw[black,very  thick] (28.5,-3) to[curve through={(31,-2)}] (33.5,-3);
\draw[black,very  thick] (28.5,3) to[curve through={(31,2)}] (33.5,3);
\draw[ gray,very thick](28.25,-2.5) ellipse (.3 and .6);
\draw[gray, very thick](28.25,2.5) ellipse (.3 and .6);
\draw[ gray,very thick](33.75,-2.5) ellipse (.3 and .6);
\draw[ gray,very thick](33.75,2.5) ellipse (.3 and .6);
\draw[ very thick](31,0) ellipse (1 and 2);

\end{tikzpicture}
\end{center}

\caption{ S and A-move }
\label{ A and S move}
\end{figure}

\begin{figure}\centering
\usetikzlibrary{backgrounds}
\begin{tikzpicture}[scale=.5]
\draw[black, very thick] (3,-2) to[curve through={(0,-3)..(-5,0)..(0,3)..(2,2.2)}] (3,2);
\draw[black, very thick] (0.5,0) to[curve through={(-0.5,1)..(-2,1)}] (-3,0);
\draw[black, very thick] (.7,.2) to[curve through={(-0.5,-.75)..(-2,-.75)}] (-3.2,.2);
\draw[gray, very thick](3,0) ellipse (.5 and 2);
\draw  node[below] at (-1,-4) {(a)};
\draw  node[below] at (-1,-3) {$\mscr{C}_1$};
\draw  node[below] at (2,-1) {$\mscr{C}_2$};
\draw  node[below] at (-5,-1) {$\mscr{C}_3$};
\draw[very  thick](-1,0) ellipse (3 and 2);
\draw[very  thick, rotate around ={35:(-1.9,0)}](-1.9,0) ellipse (2.9 and 2.2);
\draw[ very thick](-1.45,-2) ellipse (.5 and 1.05);

\draw[black,very  thick] (8,-2) to[curve through={(8.5,0)}] (7.1,2);
\draw[black,very  thick] (14,-2) to[curve through={(13.5,0)}] (14.7,1.9);
\draw[black,very  thick] (8.5,-3) to[curve through={(11,-2)}] (13.5,-3);
\draw[black,very  thick] (7.5,3) to[curve through={(9.5,3.3 )}] (10.4,4);
\draw[black,very  thick] (11.6,4) to[curve through={(13,3.1)}] (14.5,3);
\draw[gray,very  thick](8.25,-2.5) ellipse (.3 and .6);
\draw[gray,very  thick](7.25,2.5) ellipse (.3 and .6);
\draw[gray,very  thick](11,4) ellipse (.6 and .3);
\draw[gray,very  thick](13.75,-2.5) ellipse (.3 and .6);
\draw[gray,very  thick](14.75,2.5) ellipse (.3 and .6);
\draw  node[below] at (11,-4) {(b)};
\draw  node[below] at (14.65,1.8) {$\mscr{C}_1$};
\draw  node[below] at (12,-2) {$\mscr{C}_2$};
\draw  node[below] at (7.75,-.25) {$\mscr{C}_3$};
\draw  node[below] at (10,-2) {$\mscr{C}_4$};
\draw  node[below] at (7.3,1.85) {$\mscr{C}_5$};
\draw[ very thick](11,-1) ellipse (2.6 and .5);
\draw[ very  thick, rotate around={25:(10.2,2)}](10.4,2.2) ellipse (2.5 and .5);
\draw[  very thick, rotate around={-25:(11.6,2)}](11.6,2.2) ellipse (2.5 and .5);
\draw[  very thick, rotate around={25:(9.4,.5)}](9.4,.5) ellipse (.5 and 2.7);
\draw[ very  thick, rotate around={-25:(12.6,.5)}](12.6,.5) ellipse (.5 and 2.7);

\draw[black, very thick] (20,-2) to[curve through={(21,0)}] (20,2);
\draw[black, very thick] (26,-2) to[curve through={(25,0)}] (26,2);
\draw[black,very  thick] (20.5,-3) to[curve through={(23,-2)}] (25.5,-3);
\draw[black,very  thick] (20.5,3) to[curve through={(23,2)}] (25.5,3);
\draw[gray,very  thick](20.25,-2.5) ellipse (.3 and .6);
\draw[gray,very  thick](20.25,2.5) ellipse (.3 and .6);
\draw[gray,very  thick](25.75,-2.5) ellipse (.3 and .6);
\draw[gray,very  thick](25.75,2.5) ellipse (.3 and .6);
\draw  node[below] at (23,-4) {(c)};
\draw  node[below] at (23,-2) {$\mscr{C}_1$};
\draw  node[below] at (20.5,.5) {$\mscr{C}_2$};
\draw  node[below] at (26,2) {$\mscr{C}_3$};
\draw[  very thick](23,0) ellipse (.5 and 2);
\draw[ very  thick, rotate around={95:(23,0)}](23,0) ellipse (.5 and 2);
\draw[ very  thick, rotate around={133:(23,0)}](23,0) ellipse (1 and 3.3);
\end{tikzpicture}
\caption{Three relations among sequences of elementary moves.}
\label{fig:Hatch2}
\end{figure}

\newl The building blocks of any hyperbolic surface are pairs of pants. There are infinitely-many  ways that one can decompose a hyperbolic surface $\mathcal{R}$ into pairs of pants unless $\mathcal{R}$ is a pair of pants itself. In fact, there are infinitely-many isotopy classes of pants decomposition for a given hyperbolic surface. These classes are related to each other by the so-called elementary moves \cite{HatcherThurston198003}.  These are moves in which one changes one close curve at a time. Consider a pants decomposition $P$ such that removing a circle $\mscr{C}$ produces complementary components of type $(0,4)$ (a sphere with four boundaries) or $(1,1)$ (a torus with one boundary). This is equivalent to saying that $P$ contains another circle $\mathscr{C}'$ which intersects $\mscr{C}$ transversely in one or two points, respectively, and furthermore, it is disjoint from all other curves contained in $P$. One can then obtain another pants decomposition $P'$ by replacing $\mscr{C}$ with $\mscr{C}'$. These are called associativity or A-moves, and simple or S-moves, respectively. Both of these moves are their own inverse.

\begin{figure}
\begin{center}
\usetikzlibrary{backgrounds}
\begin{tikzpicture}[scale=.4]
\draw[black, very thick] (2.5,-4) to[curve through={(0,-2.5)..(-5,0)..(0,2.5)}] (2.5,4);
\draw[black,very  thick] (0.5,0) to[curve through={(-0.5,.5)..(-2,.5)}] (-3,0);
\draw[black,very  thick] (.7,.2) to[curve through={(-0.5,-.5)..(-2,-.5)}] (-3.2,.2);
\draw[black, very thick] (3,-2.5) to[curve through={(2.5,-1)..(2.5,1)}] (3,2.5);
\draw[gray, very thick](2.75,3.25) ellipse (.5 and .75);
\draw[gray,very thick](2.75,-3.25) ellipse (.5 and .75);
\draw  node[below] at (-3,-.5) {$\mscr{C}_3$};
\draw  node[below] at (1,-3) {$\mscr{C}'_1$};
\draw[very  thick](1.4,0) ellipse (.5 and 3);
\draw[ very thick](-4,0) ellipse (.95 and .8);
\draw[very thick,<->](3.5,0)--(6.5,0);
\draw[very thick,<->](0,-4)--(0,-7);
\draw  node[below] at (5,0) {\bf A};
\draw  node[right] at (0,-5.5) {\bf S};
\draw[black, very thick] (14.5,-4) to[curve through={(12,-2.5)..(7,0)..(12,2.5)}] (14.5,4);
\draw[black, very thick] (12.5,0) to[curve through={(11.5,.5)..(10,.5)}] (9,0);
\draw[black, very thick] (12.7,.2) to[curve through={(11.5,-.5)..(10,-.5)}] (8.8,.2);
\draw[black, very thick] (15,-2.5) to[curve through={(14.5,-1)..(14.5,1)}] (15,2.5);
\draw[gray,very  thick](14.75,3.25) ellipse (.5 and .75);
\draw[gray, very thick](14.75,-3.25) ellipse (.5 and .75);
\draw  node[below] at (9,-.6) {$\mscr{C}_3$};
\draw  node[below] at (15.5,0) {$\mscr{C}_1$};
\draw[very  thick](13.5,0) ellipse (.95 and .8);
\draw[ very thick](8,0) ellipse (.95 and .8);

\draw[very thick,<->](16.5,0)--(19.5,0);
\draw  node[below] at (18,0) {\bf A};
\draw[black,very  thick] (27.5,-4) to[curve through={(25,-2.5)..(20,0)..(25,2.5)}] (27.5,4);
\draw[black, very thick] (25.5,0) to[curve through={(24.5,.5)..(23,.5)}] (22,0);
\draw[black, very thick] (25.7,.2) to[curve through={(24.5,-.5)..(23,-.5)}] (21.8,.2);
\draw[black, very thick] (28,-2.5) to[curve through={(27.5,-1)..(27.5,1)}] (28,2.5);
\draw[gray, very thick](27.75,3.25) ellipse (.5 and .75);
\draw[gray, very thick](27.75,-3.25) ellipse (.5 and .75);
\draw  node[below] at (22,-1.9) {$\mscr{C}'_3$};
\draw  node[below] at (29,.5) {$\mscr{C}_1$};
\draw[very thick,<->](26,-4)--(26,-7);
\draw  node[left] at (26,-5.5) {\bf S};
\draw[very  thick](26.5,0) ellipse (.95 and .3);
\draw[ very thick] (27.5,-1.5) to[curve through={(26.5,-2.25)..(25,-1.75)..(20.5,0)..(25,1.75)..(26.5,2.25)}] (27.5,1.5);
\draw[ very thick] (27.55,-1.5) to[curve through={(27.1,-.9)..(25,-.75)..(21.5,0)..(25,.75)..(27.1,.9)}] (27.55,1.5);
\draw[black,very  thick] (2.5,-16) to[curve through={(0,-14.5)..(-5,-12)..(0,-9.5)}] (2.5,-8);
\draw[black,very  thick] (0.5,-12) to[curve through={(-0.5,-11.5)..(-2,-11.5)}] (-3,-12);
\draw[black, very thick] (.7,-11.8) to[curve through={(-0.5,-12.5)..(-2,-12.5)}] (-3.2,-11.8);
\draw[black,very  thick] (3,-14.5) to[curve through={(2.5,-13)..(2.5,-11)}] (3,-9.5);
\draw[gray,very  thick](2.75,-8.75) ellipse (.5 and .75);
\draw[gray, very thick](2.75,-15.25) ellipse (.5 and .75);
\draw  node[below] at (-3,-12.5) {$\mscr{C}_3$};
\draw  node[below] at (1,-15) {$\mscr{C}'_1$};
\draw[very  thick](1.6,-12) ellipse (.3 and 3.1);
\draw[ very thick](-1.2,-12) ellipse (2.25 and .8);
\draw[very thick,<->](3.5,-12)--(6.5,-12);
\draw  node[below] at (5,-12) {\bf A};
\draw[black,very  thick] (14.5,-16) to[curve through={(12,-14.5)..(7,-12)..(12,-9.5)}] (14.5,-8);
\draw[black, very thick] (12.5,-12) to[curve through={(11.5,-11.5)..(10,-11.5)}] (9,-12);
\draw[black, very thick] (12.7,-11.8) to[curve through={(11.5,-12.5)..(10,-12.5)}] (8.8,-11.8);
\draw[black, very thick] (15,-14.5) to[curve through={(14.5,-13)..(14.5,-11)}] (15,-9.5);
\draw[gray, very thick](14.75,-8.75) ellipse (.5 and .75);
\draw[gray, very thick](14.75,-15.25) ellipse (.5 and .75);
\draw  node[below] at (9,-12.35) {$\mscr{C}_2$};
\draw  node[below] at (15.5,-10) {$\mscr{C}'_2$};
\draw[ very thick, rotate around={25:(11.3,-11.5)}](11.3,-11.55)  ellipse (3.5 and 1.7);
\draw[ very thick](10.8,-12)  ellipse (2.25 and .8);

\draw[very thick,<->](16.5,-12)--(19.5,-12);
\draw  node[below] at (18,-12) {\bf A};
\draw[black, very thick] (27.5,-16) to[curve through={(25,-14.5)..(20,-12)..(25,-9.5)}] (27.5,-8);
\draw[black, very thick] (25.5,-12) to[curve through={(24.5,-11.5)..(23,-11.5)}] (22,-12);
\draw[black, very thick] (25.7,-11.8) to[curve through={(24.5,-12.5)..(23,-12.5)}] (21.8,-11.8);
\draw[black, very thick] (28,-14.5) to[curve through={(27.5,-13)..(27.5,-11)}] (28,-9.5);
\draw[gray,very  thick](27.75,-8.75) ellipse (.5 and .75);
\draw[gray,very  thick](27.75,-15.25) ellipse (.5 and .75);
\draw  node[below] at (22,-14) {$\mscr{C}'_3$};
\draw  node[below] at (24,-11.25) {$\mscr{C}_2$};
\draw[very  thick](23.75,-12) ellipse (2.25 and .7);
\draw[ very thick] (27.5,-13.5) to[curve through={(26.5,-14.25)..(25,-13.75)..(20.5,-12)..(25,-10.25)..(26.5,-9.75)}] (27.5,-10.5);
\draw[ very thick] (27.55,-13.5) to[curve through={(27.1,-12.9)..(25,-12.85)..(21,-12)..(25,-11.2)..(27.1,-11.1)}] (27.55,-10.5);
\end{tikzpicture}
\end{center}

\caption{ }
\label{fig:Hatch3}
\end{figure}

\newl Certain relations hold between sequences of elementary moves. Consider a pants decomposition $P$
\begin{itemize}
    \item[(1)] If removing a circle from $P$ produces a component of type $(0,4)$, then that component contains circles $\mscr{C}_i,i=1,2,3$, as in Figure \ref{fig:Hatch2}(a), such that
    \begin{equation}
        \mscr{C}_1\xrightarrow{\text{S-move}}\mscr{C}_2\xrightarrow{\text{S-move}}\mscr{C}_3\xrightarrow{\text{S-move}}\mscr{C}_1.
    \end{equation}

    \item[(2)] If removing two circles from $P$ produces a component of type $(0,5)$, then that component contains circles $\mscr{C}_i,i=1,\ldots,5$, as in Figure \ref{fig:Hatch2}(b), such that 
    \begin{equation}
        \begin{aligned}
        &\{\mscr{C}_1,\mscr{C}_3\}\xrightarrow{\text{A-move}}\{\mscr{C}_1,\mscr{C}_4\}\xrightarrow{\text{A-move}}\{\mscr{C}_2,\mscr{C}_4\}\xrightarrow{\text{A-move}}
        \\
        \xrightarrow{\text{A-move}}&\{\mscr{C}_2,\mscr{C}_5\}\xrightarrow{\text{A-move}}\{\mscr{C}_3,\mscr{C}_5\}\xrightarrow{\text{A-move}}\{\mscr{C}_3,\mscr{C}_1\}.
        \end{aligned}
    \end{equation}

    \item[(3)] If removing a circle from $P$ produces a component of type $(1,1)$, then that component contains circles $\mscr{C}_i,i=1,2,3$, as in Figure \ref{fig:Hatch2}(c), such that
    \begin{equation}
        \mscr{C}_1\xrightarrow{\text{A-move}}\mscr{C}_2\xrightarrow{\text{A-move}}\mscr{C}_3\xrightarrow{\text{A-move}}\mscr{C}_1.
    \end{equation}

    \item[(4)] If removing two circles from $P$ produces a component of type $(1,2)$, then that component contains circles $\mscr{C}_i,i=1,2,3$ and $\mscr{C}'_i,i=1,2,3$, as in Figure \ref{fig:Hatch3}, such that
    \begin{equation}
        \begin{aligned}
        &\{\mscr{C}_1,\mscr{C}_3\}\xrightarrow{\text{A-move}}\{\mscr{C}_1,\mscr{C}'_3\}\xrightarrow{\text{S-move}}\{\mscr{C}_2,\mscr{C}'_3\}\xrightarrow{\text{A-move}}
        \\
        \xrightarrow{\text{A-move}}&\{\mscr{C}_2,\mscr{C}'_2\}\xrightarrow{\text{A-move}}\{\mscr{C}_2,\mscr{C}'_1\}\xrightarrow{\text{S-move}}\{\mscr{C}_3,\mscr{C}'_1\}\xrightarrow{\text{A-move}}\{\mscr{C}_3,\mscr{C}_1\}.
        \end{aligned}
    \end{equation}

    \item[(5)] If the support of two moves of either type is disjoint in $\mathcal{R}$, then they commute. The commutator of such moves is a cycle containing four elementary moves. 
\end{itemize}

We refer to \cite{HatcherThurston198003,Hatcher199906} for further details on relations between pants decomposition and mapping-class groups of hyperbolic surfaces.

%% file: Appendices/Matrices_for_Fenchel-Nielsen_Coordinates.tex
   In this section we explain how the generates of a Fuchsian group can be written in terms of Fenchel-Nielsen coordinates for the Teichm\"uller space of $\Gamma$ following Maskit \cite{Maskit2001} (see also \cite{Maskit199905}).  These results should be relevant for the computation of correlation functions of CFTs on a given Riemann surface $\mcal{R}$ in terms of Fenchel-Nielsen coordinates. The reason is that these correlation functions are certain determinants that can be expressed in terms of generators of the Fuchsian group associated with $\mathcal{R}$ through quantities like Selberg's zeta function \cite{Kierlanczyk198609,DHokerPhong198612,Sarnak198703}.

\subsection {The Fenchel-Nielsen System}
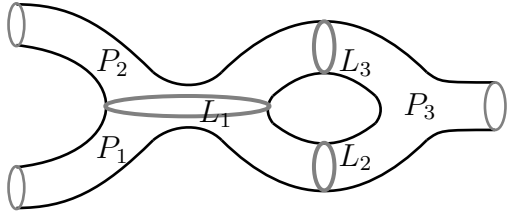
\begin{figure}
\begin{center}
\usetikzlibrary{backgrounds}
\begin{tikzpicture}[scale=.9]
\draw[line width=1pt] (1,1) .. controls (1.75,1) and (2.25,.75)  ..(3,0);
\draw[line width=1pt] (1,-2) .. controls(1.75,-2) and (2.25,-1.75)  ..(3,-1);
\draw[line width=1pt] (1,.4) .. controls(2.75,.2)  and (2.75,-1.2) ..(1,-1.4);
\draw[gray,line width=1pt] (1,.7) ellipse (.115 and .315);
\draw[gray,line width=1pt] (1,-1.7) ellipse (.115 and .315);
\draw[line width=1pt] (3,0) .. controls(3.3,-.25) and (3.75,-.25) ..(4,0);
\draw[line width=1pt] (3,-1) .. controls(3.3,-.75) and (3.75,-.75) ..(4,-1);
\draw[line width=1pt] (4,0) .. controls(5,1) and (6,1) ..(7,0);
\draw[line width=1pt] (4,-1) .. controls(5,-2) and (6,-2) ..(7,-1);
\draw[line width=1pt] (7,0) .. controls(7.2,-.15)  ..(8,-.15);
\draw[line width=1pt] (7,-1) .. controls(7.2,-.85)  ..(8,-.85);
\draw[gray,line width=1pt] (8,-.5) ellipse (.15 and .35);
\draw[line width=1pt] (4.75,-.75) .. controls(5.25,-1.15) and (5.75,-1.15) ..(6.25,-.75);
\draw[line width=1pt] (4.75,-.35) .. controls(5.25,0.1) and (5.75,0.1) ..(6.25,-.35);
\draw[line width=1pt] (4.75,-.35) .. controls(4.65,-.475) and (4.65,-.625) ..(4.75,-.75);
\draw[line width=1pt] (6.25,-.35) .. controls(6.35,-.475) and (6.35,-.625) ..(6.25,-.75);
\draw[line width =1.5pt, color=gray] (3.5,-.5) ellipse (1.2 and .15);
\draw[line width =1.5pt, color=gray] (5.5,.375) ellipse (.15 and .375);
\draw[line width =1.5pt, color=gray] (5.5,-1.39) ellipse (.15 and .35);
\draw (2,.5) node[below right] {$P_2$}  (2,-1.5) node[above right] {$P_1$} (6.5,-.5)node [ right ] {$P_3$}  (3.5,-.25) node  [below right ] {$L_1$} (5.55,-1.25) node  [ right ] {$L_2$} (5.55,0.5) node  [below right ] {$L_3$};
\end{tikzpicture}
\end{center}

\caption{Ordered coordinate geodesics and pairs of pants. Thick gray curves are the coordinate-geodesics. $L_1$ and $L_2$ are attaching  geodesics and $L_3$ is a handle geodesic.}
\label{orderedpair1}
\end{figure}

  Consider a genus-$g$ hyperbolic surface $\mathcal{R}$ with $n$ punctures, corresponding to the Fuchsian group $\Gamma$, so that $\mathcal{R}=\mathbb{H}/\Gamma$. There are at most $p=3g-3+n$ distinct  disjoint simple closed geodesics $L_1,\ldots, L_p$ on $\mathcal{R}$.  We shall call these $p$ simple geodesics the {\it coordinate geodesics}. The coordinate geodesics divide $\mathcal{R}$ into $q=2g-2+n$ pairs of pants, $P_1,\ldots,P_q$.  Each coordinate geodesic corresponds to  either a boundary element of two distinct pairs of pants or corresponds to two boundary elements of the same pair of pants.  Let us order the coordinate geodesics  and  the pairs of pants with the following set of rules, although there is no canonical way to do this.
\begin{itemize}
\item If there is a  coordinate geodesic, then $L_1$ is dividing and lies between $P_1$ and $P_2$. 

\item If $q\geq 3$, then $L_2$ lies between $P_1$ and $P_3$ (see Figure \ref{orderedpair1}).

\item The first $q-1$ coordinate geodesics, and the $q$ pairs of pants, $P_1,\ldots,P_q$, are ordered so that, for every $j=3,\ldots,q-1$, there is an $i=i(j)$, with $1\leq i(j)<j$, so that $L_j$ lies on the common boundary of $P_i$ and $P_j$. 

\item The coordinate geodesics $L_1,\ldots, L_{q-1}$ are called the attaching geodesics and coordinate geodesics $L_q,\ldots,L_p$ are called the handle geodesics.  

\end{itemize}

We shall call this ordered set as a Fenchel-Nielsen system (FN system). Denote  the Fenchel-Nielsen coordinates  defined with respect to this FN system as
\begin{equation}\label{FNcord}
\Phi=(l_1,\ldots,l_{p},\tau_1,\ldots,\tau_p)\in (\mathbb{R^+})^{p}\times\mathbb{R}^p,
\end{equation}
where $l_i,~i=1,\ldots,p$ are the geodesic lengths of curves $L_i,~i=1,\ldots,p$ and $\tau_i,~i=1,\ldots,p$ are the twist about $L_i,~i=1,\ldots,p$. The length of a geodesic is an absolute concept, but the twist about a geodesic is not an absolute quantity. It requires a reference surface. Let us call this reference surface $\mathcal{R}_0$. For convenience, let us introduce the parameters:
\begin{eqaligned}
    s_i&\equiv\mathrm{sinh}~\frac{l_i}{2}, &\qquad i&=1,\ldots,p,
    \\
    t_i&\equiv\mathrm{sinh}~\frac{\tau_i}{2}, &\qquad i&=1,\ldots,p.
\end{eqaligned}

\paragraph{The Ordering for the Boundary Components of the Pairs of Pants:} The FN system contains an ordered set of geodesics on $\mathcal{R}$. Let us assign an ordering for the boundary components of the pairs of pants obtained from $\mathcal{R}_0$ as a result of  a pair of pants decomposition of $\mathcal{R}$ along the coordinate geodesics of the FN system. In most cases, the ordering of the coordinate geodesics and boundary elements of $\mathcal{R}_0$ induces an ordering of the boundaries of the pairs of pants. There are two exceptional cases in which there are two boundary elements of a pair of pants $P$ corresponding to just one coordinate geodesic of $\mathcal{R}_0$. They are the following. 

\begin{itemize}
\item[(1)] $\mathcal{R}_0$ is a torus with one  puncture. In this case, two of the boundary elements of the pair of pants $P$ are hyperbolic, necessarily of the same size, and the third one is parabolic. It is not possible to differentiate between the two boundary elements of $P$ corresponding to one coordinate geodesic on $\mathcal{R}_0$. 

\item[(2)] The pair of pants $P\subset \mathcal{R}_0$ with one boundary element, $ b_1$, corresponds to an attaching geodesic on $\mathcal{R}_0$. While the other two boundaries $b_2, b_3$ both correspond to the same handle geodesic. Which boundary element of $P$ we call $b_2$ or $b_3$ is arbitrary. 
\end{itemize} 

 If the surface is not an example of an exceptional case then we shall label the  boundary elements of the pair of pants $P_i$ as $b_{i,1},b_{i,2},b_{i,3}$, in the following order: {\it hyperbolic boundary elements come before the parabolic boundary elements}. 

\paragraph{Directing the Coordinate and the Boundary Geodesics:} Since the twist parameter around a curve depends on the orientation, we need to assign a direction for all coordinate geodesics on $\mathcal{R}_0$. We say that  two geodesics on the boundary of some pair of pants $P$ are consistently oriented with respect to $P$, if $P$ lies on the right as we traverse either geodesic in the positive direction. Assume that we already directed geodesics $L_1,\ldots,L_j,~j\geq1$. If $L_{j+1}$ lies on the boundary of two distinct pairs of pants or is a boundary geodesic, then there is the lowest index $i$, so that $L_{j+1}$ lies on the boundary of $P_i$. Since $j+1>1,~L_{j+1}$ corresponds to either $b_{i,2}$ or $b_{i,3}$, for $b_{i,1}$ must correspond to some attaching geodesic, $L_{j'},~j'\leq j$. We direct $L_{j+1}$ so that $L_{j+1}$ and $L_{j'}$ are consistently oriented as boundary elements of $L_i$. If $L_{j+1}$ is a handle geodesic, with the same pair of pants $P_i$ on both sides of $L_{j+1}$, then its direction is more complicated. We shall explain this case when we encounter such a situation.

\subsection{From the FN System to the Generators of the Fuchsian Group}

 Assume that we are given an FN system on  $\mathcal{R}_0$, where the coordinate geodesic $L_1,\ldots,L_p$, the pairs of pants $P_1,\ldots,P_q$  are ordered and the coordinate geodesics are directed as already discussed. We also assume that we are given a point 
\begin{equation}\label{pointfn}
\Phi=(l_1,\ldots,l_{p},\tau_1,\ldots,\tau_p)\in (\mathbb{R}^+)^{p}\times \mathbb{R}^p,
\end{equation}
in the corresponding space of FN coordinates. Now we shall describe a canonical procedure for writing down a set of generators for the corresponding Fuchsian group.  We write the generators in the following order. The first $2p-q+1$ generators are hyperbolic, $p$ coordinate geodesics followed by $p-q+1$ handle-closers. The remaining $n$ generators are  parabolic boundary elements. In general, the total number of generators, $d=2q+1=4g-3+2n$, is  far from minimal, which is $2g+n$. 

\paragraph{The Normalization:} Let $a_1,\ldots,a_d$ be a set of generators for a Fuchsian group. Assume that $a_1$ and $a_2$ are as follows.
\begin{itemize}
\item The generators $a_1$ and $a_2$ are both hyperbolic with disjoint axes.\footnote{Recall that the axis $A_{\msf{g}}$ of a hyperbolic  element $\msf{g}\in\Gamma$ is the geodesic on $\mathbb{H}$ that connects the fixed points of $\msf{g}$-action on $\mbb{H}$.}

\item The axis of $a_1$ denoted by $A_1$ lie on the imaginary axis, pointing towards $\infty$; the axis of $a_2$ denoted by $A_2$ lies in the right half-plane, with the attracting fixed-point smaller than the repelling  one.

\item The common orthogonal between $A_1$ and $A_2$ lies on the unit circle.
\end{itemize}

\subsubsection*{The Basic Building Blocks} 

Let us label  the three distinct boundary elements of each pair of pants $P_i$  as $b_{i,1},b_{i,2},b_{i,3}$. The order of the boundary elements is determined by the order of the coordinates geodesics and boundary elements of the FN system, except for the cases where we have not yet distinguished between $b_{i,2}$ and $b_{i,3}$. The size of each hyperbolic boundary element is specified by the FN coordinate $\Phi$ given in \eqref{pointfn}. Let $P_i'$ be the Nielsen extension of $P_i$. For each $i=1,\ldots,q$ there is a unique normalized pants group $\Gamma_i$ representing $P_i'$, i.e., $\mathbb{H}/\Gamma_i\simeq P_i'$. $\Gamma_i$ has three distinguished generators $a_{i,1},a_{i,2},a_{i,3}$, where $a_{i,1}a_{i,2}a_{i,3}=1$. The axis of $a_{i,k}$, $A_{i,k}$, projects onto $b_{i,k},~k=1,2,3$.

\paragraph{Sphere with Holes:} Consider subspace $Q_j$ of  $\mathcal{R}_0$, defined as the interior of the closure of the union of the $P_i,~i\leq j$ for $j=1,\ldots,q-1$. Let $Q_j'$ be the Nielsen extension of $Q_j$. Each subspace $Q_j'$ has a naturally-defined FN system, where the coordinate geodesics are $L_1,\ldots,L_{j-1}$ and the pairs of pants are $P_1,\ldots,P_j$. Say, $\Gamma_1=J_1$ be the normalized pants group representing $P_1$. Assume that $a_{1,1}$ and $a_{1,2}$ generate  the group $\Gamma_1$.  We can choose the normalizations for different cases as follows 
\begin{itemize}
\item If $a_{1,1}$ is hyperbolic, then $A_1$, the axis of $a_{1,1}$, is the imaginary axis, pointing towards $\infty$.  $A_2$, the axis of $a_{1,2}$, lies in the right half-plane and $M_3$, the common orthogonal between $A_1$ and $A_2$, lies on the unit circle.  
\item If $a_{1,1}$ is parabolic, then $A_1$ is the point at infinity, and $M_3$ lies on the imaginary axis. If $a_{1,2}$ is also parabolic, then the axis  $A_2$ is necessarily at $0$. 

\end{itemize}

The imaginary axis, which is the axis of $a_{1,1}$, projects to $L_1$ and the positive direction of the axis projects to the positive direction $L_{1}$. Consider $\Gamma_2$, the normalized pants group representing $P_2$. Let $c_2$ be the hyperbolic isometry that maps the right half-plane onto the left half-plane while introducing a twist of $\tau_1$ in the positive direction on $L_1$; i.e., 
 \begin{equation}
 c_2(0)=\infty,~c_2(\infty)=0,~c_2(\mathrm{i})=e^{\tau_1}\mathrm{i}.
 \end{equation}
  Let $\Gamma_2=c_2\Gamma_2c_2^{-1}$. Then, 
   \begin{equation}
   a_{1,1}=\wh a_{1,1}=c_2a_{2,1}c_2^{-1}=\wh a_{2,1}.
   \end{equation}

It follows that one can use the {\it amalgamated free product (AFP) combination theorem} to amalgamate $\wh\Gamma_2$ to $\Gamma_1$. The AFP combination theorem  asserts that 
\begin{enumerate}
\item $J_2=\langle \Gamma_1,\wh \Gamma_2\rangle$ is Fuchsian.

\item $J_2$ is generated by $a_{1,1},a_{1,2},a_{1,3},\wh a_{2,2},\wh a_{2,3}$. In addition to the defining relations of $\Gamma_1$ and $\Gamma_2$, these satisfy the additional relation: $a_{1,1}=\wh a_{2,2}\wh a_{2,3}$.

\item $\mathbb{H}/J_2$ is a sphere with 4 boundaries; the corresponding boundary subgroups are generated by the above four generators.
\end{enumerate}

We can now identify $\mathbb{H}/J_2$ with $Q_2'$. This imposes a new order on the boundary elements of $\mathbb{H}/J_2$ as follows. 
\begin{itemize}

\item If $b$ and $b'$ are boundary elements of $Q_2$, where $b$ precede $b'$ as coordinate geodesic or as boundary elements of $\mathcal{R}_0$, then $b$ precede $b'$ as a boundary element of $Q_2$. 

\item If $b$ and $b'$ both corresponds to the same coordinate geodesic on $\mathcal{R}_0$, and $b$ corresponds to a boundary geodesic on $P_1$, while $b'$ corresponds to a boundary geodesic on $P_2$, then as boundary elements of $Q_2$, $b$ precede $b'$. 

\item If $b$ and $b'$ both correspond to boundary elements of either $P_1$ or $P_2$, then $b$ precede $b'$ if $b$ corresponds to $A_{i,2}$ and $b'$ corresponds to $A_{i,3}$.

\item  In the case that $P_i$ has two boundary elements corresponding to the same handle geodesic, $L$, we direct $L$ so that the positive direction of $A_{i,2}$ projects onto the positive direction of $L$.
\end{itemize}

 This way we can order the boundary elements of $Q_2$ and direct them. Introducing a new ordered set of generators for $J_2$ as $a_1^2,\ldots,a_5^2$, where $a_1^2=a_{1,1}=\wh a_{2,1}^{-1}$ and $a_2^2,\ldots,a_5^2$ are the generators $a_{1,2},a_{1,3},\wh a_{2,3}$. $A_j^2$ projects onto $b_{j-1}^2$. Those $A_j^2$ that are hyperbolic are all directed so that the attracting fixed point of $a_j^2$ is smaller than the repelling fixed point.  Each $A_j^2$ has a canonical base point on it. For $i=1,\ldots, q$ the point $i$ is the canonical base point on $A_{i,1}$. For $i=1,\ldots,q$ and for $k=2,3$ the base point on $A_{i,k}$ is the point of intersection of $A_{i,k}$ with the common orthogonal between $A_{i,1}$ and $A_{i,k}$. In the case that $a_j^2=a_{1,i}$, the base point is the canonical base point for $a_{1,i}$. In the case that $a_j^2=c_2a_{2,i}c_2^{-1}$, the canonical base point is the $c_2$ image of the canonical base point for $a_{2,i}$.

\newl We should now iterate the above process. Assume that we have found $J_k$ representing $Q'_k$, where $k<q$, with distinguished subgroups, $\wh \Gamma_1,\ldots,\wh \Gamma_k$. Assume that we have ordered boundary elements of $Q_k$, and that we have  found the distinguished generators $a_1^k,\ldots,a_{2k+1}^k$ for $J_k$, where each distinguished generator lies in some distinguished subgroup, so that, for $i=1,\ldots,k$, $A_i^k$ projects onto the boundary element $b^k_{k-i}$ of $Q_k$. Assume that we have assigned a canonical base point for each geodesic $A_i^k,~i=k+1,\ldots,2k+1$. Also, assume that all the hyperbolic boundary generators of $J_k$ are directed so that the attracting fixed point is smaller than the repelling point. Then, there is some $j$ so that $A^k_{k+j}$  projects onto $L_{k+1}$. We need to renormalize $\Gamma_{k+1}$ so that the renormalized $A_{k+1,1}$ agrees with $A^k_{k+j}$, but with opposite direction and with an appropriate twist.

\newl Let us define the conjugator $c_{k+1}$ to be the unique orientation-preserving hyperbolic isometry mapping the left half-plane onto the action half-plane of $A^k_{k+j}$\footnote{See below for the definition of action half-plane.} while mapping the base point $i$ onto the point whose distance from the base point on $A^k_{k+j}$, measured in the positive direction along $A^k_{k+j}$, is exactly $\tau_k$. The AFP combination theorem assures that
\begin{enumerate} 
\item $J_{k+1}=\langle J_k,c_{k+1}\Gamma_{k+1}c^{-1}_{k+1}\rangle$ is Fuchsian.

\item $J_{k+1}=\langle a^k_1,\ldots, a^k_{2k+1},c_{k+1}a_{k+1}c^{-1}_{k+1},c_{k+1}a_{k+1,2}c_{k+1}^{-1},c_{k+1}a_{k+1,3}c_{k+1}^{-1}\rangle$. These satisfy the defining relations of $J_k$ and $\Gamma_{k+1}$ together with the additional defining relation: $a^k_{k+j}=(c_{k+1}a_{k+1,1}c^{-1}_{k+1})^{-1}$.

\item $\mathbb{H}/J_{k+1}$ is a sphere with $2k+1$ holes.
\end{enumerate}

As above, we rewrite the generators of $J_{k+1}$ as $a_1^{k+1},\ldots,a_{2k+3}^{k+1}$, where the first $k$ generators correspond to the $k$ coordinate geodesics of $Q_{k+1}$, and the remaining generators correspond to the boundary elements of $Q_{k+1}$ in the following order.  Each boundary generator $a_i$ of $J_{k+1}$ corresponds to either a coordinate geodesic $L_j$ on $\mathcal{R}_0$ or it corresponds to a boundary element $b_j$ of $\mathcal{R}_0$. Pulling back the order of the coordinate geodesic and boundary elements for $\mathcal{R}_0$ imposes a partial order on the boundary generators of $J_{k+1}$. The only ambiguities occur  when the generators $a_i$ and $a_i'$ both correspond to the same coordinate geodesic. If $a_i$, respectively , $a_i'$, lies in the distinguished subgroups $\wh \Gamma_j$, respectively, $\wh \Gamma_{j'}$, where $j<j'$, then $a_i$ precedes $a_i'$. If $a_i$ and $a_i'$ both lie in the same distinguished subgroup, $\wh \Gamma_j$, then $\wh a_{j,2}$ precedes $\wh a_{j,3}$.

\newl The hyperbolic boundary generators of $J_k$ all have distinguished base points. We assign the $c_k$ image of the distinguished point on $a_{k_1,2}$ and $a_{k+1,3}$ as the distinguished base point on the new boundary generators of $J_{k+1}$. We also observe that the hyperbolic boundary generators of $J_{k+1}$ are all directed so that their attracting fixed points are smaller than  their repelling ones. When $k=q-1$, we reach the group $J_q$, representing $Q_q'$. 

\paragraph{Closing the Handles:} Let us rename the group $J_q$ and call it $K_0$ and also rename its ordered set of generators and call them $a_1^0,\ldots,a^0_{2q+1}$. The first $q-1$ of these generators correspond to the attaching coordinate geodesics of $\mathcal{R}_0$; the next $2(p-q+1)$ generators correspond to the boundary elements of $Q_q$ that are handle-geodesics on $\mathcal{R}_0$.  We define the first handle-closer $d_1$ as the orientation preserving hyperbolic isometry mapping the action half-plane of $a_q^0$ onto the boundary half-plane of $a^0_{q+1}$ while mapping the point at distance $-\tau_q$ from the base point on $A^0_q$ to the base point on $A^0_{q+1}$. Note that $d_1$ conjugates $a^0_q$ onto $(a^0_{q+1})^{-1}$. Then the HNN combination theorem asserts that
\begin{enumerate}
\item $K_1=\langle K_0,d_1\rangle$ is Fuchsian. 

\item $K_1$ is generated by $a^0_1,\ldots,a^0_{2q+1},d_1$. These satisfy the defining relations of $K_1$ together with the additional relation: $a^0_{q+1}=d_1(a^0_q)^{-1}d_1^{-1}$.

\item $\mathbb{H}/K_1$ is a genus-one hyperbolic surface and $q$ boundaries.
\end{enumerate}

The distinguished subgroups of $K_1$ are the distinguished subgroups of $K_0$. The generators of $K_1$ are generators of $K_0$, in the same order, but with $a^0_{q+1}$ deleted, and $d_1$ added to the list. In the list of generators for $K_1$, $d_1$ appears after all the generators corresponding to coordinate geodesics on $\mathcal{R}_0$, and before the first generator corresponding to a boundary element of $\mathcal{R}_0$. After $g$ iterations, we reach the discrete group $\Gamma=K_g$, where $\mathbb{H}/\Gamma=\mathcal{R}_0$. Furthermore, $\Gamma$ has $q$ distinguished subgroups, representing in order the $q$ pairs of pants, $P_1,\ldots,P_q$. $\Gamma$ has $2q+1$ distinguished generators, $a_1,\ldots, a_{2p-q+1}$. For $i=1,\ldots,p$ the axis $A_i$ projects onto the coordinate geodesic $L_i$; the generators $a_{p+1},\ldots, a_{2p-q+1}$ are handle-closers and the axes of the remaining generators project, in order, onto the boundaries of $\mathcal{R}_0$.

\subsection{The Explicit Matrices}

In this section, we shall describe an algorithm for writing the generators of the Fuchsian group in terms of FN coordinates. 

\subsubsection*{Reduction to Primitive Conjugators}
Let $\Gamma_i$ and $\Gamma_{j+1}$ be the normalized pants groups representing the pairs of pants, $P_i$ and $P_{j+1}$, respectively. We assume that $i<j+1$, and that there is a $k,~1\leq k\leq 3$, so that $a_{j+1,1}$ and $a_{i,k}$ represent the same attaching geodesic, $L_j$ on $\mathcal{R}_0$, but with reverse orientations. Then $|\mathrm{tr}(\wt a_{j+1,1})|=|\mathrm{tr}(\wt a_{i,k})|$. The {\it elementary conjugator $e_{i,k}$} maps the boundary half-plane of $a'=a_{j+1,1}$ onto the action half-plane of $a=a_{i,k}$ while introducing a twist of $\tau_j$.  The {\it untwisted elementary conjugator, $e^0_{i,k}$}, which is independent of the index $j$, also maps the left half-plane onto the action half-plane of $a$, but maps the base point on $A_{j+1,1}$ to the base point on $A_{i,k}$. If $a$ generates a  boundary subgroup of $G$, then the corresponding axis $A$ separates the upper half-plane  into two half-planes. The boundary half-plane is precisely invariant under $\langle a\rangle$ in $G$. The other half-plane, which is not precisely invariant (unless $G$ is elementary), is called the {\it action half-plane}.

\newl Once we have found matrices for the elementary conjugators, then we can inductively find matrices for all conjugators. Let $P_i$ and $P_{j+1}$ be the adjacent pairs of pants in the FN system on $\mathcal{R}_0$, with $i<j+1$. Let, as above, $b_{j+1,1}$ attached to $b_{i,k}$. Once we have found the matrix for the elementary conjugator, $e_{i,k}$, the matrix for the {\it conjugator $c_{j+1}$} is given by 
\begin{equation}\label{econj}
\wt c_{j+1}=\wt c_i\wt e_{i,k}.
\end{equation}
For each $j=q,\ldots, p$, the handle closer $d_j$ conjugates one distinguished boundary generator of $K_0$ onto the inverse of another distinguished boundary generator. These two boundary generators either lie in the same distinguished subgroup or they lie in different ones. If the two boundary generators lie in the same distinguished subgroup, $\wh \Gamma_i$, then there is a conjugator $c_i$, and there is normalized pants subgroup $\Gamma_i$ such that that $\wh \Gamma_i=c_i\Gamma_ic^{-1}_i$. In this case, we can choose the matrix for the corresponding handle-closer as 
\begin{equation}\label{hc}
\wt d_j=\wt c_i\wt d\wt c_i^{-1},
\end{equation}
where $\wt d$ is the matrix for the handle closer. If these two boundary generators lie in distinct distinguished subgroups, say $\wh \Gamma_i$ and $\wh \Gamma_{i'}$, where $i<i'$. Assume that the first boundary generator corresponds to $\wh a_{i,k}$ and the second corresponds to $\wh a_{i',k'}$. We can write $d_j$ as a product of four transformations: First, the twist along the axis $\wh A_{i,k}$, then map the boundary half-plane of $a_{i,k}$ onto the right half-plane; then interchange left and right half-planes, and last map the right half-plane onto the boundary half-plane of $\wh a_{i',k'}$. 

\begin{itemize}
\item[(1)] The twist transformation preserves both sides of $\wh A_{i,k}$ and maps the point on $\wh A_{i,k}$
, whose distance from the base point is $-\tau_j$, to the base point on $\wh A_{i,k}$. We can write the matrix for this transformation as 
\begin{equation}\label{twist}
(\wt c_i\wt e^0_{i,k})\wt f_{\tau_j}(\wt c_i\wt e^0_{i,k})^{-1},
\end{equation}
where $f_{\tau}$ is the universal twist map, $f_{\tau}(z)=e^{-\tau}z$. 

\item[(2)] The second transformation maps the boundary half-plane of $\wh a_{i,k}$ onto the right half-plane, and maps the base point on $\wh A_{i,k}$ to the base point on the imaginary axis.  The matrix for this transformation is 
\begin{equation}\label{brmap}
(\wt c_i\wt e^0_{i,k})^{-1}.
\end{equation}

\item[(3)] The interchange transformation $h$ interchanges left and right half-planes and preserves the base point on the imaginary axis,  that is 
\begin{equation}\label{in}
h(z)=-\frac{1}{z}.
\end{equation}

\item[(4)] The final transformation maps the left half-plane onto the boundary half-plane of $\wh a_{i',k'}$. The matrix representation of this transformation can be written as 
\begin{equation}\label{final}
\wt c_{i'}\wt e^0_{i',k'}.
\end{equation}
\end{itemize}

Combining all we obtain
\begin{equation}\label{comb}
\wt d_j=\wt c_{i'}\wt e^0_{i',k'}\wt g \wt f_{\tau_j}(\wt c_i\wt e^0_{i,k})^{-1}.
\end{equation}

\hypertarget{NormalizedPantsGroup}{\subsubsection*{The Normalized Pants Groups:}}

Given three numbers $\lambda_1,\lambda_2$ and $\lambda_3$, it is possible to  write down matrices $\wt a_1$ and $\wt a_2$, corresponding to the geometric generators for the pants group, such that  $|\mathrm{tr}(\wt a_1)|=2~\mathrm{cosh}~\lambda_1, |\mathrm{tr}(\wt a_2)|=2~\mathrm{cosh}~\lambda_2$ and $|\mathrm{tr}(\wt a_3)|=|\mathrm{tr}(\wt a_1\wt a_2)^{-1}|=2~\mathrm{cosh}~\lambda_3$. We can assume that $\lambda_i$ are given so that $a_i$ are in the non-increasing order. Without loss of generality, we assume that $\mathrm{tr}(\wt a_1)\geq 0$ and $\mathrm{tr}(\wt a_2)\geq 0$. Let us choose normalizations for different cases as follows

\begin{itemize}
\item $a_1$ is hyperbolic:  $A_1$ is the imaginary axis, pointing towards $\infty$.  $A_2$ lies in the right half-plane and $M_3$, the common orthogonal between $A_1$ and $A_2$, lies on the unit circle. 

\item $a_1$ is parabolic: $A_1$ is the point at infinity, and $M_3$ lies on the imaginary axis. If $a_2$ is also parabolic, then $A_2$ is necessarily at $0$.  

\end{itemize}

Now we shall list the generators of all possible pants groups satisfying the above normalization.

\paragraph{Three Hyperbolic Elements:} In this case $\lambda_i>0,~i=1,2,3$. We need to find matrices $\wt a_1$ and $\wt a_2$ with $\mathrm{tr}(\wt a_1)=2~\mathrm{cosh}~(\lambda_1);~\mathrm{tr}(\wt a_2)=2~\mathrm{cosh}~(\lambda_2)$ and $\mathrm{tr}(\wt a_1\wt a_2)=-2~\mathrm{cosh}~(\lambda_3)$. $a_1$ and $a_2$ should be appropriately oriented; hence we write our matrices so that the repelling fixed point of $a_2$ is greater than 1, while its attracting fixed point is less than 1. Define $\mu$ and $\nu$ by

\begin{eqaligned}
    \label{mu}
\mathrm{coth}~\mu&=\frac{\mathrm{cosh}~\lambda_1\mathrm{cosh}~\lambda_2+\mathrm{cosh}~\lambda_3}{\mathrm{sinh}~\lambda_1\mathrm{sinh}~\lambda_2},\qquad \mu>0,
\\
\mathrm{coth}~\nu&=\frac{\mathrm{cosh}~\lambda_1\mathrm{cosh}~\lambda_3+\mathrm{cosh}~\lambda_2}{\mathrm{sinh}~\lambda_1\mathrm{sinh}~\lambda_3},\qquad \nu>0.
\end{eqaligned}
We can then write the generators $\wt a_1, \wt a_2$ and $\wt a_3=-\wt a_2^{-1}\wt a_2^{-1}$ as follows
\begin{eqgathered}
\label{3hypg}
\wt a_1=\left(\begin{array}{cc}e^{\lambda_1} & 0 \\0 & e^{-\lambda_1}\end{array}\right),
\\
\begin{aligned}
    \wt a_2&=\frac{1}{\mathrm{sinh}~\mu}\left(\begin{array}{cc}\mathrm{sinh}~(\mu-\lambda_2)& \mathrm{sinh}~\lambda_2 \\-\mathrm{sinh}~\lambda_2 & \mathrm{sinh}~(\mu+\lambda_2)\end{array}\right),
    \\
\wt a_3&=\frac{1}{\mathrm{sinh}~\nu}\left(\begin{array}{cc}\mathrm{sinh}~(\nu-\lambda_3)& e^{\lambda_1}\mathrm{sinh}~\lambda_3 \\-e^{\lambda_1}\mathrm{sinh}~\lambda_3 & \mathrm{sinh}~(\nu+\lambda_3)\end{array}\right).    
\end{aligned}
\end{eqgathered}
$\mu$ is related to $\delta$, the distance between $A_1$ and $A_2$, by $\mathrm{coth}~\mu=\mathrm{cosh}~\delta$.

\paragraph{Exactly One Hyperbolic and at Least One Parabolic Element:}

In this case $\lambda_1>0,~\lambda_2=0$. The matrices are
\begin{eqgathered}
    \label{ohop}
 \wt a_1=\left(\begin{array}{cc}e^{\lambda_1} & 0 \\0 & e^{-\lambda_1}\end{array}\right), \qquad \wt a_2=\left(\begin{array}{cc}1+\beta & -\beta\\\beta & 1-\beta\end{array}\right),
 \\
\beta\equiv -\frac{\text{cosh}~\lambda_1+\text{cosh}~\lambda_3}{\text{sinh}~\lambda_1}.
\end{eqgathered}
 
\paragraph{Closing a Handle on a Pair of Pants:} Consider three hyperbolic elements $\wt a_1,\wt a_2$ and $\wt a_3$ with $\lambda_2=\lambda_3$. We need to find the matrix corresponding to the handle-closer $d$ which identifies the action of $a_2$ and $a_3$, while twisting by $\tau$ in the positive direction of $A_2$. The matrix corresponding to $d$ is given 
\begin{eqgathered}
\label{hclose}
\widetilde d=\wt r\wt r_2\wt e_{\tau},
\\
\widetilde r\equiv 
\begin{pmatrix}
0 & e^{\frac{\lambda_1}{2}} 
\\
e^{-\frac{\lambda_1}{2}} & 0
\end{pmatrix},
\qquad 
\widetilde r_2\equiv\frac{1}{\mathrm{sinh}}
\begin{pmatrix}
\cosh\mu & 1
\\-1& \cosh\mu
\end{pmatrix},
\\
\widetilde{e}_{\tau}\equiv\frac{1}{\sinh\mu}
\begin{pmatrix}
\sinh(\mu-\frac{\tau}{2}) & \sinh\frac{\tau}{2}
\\
-\sinh~\frac{\tau}{2}& \sinh\left(\mu+\frac{\tau}{2}\right)
\end{pmatrix}.    
\end{eqgathered}
Here $e_{\tau}$ is the hyperbolic motion with the same fixed point as $a_2$ whose trace is equal to $2~\mathrm{cosh} ~\frac{\tau}{2}$. $a_2$ and $e_{\tau}$ have the same attracting fixed point if $\tau>0$, and have opposite attracting fixed points if $\tau<0$. $r_2$ is the reflection in $A_2$ and $r$ is the reflection in the line half-way between $A_2$ and $A_3$.

\paragraph{Torus with One Puncture:}

We require $a_1$ to be parabolic.  Then $\lambda_2=\lambda_3>0$. We normalize so that $A_1$ lies on the positive imaginary axis so that the unit circle is the common orthogonal between $A_2$ and $A_3$. Here the fixed points of $a_3$ are positive, with the repelling fixed point larger than the attracting one and the fixed point of $a_2$ being negative. Then $\wt a_2,\wt a_3$ is given as follows

\begin{eqgathered}
\label{toop}
  \wt a_2=\frac{1}{\mathrm{sinh}~\mu}\left(\begin{array}{cc}\mathrm{sinh}~(\mu+\lambda_2) & \mathrm{sinh}~\lambda_2 \\-\mathrm{sinh}~\lambda_2 & \mathrm{sinh}~(\mu-\lambda_2)\end{array}\right),
  \\
  \wt a_3=\frac{1}{\mathrm{sinh}~\mu}\left(\begin{array}{cc}\mathrm{sinh}~(\mu-\lambda_2) & \mathrm{sinh}~\lambda_2 \\-\mathrm{sinh}~\lambda_2 & \mathrm{sinh}~(\mu+\lambda_2)\end{array}\right),
  \\
  \mathrm{sinh}^2~\mu = \frac{2\mathrm{sinh}^2~\lambda_2}{\mathrm{cosh}~\lambda_1+1},\qquad \mu>0.
\end{eqgathered}
The handle-closer $d$ which conjugates $a_2$ onto $a_3^{-1}$ while introducing a twist of $\tau$ along $A_2$ is given by 
 \begin{equation}
 d=r_0f_{\tau}r_2.
 \end{equation}
  Here $r_2$ is the reflection in $A_2$, $f_{\tau}$ is the twist by $\tau$ along the imaginary axis  and $r_0$ is the reflection in the imaginary axis. The corresponding matrices are given by 
    \begin{eqgathered}
        \label{toop1}
        \wt r_0= \left(\begin{array}{cc}1 & 0 \\0 & -1\end{array}\right),\qquad \wt r_2= \frac{1}{\mathrm{sinh}~\mu}\left(\begin{array}{cc}\mathrm{cosh}~\mu & 1 \\-1 & -\mathrm{cosh}~\mu\end{array}\right),
        \\
        \wt f_{\tau}= \left(\begin{array}{cc}e^{\frac{\tau}{2}} & 0 \\0 & e^{-\frac{\tau}{2}}\end{array}\right).
    \end{eqgathered}
 
 \subsubsection*{Matrices for the Universal Twist Map, Interchange, and Elementary Conjugators}
 
 The explicit matrices for universal twist map, interchange, and elementary conjugators are as follows.

 \paragraph{The Universal Twist Map and The Interchange:}
The explicit matrices generating the twist and interchange are given by
\begin{equation}
    \label{utmi}
\wt f_{\tau}=\left(\begin{array}{cc}e^{-\frac{\tau}{2}} & 0 \\0 & e^{\frac{\tau}{2}}\end{array}\right),
\qquad 
\wt h=\left(\begin{array}{cc}0 & -1 \\1 & 0\end{array}\right).
\end{equation}

\paragraph{Elementary Conjugator for $a_{i,1}$}

This case occurs only for $i=j=1$. We can choose $\wt h$ as the matrix for untwisted elementary conjugator $\wt e_{2,1}^0$. The matrix for the twisted elementary conjugator is given by
\begin{equation}
    \label{ecai1}
\wt e_{2,1}=\wt e^0_{2,1}\wt f_{\tau}=\left(\begin{array}{cc}0 & -e^{\frac{\tau_1}{2}} \\e^{-\frac{\tau_1}{2}} & 0\end{array}\right).
\end{equation}

\paragraph{Elementary Conjugator for $a_{i,2}$:}
This matrix is given by
\begin{equation}
\label{tecai2}
\wt e^0_{i,2}=\frac{1}{\sqrt{2\mathrm{sinh}~\mu_i}}\left(\begin{array}{cc}e^{\frac{1}{2}\mu_i} & e^{-\frac{1}{2}\mu_i} \\e^{-\frac{1}{2}\mu_i} & e^{\frac{1}{2}\mu_i}\end{array}\right),
\end{equation} 
where $\mu_i$ defined in \eqref{mu}  or \eqref{ohop}, depending on the type of $a_{i,3}$. Furthermore,

\begin{equation}\label{ecai2}
\wt e_{i,2}=\frac{1}{\sqrt{2\mathrm{sinh}~\mu_i}}\left(\begin{array}{cc}e^{\frac{1}{2}(\mu_i-\tau_j)} & e^{-\frac{1}{2}(\mu_i-\tau_j)} \\e^{-\frac{1}{2}(\mu_i+\tau_j)} & e^{\frac{1}{2}(\mu_i+\tau_j)}\end{array}\right).
\end{equation}

\paragraph{Primitive conjugator for $a_{i,3}$:} This matrix is given by

\begin{equation}
    \label{tecai3}
\wt e^0_{i,3}=\frac{1}{\sqrt{2\mathrm{sinh}~\nu_i}}\left(\begin{array}{cc}e^{\frac{1}{2}\nu_i+\lambda_i} & e^{-\frac{1}{2}\mu_i} \\e^{-\frac{1}{2}\mu_i}& e^{\frac{1}{2}\mu_i-\lambda_i}\end{array}\right),
\end{equation}
where $\nu_i$ defined by \eqref{mu}. $\lambda_i$ is the size of $a_{i,1}$, i.e., $|\mathrm{tr}(a_i,1)|=\mathrm{2 ~cosh}~\lambda_i$. Furthermore, 

\begin{equation}\label{ecai3}
\wt e_{i,3}=\frac{1}{\sqrt{2\mathrm{sinh}~\mu_i}}\left(\begin{array}{cc}e^{\frac{1}{2}(\mu_i-\tau_j)+\lambda_i} & e^{-\frac{1}{2}(\mu_i-\tau_j)} \\e^{-\frac{1}{2}(\mu_i+\tau_j)} & e^{\frac{1}{2}(\mu_i+\tau_j)-\lambda_i}\end{array}\right).
\end{equation}

\subsection{The Algorithm}

We now summarize the algorithm for constructing the generators of a Fuchsian group as a function of FN coordinates. An explicit FN system on a hyperbolic surface can be represented by a table called {\it the pairing table for the  FN system}. A pairing table has $q$ rows, one for each pair of pants $P_i$. The entry in the $i$\textsuperscript{th} row and $k$\textsuperscript{th} column identifies the boundary element $b_{i,k}$ as corresponding to either a coordinate geodesic $L_j$ or a boundary element $b_j$ (cusp)  of $\mathcal{R}_0$. The following algorithm then gives the explicit matrices

\begin{enumerate}

\item For each $i=1,\ldots,q$ and for $k=1,2,3$, we read off from the pairing table whether $b_{i,k}$ corresponds to a coordinate geodesic or to a boundary element of $\mathcal{R}_0$. If $b_{i,k}$ corresponds to coordinate geodesic $L_j$, then we read off the size of $a_{i,k}$ from $j^{\mathrm{th}}$ entry in $\Phi$ (see \eqref{pointfn}). Due to the ordering, each $b_{i,1}$ corresponds to an attaching geodesic; we write the size of this geodesic as $\lambda_i$.  Use the constructions of \hyperlink{NormalizedPantsGroup}{The Normalized Pants Group} We can write down the matrices $\wt a_{i,k}$.

\item The first conjugator $c_1$ is identity. The second conjugator is the elementary conjugator $\wt e_{2,1}$. Continuing inductively, assume that we have found matrices for the conjugators $c_1,\ldots,c_{j-1},~j\leq q$. The attaching geodesic $L_j$ appears in the pairing table once as $b_{j+1,1}$ and once as $b_{i,k},~i<j+1,~k>1$. \eqref{econj} gives the formula for $\wt c_j$. 

\item Write the matrices for the first  $q-1$ generators. These are the generators corresponding to attaching geodesics. They are given as,
\begin{equation}\label{q-1g}
\wt a_1=\wt a_{1,1},\quad \wt a_2=\wt c_3\wt a_{3,1}^{-1},\quad\ldots,\quad\wt a_{q-1}=\wt c_q\wt a_{q,1}^{-1}\wt c_q^{-1}. 
\end{equation} 

\item Find matrices for generators corresponding to the handle geodesics; these are the generators $a_q,\ldots,a_p$. Each $L_j,~ q\leq j \leq p$, appears twice in the pairing table. Either there is some $i$ so that $L_j$ appears as both $b_{i,2}$ and $b_{i,3}$ or there are two distinct rows $i<i'$, so that $L_j$ appears as $b_{i,k}$ and $b_{i',k'}$. In the first case, we write the matrix for the handle generator as $\wt a_j=\wt c_i\wt a_{i,2}\wt c_i^{-1}$. In the second case, we write the matrix for the handle generator as $\wt a_j=\wt c_i \wt a_{i,k}\wt c_i^{-1}$.

\item Find matrices for handle-closing generators $a_{p+1},\ldots, a_{2p-q+1}$. If there is some $j,~q\leq j\leq p,$ so that $L_j$ appears as both $b_{i,2}$ and $b_{i,3}$ in the pairing table, then we write $\wt a_{p+j}=\wt d_j$, where $\wt d_j$ is given by \ref{hc}. If the two entries of $L_j$ in the pairing table appear as both $b_{i,k}$ and $b_{i',k'}$, where $i<i'$, then we write  $\wt a_{p+j}=\wt d_j$, where $\wt d_j$ is given by \eqref{comb}.

\item Write down the matrices corresponding to the boundary elements of $\mathcal{R}_0$. Each $b_j,~j=1,\ldots,m+n$, appears exactly once in the pairing table. If $b_j$ corresponds to $b_{i,k}$, then the matrix for the corresponding generator is given by $\wt a_{2p+q+j}=\wt c_i \wt a_{i,k}\wt c_i^{-1}$. 

\end{enumerate}

This concludes our brief recap of \cite{Maskit2001}.

%% file: Main.bbl
\providecommand{\href}[2]{#2}\begingroup\raggedright\begin{thebibliography}{10}

\bibitem{Polyakov1981a}
A.~M. Polyakov, \emph{{Quantum Geometry of Bosonic Strings}},
  \href{https://doi.org/10.1016/0370-2693(81)90743-7}{\emph{Phys Lett.}
  {\bfseries B103} (1981) 207}.

\bibitem{Polyakov1981b}
A.~M. Polyakov, \emph{{Quantum Geometry of Fermionic Strings}},
  \href{https://doi.org/10.1016/0370-2693(81)90744-9}{\emph{Phys Lett.}
  {\bfseries B103} (1981) 211}.

\bibitem{FriedanMartinecShenker1986}
D.~Friedan, E.~J. Martinec and S.~Shenker, \emph{{Conformal Invariance,
  Supersymmetry and String Theory}},
  \href{https://doi.org/10.1016/S0550-3213(86)80006-2}{\emph{Nucl. Phys.}
  {\bfseries B271} (1986) 93}.

\bibitem{Berera199301}
A.~Berera, \emph{{Unitary String Amplitudes}},
  \href{https://doi.org/https://doi.org/10.1016/0550-3213(94)90057-4}{\emph{Nucl.
  Phys. B} {\bfseries 411} (1994) 157}.

\bibitem{Berera199406}
A.~{Berera}, \emph{{All-Order Mass Renormalization of String Theory}},
  \href{https://doi.org/10.1103/PhysRevD.49.6674}{\emph{Phys. Rev.} {\bfseries
  D49} (1994) 6674}.

\bibitem{Belopolsky1997a}
A.~{Belopolsky}, \emph{{De Rham Cohomology of the Supermanifolds and
  Superstring BRST Cohomology}},
  \href{https://doi.org/10.1016/S0370-2693(97)00445-0}{\emph{Phys. Lett.}
  {\bfseries B403} (1997) 47}
  [\href{https://arxiv.org/abs/hep-th/9609220}{{\ttfamily hep-th/9609220}}].

\bibitem{Belopolsky1997b}
A.~{Belopolsky}, \emph{{New Geometrical Approach to Superstrings}},
  {\emph{ArXiv e-prints} (1997) }
  [\href{https://arxiv.org/abs/hep-th/9703183}{{\ttfamily hep-th/9703183}}].

\bibitem{Belopolsky1997c}
A.~{Belopolsky}, \emph{{Picture Changing Operators in Supergeometry and
  Superstring Theory}}, {\emph{ArXiv e-prints} (1997) }
  [\href{https://arxiv.org/abs/hep-th/9706033}{{\ttfamily hep-th/9706033}}].

\bibitem{Witten2012a}
E.~{Witten}, \emph{{Notes On Supermanifolds and Integration}}, {\emph{ArXiv
  e-prints} (2012) } [\href{https://arxiv.org/abs/1209.2199}{{\ttfamily
  1209.2199}}].

\bibitem{Witten2012b}
E.~{Witten}, \emph{{Notes On Super Riemann Surfaces and Their Moduli}},
  {\emph{ArXiv e-prints} (2012) }
  [\href{https://arxiv.org/abs/1209.2459}{{\ttfamily 1209.2459}}].

\bibitem{Witten2012c}
E.~{Witten}, \emph{{Superstring Perturbation Theory Revisited}}, {\emph{ArXiv
  e-prints} (2012) } [\href{https://arxiv.org/abs/1209.5461}{{\ttfamily
  1209.5461}}].

\bibitem{Witten201306}
E.~Witten, \emph{{Notes On Holomorphic String And Superstring Theory Measures
  Of Low Genus}},  \href{https://arxiv.org/abs/1306.3621}{{\ttfamily
  1306.3621}}.

\bibitem{DonagiWitten201304}
R.~Donagi and E.~Witten, \emph{{Supermoduli Space Is Not Projected}},
  {\emph{Proc. Symp. Pure Math.} {\bfseries 90} (2015) 19}
  [\href{https://arxiv.org/abs/1304.7798}{{\ttfamily 1304.7798}}].

\bibitem{DonagiWitten201404}
R.~Donagi and E.~Witten, \emph{{Super Atiyah Classes and Obstructions to
  Splitting of Supermoduli Space}},
  \href{https://doi.org/10.4310/PAMQ.2013.v9.n4.a5}{\emph{Pure Appl. Math.
  Quart.} {\bfseries 09} (2013) 739}
  [\href{https://arxiv.org/abs/1404.6257}{{\ttfamily 1404.6257}}].

\bibitem{Sen201408}
A.~Sen, \emph{{Off-Shell Amplitudes in Superstring Theory}},
  \href{https://doi.org/10.1002/prop.201500002}{\emph{Fortsch. Phys.}
  {\bfseries 63} (2015) 149} [\href{https://arxiv.org/abs/1408.0571}{{\ttfamily
  1408.0571}}].

\bibitem{SenWitten201504}
A.~Sen and E.~Witten, \emph{{Filling the Gaps with PCOs}},
  \href{https://doi.org/10.1007/jhep09(2015)004}{\emph{J. High Energ. Phys.}
  {\bfseries 2015} (2015) } [\href{https://arxiv.org/abs/1504.00609}{{\ttfamily
  1504.00609}}].

\bibitem{PiusRudraSen201410}
R.~{Pius}, A.~{Rudra} and A.~{Sen}, \emph{{String Perturbation Theory around
  Dynamically-Shifted Vacuum}},
  \href{https://doi.org/10.1007/JHEP10(2014)070}{\emph{J. High Energ. Phys.}
  {\bfseries 2014} (2014) 70}
  [\href{https://arxiv.org/abs/1404.6254}{{\ttfamily 1404.6254}}].

\bibitem{Sen201411}
A.~Sen, \emph{{Gauge Invariant 1PI Effective Action for Superstring Field
  Theory}}, \href{https://doi.org/10.1007/JHEP06(2015)022}{\emph{J. High Energ.
  Phys.} {\bfseries 2015} (2015) 22}
  [\href{https://arxiv.org/abs/1411.7478}{{\ttfamily 1411.7478}}].

\bibitem{Sen201501}
A.~Sen, \emph{{Gauge Invariant 1PI Effective Superstring Field Theory:
  Inclusion of the Ramond Sector}},
  \href{https://doi.org/10.1007/JHEP08(2015)025}{\emph{J. High Energ. Phys.}
  {\bfseries 2015} (2015) 25}
  [\href{https://arxiv.org/abs/1501.00988}{{\ttfamily 1501.00988}}].

\bibitem{Sen201508a}
A.~{Sen}, \emph{{Supersymmetry Restoration in Superstring Perturbation
  Theory}}, \href{https://doi.org/{10.1007/JHEP12(2015)075}}{\emph{J. High
  Energ. Phys.} {\bfseries 12} (2015) 75}
  [\href{https://arxiv.org/abs/1508.02481}{{\ttfamily 1508.02481}}].

\bibitem{Sen201508b}
A.~Sen, \emph{{BV Master Action for Heterotic and Type-II String Field
  Theories}}, \href{https://doi.org/10.1007/JHEP02(2016)087}{\emph{J. High
  Energ. Phys.} {\bfseries 2016} (2016) 87}
  [\href{https://arxiv.org/abs/1508.05387}{{\ttfamily 1508.05387}}].

\bibitem{PiusSen1604}
R.~{Pius} and A.~{Sen}, \emph{{Cutkosky Rules for Superstring Field Theory}},
  \href{https://doi.org/10.1007/JHEP10(2016)024}{\emph{J. High Energ. Phys.}
  {\bfseries 10} (2016) 24} [\href{https://arxiv.org/abs/1604.01783}{{\ttfamily
  1604.01783}}].

\bibitem{Sen201610}
A.~Sen, \emph{{Equivalence of Two Contour Prescriptions in Superstring
  Perturbation Theory}},
  \href{https://doi.org/10.1007/JHEP04(2017)025}{\emph{J. High Energ. Phys.}
  {\bfseries 04} (2017) 025}
  [\href{https://arxiv.org/abs/1610.00443}{{\ttfamily 1610.00443}}].

\bibitem{Sen201606}
A.~Sen, \emph{{Reality of Superstring Field Theory Action}},
  \href{https://doi.org/10.1007/JHEP11(2016)014}{\emph{J. High Energ. Phys.}
  {\bfseries 11} (2016) 014}
  [\href{https://arxiv.org/abs/1606.03455}{{\ttfamily 1606.03455}}].

\bibitem{Sen201607}
A.~{Sen}, \emph{{Unitarity of Superstring Field Theory}},
  \href{https://doi.org/10.1007/JHEP12(2016)115}{\emph{J. High Energ. Phys.}
  {\bfseries 12} (2016) 115}
  [\href{https://arxiv.org/abs/1607.08244}{{\ttfamily 1607.08244}}].

\bibitem{Sen201609}
A.~Sen, \emph{{Wilsonian Effective Action of Superstring Theory}},
  \href{https://doi.org/10.1007/jhep01(2017)108}{\emph{J. High Energ. Phys.}
  {\bfseries 2017} (2017) }
  [\href{https://arxiv.org/abs/arXiv:1609.00459}{{\ttfamily
  arXiv:1609.00459}}].

\bibitem{deLacroixErbinKashyapSenVerma1703}
C.~{de Lacroix}, H.~{Erbin}, S.~P. {Kashyap}, A.~{Sen} and M.~{Verma},
  \emph{{Closed Superstring Field Theory and its Applications}},
  \href{https://doi.org/10.1142/S0217751X17300216}{\emph{Int. J. Mod. Phys.}
  {\bfseries A32} (2017) 1730021}
  [\href{https://arxiv.org/abs/1703.06410}{{\ttfamily 1703.06410}}].

\bibitem{DeligneMumford1969a}
P.~Deligne and D.~Mumford, \emph{{The Irreducibility of the Space of Curves of
  Given Genus}}, \href{https://doi.org/10.1007/bf02684599}{\emph{Publications
  Math\'ematiques de l\'IH\'ES} {\bfseries 36} (1969) 75}.

\bibitem{Abikoff198110}
W.~Abikoff, \emph{{The Uniformization Theorem}},
  \href{https://doi.org/10.2307/2320507}{\emph{Am. Math. Mon} {\bfseries 88}
  (1981) 574}.

\bibitem{FenchelNielsen2002}
W.~Fenchel and J.~Nielsen,
  \emph{\href{https://www.degruyter.com/document/doi/10.1515/9783110891355/html?lang=en}{Discontinuous
  Groups of Isometries in the Hyperbolic Plane}}, de Gruyter Studies in
  Mathematics. De Gruyter, 2002.

\bibitem{ImayoshiTaniguchi1992}
Y.~Imayoshi and M.~Taniguchi,
  \emph{\href{https://link.springer.com/book/10.1007/978-4-431-68174-8}{An
  Introduction to Teichm\"uller Spaces}}. Springer Japan, 1992.

\bibitem{Hubbard2006}
J.~H. {Hubbard},
  \emph{\href{https://matrixeditions.com/TeichmullerVol1.html}{Teichm\"uller
  Theory and Applications to Geometry, Topology, and Dynamics}}. Matrix
  Editions, 2006.

\bibitem{Maskit2001}
B.~Maskit,
  \emph{\href{https://www.acadsci.fi/mathematica/Vol26/maskit.pdf}{Matrices for
  Fenchel-Nielsen Coordinates}}, {\emph{Ann. Acad. Sci. Fenn. Math.} {\bfseries
  26} (2001) 267}.

\bibitem{Mirzakhani200610}
M.~Mirzakhani, \emph{{Simple Geodesics and Weil-Petersson Volumes of Moduli
  Spaces of Bordered Riemann Surfaces}},
  \href{https://doi.org/10.1007/s00222-006-0013-2}{\emph{Inventiones
  Mathematicae} {\bfseries 167} (2006) 179}.

\bibitem{BaranovSchwarz198510}
M.~A. Baranov and A.~Schwarz,
  \emph{\href{http://jetpletters.ru/ps/1436/article_21849.pdf}{Multiloop
  Contribution to String Theory}}, {\emph{JETP Lett.} {\bfseries 42} (1986)
  419}.

\bibitem{Friedan1986a}
D.~Friedan,
  \emph{\href{https://lib-extopc.kek.jp/preprints/PDF/2000/0038/0038836.pdf}{Notes
  on String Theory and Two-Dimensional Conformal Field Theory}}, {\emph{Unified
  String Theories} (1986) 162}.

\bibitem{RoslySchwarzVoronov1988a}
A.~A. Rosly, A.~S. Schwarz and A.~A. Voronov, \emph{{Geometry of Superconformal
  Manifolds}}, \href{https://doi.org/10.1007/bf01218264}{\emph{Comm. Math.
  Phys.} {\bfseries 119} (1988) 129}.

\bibitem{VerlindeVerlinde198702}
E.~{Verlinde} and H.~{Verlinde}, \emph{{Multiloop Calculations in Covariant
  Superstring Theory}},
  \href{https://doi.org/10.1016/0370-2693(87)91148-8}{\emph{Phys. Lett.}
  {\bfseries B192} (B1987) 95}.

\bibitem{ErlerKonopkaSachs201312}
T.~Erler, S.~Konopka and I.~Sachs, \emph{{Resolving Witten`s Superstring Field
  Theory}}, \href{https://doi.org/10.1007/JHEP04(2014)150}{\emph{J. High Energ.
  Phys.} {\bfseries 04} (2014) 150}
  [\href{https://arxiv.org/abs/1312.2948}{{\ttfamily 1312.2948}}].

\bibitem{DineSeibergWitten198701}
M.~{Dine}, N.~{Seiberg} and E.~{Witten}, \emph{{Fayet-Iliopoulos Terms in
  String Theory}},
  \href{https://doi.org/10.1016/0550-3213(87)90395-6}{\emph{Nucl. Phys.}
  {\bfseries B289} (1987) 589}.

\bibitem{AtickDixonSen198702}
J.~J. {Atick}, L.~J. {Dixon} and A.~{Sen}, \emph{{String Calculation of
  Fayet-Iliopoulos D-Terms in Arbitrary Supersymmetric Compactifications}},
  \href{https://doi.org/10.1016/0550-3213(87)90639-0}{\emph{Nucl. Phys.}
  {\bfseries B292} (1987) 109}.

\bibitem{DineIchinoseSeiberg198711}
M.~{Dine}, I.~{Ichinose} and N.~{Seiberg}, \emph{{F-Terms and D-Terms in String
  Theory}}, \href{https://doi.org/10.1016/0550-3213(87)90072-1}{\emph{Nucl.
  Phys.} {\bfseries B293} (1987) 253}.

\bibitem{GreenSeiberg198804}
M.~B. {Green} and N.~{Seiberg}, \emph{{Contact Interactions in Superstring
  Theory}}, \href{https://doi.org/10.1016/0550-3213(88)90549-4}{\emph{Nucl.
  Phys.} {\bfseries B299} (1988) 559}.

\bibitem{D'HokerPhong2002a}
E.~{D'Hoker} and D.~H. {Phong}, \emph{{Two-Loop Superstrings I. Main
  Formulas}}, \href{https://doi.org/10.1016/S0370-2693(02)01255-8}{\emph{Phys.
  Lett.} {\bfseries B529} (2002) 241}
  [\href{https://arxiv.org/abs/hep-th/0110247}{{\ttfamily hep-th/0110247}}].

\bibitem{D'HokerPhong2002b}
E.~{D'Hoker} and D.~H. {Phong}, \emph{{Two-Loop Superstrings II. The Chiral
  Measure on Moduli Space}},
  \href{https://doi.org/10.1016/S0550-3213(02)00431-5}{\emph{Nucl. Phys.}
  {\bfseries B636} (2002) 3}
  [\href{https://arxiv.org/abs/hep-th/0110283}{{\ttfamily hep-th/0110283}}].

\bibitem{D'HokerPhong2002c}
E.~{D'Hoker} and D.~H. {Phong}, \emph{{Two-Loop Superstrings III.
  Slice-Independence and Absence of Ambiguities}},
  \href{https://doi.org/10.1016/S0550-3213(02)00432-7}{\emph{Nucl. Phys.}
  {\bfseries B636} (2002) 61}
  [\href{https://arxiv.org/abs/hep-th/0111016}{{\ttfamily hep-th/0111016}}].

\bibitem{D'HokerPhong2002d}
E.~{D'Hoker} and D.~H. {Phong}, \emph{{Two-Loop Superstrings IV. The
  Cosmological Constant and Modular Forms}},
  \href{https://doi.org/10.1016/S0550-3213(02)00516-3}{\emph{Nucl. Phys.}
  {\bfseries B639} (2002) 129}
  [\href{https://arxiv.org/abs/hep-th/0111040}{{\ttfamily hep-th/0111040}}].

\bibitem{D'HokerPhong2002e}
E.~{D'Hoker} and D.~H. {Phong}, \emph{{Two-Loop Superstrings V. Gauge-Slice
  Independence of the N-Point Function}},
  \href{https://doi.org/10.1016/j.nuclphysb.2005.02.042}{\emph{Nucl. Phys.}
  {\bfseries B715} (2005) 91}
  [\href{https://arxiv.org/abs/hep-th/0501196}{{\ttfamily hep-th/0501196}}].

\bibitem{D'HokerPhong2002f}
E.~{D'Hoker} and D.~H. {Phong}, \emph{{Two-Loop Superstrings VI.
  Nonrenormalization Theorems and the 4-Point Function}},
  \href{https://doi.org/10.1016/j.nuclphysb.2005.02.043}{\emph{Nucl. Phys.}
  {\bfseries B715} (2005) 3}
  [\href{https://arxiv.org/abs/hep-th/0501197}{{\ttfamily hep-th/0501197}}].

\bibitem{D'HokerPhong2002g}
E.~{D'Hoker} and D.~H. {Phong}, \emph{{Two-Loop Superstrings VII. Cohomology of
  Chiral Amplitudes}},
  \href{https://doi.org/10.1016/j.nuclphysb.2008.04.030}{\emph{Nucl. Phys.}
  {\bfseries B804} (2008) 421}
  [\href{https://arxiv.org/abs/0711.4314}{{\ttfamily 0711.4314}}].

\bibitem{PiusRudraSen201311}
R.~Pius, A.~Rudra and A.~Sen, \emph{{Mass Renormalization in String Theory:
  Special States}}, \href{https://doi.org/10.1007/jhep07(2014)058}{\emph{J.
  High Energ. Phys.} {\bfseries 2014} (2014) }
  [\href{https://arxiv.org/abs/arXiv:1311.1257}{{\ttfamily arXiv:1311.1257}}].

\bibitem{PiusRudraSen201401}
R.~Pius, A.~Rudra and A.~Sen, \emph{{Mass Renormalization in String Theory:
  General States}}, \href{https://doi.org/10.1007/jhep07(2014)062}{\emph{J.
  High Energ. Phys.} {\bfseries 2014} (2014) }
  [\href{https://arxiv.org/abs/arXiv:1401.7014}{{\ttfamily arXiv:1401.7014}}].

\bibitem{Wolpert198203}
S.~Wolpert, \emph{{The Fenchel-Nielsen Deformation}},
  \href{https://doi.org/10.2307/2007011}{\emph{Ann. Math.} {\bfseries 115}
  (1982) 501}.

\bibitem{Wolpert198302}
S.~Wolpert, \emph{{On the Symplectic Geometry of Deformations of a Hyperbolic
  Surface}}, \href{https://doi.org/10.2307/2007075}{\emph{Ann. Math.}
  {\bfseries 117} (1983) 207}.

\bibitem{Wolpert198508}
S.~A. Wolpert, \emph{{On the Weil-Petersson Geometry of the Moduli Space of
  Curves}}, \href{https://doi.org/10.2307/2374363}{\emph{American Journal of
  Mathematics} {\bfseries 107} (1985) 969}.

\bibitem{Wolpert199002}
S.~A. Wolpert, \emph{{The Hyperbolic Metric and the Geometry of the Universal
  Curve}}, \href{https://doi.org/10.4310/jdg/1214444322}{\emph{J. Differential
  Geom.} {\bfseries 31} (1990) 417}.

\bibitem{ObitsuWolpert200704}
K.~Obitsu and S.~A. Wolpert, \emph{{Grafting Hyperbolic Metrics and Eisenstein
  Series}}, \href{https://doi.org/10.1007/s00208-008-0210-y}{\emph{Math. Ann.}
  {\bfseries 341} (2008) 685}
  [\href{https://arxiv.org/abs/arXiv:0704.3169}{{\ttfamily arXiv:0704.3169}}].

\bibitem{MelroseZhu201606}
R.~Melrose and X.~Zhu, \emph{{Boundary Behaviour of Weil-Petersson and Fibre
  Metrics for Riemann Moduli Spaces}},
  \href{https://doi.org/10.1093/imrn/rnx264}{\emph{Int. Math. Res. Not}
  {\bfseries 2019} (2017) 5012}
  [\href{https://arxiv.org/abs/arXiv:1606.01158v5}{{\ttfamily
  arXiv:1606.01158v5}}].

\bibitem{Randol197910}
B.~Randol, \emph{{Cylinders in Riemann Surfaces}},
  \href{https://doi.org/10.1007/bf02566252}{\emph{Comment. Math. Helv.}
  {\bfseries 54} (1979) 1}.

\bibitem{Wolpert1987a}
S.~A. Wolpert, \emph{{Asymptotics of the Spectrum and the Selberg Zeta-Function
  on the Space of Riemann Surfaces}},
  \href{https://doi.org/10.1007/bf01217814}{\emph{Comm. Math. Phys.} {\bfseries
  112} (1987) 283}.

\bibitem{McShane199105}
G.~McShane, \emph{\href{http://wrap.warwick.ac.uk/4008/}{A Remarkable Identity
  for Lengths of Curves}}, Ph.D. thesis, {\it University of Warwick}, 1991.

\bibitem{McShane199805}
G.~McShane, \emph{{Simple Geodesics and a Series Constant over Teichm\"uller
  Space}}, \href{https://doi.org/10.1007/s002220050235}{\emph{Invent. Math.}
  {\bfseries 132} (1998) 607}.

\bibitem{Pius201808}
R.~Pius, \emph{{Quantum Closed Superstring Field Theory and Hyperbolic Geometry
  I: Construction of String Vertices}},
  \href{https://arxiv.org/abs/1808.09441}{{\ttfamily 1808.09441}}.

\bibitem{ErlerKonopka201710}
T.~Erler and S.~Konopka, \emph{{Vertical Integration from the Large Hilbert
  Space}}, \href{https://doi.org/10.1007/JHEP12(2017)112}{\emph{J. High Energ.
  Phys.} {\bfseries 12} (2017) 112}
  [\href{https://arxiv.org/abs/1710.07232}{{\ttfamily 1710.07232}}].

\bibitem{MoosavianPius201706}
S.~F. Moosavian and R.~Pius, \emph{{Hyperbolic Geometry and Closed
  Bosonic-String Field Theory I: The String Vertices via Hyperbolic Riemann
  Surfaces}}, \href{https://doi.org/10.1007/JHEP08(2019)157}{\emph{J. High
  Energy Phys} {\bfseries 08} (2019) 157}
  [\href{https://arxiv.org/abs/1706.07366}{{\ttfamily 1706.07366}}].

\bibitem{MoosavianPius201708}
S.~F. Moosavian and R.~Pius, \emph{{Hyperbolic Geometry and Closed
  Bosonic-String Field Theory II: The Rules for Evaluating the Quantum BV
  Master Action}}, \href{https://doi.org/10.1007/JHEP08(2019)177}{\emph{J. High
  Energy Phys} {\bfseries 08} (2019) 177}
  [\href{https://arxiv.org/abs/1708.04977}{{\ttfamily 1708.04977}}].

\bibitem{CostelloZwiebach201909}
K.~Costello and B.~Zwiebach, \emph{{Hyperbolic String Vertices}},
  \href{https://doi.org/10.1007/JHEP02(2022)002}{\emph{J. High Energ. Phys.}
  {\bfseries 02} (2022) 002}
  [\href{https://arxiv.org/abs/1909.00033}{{\ttfamily 1909.00033}}].

\bibitem{Cho201912}
M.~Cho, \emph{{Open-Closed Hyperbolic String Vertices}},
  \href{https://doi.org/10.1007/JHEP05(2020)046}{\emph{J. High Energ. Phys.}
  {\bfseries 05} (2020) 046}
  [\href{https://arxiv.org/abs/1912.00030}{{\ttfamily 1912.00030}}].

\bibitem{Firat202102}
A.~H. F\i{}rat, \emph{{Hyperbolic Three-String Vertex}},
  \href{https://doi.org/10.1007/JHEP08(2021)035}{\emph{J. High Energ. Phys.}
  {\bfseries 08} (2021) 035}
  [\href{https://arxiv.org/abs/2102.03936}{{\ttfamily 2102.03936}}].

\bibitem{SchifferSpencer}
M.~Schiffer and D.~C. Spencer,
  \emph{\href{https://press.princeton.edu/books/hardcover/9780691653167/functionals-of-finite-riemann-surfaces}{Functionals
  of Finite Riemann Surfaces}}. Princeton University Press, 2015.

\bibitem{Gardiner197501}
F.~P. Gardiner, \emph{{Schiffer's Interior Variation and Quasiconformal
  Mapping}}, \href{https://doi.org/10.1215/s0012-7094-75-04235-0}{\emph{Duke
  Math. J.} {\bfseries 42} (1975) }.

\bibitem{Zwiebach199206}
B.~Zwiebach, \emph{{Closed-String Field Theory: Quantum Action and the BV
  Master Equation}},
  \href{https://doi.org/10.1016/0550-3213(93)90388-6}{\emph{Nucl. Phys.}
  {\bfseries B390} (1993) 33}
  [\href{https://arxiv.org/abs/hep-th/9206084}{{\ttfamily hep-th/9206084}}].

\bibitem{Nelson198902}
P.~{Nelson}, \emph{{Covariant Insertion of General Vertex Operators}},
  \href{https://doi.org/10.1103/PhysRevLett.62.993}{\emph{Phys. Rev. Lett.}
  {\bfseries 62} (1989) 993}.

\bibitem{Polchinskivol011998}
J.~Polchinski,
  \emph{\href{https://www.cambridge.org/core/books/string-theory/30409AF2BDE27D53E275FDA395AB667A}{String
  Theory, Vol. 1 - An Introduction to the Bosonic String}}. Cambridge
  University Press, 1998.

\bibitem{KeenLakic2007}
L.~Keen and N.~Lakic,
  \emph{\href{https://www.cambridge.org/core/books/hyperbolic-geometry-from-a-local-viewpoint/A04F32E52B19DFE6634A40105467BFD4}{Hyperbolic
  Geometry from a Local Viewpoint}}, London Mathematical Society Student Texts.
  Cambridge University Press, 2007.

\bibitem{Wolpert201001}
S.~A. Wolpert, \emph{\href{https://bookstore.ams.org/cbms-113}{Families of
  Riemann Surfaces and Weil-Petersson Geometry}}. American Mathematical
  Society, 2010.

\bibitem{Keen1974}
L.~Keen, \emph{{On Fundamental Domains and the Teichm\"uller Modular Group}},
  in
  \emph{\href{https://www.sciencedirect.com/book/9780120448500/contributions-to-analysis}{Contributions
  to Analysis: A Collection of Papers Dedicated to Lipman Bers}}, pp.~185--194.
\newblock Elsevier, 1974.

\bibitem{Keen197410}
L.~Keen, \emph{{Collars on Riemann Surface}},  in
  \emph{\href{https://www.degruyter.com/document/doi/10.1515/9781400881642/html}{Discontinuous
  Groups and Riemann Surfaces}} (L.~Greenberg, ed.), pp.~263--268.
\newblock Princeton University Press, dec, 1974.

\bibitem{Matelski197610}
J.~P. Matelski, \emph{{A Compactness Theorem for Fuchsian Groups of the Second
  Kind}}, \href{https://doi.org/10.1215/s0012-7094-76-04364-7}{\emph{Duke Math.
  J.} {\bfseries 43} (1976) 829}.

\bibitem{ChavelFeldman197812}
I.~Chavel and E.~A. Feldman, \emph{{Cylinders on Surfaces}},
  \href{https://doi.org/10.1007/bf02566089}{\emph{Comment. Math. Helv.}
  {\bfseries 53} (1978) 439}.

\bibitem{Mirzakhani200603}
M.~Mirzakhani, \emph{{Weil-Petersson Volumes and Intersection Theory on the
  Moduli Space of Curves}},
  \href{https://doi.org/10.1090/s0894-0347-06-00526-1}{\emph{J. Am. Math. Soc.}
  {\bfseries 20} (2006) 1}.

\bibitem{HatcherThurston198003}
A.~Hatcher and W.~Thurston, \emph{{A Presentation for the Mapping-Class Group
  of a Closed Orientable Surface}},
  \href{https://doi.org/10.1016/0040-9383(80)90009-9}{\emph{Topology}
  {\bfseries 19} (1980) 221}.

\bibitem{Hatcher199906}
A.~{Hatcher}, \emph{{Pants Decompositions of Surfaces}}, {\emph{ArXiv e-prints}
  (1999) } [\href{https://arxiv.org/abs/math/9906084}{{\ttfamily
  math/9906084}}].

\bibitem{GrossHarveyMartinecRohm1985a}
D.~J. Gross, J.~A. Harvey, E.~J. Martinec and R.~Rohm, \emph{{Heterotic
  String}}, \href{https://doi.org/10.1103/PhysRevLett.54.502}{\emph{Phys. Rev.
  Lett.} {\bfseries 54} (1985) 502}.

\bibitem{GrossHarveyMartinecRohm1985b}
D.~J. Gross, J.~A. Harvey, E.~J. Martinec and R.~Rohm, \emph{{Heterotic String
  Theory I. The Free Heterotic String}},
  \href{https://doi.org/10.1016/0550-3213(85)90394-3}{\emph{Nucl. Phys.}
  {\bfseries B256} (1985) 253}.

\bibitem{Lechtenfeld198911}
O.~Lechtenfeld, \emph{{Superconformal Ghost Correlations on Riemann Surfaces}},
  \href{https://doi.org/10.1016/0370-2693(89)91686-9}{\emph{Phys. Lett.}
  {\bfseries B232} (1989) 193}.

\bibitem{Morozov199001}
A.~{Morozov}, \emph{{A Straightforward Proof of Lechtenfeld's Formula for the
  $\beta\gamma$ Correlator}},
  \href{https://doi.org/10.1016/0370-2693(90)91993-L}{\emph{Phys. Lett.}
  {\bfseries B234} (1990) 15}.

\bibitem{Witten199001}
E.~Witten, \emph{{Two-Dimensional Gravity and Intersection Theory on Moduli
  Space}}, \href{https://doi.org/10.4310/sdg.1990.v1.n1.a5}{\emph{Surveys Diff.
  Geom.} {\bfseries 1} (1990) 243}.

\bibitem{Kontsevich199206}
M.~Kontsevich, \emph{{Intersection Theory on the Moduli Space of Curves and the
  Matrix Airy Function}},
  \href{https://doi.org/10.1007/bf02099526}{\emph{Commun. Math. Phys}
  {\bfseries 147} (1992) 1}.

\bibitem{DijkgraafVerlindeVerlinde199011}
R.~Dijkgraaf, H.~L. Verlinde and E.~P. Verlinde,
  \emph{\href{https://lib-extopc.kek.jp/preprints/PDF/1991/9103/9103452.pdf}{Notes
  on Topological String Theory and 2D Quantum Gravity}},  in \emph{{Cargese
  Study Institute: Random Surfaces, Quantum Gravity and Strings}}, 11, 1990.

\bibitem{Maskit199905}
B.~Maskit, \emph{{New Parameters for Fuchsian Groups of Genus $2$}},
  \href{https://doi.org/10.1090/s0002-9939-99-04973-4}{\emph{Proc. Am. Math.
  Soc.} {\bfseries 127} (1999) 3643}.

\bibitem{Kierlanczyk198609}
M.~Kierlanczyk,
  \emph{\href{http://dspace.mit.edu/bitstream/handle/1721.1/87802/16113111-MIT.pdf?sequence=2}{Determinants
  of Laplacians}}, Ph.D. thesis, {\it Massachusetts Institute of Technology},
  1986.

\bibitem{DHokerPhong198612}
E.~D'Hoker and D.~H. Phong, \emph{{On Determinants of Laplacians on Riemann
  Surfaces}}, \href{https://doi.org/10.1007/BF01211063}{\emph{Commun. Math.
  Phys.} {\bfseries 104} (1986) 537}.

\bibitem{Sarnak198703}
P.~Sarnak, \emph{{Determinants of Laplacians}},
  \href{https://doi.org/10.1007/bf01209019}{\emph{Commun. Math. Phys.}
  {\bfseries 110} (1987) 113}.

\bibitem{FarbMargalit2017}
B.~Farb and D.~Margalit,
  \emph{\href{https://press.princeton.edu/books/hardcover/9780691147949/a-primer-on-mapping-class-groups-pms-49}{A
  Primer on Mapping Class Groups}}. Princeton University Press, oct, 2017.

\end{thebibliography}\endgroup
